\documentclass[lettersize,journal,compsoc]{IEEEtran}
\usepackage{makecell}
\usepackage{multirow}
\usepackage[utf8]{inputenc} 
\usepackage{threeparttable}
\usepackage{hyperref}       
\usepackage{url}            
\usepackage{booktabs}       
\usepackage{amsfonts}       
\usepackage{nicefrac}       
\usepackage{microtype}      
\usepackage[table]{xcolor}         
\usepackage{orcidlink}
\usepackage{soul}
\usepackage{algorithmic}
\usepackage{graphicx}
\usepackage{bbm}
\usepackage[linesnumbered,ruled,vlined]{algorithm2e}
\def\BibTeX{{\rm B\kern-.05em{\sc i\kern-.025em b}\kern-.08em
    T\kern-.1667em\lower.7ex\hbox{E}\kern-.125emX}}
\usepackage{balance}

\usepackage{caption}
\usepackage{subcaption}
\usepackage{textcomp}
\usepackage{amsmath}
\usepackage{wrapfig}
\usepackage{lscape}
\usepackage[normalem]{ulem}
\newcommand{\model}{\textsc{RiGPS}}

\newcommand{\funnymodel}{\textbf{R}einforced \textbf{I}terative \textbf{G}ene \textbf{P}anel \textbf{S}election}
\DeclareMathOperator*{\argmax}{argmax} 

\title{Knowledge-Guided Gene Panel Selection for Label-Free Single-Cell RNA-Seq Data: A Reinforcement Learning Perspective}
\author{{Meng Xiao\orcidlink{0000-0001-5294-5776},~\IEEEmembership{Member,~IEEE}, Weiliang Zhang\orcidlink{0009-0006-8323-1315}, Xiaohan Huang\orcidlink{0009-0008-2066-2905}, Hengshu Zhu\orcidlink{0000-0003-4570-643X},~\IEEEmembership{Senior~Member,~IEEE,}\\ Min Wu\orcidlink{0000-0003-0977-3600},~\IEEEmembership{Senior~Member,~IEEE}, Xiaoli Li\orcidlink{0000-0002-0762-6562},~\IEEEmembership{Fellow,~IEEE,} and Yuanchun Zhou\orcidlink{0000-0003-2144-1131}}%
\IEEEcompsocitemizethanks{
\IEEEcompsocthanksitem Meng Xiao is with the Computer Network Information Center, Chinese Academy of Sciences, Beijing, and the DUKE-NUS Medical School, National University of Singapore. E-mails: meng.xiao@nus.edu.sg
\IEEEcompsocthanksitem Weiliang Zhang and Xiaohan Huang are with the Computer Network Information Center, Chinese Academy of Sciences, and the University of Chinese Academy of Sciences, Beijing.
E-mails: \{wlzhang, xhhuang\}@cnic.cn
\IEEEcompsocthanksitem Hengshu Zhu is with the Computer Network Information Center, Chinese Academy of Sciences, Beijing. E-mails: hszhu@cnic.cn
\IEEEcompsocthanksitem Min Wu is with the Institute for Infocomm Research, Agency for Science, Technology, and Research. E-mails: wumin@i2r.a-star.edu.sg
\IEEEcompsocthanksitem Xiaoli Li is with Singapore University of Technology and Design, the Information Systems Technology and Design Pillar. Email: xiaoli\_li@sutd.edu.sg
\IEEEcompsocthanksitem  Yuanchun Zhou is with Computer Network Information Center, Chinese Academy of Sciences, Beijing, University of Chinese Academy of Sciences, Beijing, and University of Science and Technology of China, Hefei. E-mails: zyc@cnic.cn
\IEEEcompsocthanksitem Corresponding author: Yuanchun Zhou (zyc@cnic.cn). Meng Xiao and Weiliang Zhang contributed equally to this study.}}
\begin{document}

\IEEEtitleabstractindextext{%
\begin{abstract}
Gene panel selection aims to identify the most informative genomic biomarkers in label-free genomic datasets. 
Traditional approaches, which rely on domain expertise, embedded machine learning models, or heuristic-based iterative optimization, often introduce biases and inefficiencies, potentially obscuring critical biological signals. To address these challenges, we present an iterative gene panel selection strategy that harnesses ensemble knowledge from existing gene selection algorithms to establish preliminary boundaries or prior knowledge, which guide the initial search space. Subsequently, we incorporate reinforcement learning (RL) through a reward function shaped by expert behavior, enabling dynamic refinement and targeted selection of gene panels. This integration mitigates biases stemming from initial boundaries while capitalizing on RL's stochastic adaptability. Comprehensive comparative experiments, case studies, and downstream analyses demonstrate the effectiveness of our method, highlighting its improved precision and efficiency for label-free biomarker discovery. Our results underscore the potential of this approach to advance single-cell genomics data analysis.
\end{abstract}}

\maketitle

\section{Introduction}
\IEEEPARstart{S}{ingle-cell} RNA sequencing (scRNA-seq) has emerged as a landmark advance in transcriptional analysis~\cite{schwartzman2015single,gawad2016single,woodworth2017building,liugut}, affording a high-resolution, cell-specific perspective on tissues, organs, and entire organisms~\cite{lee2020single,baysoy2023technological}. 
This capability enables a wide range of applications, from spatial transcriptomic analysis~\cite{huynh2024topological}, exploration of tissue-level architecture~\cite{rao2021exploring_tissue_architecture,longo2021intercellulartissuedynamics}, identification of salient cell subpopulations~\cite{sun2022identifyingphenotype-associatedsubpopulations}, to the development of large-scale domain foundation models~\cite {cui2023scgpt,theodoris2023geneformer,yang2023genecompass,qin2025scihorizon}. 
However, scRNA-seq data present substantial analytical challenges, including a lack of label, high dimensionality, sparsity, and noise, often culminating in the well-known `curse of dimensionality'~\cite{kiselev2019challenges(dimension_curse),li2021sparse}.
These issues complicate downstream analysis tasks such as biomarker discovery, making robust, scalable solutions an urgent necessity.

\begin{figure}[!h]
    \centering
\includegraphics[width=\linewidth]{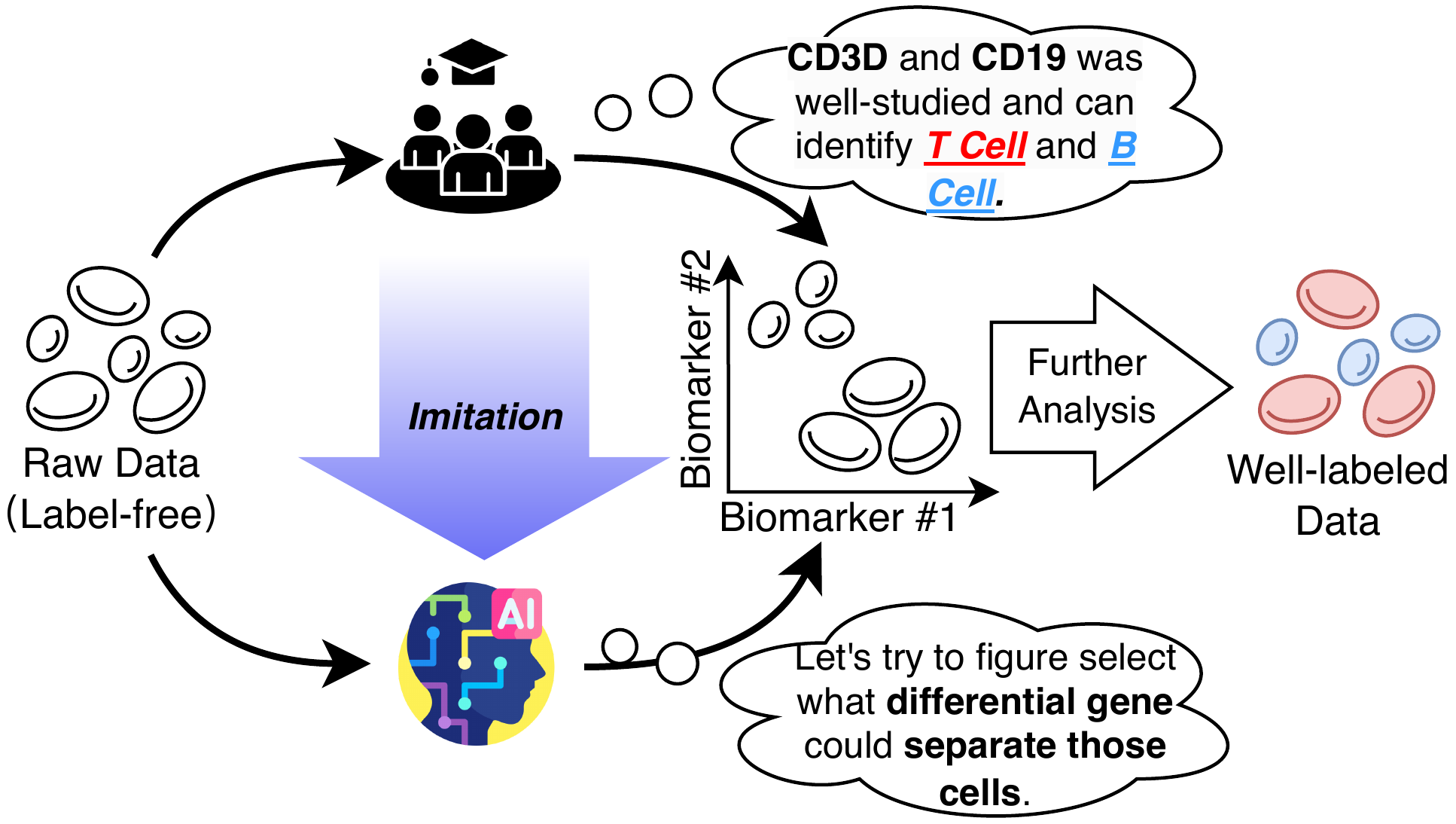}
    \caption{The Motivation of our study is the imitation of expert-driven genomic data analysis.}
    \label{motivation}
\end{figure}


Existing approaches to manage scRNA-seq complexity generally fall into three categories:
(1) Dimensional Reduction Techniques, such as PCA~\cite{mackiewicz1993principal}, t-SNE~\cite{kobak2019art}, and UMAP~\cite{becht2019dimensionality} are essential for managing the complexity of scRNA-seq data, especially for visualization. However, these methods also have several drawbacks:
These methods can result in the loss of subtle yet biologically significant information, distort the true structure of the data, and are highly dependent on the choice of parameters, such as the number of principal components or the perplexity value in t-SNE. 
(2) Statistical Methods, including the use of p-values, fold changes~\cite{dalman2012fold}, or analysis of highly variable genes (HVG)~\cite{yip2019evaluation,Seuratv3,luecken2022benchmarking}, are fundamental steps in identifying significant characteristics in scRNA-seq data analysis~\cite{hetzel2022predicting}, or domain foundation model research~\cite{cui2023scgpt,gong2024xtrimogene}. 
However, these methods often assume data normality and independence assumptions that may not hold in scRNA-seq contexts and are sensitive to the inherent noise and sparsity of the data, potentially leading to inaccuracies by either masking biological signals or amplifying artifacts. 
(3) Gene Selection Approaches, specifically tailored for genomics research, including scRNA-seq studies. 
Those approaches, whether they highly depend on well-trained embedded machine learning models~\cite{CellBRF} to identify the importance of each gene or they utilize heuristic metrics to determine biomarker~\cite{geneBasis, HRG}, are always unstable and not optimization-directed. 

To tackle these previously discussed issues, we take a moment to reassess and examine the classical expert-driven solution~\cite{pont2019single}. 
The upper portion of Figure~\ref{motivation} illustrates the current manual pipeline inherent in scRNA-seq data analysis. 
A promising avenue to automate the pipeline is to emulate expert decision-making via reinforcement learning (RL)~\cite{volk2023alphaflow}. 
Unlike supervised or unsupervised learning paradigms, RL excels at modeling sequential decision processes, in which an “agent” learns by interacting with an environment and receiving reward signals~\cite{cai2022survey,saadatmand2024many}.
By framing the gene panel selection process as a systematic path of decisions, similar to how an expert iteratively refines a candidate gene set, an RL agent can progressively converge on an optimal or near-optimal solution under varying conditions. 
The down portion of Figure~\ref{motivation} encapsulates our central idea: domain knowledge provides the foundations, but reinforcement learning operationalizes and enhances it, offering a scalable and automated system to select genomic biomarkers. 

Inspired by the discussions, we proposed an automated label-free gene panel selection pipeline, namely \funnymodel\ (\textbf{\model}) framework. 
Our approach is distinguished by its ability to ensemble prior knowledge from existing gene panel selection algorithms. 
This ensemble of knowledge serves as valuable preliminary boundaries or essential prior experiences that bootstrapped the initial phase of gene panel selection, allowing for a more directed and informed biomarker search.
Moreover, we incorporate the principles of stochastic exploration in RL and its continuous optimization capabilities through a reward-based feedback mechanism. 
This innovative combination allows our model to adjust and refine the gene panel selection process, mitigating the biases and limitations inherent in the initial boundaries set by previous algorithms. 
Ultimately, the reward function is crafted based on the imitation of the pseudo-experiment-driven pipeline, driving the framework to select the pivotal biomarker that can most effectively distinguish each cell sample unsupervised.
The contributions of the paper can be summarized as follows:
\begin{itemize}
    \item \textbf{Knowlege-Ensembled Initialization}: We derive preliminary boundaries from multiple gene selection algorithms to bootstrap the RL agent, reducing both search complexity and computational overhead.
    \item \textbf{Multi-Agent Reinforcement Learning}: Our framework coordinates multiple agents, each exploring a subset of genes, to collectively determine optimal panels under stochastic exploration.
    \item \textbf{Pseudo-Experiment-Guided Reward Design}: The reward function encodes expert-like criteria, ensuring that the agent focuses on biologically relevant distinctions in a label-free manner.
    \item \textbf{Extensive Evaluation}: We conduct comprehensive quantitative and qualitative experiments on diverse scRNA-seq datasets across multiple species and tissue types, demonstrating significant gains in both performance and efficiency compared to baseline methods.
\end{itemize}

\section{Background and Preliminary}~\label{preliminar}


\noindent\textbf{Common Tasks in Single-Cell Data Analysis.}
In the context of scRNA-seq analysis, \textit{Clustering} is often a preliminary downstream task, and gene selection plays a pivotal role in making this process effective. 
By grouping cells based on their expression profiles, clustering enables the discovery of putative cell types, states, or patterns in an unsupervised manner, thus revealing novel insights into cellular diversity. 
Beyond clustering, \textit{Visualization} techniques are equally central to interpret single-cell data. 
Dimensional reduction methods (such as t-SNE, UMAP, or PCA) project the high-dimensional gene expression matrix into two or three dimensions, enabling researchers to observe distinct cell groups and evaluate how well their chosen gene set resolves meaningful biological structures.
\textit{Heatmap Analysis} further augments clustering and visualization by highlighting gene expression patterns across cell subsets in a more interpretable, matrix-like depiction. Through a heatmap, one can assess how the chosen genes distribute their expression levels across cell clusters, either verifying known biological signatures or uncovering unexpected relationships. 
\textit{Differential Expression Analysis} is a crucial follow-up to clustering. 
It seeks to identify specific genes that are significantly up- or downregulated between identified cell clusters, conditions (e.g., treatment vs. control), or developmental stages. 
In summary, because the clustering task is more suitable for quantitative evaluation, this study adopts clustering-based metrics as the principal quantitative evaluation tools for model performance. 
Currently, to comprehensively assess the applicability of the model in real-world biological contexts, we leverage additional tasks, such as visualization, heatmap analysis, and differential expression analysis, to conduct qualitative evaluations.


\noindent\textbf{Gene Selection Problem.} 
Formally, the given scRNA-seq dataset can be denoted as $D=\{G,X\}$, where $G$ denoted the overall gene set and $X$ is the expression matrix. 
We can use $X[G']=\{x_{i,j}\}_{j\in G'}$ to denote select gene expression matrix with a gene subset $G'$, where $x_{i,j}\in X$ represents cell-$i$'s expression of the genes-$j$.
We aim to develop a generalized yet robust gene panel selection method that can identify the optimal key gene panel $G^*$ from a scRNA-seq dataset $D$, optimally preserving biology signal for various downstream analysis tasks:
\begin{equation}
    \label{objective}
    G^{*} = \argmax_{G'\subseteq G} \mathcal{E}(\mathcal{C}(X[G'])),
\end{equation}
where $\mathcal{E}$ and $\mathcal{C}$ denoted the evaluation metric and downstream analysis method, respectively. 

\section{Proposed Methodology}\label{method}
\begin{figure*}[!t]
    \centering
    
    \includegraphics[width=0.95\linewidth]{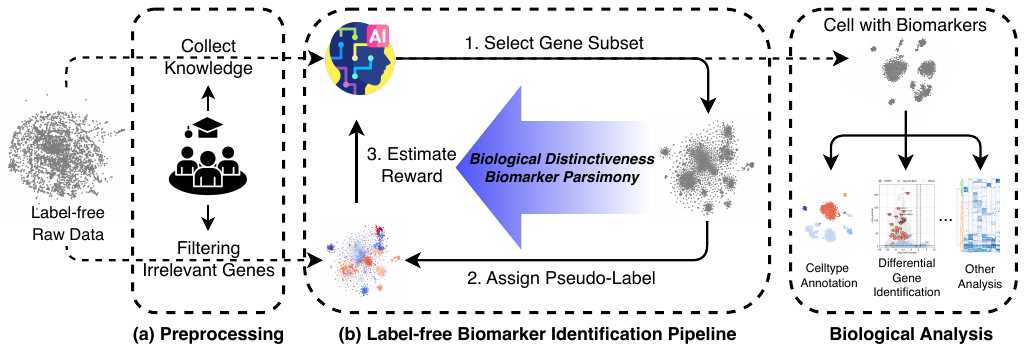}
    \caption{The overview of our framework. (a) After raw data preprocessing, we filter the irrelevant genes and collect expert knowledge from the basic methods. The streamlined gene set will then feed into \model.
    (b) The pipeline consists of three main stages: cell agents will first cooperate to select from the filtered gene set as candidate biomarkers. Then, each cell is assigned a pseudo-label based on the chosen biomarker. Finally, the model will estimate and assign the reward to each cell agent by an expert-knowledge-guided reward function.
    (c) The selected biomarker will be applied to enhance the downstream analysis.}

    \label{fig:framework}
\end{figure*}
Figure~\ref{fig:framework} illustrates the overview of \model, an iterative gene panel selection method. 
This section will begin with a brief introduction of the micro-view of \model{}, the collaborative gene agents. Then, we step deeper into the expert-knowledge-guided reward function and the whole framework.

\subsection{Gene Agents for Collaborative Gene Selection}\label{met2} 
The central panel of Figure~\ref{fig:framework} illustrates the pipeline in which multiple gene agents collaborate iteratively to select the most informative genes. 
Specifically, we construct agents with the same number as the candidate genes. 
Each gene agent consists of the following components: 

\smallskip
\noindent\textbf{Action.} The action token $a_t^i$ denoted gene $i$'s agent at $t$-th iteration is to select or discard its corresponding gene. Its candidate action space is $a_t^i\in\{select, discard\}$. 

\smallskip
\noindent\textbf{State.} The state at $t$-th iteration is a vectorized representation derived from the previously selected gene subset $G_t$. 
First, we extract each gene's descriptive statistics from the selected subset to preserve the biological signal (e.g., the standard deviation, minimum, maximum, and the first, second, and third quartile, etc.). 
Then, we flatten and concatenate all descriptive statistics vectors and feed them into an autoencoder. 
This autoencoder has a fixed $k$-length latent vector and variable input and output dimensions according to the selected gene subset. 
Its goal is to minimize the reconstruction loss between the input and output, thus compressing the information from descriptive statistics vectors into a fixed size. 
After the autoencoder converges, the hidden vector $\mathcal{S}_t$ with dimension of $k$ will be used as the state representation at the $t$-th iteration.

\smallskip
\noindent\textbf{Policy Network.} 
Each gene agent will share the state in each iteration. 
Their policy network $\pi(\cdot)$ is a feed-forward neural network with a binary classification head. 
Formally, for gene $i$, its action in $t$-th iteration is then derived by: $a_t^i=\pi^i(\mathcal{S}_t)$.

\subsection{Pseudo-Experiment Guided Label-free Reward Estimation}\label{met3}

Gene agents will decide whether to select or discard their corresponding gene in each iteration by policy network. 
By combining those decisions, we can obtain the selection in the current iteration, given as $\mathcal{A}_t = \{a^i_t\}_{i=1}^n$.
Meanwhile, the coarse boundary can be refined by $\tilde{G}\xrightarrow{\mathcal{A}_t}G_t$, where $G_t$ is the selected subset in $t$-th iteration. 
As illustrated in Figure~\ref{fig:framework}, we designed the reward function from two perspectives: 

\smallskip
\noindent\textbf{Biological Distinctiveness.} 
Visualization analysis is widely employed to explore the biological significance of single-cell transcriptomic data~\cite{pont2019single}.  
To mimic this progress, the first part of the reward function evaluates biological differentials by leveraging normalized mutual information (NMI).  
In each step, the model first clusters the cells with the expression of the currently selected gene and assigns a pseudo-label $\hat{y}$ to each cell. 
Then, the reward estimator will obtain the biological differential reward by $\hat{y}$: 
\begin{equation}
    r_t^s = \frac{2 \times I(X[G_t]; \hat{y})}{H(X[G_t]) + H(\hat{y})},
\end{equation}
where $I(X[G_t]; \hat{y})$ denotes the mutual information between the selected gene expression on each cell $X[G_t]$ and the pseudo labels $\hat{y}$.  $H(X[G_t])$ and $H(\hat{y})$ are the entropies of $X[G_t]$ and $\hat{y}$, respectively. 
These metrics reward gene agents for an unsupervised spatial separation understanding between and within each cluster. 

\smallskip

\noindent\textbf{Biomarker Parsimony.} The second perspective focuses on ensuring a compact number of genes through: 
\begin{equation}\label{reward_compact}
    r_t^c = \frac{|\tilde{G}| - |G_t|}{|\tilde{G}| + \lambda \cdot |G_t|},
\end{equation}
where $\lambda$ is a hyperparameter and $|\cdot|$ denoted the size of given set. 
This formula balances the reduction of the gene set size with the penalty for overly aggressive reduction. 
As $\lambda$ increases, the penalty for keeping too many genes (large $|G_t|$) becomes more severe, thus encouraging more substantial gene reduction. 
Conversely, a lower value of $\lambda$ relaxes the penalty against the size of $|G_t|$, suitable when minimal reduction is sufficient. 
This metric ensures that the selection process strategically reduces the number of genes. 

\smallskip
\noindent\textbf{Reward Assignment.} 
We combine two perspectives and obtain the reward in step-$t$: 
\begin{equation}\label{reward_overall_func}
    r_t = \alpha \cdot r^s_t + (1-\alpha) \cdot r^c_t,
\end{equation}
where $r_t$ is the total reward in step-$t$. 
$\alpha$ is a hyperparameter for adjusting the weight of two perspectives. 
After that, the framework will assign the reward equally to each agent. 

\subsection{Preprocessing and Knowledge Collection}\label{met1}
This section demonstrates the pre-processing of raw label-free scRNA-seq data, filtering out irrelevant genes, and gathering knowledge for the subsequent pipeline.

\smallskip
\noindent\textbf{Raw Data Preprocessing.} 
The raw single-cell RNA sequencing data will undergo a rigorous preprocessing pipeline~\cite{pre} to ensure data quality and minimize technical artifacts.  
Initially, a comprehensive quality control procedure is implemented to identify and exclude low-quality cells characterized by aberrant mitochondrial gene expression levels or an insufficient number of detected genes.  
Subsequently, the retained gene count matrix is normalized to account for differences in sequencing depth across cells.  
A logarithmic transformation is then applied to the expression matrix, which serves to stabilize variance and attenuate the impact of extreme values, thereby enhancing the signal-to-noise ratio for downstream analyses.  
Finally, we feed the processed dataset into the next step to collect knowledge. 

\smallskip
\noindent\textbf{Collect Knowledge Set from Basic Methods.} 
Formally, the basic selection method pipeline can be divided into estimating the importance of the gene, ranking and selecting the top-$k$ genes, denoted as: 
$f(D)\rightarrow\{S,G^f\}$,
where $f(\cdot)$ is the basic method (such as the high variable gene~\cite{HVG} method), $S=\{s^i\}_{i=1}^{|G|}$ is the estimated score of each gene, and $G^f$ is the selected gene subset. 
Suppose that we have a $m$ basic methods, denoted as $F=\{f_i\}_{i=1}^m$. 
Each gene in the original gene set can have its significance score calculated using the methods in $F$, represented as $\mathbf{S} = \{S_i\}_{i=1}^m$. 
We collect the selection results as the knowledge set, given as $K = \{G^f_i\}_{i=1}^m$.  

\smallskip
\noindent\textbf{Filtering Irrelevant Genes.} 
We then utilize the knowledge set and the estimated gene score to form the coarse boundary. 
To reduce the bias from the results of simple methods, we introduce the idea of meta-votes~\cite{vote} from ensemble learning to identify the boundary with high recall but low precision. 
We can first adopt the same approach by assigning a pseudo-label and estimating the reliable weights from Equation~\ref{reward_overall_func}. 
The estimated reliable weights of each component within the knowledge set can be denoted as $P=\{r_i\}_{i=1}^m$.
After that, we calculate the normalized reliable weights for each model: 
\begin{equation}
    w_i = \frac{r_i}{\sum_{r_j \in P} r_j},
\end{equation}
where $w_i$ is the reliable weight for model $f_i\in F$, and the weight of each method can be denoted by $W = \{w_i\}_{i=1}^m$. 
For gene $g_i$, its meta-vote score $\hat{s}^i$ can be obtained by weighted aggregation from the reliable weight of each method: 
\begin{equation}
    \hat{s}^i = \sum_{j\in\{1,\dots, m\}} w_j \cdot s^i_j,
\end{equation}
To identify genes whose meta-vote scores significantly deviate from the average, we first calculate the mean $\mu$ and standard deviation $\sigma$ of the scores across all genes:
\begin{equation}
\mu = \frac{1}{n} \sum_{i=1}^n \hat{s}^i,\text{   } \sigma = \sqrt{\frac{1}{n} \sum_{i=1}^n (\hat{s}^i - \mu)^2}.
\end{equation}
We can then form the coarse boundary by filtering genes based on whether their scores fall outside the range defined by two standard deviations from the mean (2-sigma): $\tilde{G} = \{g^i: \hat{s}^i > \mu + 2\sigma\}$. The filtered gene set $\tilde{G} \subseteq G$ consists of genes whose meta-vote scores are significantly higher than the mean by at least two standard deviations. 
The objective of the gene selection problem in Equation~\ref{objective} can be reformulated as:
\begin{equation}
    \label{final_objective}
    G^{*} = \argmax_{G'\subseteq \tilde{G}} \mathcal{E}(\mathcal{C}(X[G'])).
\end{equation}
By utilizing the coarse boundary $\tilde{G}$, we are able to retain the most informative genes while greatly decreasing the complexity of the overall process.

\subsection{Iteration and Optmization}\label{met4}
We then introduce the detail of \model{} iteration.
In the initialization phase, we inject the collected knowledge set $K$ into each gene agent's memory queue.
Then, the \model{} explores and refines the coarse boundary, collects memories, and injects them into the memory queue.
When the memory queue exceeds a sufficient number, the model will explore and optimize each gene agent alternately. 

\smallskip
\noindent\textbf{Knowledge Set Injection.} 
Given a collected knowledge $G^f$, for gene agent $i$, an experience of the following form is injected into its memory queue: $m^{f} = \{\mathcal{S}^{0}, a_i^{f}, r^{f}, \mathcal{S}^{f}\}$. 
Here, $a_i^f$ represents select or discard the gene $i$:
\begin{equation}
    a_i^{f} =
\begin{cases} 
Select, & \text{if } g_i \in G^f, \\
Discard, & \text{otherwise}.
\end{cases}
\end{equation}
$\mathcal{S}^{0}$ and $\mathcal{S}^{f}$ are the state representation extracted from $\tilde{G}$ and $G^f$, respectively.
$r^f$ is the reward based on the coarse boundary calculated following the reward estimation in Equation~\ref{reward_overall_func}.

\smallskip
\noindent\textbf{Pipeline Exploration.}
Each gene agent executes actions guided by their policy networks during exploration. 
These agents process the current state as input and choose to select or discard its correlated gene. 
Those actions will then affect the size and composition of the gene subset, consequently refining a newly selected gene subspace. 
The actions performed by the gene agents accumulate an overall reward that is subsequently assigned to all the participating agents in the optimization phase. 
Specifically, for gene $i$, in step-$t$, the collected experience can be denoted as: $m_t^i =\{\mathcal{S}_{t}, a_t^i, r_t^i, \mathcal{S}_{t+1}\}$.

\smallskip
\noindent\textbf{Pipeline Optimization.}
In the optimization phase, each gene agent will train their policy independently with a shared goal through the mini-batch of memory derived from the replay of the prioritized experience~\cite{schaul2015prioritized}. 
We optimized the policy based on the Actor-Critic approach~\cite{konda1999actor,sewak2019actor}, where the policy network $\pi(\cdot)$ is the actor and $V^{\pi}(\cdot)$ is its correlated critic.
The agent-$i$ seeks to maximize its expected cumulative reward:
\begin{equation}
    \max_{\pi} \mathbb{E}_{m_t^i \sim \mathcal{B}} \left[ \sum_{t=0}^{T} \gamma r_t^i \right],
\end{equation}
where $\mathcal{B}$ denotes the distribution of experiences within the prioritized replay buffer, $\gamma$ is the discount factor, and $T$ represents the temporal horizon of an episode.
To learn the advantage function required for policy updates, we define a state-action value function $Q^{\pi}(\mathcal{S}, a)$ under the policy $\pi$:
\begin{equation}
    Q^\pi(\mathcal{S}, a) = \mathbb{E}\Bigl[r + \gamma V^{\pi}(\mathcal{S}')
    \big|\mathcal{S}, a \Bigr].
\end{equation}
The training updates for the actor (A) and critic (C) networks are computed as follows:
\begin{align}
    \text{C:} & \quad L(V^\pi) = \mathbb{E}_{m_t^i \sim \mathcal{B}} \left[ V^\pi(\mathcal{S}_t) - \left( r_t^i + \gamma V^\pi(\mathcal{S}_{t+1}) \right) \right]^2, \\
    \text{A:} & \quad \nabla_{\theta} J(\pi) = \mathbb{E}_{m_t^i \sim \mathcal{B}} \left[ \nabla_{\theta} \log \pi(a_t^i | \mathcal{S}_t) A^\pi(\mathcal{S}_t, a_t^i) \right].
\end{align}
Here, $A^{\pi}(\mathcal{S}, a) = Q^{\pi}(\mathcal{S}, a) - V^{\pi}(\mathcal{S})$ represents the advantage function to estimate the gradient for policy improvement.

\section{Experiment Setting}\label{exp_setting}
\subsection{Dataset Description}\label{dataset_des}
Our research involved 24 public single-cell RNA sequencing (scRNA-seq) datasets derived from various sequencing technologies and representing diverse biological conditions. 
These datasets were collected from several public databases
~\cite{edgar2002gene, brazma2003arrayexpress, leinonen2010sequence}
, including the National Center for Biotechnology Information's Gene Expression Omnibus (GEO), ArrayExpress, and the Sequence Read Archive (SRA), etc.
The "Cao" dataset~\cite{Cao} was procured from a study utilizing the sci-RNA-seq method (single-cell combinatorial indexing RNA sequencing). 
The "Han" dataset~\cite{Han} originates from the Mouse Cell Atlas.
To test the model's robustness, we include one dataset with a batch effect, Human Pancreas~\cite{batch_effect}. 
Detailed specifics, including each dataset's origins, description, and size of cells and genes, are provided in the \textit{Supplementary Material} and our code based\footnote{
Our codes,  selected gene set of each dataset, and example dataset are publicly accessible via ~\href{https://www.dropbox.com/scl/fo/brc4neuc9o336l4ra9s24/AOGxyfVRqC_Q0CEUmI7GdtY?rlkey=77xamb700pl12xev21gsfbczk&st=isq4wp7u&dl=0}{Dropbox}.}. 

\subsection{Evaluation Metrics.}\label{metrics}
To compare the performance of these methods, we evaluate the cell-type-discriminating performance of genes via cell clustering. 
We follow the same setting as CellBRF~\cite{CellBRF} by adopting a graph-based Louvain community detection algorithm in Seurat~\cite{Seurat} as the downstream cell clustering analysis method, a commonly used software toolkit for scRNA-seq clustering. 
We adopted three widely used metrics to quantitatively assess model performance, including normalized mutual information (NMI)~\cite{NMI}, adjusted rand index (ARI)~\cite{ARI}, and silhouette index (SI)~\cite{SI}. 
All metrics range from 0 to 1, where the higher the value, the better the model performance. 
We also included accuracy, balanced accuracy, Micro-F1, and Macro-F1 for the evaluation of the cell-type annotation task. 


\subsection{Baseline Methods.}\label{baselines}
Our comparative analysis evaluated \model\ against seven widely used baselines. 
The detailed descriptions are listed as follows:
(1) \textbf{CellRanger}~\cite{CellRanger} converts scRNAseq data into a gene-barcode matrix suitable for gene selection through sample demultiplexing, barcode processing, and single-cell gene counting; 
(2) \textbf{Pearson Residuals (PR)}~\cite{PearsonResidualS} normalizes and identifies biologically variable genes by quantifying the deviation of observed gene expression counts from an expected model of constant expression across cells; 
(3) \textbf{Seurat v3}~\cite{Seuratv3} performs gene selection by applying a variance-stabilizing transformation to account for the mean-variance relationship inherent in scRNAseq data, then identifying the top genes with the highest variance after standardization;
(4) \textbf{HRG}~\cite{HRG} utilizes a graph-based approach to identify genes that exhibit regional expression patterns within a cell-cell similarity network;
(5) \textbf{geneBasis}~\cite{geneBasis} aims to select a small, targeted panel of genes from scRNA-seq datasets that can effectively capture the transcriptional variability present across different cells and cell types; 
(6) \textbf{CellBRF}~\cite{CellBRF}, selects the most significant gene subset evaluated using Random Forest.
(7) \textbf{gpsFISH}~\cite{gpsFISH} utilizes a genetic algorithm to optimize gene panel selection for targeted spatial transcriptomics by accounting for platform effects and incorporating cell type hierarchies and custom gene preferences.
(8) \textbf{scGIST}~\cite{scGIST} appies a deep learning-based approach for spatial transcriptomics that prioritizes user-specified genes while maintaining accuracy in cell type detection, allowing for more comprehensive analysis within the constraints of limited panel sizes.

\subsection{Hyperparameter Setting and Reproducibility}\label{hyper}
For all experiments and datasets, we ran 400 epochs for exploration and optimization. The memory size is set to 400. 
The basic methods for the gene pre-filtering module consist of Random Forest~\cite{rf}, SVM~\cite{svm}, RFE~\cite{rfe}, geneBasis, and KBest~\cite{KBest}.
We adopt the Louvain community detection algorithm to generate pseudo-labels for reward estimation and those supervised selection methods, as same as the downstream clustering method. 
The gene state representation component consists of an autoencoder, which includes two structurally mirrored three-layer feed-forward networks. 
The first network serves as the encoder, with the first layer containing 256 hidden units, the second layer containing 128 hidden units, and the third layer containing 64 hidden units, progressively compressing the data to capture its intrinsic features. 
The second network acts as the decoder, mirroring the encoder structure. 
The training epochs in each step for the gene subset state representation component are set to 10. 
For the knowledge injection setting, we adopt the gene subsets selected by CellBRF, geneBasis, and HRG as our prior knowledge.
In the reinforcement iteration, we set each gene agent's actor and critic network as a two-layer neural network with 64 and 8 hidden sizes in the first and second layers, respectively. 
Following the hyperparameter study, we set $\alpha$ (the trade-off between spatial coefficient and quantity suppression in  Equation~\ref{reward_overall_func}) to 0.5, so each part of the reward function has a balanced weight.
To train the policy network in each gene agent, we set the minibatch size to 32 and used the Adam optimizer with a learning rate of 1e-5. 
The parameter settings of all baselines follow the original papers.

\subsection{Experiment Platform Settings}\label{platform} All experiments were ran on the Ubuntu 18.04.6 LTS operating system, Intel(R) Xeon(R) Gold 6338 CPU, and 4 NVIDIA V100 GPUs, with the framework of Python 3.11.5 and PyTorch 2.1.1.


\begin{figure*}[!t]
    \centering
    \scalebox{1}{
    \begin{subfigure}[b]{0.32\textwidth} 
        \includegraphics[width=\textwidth]{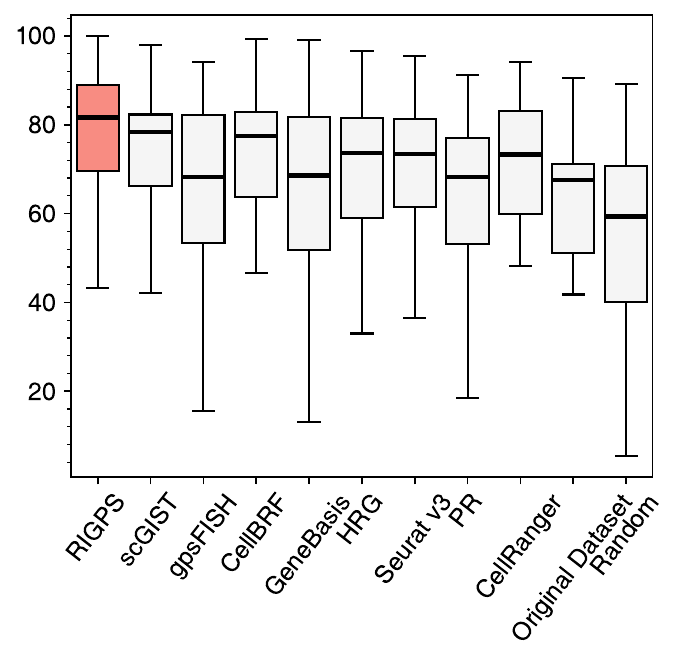}
        \caption{NMI}
    \end{subfigure}
    \begin{subfigure}[b]{0.32\textwidth} 
        \includegraphics[width=\textwidth]{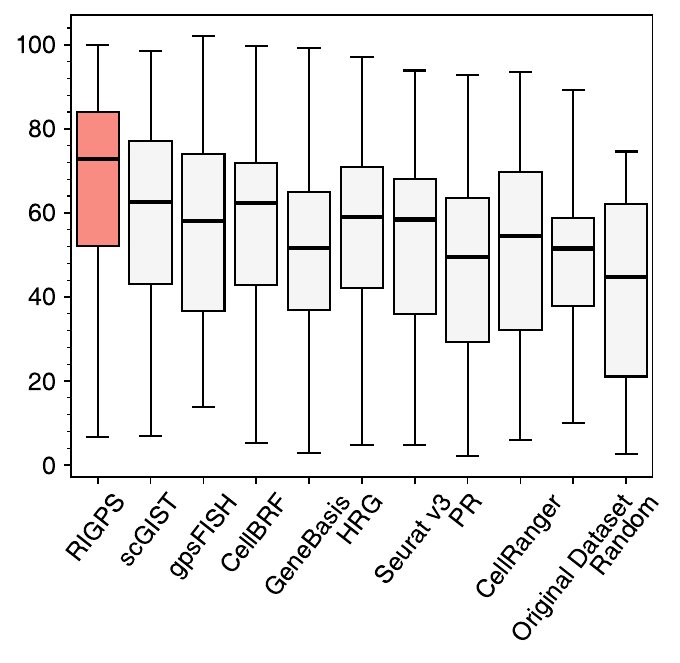}
        \caption{ARI}
    \end{subfigure}
    \begin{subfigure}[b]{0.32\textwidth} 
\includegraphics[width=\textwidth]{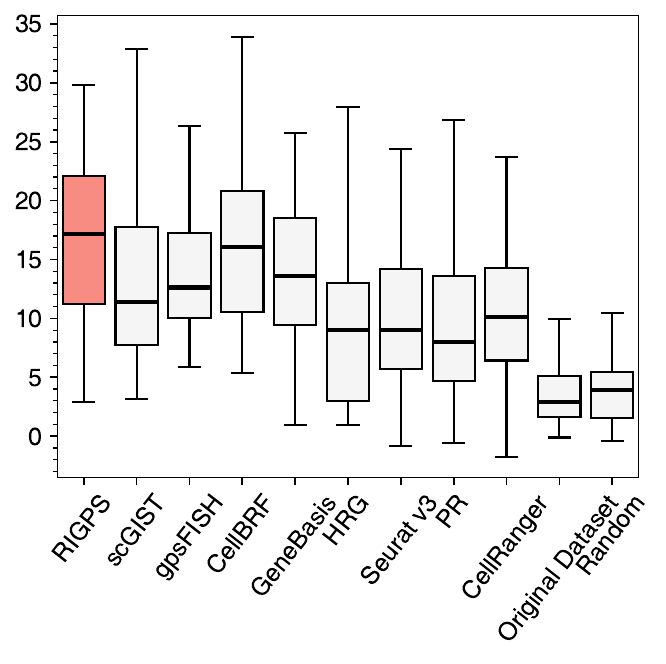}
        \caption{SI}
    \end{subfigure}}
    \scalebox{1}{
    \begin{subfigure}[b]{0.32\textwidth} 
        \includegraphics[width=\textwidth]{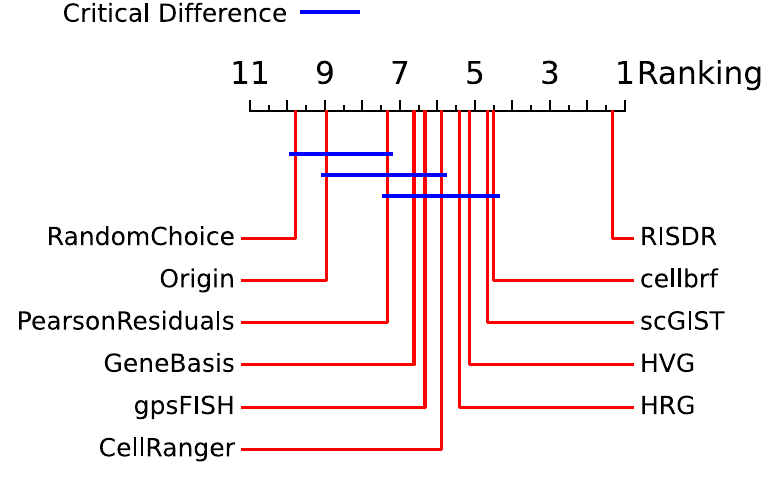}
        \caption{NMI}
    \end{subfigure}
    \begin{subfigure}[b]{0.32\textwidth} 
        \includegraphics[width=\textwidth]{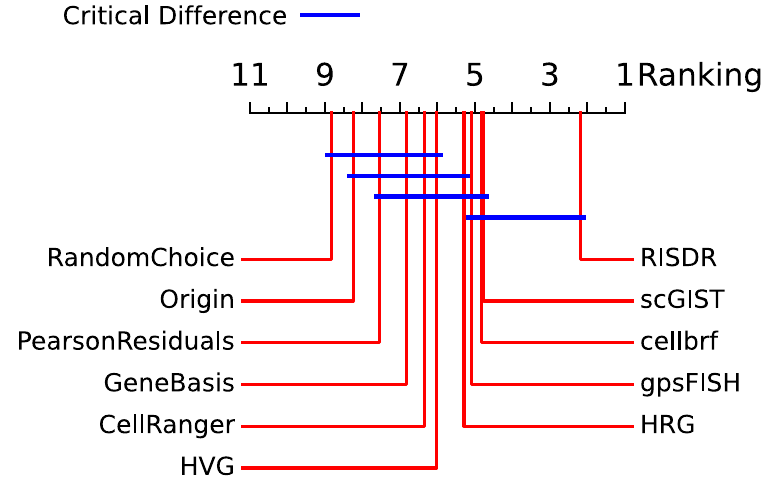}
        \caption{ARI}
    \end{subfigure}
    \begin{subfigure}[b]{0.32\textwidth} 
\includegraphics[width=\textwidth]{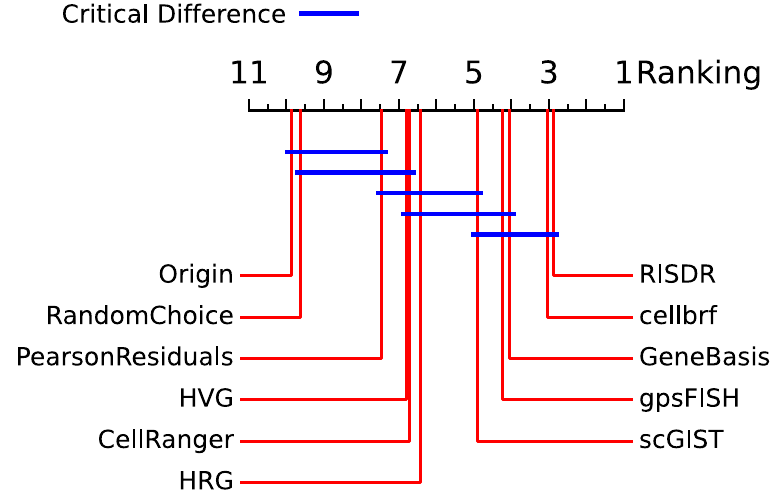}
        \caption{SI}
    \end{subfigure}}
    \scalebox{1}{
    \hspace{-0.5cm}\begin{subfigure}[t]{0.56\textwidth} 
    \includegraphics[width=\textwidth]
    {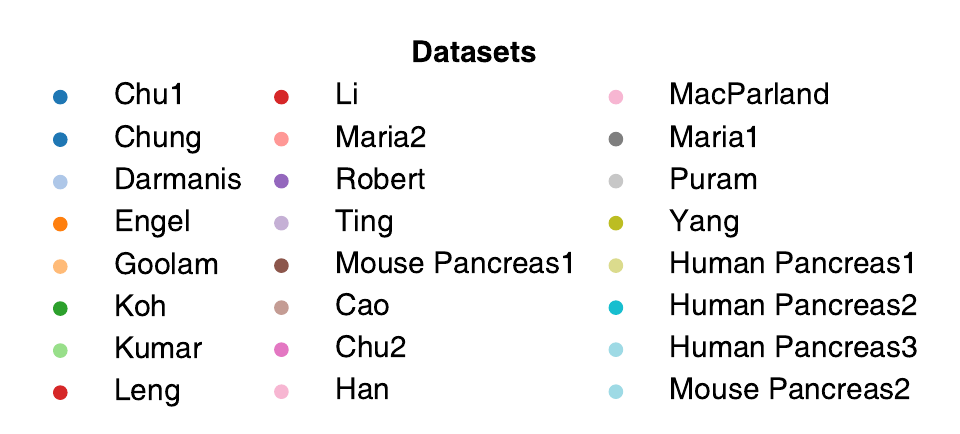}
    \end{subfigure}
    \hspace{0.2cm}
    \begin{subfigure}[t]{0.4\textwidth} %
    \includegraphics[width=\textwidth]{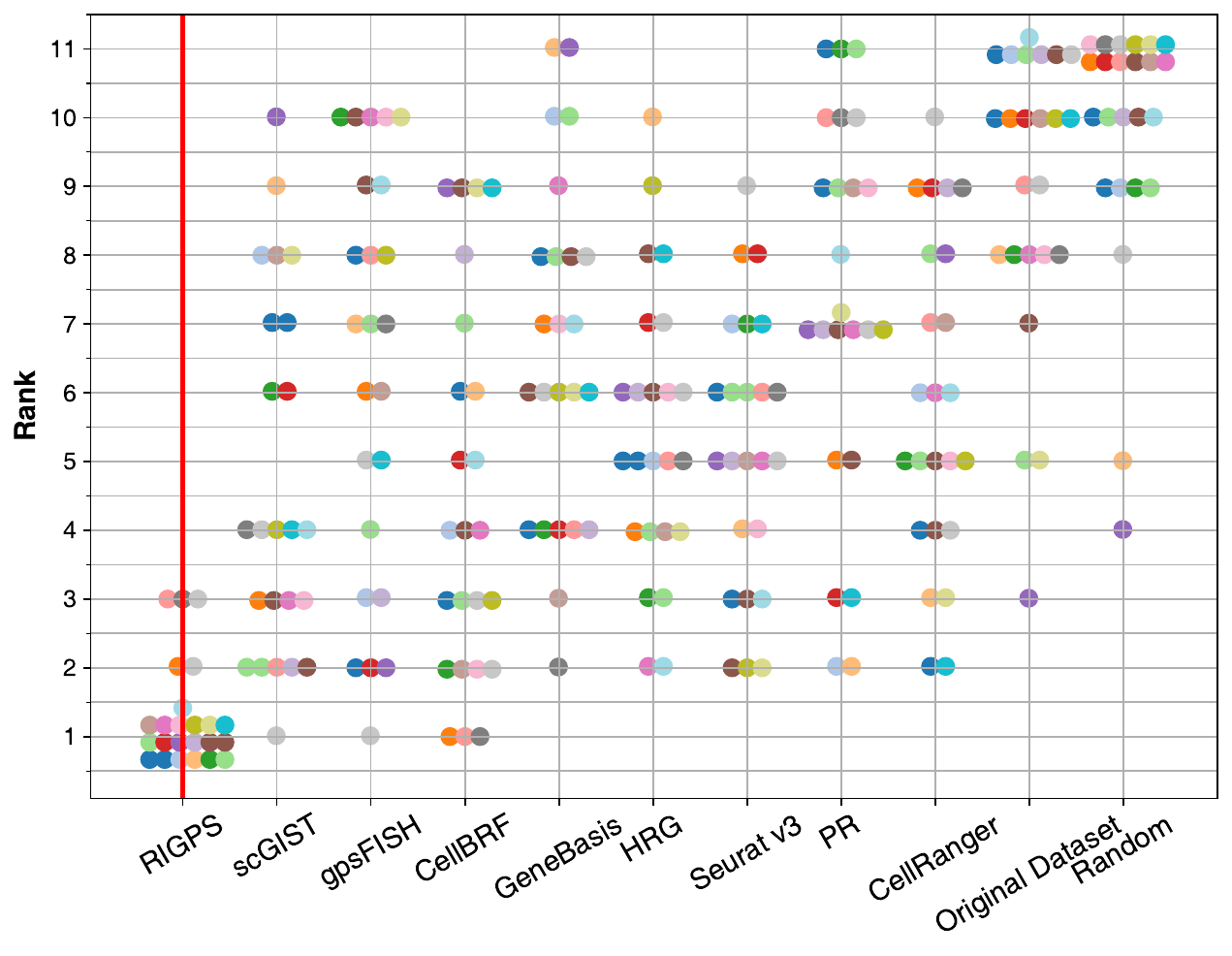}
        \caption{Performance Rank}
        \label{fig:sub4}
    \end{subfigure}
    }
    \caption{Overall quantitative analysis on downstream clustering task performance comparison: (a-c) Comparison of \model\ with seven state-of-the-art gene panel selection methods for single-cell clustering in NMI, ARI,  and SI. 
    (d-f) Evaluating each baseline with the Nemenyi test based on NMI, ARI, and SI. (g) Performance Ranking in NMI.
    }
    \label{fig:main_figure}
\end{figure*}

\section{Experimental Results}\label{exp}
This section reports the quantitative evaluation of \model\ against other baselines and ablation variations. 
To thoroughly analyze the multiple characteristics of \model, we also analyze the model robustness under batch effect, the convergence speed of optimization, the hyperparameter of the reward function, the time/space scalability, coarse boundary setting, knowledge injection setting, and the size of selected biomarkers.

\begin{figure*}[!ht]
    \centering
    \begin{subfigure}[b]{0.48\textwidth} 
        \includegraphics[width=\textwidth]{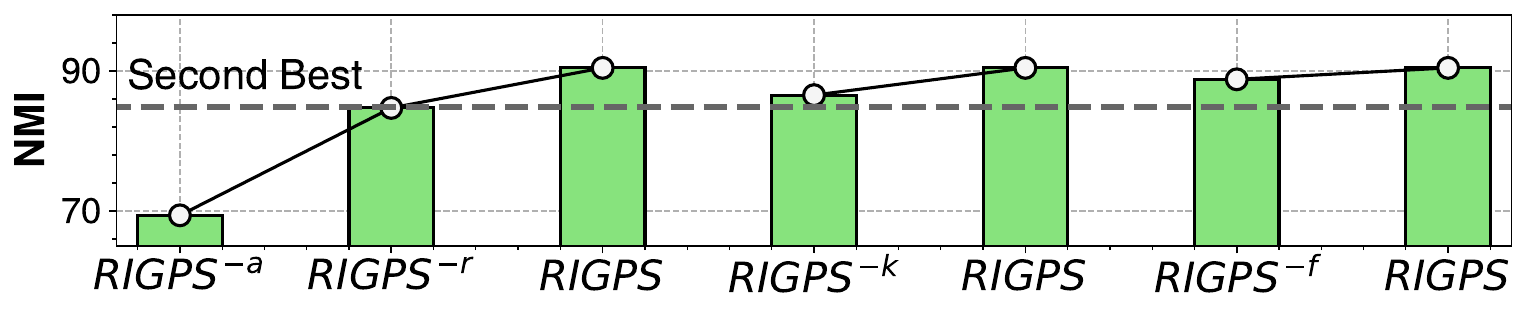}
        \caption{Chu1}
    \end{subfigure}
    \begin{subfigure}[b]{0.48\textwidth} 
        \includegraphics[width=\textwidth]{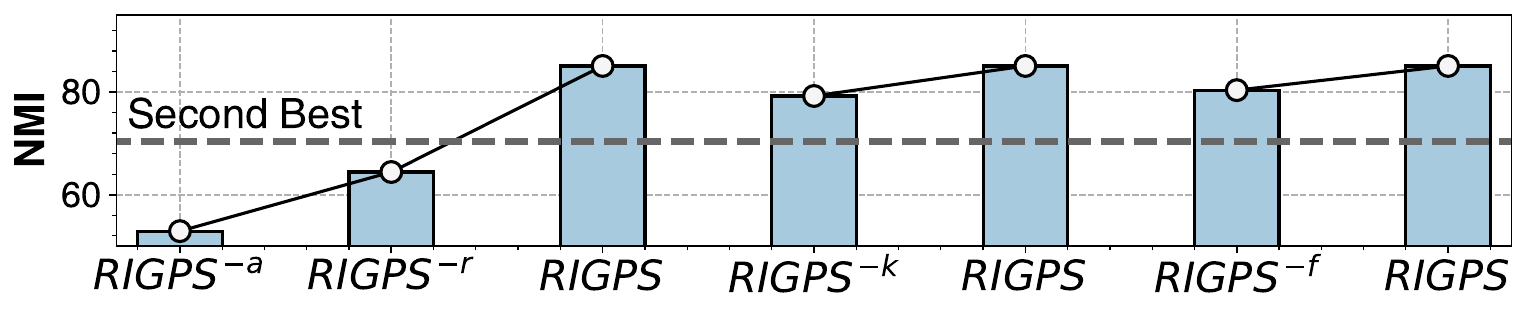}
        \caption{Leng}
    \end{subfigure}
    \begin{subfigure}[b]{0.48\textwidth} 
        \includegraphics[width=\textwidth]{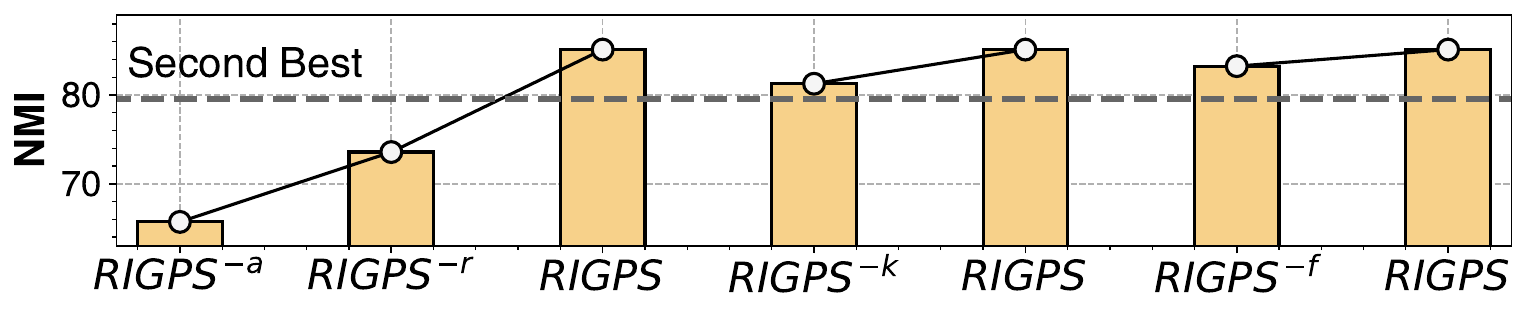}
        \caption{Puram}
    \end{subfigure}
    \begin{subfigure}[b]{0.48\textwidth} 
        \includegraphics[width=\textwidth]{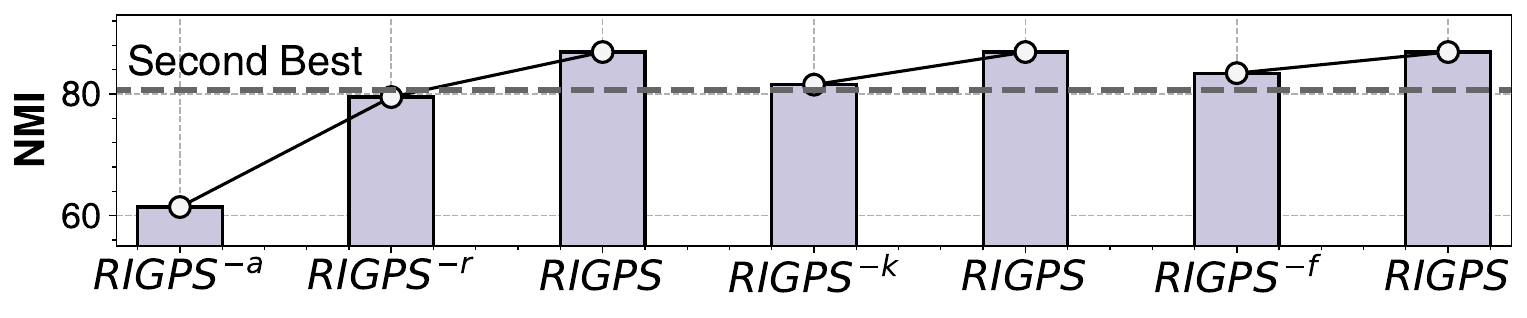}
        \caption{Mouse Pancreas1}
    \end{subfigure}
    \caption{
    Ablation studies of \model\ in terms of NMI. The comparison between \model${^{-a}}$, \model${^{-r}}$, and \model\ shows the impact of reinforced optimization. The comparison between \model${^{-k}}$ and \model\ shows the impact of the knowledge injection. The comparison between \model${^{-f}}$ and \model\ shows the impact of the pre-filtering component. 
    }
    \label{ablation_fig}
\end{figure*}
\begin{figure*}[!h]
    \centering
    \begin{subfigure}[b]{0.24\textwidth} 
        \includegraphics[width=\textwidth]{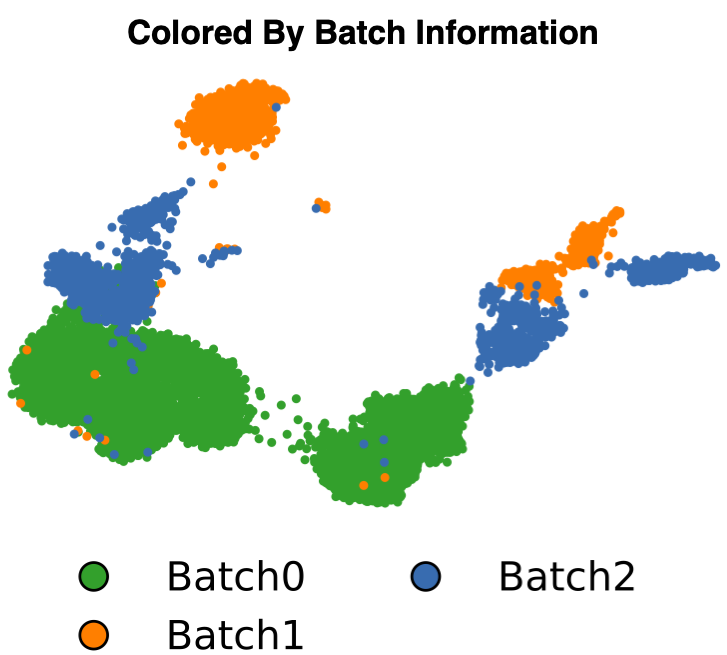}
        \caption{The original dataset.}
    \end{subfigure}
    \begin{subfigure}[b]{0.24\textwidth} 
        \includegraphics[width=\textwidth]{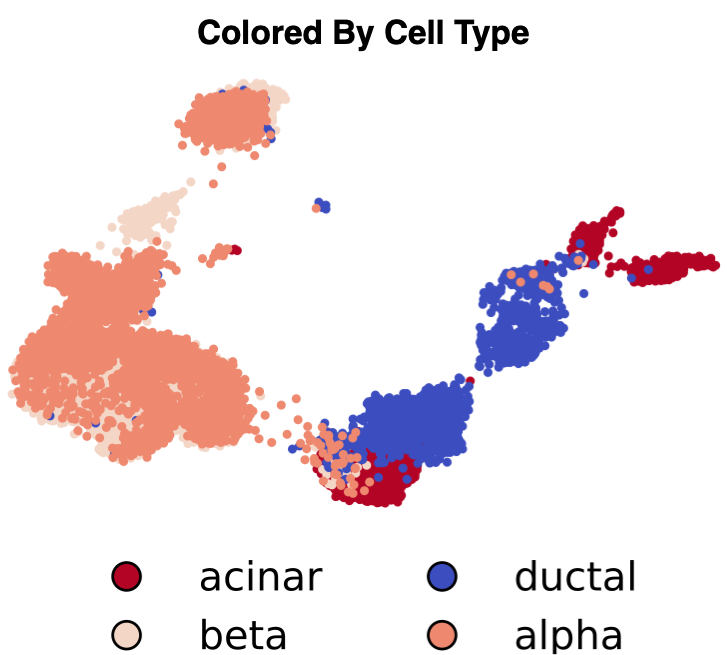}
        \caption{The original datasets.}
    \end{subfigure}
    \begin{subfigure}[b]{0.24\textwidth} 
\includegraphics[width=\textwidth]{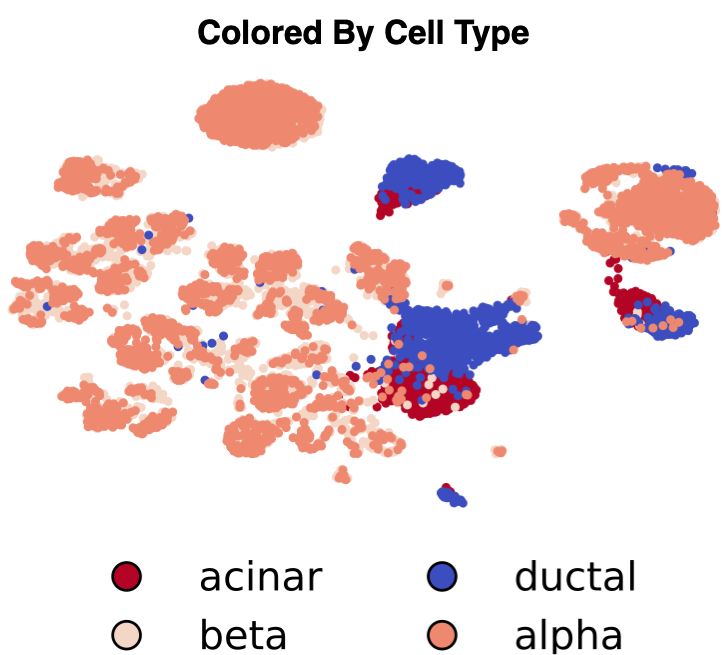}
        \caption{CellBRF Selected.}
    \end{subfigure}
    \begin{subfigure}[b]{0.24\textwidth} 
    \includegraphics[width=\textwidth]{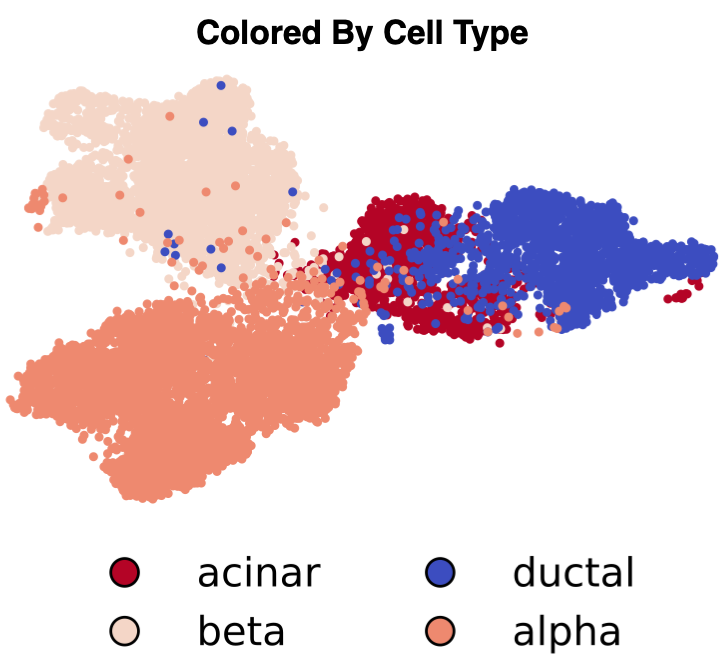}
        \caption{\model{} Selected.}
    \end{subfigure}
    \caption{ 
 Model robustness checks on the Human Pancreas dataset under batch effect circumstances. (a) The visualization of the original dataset is colored by batch number. (b-d) Visualizing the original dataset, the optimized dataset via gene panel selection methods, i.e., CellBRF and \model{}, colored by cell type.}
\label{batch_effect_fig}
\end{figure*}

\subsection{Overall Comparison}\label{overall_com}

This experiment aims to answer: \textit{Is \model\ capable of effectively identifying the biomarkers?}
Figure~\ref{fig:main_figure} (a-c) compares \model\ with ten gene panel selection methods for single-cell clustering on all datasets regarding NMI, ARI, and SI. 
We also reported the Nemenyi test in Figure~\ref{fig:main_figure} (d-f) and performance ranking visualization regarding NMI in Figure~\ref{fig:main_figure} (g). 
We observed that the average performance of \model\ outperforms all the baseline methods. 
Additionally, \model\ achieves the highest rank on 19 out of 24 datasets and ranks within the top 3 for all datasets in terms of NMI.
The underlying driver for this observation is that \model\ eliminates redundant genes through gene pre-filtering and then effectively selects the most vital gene panel by reinforcement-optimized strategy. 
Overall, this experiment demonstrates that \model\ is effective and robust across diverse datasets, encompassing various species, tissues, and topic-related complexities, underscoring its broad applicability for single-cell genomic data analysis tasks.
The results of the \textit{Numerical Comparison} regarding NMI, ARI, and SI are provided in the \textit{Supplementary Material}.

\subsection{Study of the Impact of Each Technical Component}\label{ablation}
This experiment aims to answer: 
\textit{How does each technical component of \model\ affect its performance?}
We developed four variants of \model\ to validate the impact of each technical component.
(i) \textbf{$\model^{-r}$} uses the gene subset obtained by pre-filtering as the final gene panel without the reinforced optimization; its results will depend on random exploration.
(ii) \textbf{$\model^{-k}$} reinforced optimize the whole pipeline without the knowledge injection, which results in a random-initialized exploration start point.
(iii) \textbf{$\model^{-f}$} reinforced optimize the whole pipeline without the pre-filtering component, which results in a vast search space.
(iv) \textbf{$\model^{-a}$} ablated all components, i.e., the performance on the original dataset, which will use all genes to cluster. 
Figure~\ref{ablation_fig} illustrates the results on Chu1, Leng, Puram, and Mouse Pancreas1 datasets.
We observed that \model\ significantly outperforms $\model^{-r}$ and $\model^{-a}$ in terms of performance.
The underlying driver is that reinforcement iteration has a powerful learning ability to screen the key gene panel from the pre-filter gene subset through iterative feedback with the reward estimation.
We also observed that \model\ is superior to $\model^{-k}$ 
in all cases.
The underlying driver is that prior knowledge injection provides a better starting point for reinforcement optimization. 
Then, RL's stochastic nature will explore and enhance them to a higher-performance gene subset.
Moreover, We found that \model\ surpasses $\model^{-f}$.
The underlying driver is that gene pre-filtering integrates multiple gene importance evaluation methods to ensure it removes the most redundant genes. 
It obtains a modest set of genes, reducing the complexity of the gene panel selection problem and helping the reinforcement iteration to find a gene panel with even better performance. 
In summary, this experiment validates that the individual components of \model\ can greatly enhance performance.

\begin{figure*}[!t]
    \centering
    \begin{subfigure}[b]{0.245\textwidth} 
        \includegraphics[width=\textwidth]{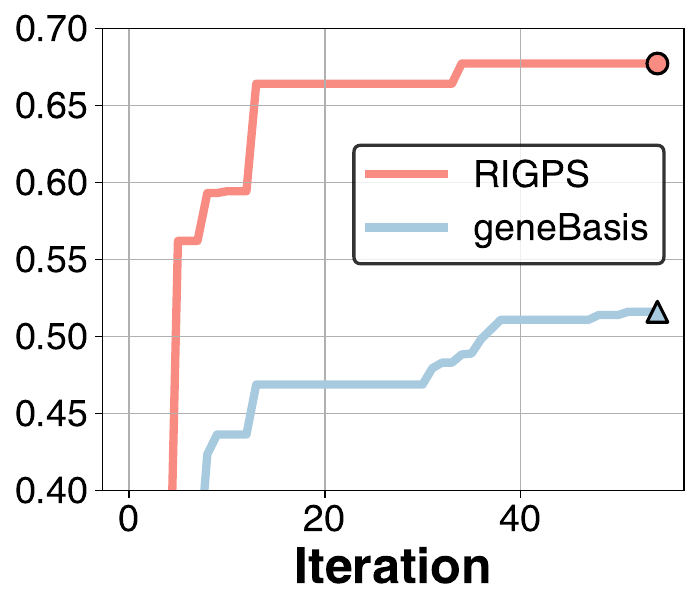}
        \caption{Cao}
    \end{subfigure}
    \begin{subfigure}[b]{0.24\textwidth} 
        \includegraphics[width=\textwidth]{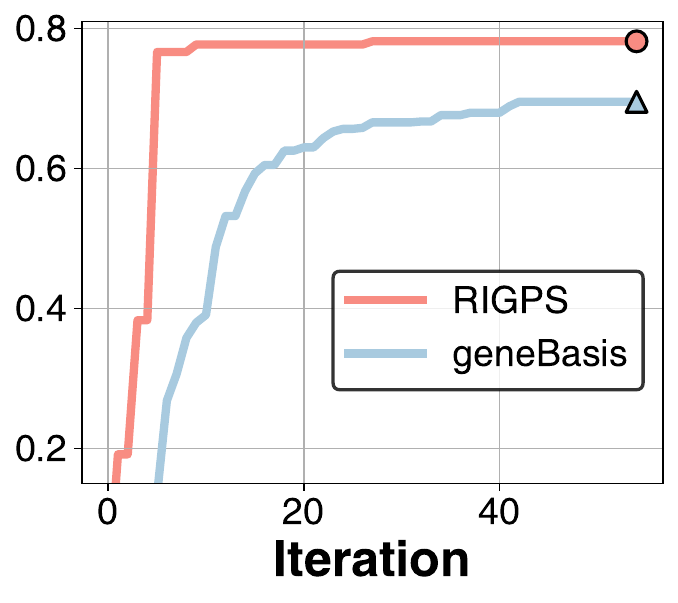}
        \caption{Han}
    \end{subfigure}
    \begin{subfigure}[b]{0.24\textwidth} 
        \includegraphics[width=\textwidth]{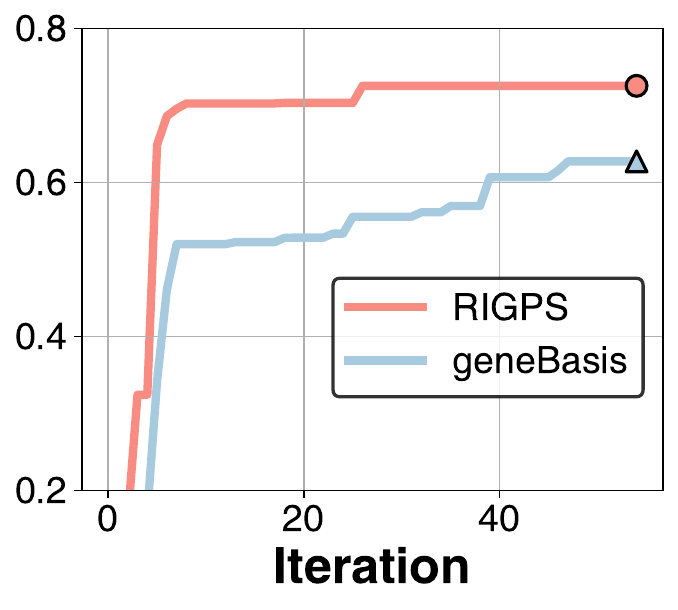}
        \caption{Yang}
    \end{subfigure}
    \begin{subfigure}[b]{0.24\textwidth} 
        \includegraphics[width=\textwidth]{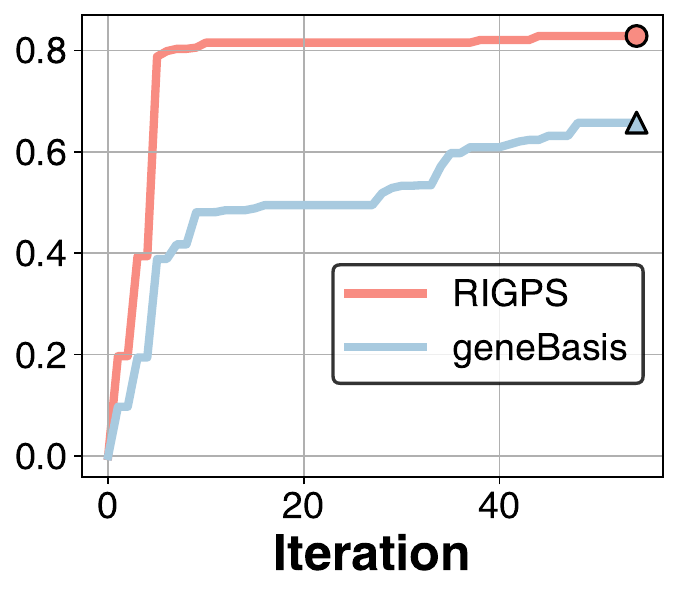}
        \caption{Puram}
    \end{subfigure}
    \caption{
        Iterative convergence speed comparison between \model\ (reinforced optimized) and geneBasis (heuristic optimized).
    }
    \label{iteration_speed_fig}
\end{figure*}

\subsection{Study of the Robustness under Batch Effect}
This experiment aims to answer: \textit{can \model{} robustly find critical genes even under the batch effect?} 
To answer the question, we utilized the Human Pancreas dataset, a benchmark composed of three single-cell RNA sequencing (scRNA-seq) datasets collected in three batches, making it ideal for evaluating batch effects.
As visualized in Figure~\ref{batch_effect_fig}(a), cells from different batches are distinctly separated when projected into two-dimensional space due to the batch effect.  
Conversely, in Figure~\ref{batch_effect_fig}(b), these cells are intermixed, particularly the alpha and beta cells, which are two cell types with markedly different gene expression profiles.
Figures~\ref{batch_effect_fig}(c) and (d) present the two-dimensional visualizations of the processed dataset using the second-best method, CellBRF, and our proposed \model{}, respectively.
When comparing the results, it is evident that the genes selected by \model{} lead to a better clustering of cells according to their types. 
Specifically, the batch effect significantly impacts CellBRF, resulting in the intermixing of alpha and beta cell clusters. 
This fusion indicates that CellBRF is not robust enough to preserve biological signals while handling batch effects. 
In contrast, the gene selected by \model{} could effectively distinguish between alpha and beta cells, as well as between acinar and ductal cells. 
This enhanced distinction is attributed to the characteristic of \model{}, where the reinforcement learning (RL) agents utilize descriptive statistics as state representations instead of directly processing raw gene expression data. 
By employing descriptive statistics, \model{} introduces less noise into the analysis and better preserves the underlying biological signals. 
This methodology allows for a more robust identification of critical genes, leading to improved differentiation of cell types even in batch effects.

\subsection{Study of Convergence Speed between Reinforced Iteration and Heuristic Iteration}\label{reinforce}

This experiment aims to answer: \textit{Will the rules learned by \model\ outperform heuristic iteration?} 
Figure~\ref{iteration_speed_fig} shows the performance (NMI) of genes selected in the first 50 iterations of \model\ and geneBasis (a commonly used iteration-based gene selection method by optimizing and selecting the gene that can minimize Minkowski distances in each step) on Cao, Han, Yang, and Puram datasets.
We found that the speed of convergence and the performance of \model\ at convergence are far better than geneBasis. 
This observation indicates that our reinforcement iteration can quickly and accurately find the best-performing gene subset by interacting with the environment through the rewards of each iteration, compared to geneBasis, which considers maximizing statistical metrics at each iteration. 
This demonstrates that the reinforcement iteration possesses strong learning capabilities and robustness. 
Therefore, this experiment proves that \model\ is superior to existing methods both in terms of the speed of iterative convergence and the performance of the gene subset obtained after convergence.

\subsection{Study of Expert Knowledge-Guided Optmized Result}\label{more} 

\begin{figure}[!t]
    \centering
\includegraphics[width=0.45\textwidth]{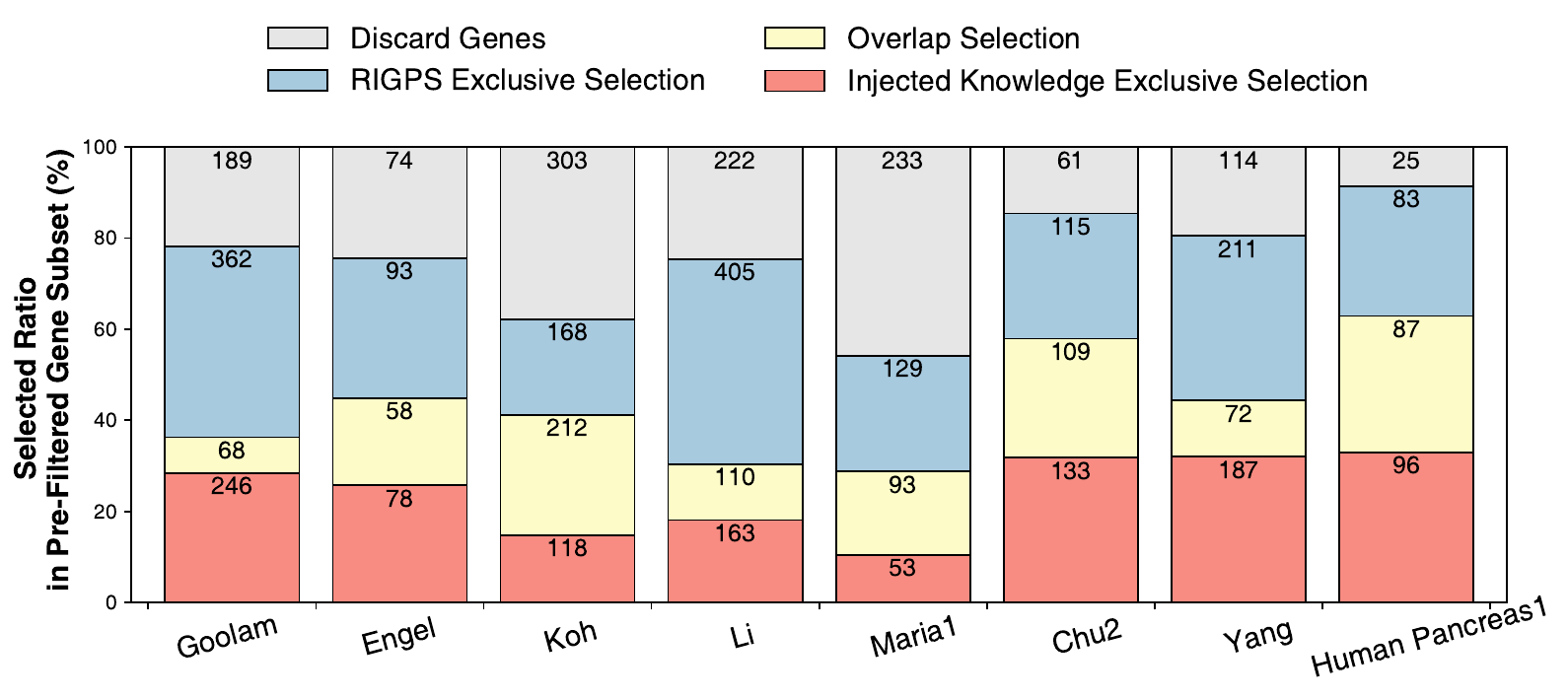}
\caption{The comparison of the selected result in the coarse boundary gene subset by \model\ exclusive selection, overlap selection, and injected knowledge exclusive selection.}
    \label{fig:refinement}
\end{figure}

This experiment aims to answer: \textit{is \model\ more than just an ensemble of other methods?} 
Figure \ref{fig:refinement} shows the comparison of the selected ratio in the coarse boundary by \model\ \textit{exclusive selection}, \textit{overlap selection}, and \textit{injected knowledge exclusive selection} on eight datasets.
From the figure, we can first observe the overlap (colored in yellow) between the injected gene set and the RL-refined gene set in a relatively small proportion. 
We also found that the gene panel selected by \model\ is substantially varied from prior knowledge.
This illustrates that reinforcement iteration with prior knowledge does not simply repeat the injected selection pattern. 
In contrast, prior knowledge will help reinforcement iteration to get a better starting point while allowing the framework to refine the selection and search for a more streamlined biomarker set.
In summary, the experiments validate that ensemble diverse gene selection methods as prior knowledge and the stochastic nature of reinforcement learning contribute significantly to the superior performance and robustness of \model.

\subsection{Study of Trade-off in Reward Function}\label{hyper_study}
\begin{figure}[!t]
    \centering
    \begin{subfigure}[b]{0.23\textwidth} 
    \includegraphics[width=\textwidth]{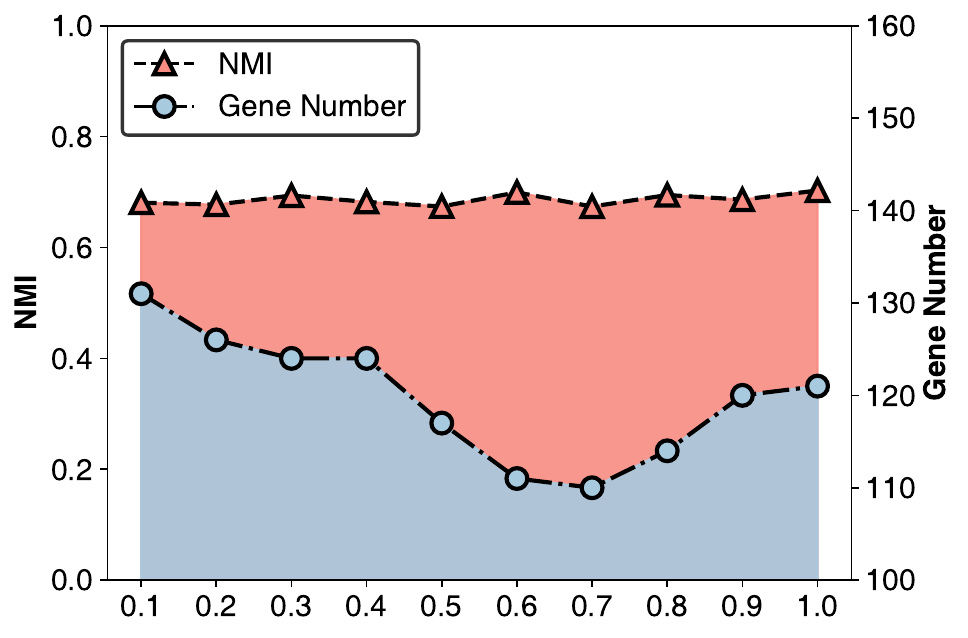}
    \caption{$\lambda$ in Equation~\ref{reward_compact}}
    \end{subfigure}
    \begin{subfigure}[b]{0.23\textwidth} 
    \includegraphics[width=\textwidth]{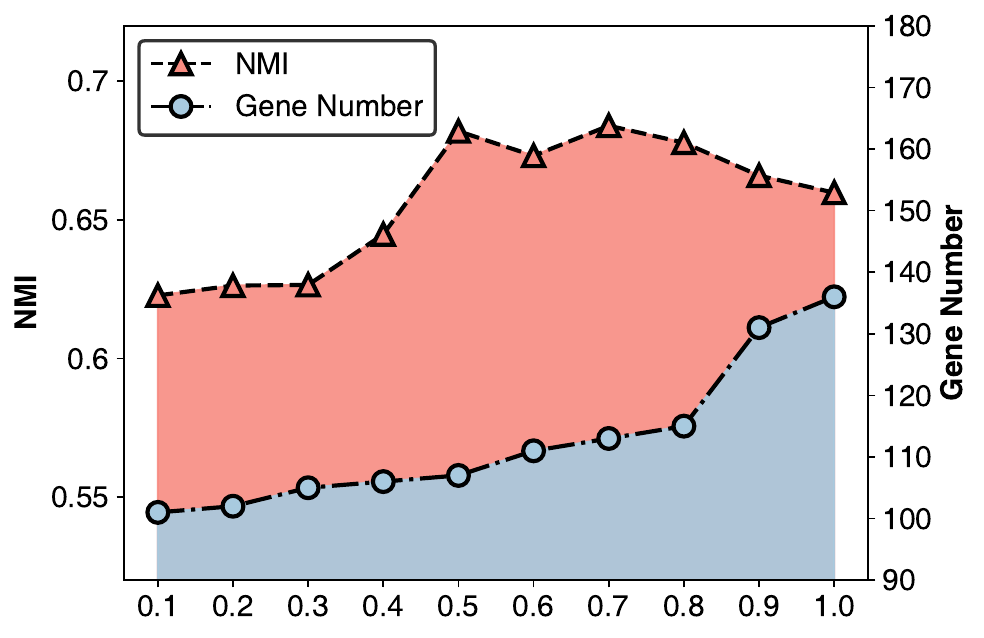}
        \caption{$\alpha$ in Equation~\ref{reward_overall_func}}
    \end{subfigure}
    \caption{
     The result of the hyperparameter sensitivity test on Cao.
    }
    \label{hyper_fig}
    \vspace{-0.4cm}
\end{figure}

\begin{figure*}[!htbp]
    \centering
    \begin{subfigure}[b]{0.45\textwidth} 
        \includegraphics[width=\textwidth]{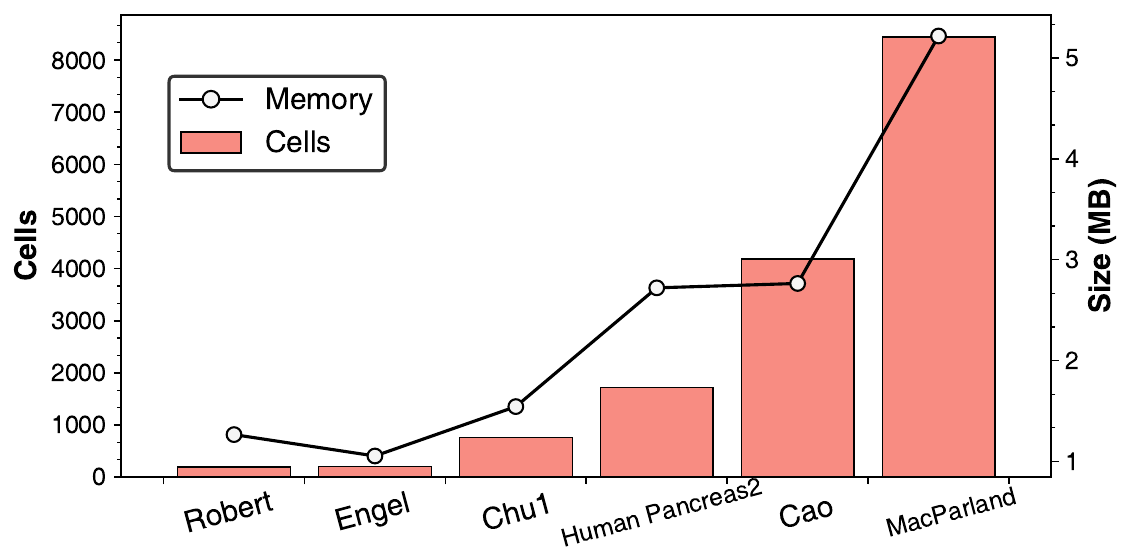}
        \caption{Parameter Size}
    \end{subfigure}
    \begin{subfigure}[b]{0.45\textwidth} 
        \includegraphics[width=\textwidth]{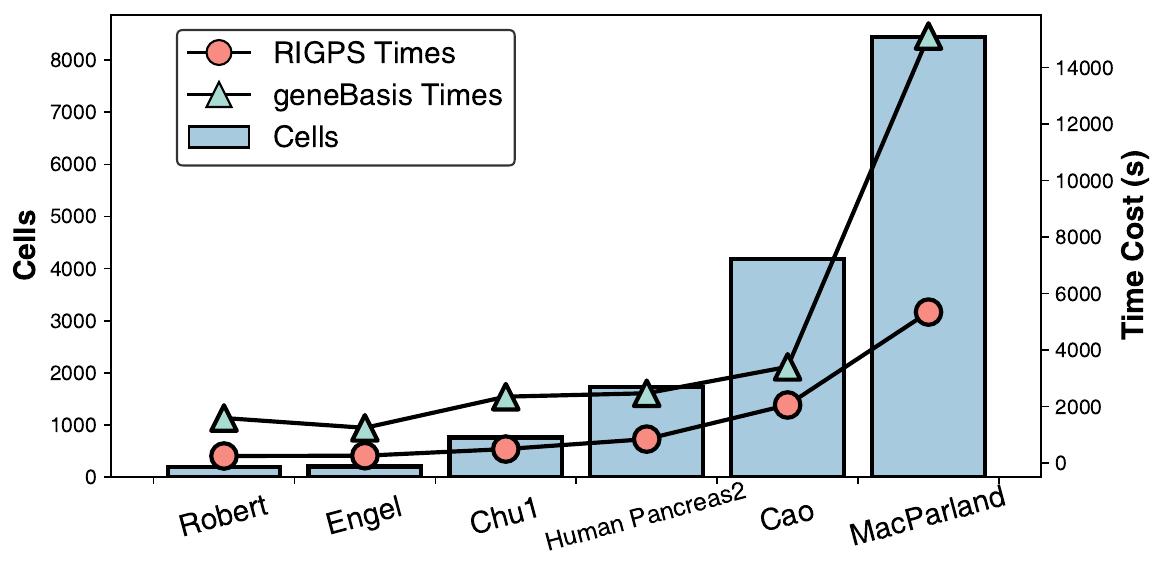}
        \caption{Train Time}
    \end{subfigure}
    \caption{
      Scalability check of \model\ regarding parameter size and training time.
    }
    \label{time_space_efficiency_fig}
\end{figure*}
\begin{figure*}[!htbp]
    \centering
    \begin{subfigure}[b]{0.245\textwidth} 
        \includegraphics[width=\textwidth]{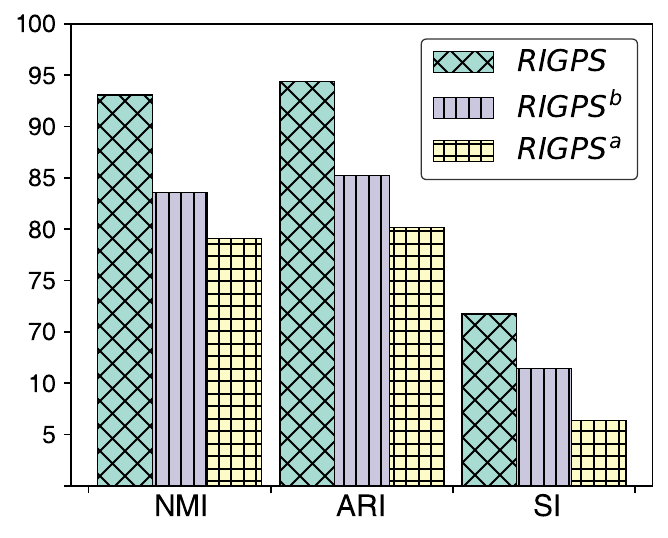}
        \caption{Leng}
    \end{subfigure}
    \begin{subfigure}[b]{0.24\textwidth} 
        \includegraphics[width=\textwidth]{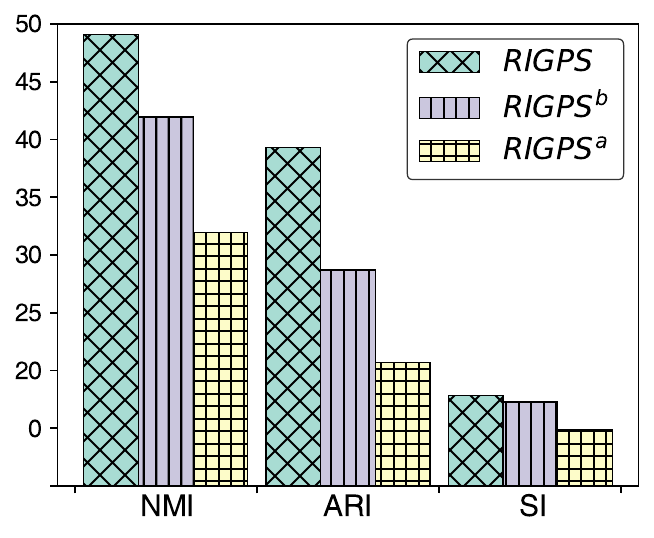}
        \caption{Maria2}
    \end{subfigure}
    \begin{subfigure}[b]{0.24\textwidth} 
        \includegraphics[width=\textwidth]{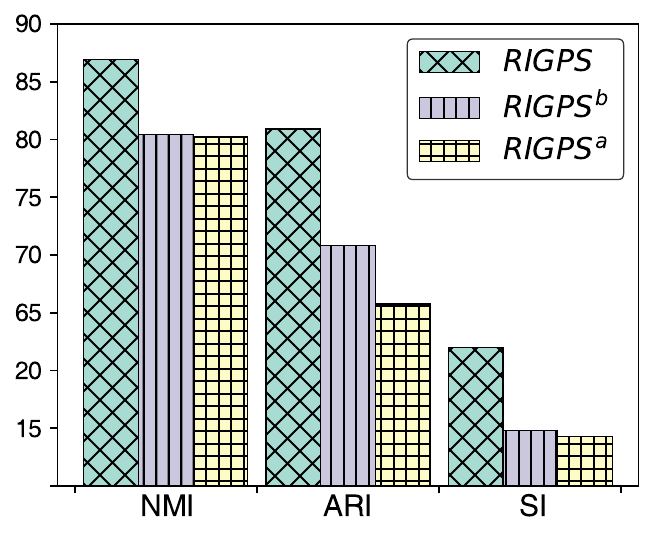}
        \caption{Mouse Pancreas1}
    \end{subfigure}
    \begin{subfigure}[b]{0.24\textwidth} 
        \includegraphics[width=\textwidth]{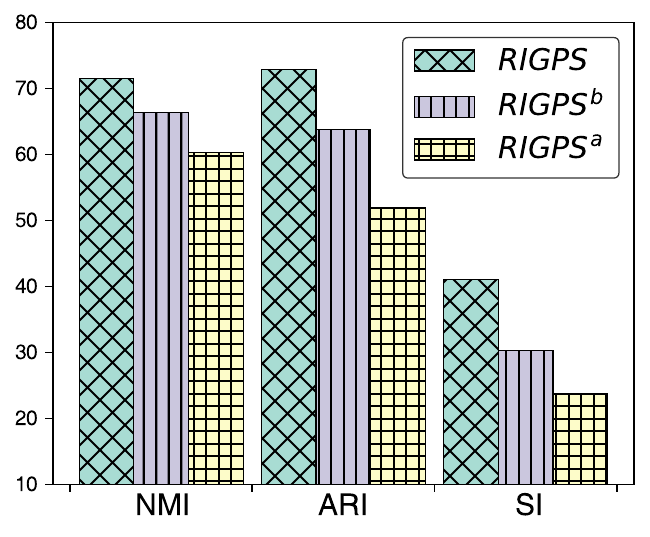}
        \caption{Robert}
    \end{subfigure}
    \caption{
    Performance comparison between three pre-filtering strategies. 
     $\model^{a}$ adopts Random Forest in the gene pre-filtering module, and $\model^{b}$ adopts RandomForest, SVM, and RFE in the gene pre-filtering module.
    }
    \label{gene_pre_filter_fig}
    \vspace{-0.4cm}
\end{figure*}

This experiment aims to answer:
\textit{How do the reward function's hyperparameters affect the model's performance and the number of genes selected?} 
This experiment investigates how the reward function's hyperparameters, $\alpha$ and $\lambda$, influence model performance and selected biomarker set size.  
In Equation~\ref{reward_overall_func}, higher $\alpha$ prioritizes performance over compactness, while in Equation~\ref{reward_compact}, higher $\lambda$ favors compact gene selection.  
We varied $\alpha$ and $\lambda$ from 0.1 to 1.0 using the Cao dataset, with results shown in Figure~\ref{hyper_fig}.  
Increasing $\lambda$ reduces the number of selected genes initially but causes it to rise at higher values, as the suppression effect weakens when $r^{c}_t$ variation narrows with an increasing $k$.  
In contrast, higher $\alpha$ initially improves performance but eventually degrades it while consistently increasing gene selection.  
This is due to reduced suppression of gene quantity and the introduction of redundancy from selecting too many genes.  
These findings confirm that $\alpha$ and $\lambda$ significantly affect both performance and gene selection.  
For balanced performance, we set $\alpha = 0.5$ and $\lambda = 0.7$.

\subsection{Study of the Time/Space Efficiency}\label{scalable_check}
This experiment aims to answer the following question:
\textit{is  \model\ excels in both temporal (time efficiency) and spatial (memory usage)?}
To this end, we selected six scRNA-seq datasets varying in cell count—Robert, Engel, Chu1, Human Pancreas2, Cao, and MacParland—ranging from small to large to provide a comprehensive evaluation. 
Figure \ref{time_space_efficiency_fig} illustrates the comparison results regarding model parameter size and training time. Our analysis revealed the following key insights: 

\smallskip
\noindent\textbf{Parameter Size Efficiency.} 
We observed that the parameter size of \model\ increases proportionally with the number of cells. This indicates that the state representation component of the reinforcement iteration, specifically the autoencoder, compresses the gene panel into a k-length latent vector. This transformation significantly reduces the parameter size compared to models that might not leverage such efficient encoding mechanisms, thus demonstrating spatial efficiency. 

\smallskip
\noindent\textbf{Training Time Efficiency.} 
The training time exhibited a linear relationship with the number of cells. 
This linear scalability suggests that \model\ maintains consistent training durations relative to dataset size, indicative of robust learning capabilities. 
Through reinforcement iteration, \model\ successfully pinpoints the optimal gene panel in a fair number of iterations and runtime. 
It outperforms geneBasis, a comparable iterative method, by demonstrating superior temporal efficiency.


\begin{figure*}[!t]
    \centering
        \begin{subfigure}[b]{0.24\textwidth} 
        \includegraphics[width=\textwidth]{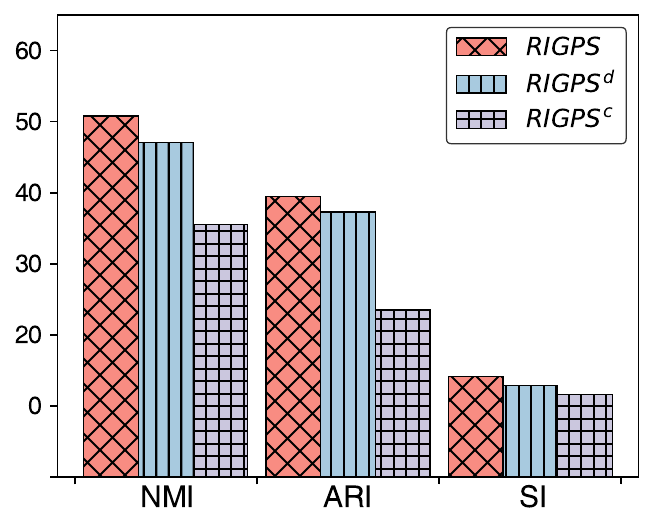}
        \caption{Maria2}
    \end{subfigure}
    \begin{subfigure}[b]{0.24\textwidth} 
        \includegraphics[width=\textwidth]{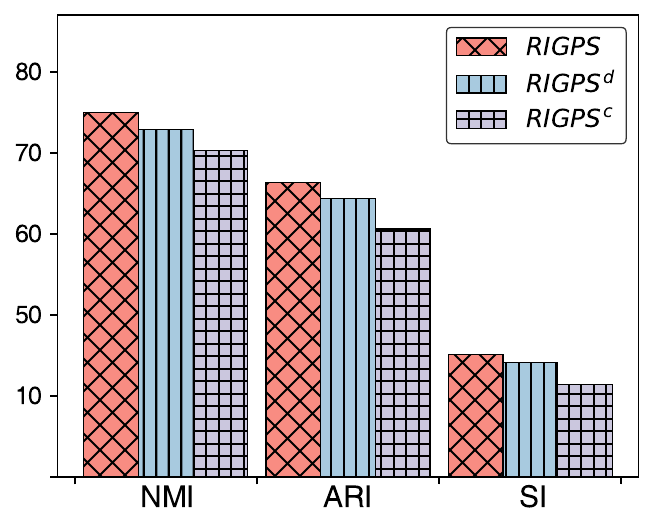}
        \caption{Cao}
    \end{subfigure}
    \begin{subfigure}[b]{0.24\textwidth} 
        \includegraphics[width=\textwidth]{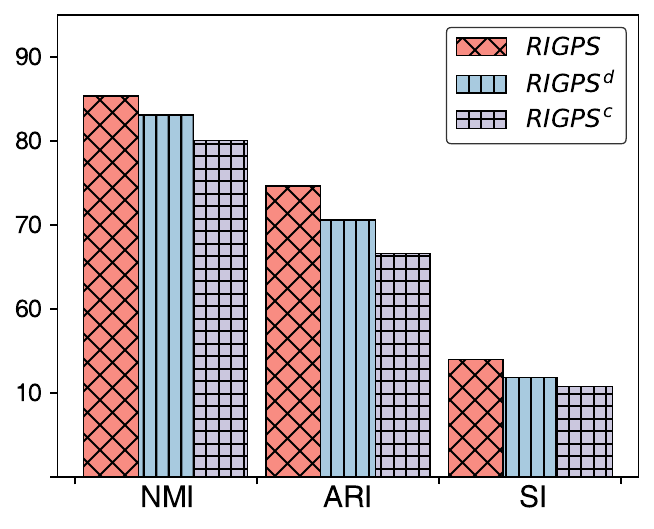}
        \caption{Puram}
    \end{subfigure}
    \begin{subfigure}[b]{0.245\textwidth} 
        \includegraphics[width=\textwidth]{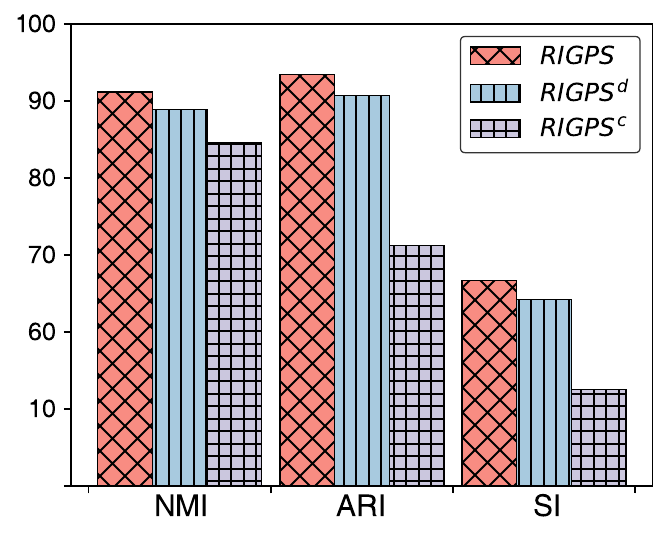}
        \caption{Human Pancreas2}
    \end{subfigure}
    \caption{
    Comparison with different knowledge injection settings. 
    (a-d) The performance of $\model$, $\model^{c}$, and $\model^{d}$ on Maria2, Cao, Puram, Human Pancreas2 datasets. 
    }
    \vspace{-0.4cm}
    \label{prior_knowledge_fig}
\end{figure*}

\subsection{Study of Coarse Boundary Settings}\label{gpf}

This experiment aims to answer: 
\textit{How do different basic method combinations in preprocessing affect the performance of \model?}
To examine the impact of different preprocessing settings, we developed two model variants of \model:
(i) $\model^{a}$: adopting \textit{Random Forest} as the basic method in the gene pre-filtering module.
(ii) $\model^{b}$: adopting \textit{Random Forest}, \textit{SVM}, and \textit{RFE} as the basic methods in the gene pre-filtering module. 
(iii) $\model$: as introduced in Section~\ref{hyper}, the basic methods in our method consist of \textit{Random Forest}, \textit{SVM}, \textit{RFE}, \textit{geneBasis}, and \textit{KBest}.
The comparative analysis of these variants was conducted using datasets from Leng, Maria2, Mouse Pancreas1, and Robert, with the results depicted in Figure \ref{gene_pre_filter_fig}. 
The findings from this study are as follows:
We found that the performance of downstream clustering tasks correlates with the number of basic methods; the more basic methods there are, the better the clustering effect. 
The underlying driver is that introducing more basic methods would reduce the overall bias and raise the recall of select informative biomarkers. 
Further, with many basic methods from multiple perspectives to identify a more comprehensive and form the coarse boundary, the reinforcement iteration is more likely to converge in a gene subset with superior performance. 
In summary, the pre-filtering module options a larger and more comprehensive subset of vital genes and avoids the problem of missing key information, which is highly critical and correlated with the performance of \model.

\subsection{Study of Injected Start Points}\label{ki} 
This experiment aims to answer:
\textit{Will different knowledge set affect the model performance?} 
To validate the effectiveness and extensibility of knowledge injection, we developed two model variants to establish the control group: 
(i) $\model^{c}$, we injected the gene panel selected by \textit{CellBRF} as the prior knowledge. 
(ii) $\model^{d}$, we injected the gene panels selected by \textit{CellBRF}, \textit{geneBasis}, and \textit{HRG} as the prior knowledge.
(iii) $\model$, as introduced in Section~\ref{hyper}, we injected the gene panels selected by \textit{CellBRF}, \textit{geneBasis}, and \textit{HRG} as the prior knowledge. 
Figure~\ref{prior_knowledge_fig} (a-d) shows the comparison results on Maria2, Cao, Puram, and Human Pancreas2. 
We found that as prior knowledge increases, the gene panel obtained by reinforcement iterations becomes increasingly effective. 
This illustrates that increasing prior knowledge injection allows the reinforcement iteration module to attain high-quality starting points, leading to a more informative biomarker.
While models such as CellBRF, which uses a single classical machine learning method, and geneBasis, which iterates using artificial statistical metrics, both have limitations in the biomarker identification, \model\ can ensemble their knowledge through the knowledge injection and identify a superior gene panel in performance.
In summary, prior knowledge injection does help \model\ to find unique and enhanced biomarkers.

\subsection{Study of the Selected Gene Panel Size}
\begin{figure}[!t]
    \centering
\includegraphics[width=0.45\textwidth]{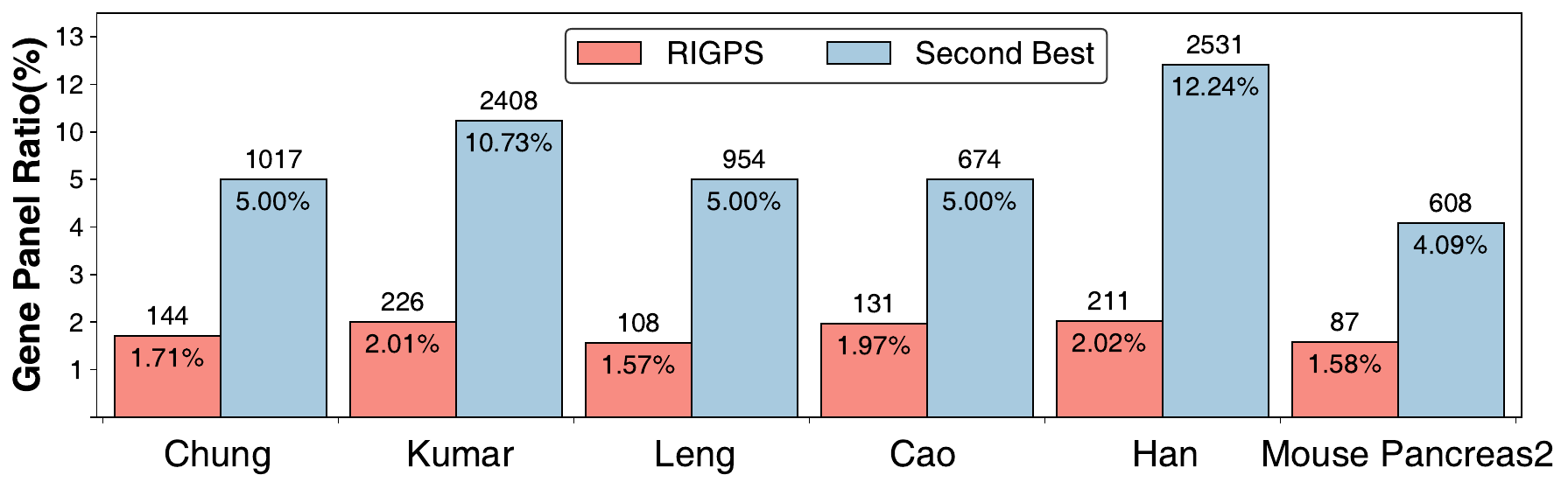}
    \caption{
     Comparison between \model\ and the second-best baseline regarding the selected gene panel size.
    }
    \label{compress_fig}
\end{figure}

This experiment aims to answer this question:
\textit{Is our proposed model capable of selecting a small, yet effective, biomarker set? }
We illustrate the selected gene panel ratio between \model\ and the second-best baseline model on six datasets in Figure~\ref{compress_fig}.
We found that the gene panel obtained by \model\ is significantly more compact than the second best while still outperforming it.
We speculate the underlying driver for this observation is that gene pre-filtering will remove many redundant genes. 
Then, our reinforcement iteration carried out further screening to obtain a compact but effective gene subset. 
Furthermore, this experiment demonstrates that the gene panel selected by \model\ can effectively decrease computational expenses with better performance.

\section{Downstream Biological Analysis}\label{vis}
This section reports four common downstream biological analyses to evaluate selected gene panels qualitatively. 
The results of \textit{Biological Analyses} on all datasets are provided in the \textit{Supplementary Material}. 

\begin{figure}[!h]
    \centering
    \begin{subfigure}[b]{0.44\linewidth} 
        \includegraphics[width=\textwidth]{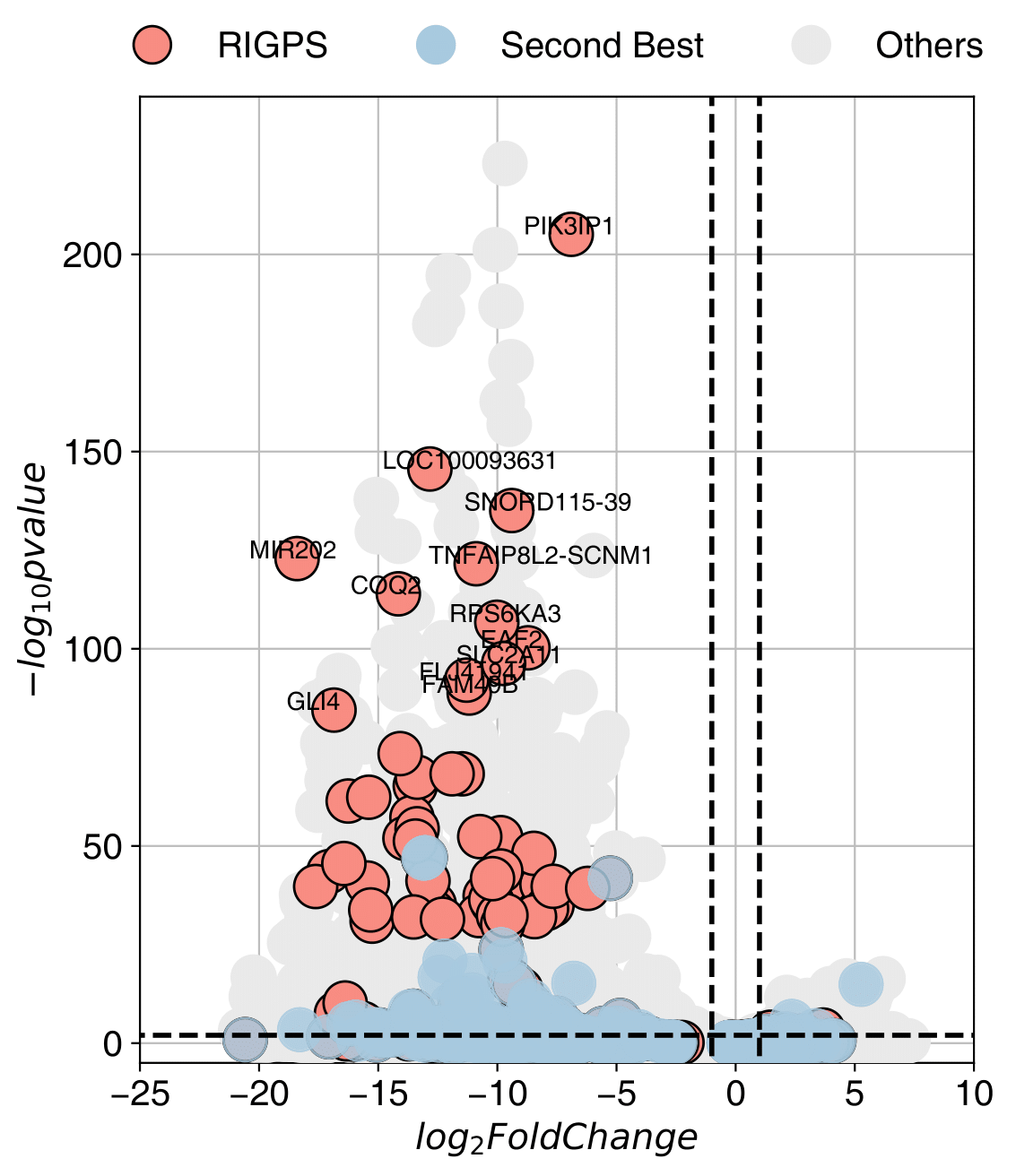}
        \caption{Puram}
    \end{subfigure}
    \begin{subfigure}[b]{0.44\linewidth} 
        \includegraphics[width=\textwidth]{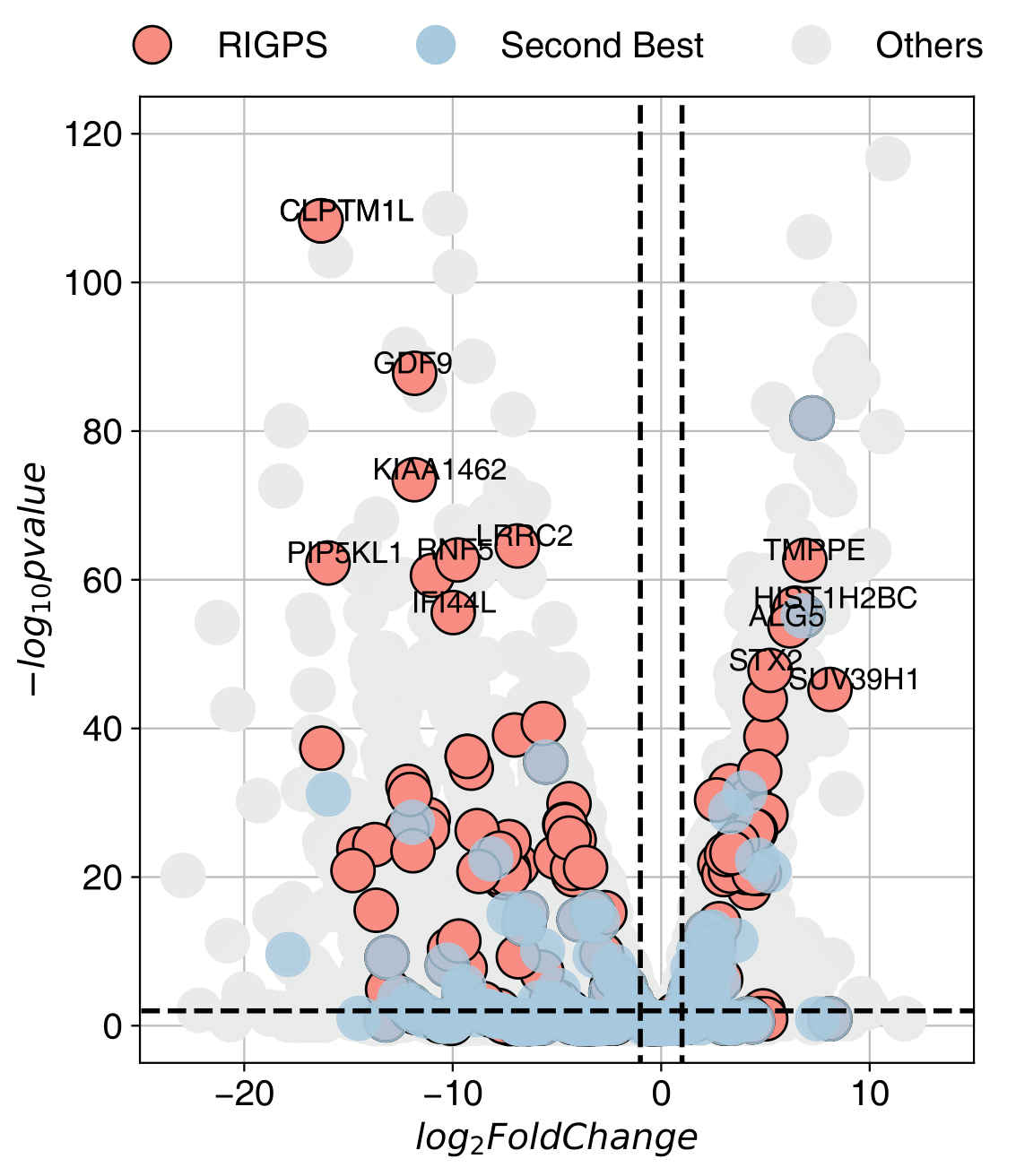}
        \caption{Chu1}
    \end{subfigure}
    \caption{Differential expression analysis. We highlight the selected gene by \model\ (in red) and by the second-best method, CellBRF (in blue).}
    \label{visualization_d}
    \vspace{-0.4cm}
\end{figure}


\subsection{Differential Expression Analysis} Figure~\ref{visualization_d} displays two sets of volcano plots for differential expression analysis, corresponding to the Puram dataset (a) and the Chu1 dataset (b). 
In these plots, red dots represent genes selected by the \model\, blue dots represent genes selected by the second-best method, CellBRF, and grey dots represent other genes.
We can observe that red dots (genes selected by \model) are located in the upper left and upper right regions, indicating significant upregulation or downregulation under experimental conditions. 
Compared with the second-best method (blue dots), genes selected by the \model\ show high significance and fold change in both datasets, demonstrating the method's effectiveness.

\subsection{Visualization Analysis.} 

\begin{figure}[!t]
    \centering
    \begin{subfigure}[b]{0.44\linewidth} 
        \includegraphics[width=\textwidth]{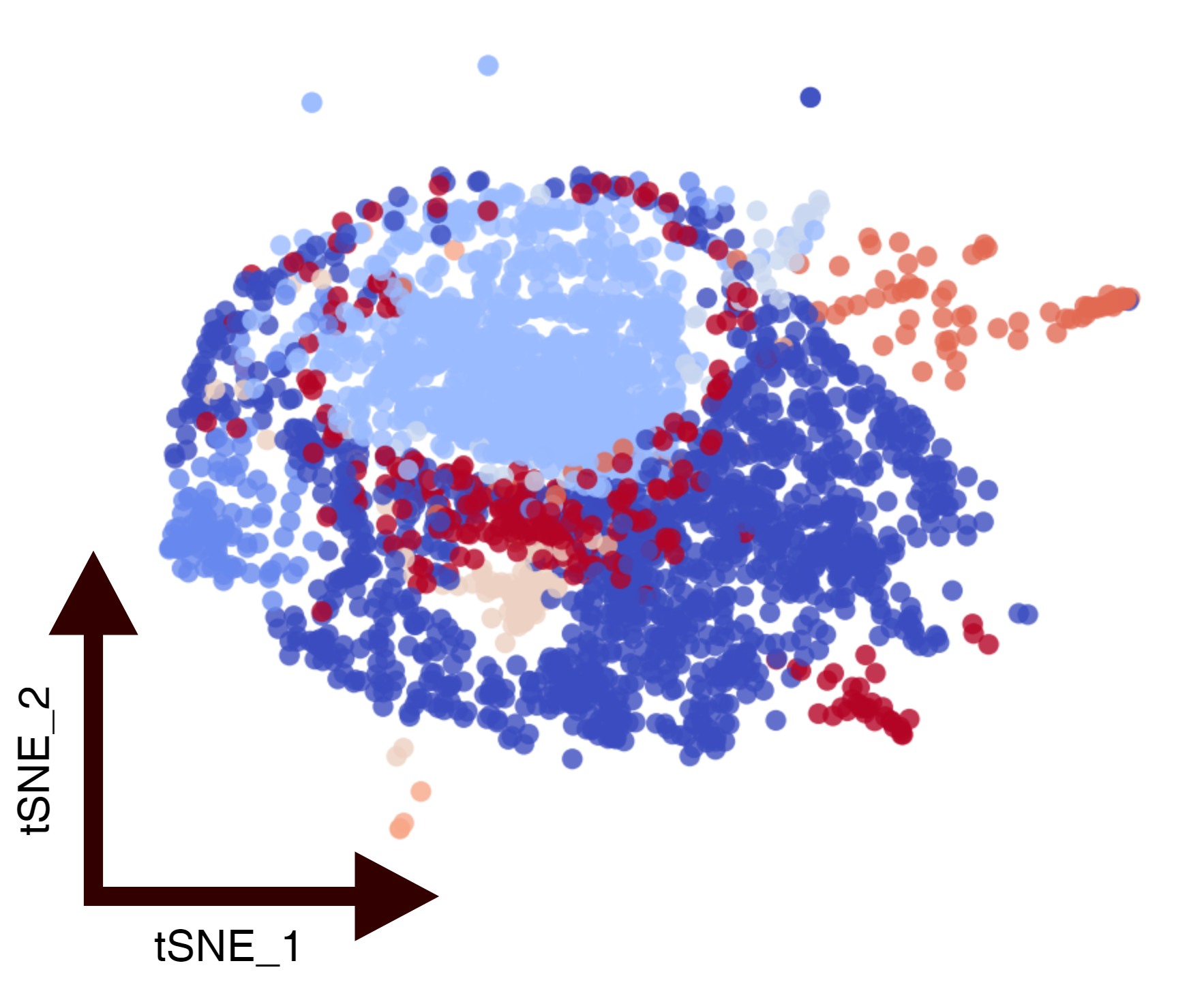}
        \caption{The original dataset.}
    \end{subfigure}
    \begin{subfigure}[b]{0.44\linewidth} 
        \includegraphics[width=\textwidth]{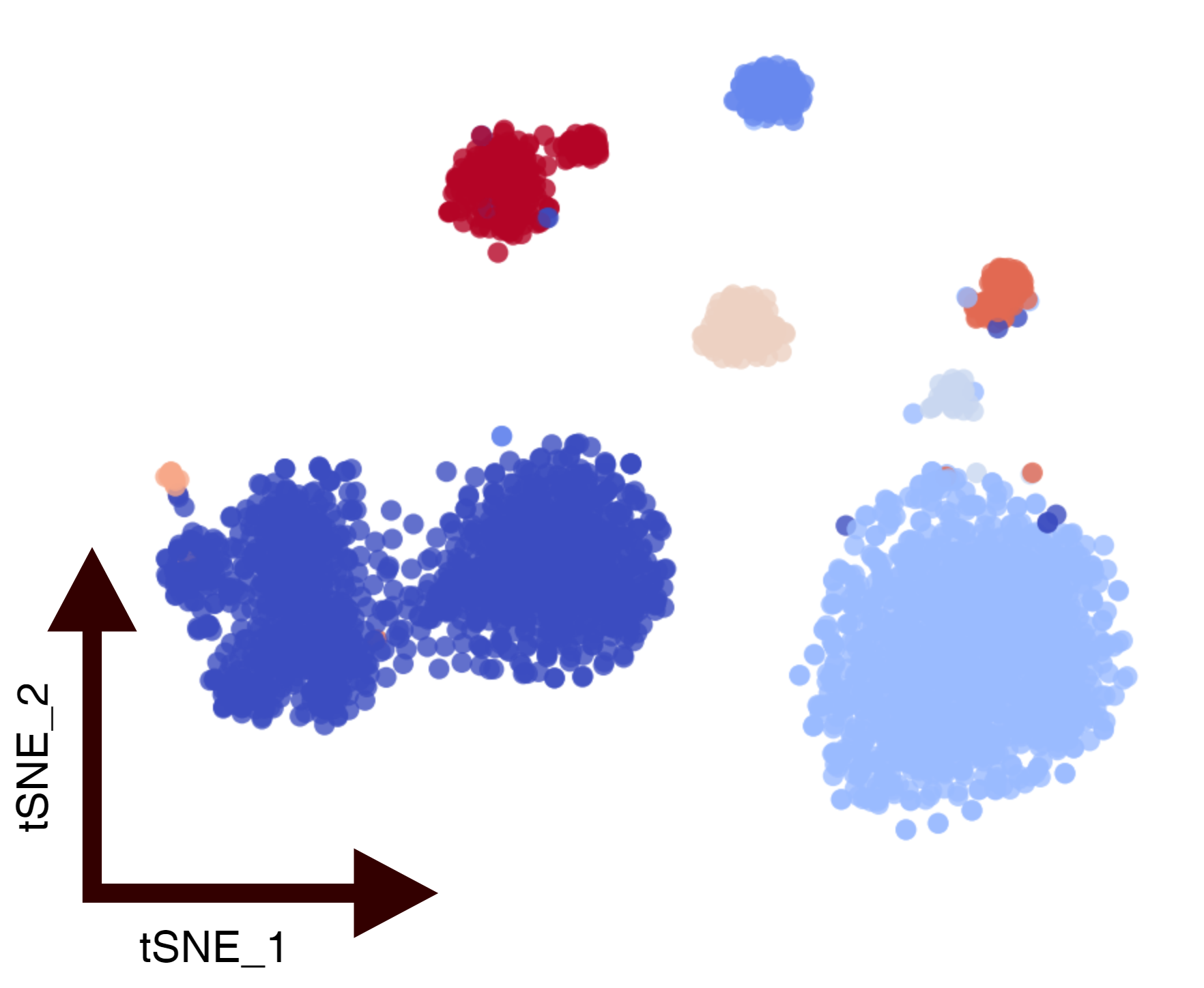}
        \caption{The optimized dataset.}
    \end{subfigure}
    \caption{Visualization analysis of the Puram dataset. (a) t-SNE visualization of the original dataset; (b) t-SNE visualization of \model\ optimized dataset.}
    \label{visualization_v}
\end{figure}

Figure~\ref{visualization_v} applies t-SNE to visualize the Puram dataset with the original genes and the gene panel selected by \model. 
We can observe that cells with the gene subset selected by \model\ self-grouped into distinct groups according to their type. 
In contrast, cells with original genes are jumbled, making identifying their types impossible. 
This visualization analysis shows that our reward function can guide RL agents in selecting the gene set that most distinguishes the cell type.

\subsection{Heatmap Analysis.} 

\begin{figure}[!t]
    \centering
    \begin{subfigure}[b]{0.44\linewidth} 
        \includegraphics[width=\textwidth]{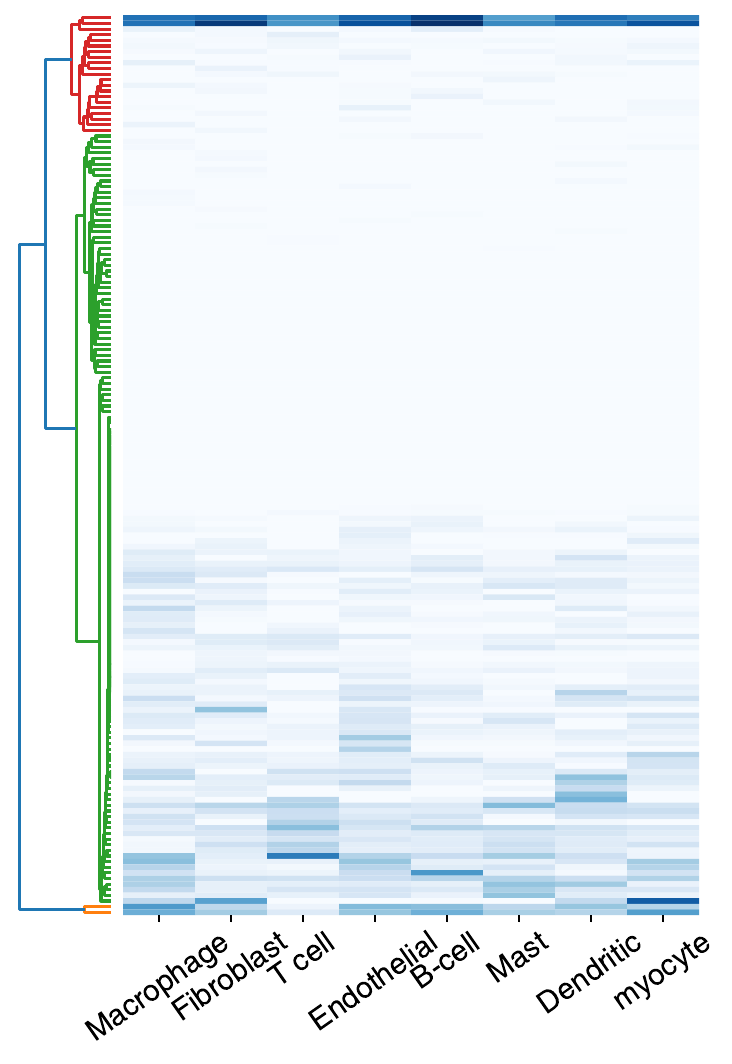}
        \caption{The original dataset.}
    \end{subfigure}
    \begin{subfigure}[b]{0.44\linewidth} 
        \includegraphics[width=\textwidth]{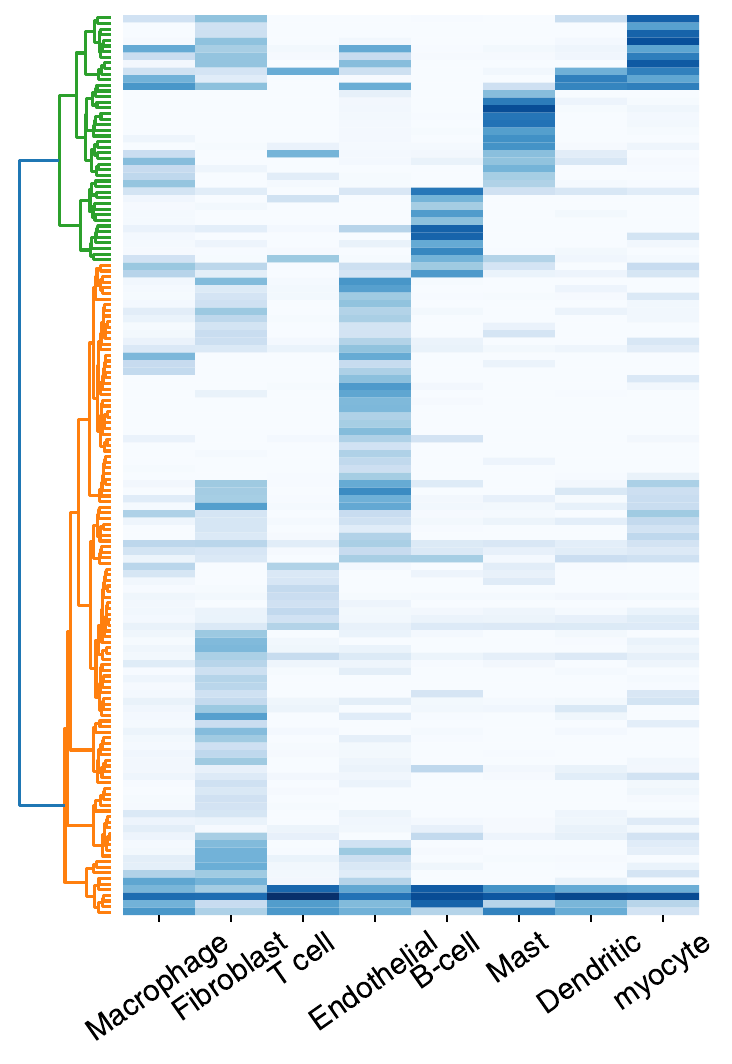}
        \caption{The optimized dataset.}
    \end{subfigure}
    \caption{Heatmap analysis of the Puram dataset. (a) expression heatmap of genes on the original dataset; (b) expression heatmap of genes selected by \model.}
    \vspace{-0.4cm}
    \label{visualization_h}
\end{figure}

Figure~\ref{visualization_h} shows the expression heatmap for the original genes and the gene subset chosen by \model, with the horizontal and vertical axes indicating various cells and genes, respectively. 
The intensity of the gene color increases with the level of gene expression. 
We found that the genes selected by \model\ expressed different patterns between each cell type.
In contrast, the gene expression patterns from the original dataset are extremely similar and difficult to distinguish.
Those observations indicate that by following a spatial separability-based reward function, \model\ can spontaneously find the key genes that most determine cell type, resulting in a visible improvement in these visualizations. 
This finding corresponds with the visualization analysis. 


\begin{table}[!t]
\centering
\caption{Three classification metrics of \model\ against baseline methods on 24 dataset}
\label{cell_type_classificaction_table}
\begin{tabular}{ccccc}
\toprule
\textit{Metrics} & \textbf{Acc$_{avg}$} & \textbf{Acc$^{b}_{avg}$} & \textbf{MaF1$_{avg}$} & \textbf{MiF1$_{avg}$} \\
\midrule
\textbf{\model\ (Ours)} & \textbf{85.90} & \textbf{80.50} & \textbf{77.87} & \textbf{87.88} \\
scGIST & 77.83 & 71.58 & 64.62 & 67.88 \\
gpsFISH & 69.05 & 66.67 & 68.86 & 83.99 \\
CellBRF & 82.55 & 76.19 & 72.72 & 82.13 \\
GeneBasis & 83.24 & 78.27 & 76.07 & 81.90 \\
HRG & 76.79 & 72.31 & 68.33 & 77.71 \\
Seurat v3 & 84.48 & 73.71 & 71.25 & 77.47 \\
PR & 79.39 & 73.72 & 71.19 & 73.60 \\
CellRanger & 81.90 & 74.19 & 71.21 & 77.83 \\
Original Dataset & 72.51 & 71.31 & 68.49 & 61.43 \\
Random & 67.84 & 66.78 & 63.40 & 64.46 \\
\bottomrule
\end{tabular}
\end{table}


\subsection{Cell Type Annotation}
\label{supervised_performance}
This experiment aims to answer the following question: \textit{Does \model\ work as well for supervised tasks?} 
We evaluate the effectiveness of our approach on supervised cell type annotation by feeding the selected genes into a three-layer neural network (with 256 and 64 hidden units in the first two layers, and the number of cell types in the output layer). The first two layers use ReLU activations, while the output employs a SoftMax function. For all datasets, we split data into training and testing sets in a 60\%/40\% ratio, with a batch size of 64, learning rate of 0.01, and trained for 30 epochs.
Given the imbalanced nature of cell types, we report Accuracy (Acc¥$_{avg}$), Balanced Accuracy (Acc$_{avg}^{b}$), Macro-F1 (MaF1), and Micro-F1 (MiF1) as evaluation metrics. Table~\ref{cell_type_classificaction_table} summarizes the comparative performance of \model\ and ten baseline methods across 24 datasets.
Across all four metrics, \model\ (Ours) consistently surpasses all other methods by a clear margin. This demonstrates that the gene selection strategy empowered by reinforced iteration not only enhances unsupervised clustering but also excels in downstream supervised annotation tasks.
Moreover, the observed discrepancies between Acc$_{avg}$ and Acc$^{b}_{avg}$ for all methods (ranging from 2.5\% to over 10\%) highlight pronounced class imbalance in the datasets. This imbalance is further evidenced by the consistent gap between MiF1$_{avg}$ and MaF1$_{avg}$ (often 5–20 points), indicating that performance on the majority classes tends to dominate overall metrics. 
Notably, \model\ exhibits one of the smallest gaps between standard and balanced accuracy (Acc$_{avg}$ vs. Acc$^{b}_{avg}$, 4.49\%), suggesting robust and balanced predictive capability across both majority and minority cell types. 
Overall, these results confirm that \model\ provides a substantial and consistent improvement in supervised cell type annotation, particularly under challenging imbalanced conditions, and outperforms existing baseline methods by a wide margin.

\section{Related Work}\label{related}

\smallskip
\noindent\textbf{Gene Panel Selection.}
Gene panel selection can be broadly categorized by selection strategies based on the statistical measure of the individual gene, the correlation among genes, or the relevance of genes and cell type.
Initial studies~\cite{yip2019evaluation,luecken2022benchmarking} often employ simple statistical metrics such as variance and mean to select genes. 
However, such methods can be suboptimal as genes with random expression across cell types may also display high variance, rendering them only marginally better than random selection~\cite{M3Drop}. 
More recent efforts have shifted towards exploring the correlation among genes. 
geneBasis~\cite{geneBasis} utilizes a k-nearest neighbor (k-NN) graph to select genes that maximize discrepancies within the graph iteratively. 
Despite their utility, these approaches often overlook the noise in gene expression-based correlation, resulting in a suboptimal performance.
Concurrently, there has been an increasing focus on the relevance of genes to specific cell types. 
These methods~\cite{FEATS, FEAST, NS-forest, gpsFISH} are generally more effective for tasks directly related to cell type. 
However, their performance may falter in applications less tied to cell typology.
Specifically, CellBRF~\cite{CellBRF} employs RandomForest to model cell clustering tasks, thereby selecting genes based on their discriminative power in tree partitioning. 
Unlike these studies, our framework raises a new perspective on gene panel selection, ensembles the knowledge from other basic gene panel selection algorithms, and then employs expert knowledge-guided reinforced iteration to determine the optimal gene panel.

\smallskip
\noindent\textbf{Reinforcement Learning.}
Reinforcement Learning~\cite{sutton2018reinforcement,gu2024review} (RL), where an agent learns through interactions with its environment under a specific policy, has demonstrated remarkable versatility in addressing a range of complex tasks~\cite{li2021structured}, including autonomous driving~\cite{he2023fear}, cloud computing~\cite{zhang2022efficient}, recommendation~\cite{guo2022reinforcement,du2022learning}, and multi-agent system~\cite{zhang2024large}. 
More recently, there has been growing interest in multi-agent reinforcement learning (MARL)\cite{wen2022multi,liu2021automated}, which seeks to solve intricate problems via collaborative or competitive interactions among multiple agents\cite{hu2023mo}. 
Within the feature engineering domain, GRFG~\cite{xiao2022self} proposed a self-optimizing MARL framework designed for feature transformation, bypassing prevalent limitations in traditional feature engineering. 
Meanwhile, HRLFS\cite{zhang2025comprehend} showcases the application of MARL to feature selection, where agents cooperatively determine the most critical subset of features. 
These developments inspired our approach, wherein multiple gene agents operate within an RL framework to identify optimal biomarkers. 
Nonetheless, the direct application of MARL to gene panel selection remains non-trivial, given the absence of ground-truth labels, the complexities of reward function design, and the inherent scalability challenges associated with high-dimensional biological data.

\section{Conclusion Remarks}\label{limitation}
This paper aims to address the challenges inherent in the single-cell genomic data analysis pipeline, which are compounded by issues such as high dimensionality in sequential modeling, sparsity of informative biological signals, and noise accompanying the batch effect. 
To overcome those challenges, we proposed a gene panel selection method with knowledge-ensembled multi-agent reinforcement learning.
Specifically, we reformulated the gene panel selection problem through the pre-filtering strategy, knowledge injection, and the iterative reinforced optimization pipeline, guided by an expert knowledge-based reward function. 
We conducted comprehensive quantitative and qualitative evaluations of \model, demonstrating its robustness under many challenging scenarios and superior performance in various scRNA-seq datasets of different species and tissues. 
The most significant discovery from the research shows that \model, by utilizing numerous cooperating gene agents, independently formulates a more efficient gene selection strategy compared to conventional heuristic-based approaches. 

\section{Acknowledgement} 
This study is supported by grants from the Strategic Priority Research Program of the Chinese Academy of Sciences XDA0460101, the National Natural Science Foundation of China (No.92470204), the Beijing Natural Science Foundation (No.4254089), and the SciHorizon Platform via the Fundamental Research Project of CNIC (No.E455230401).

\bibliographystyle{IEEEtran}
\bibliography{ref}

\begin{thebibliography}{10}
\providecommand{\url}[1]{#1}
\csname url@samestyle\endcsname
\providecommand{\newblock}{\relax}
\providecommand{\bibinfo}[2]{#2}
\providecommand{\BIBentrySTDinterwordspacing}{\spaceskip=0pt\relax}
\providecommand{\BIBentryALTinterwordstretchfactor}{4}
\providecommand{\BIBentryALTinterwordspacing}{\spaceskip=\fontdimen2\font plus
\BIBentryALTinterwordstretchfactor\fontdimen3\font minus \fontdimen4\font\relax}
\providecommand{\BIBforeignlanguage}[2]{{%
\expandafter\ifx\csname l@#1\endcsname\relax
\typeout{** WARNING: IEEEtran.bst: No hyphenation pattern has been}%
\typeout{** loaded for the language `#1'. Using the pattern for}%
\typeout{** the default language instead.}%
\else
\language=\csname l@#1\endcsname
\fi
#2}}
\providecommand{\BIBdecl}{\relax}
\BIBdecl

\bibitem{schwartzman2015single}
O.~Schwartzman and A.~Tanay, ``Single-cell epigenomics: techniques and emerging applications,'' \emph{Nature Reviews Genetics}, vol.~16, no.~12, pp. 716--726, 2015.

\bibitem{gawad2016single}
C.~Gawad, W.~Koh, and S.~R. Quake, ``Single-cell genome sequencing: current state of the science,'' \emph{Nature Reviews Genetics}, vol.~17, no.~3, pp. 175--188, 2016.

\bibitem{woodworth2017building}
M.~B. Woodworth, K.~M. Girskis, and C.~A. Walsh, ``Building a lineage from single cells: genetic techniques for cell lineage tracking,'' \emph{Nature Reviews Genetics}, vol.~18, no.~4, pp. 230--244, 2017.

\bibitem{liugut}
Y.~Liu, L.~Yang, M.~Meskini, A.~Goel, M.~Opperman, S.~S. Shyamal, A.~Manaithiya, M.~Xiao, R.~Ni, Y.~An \emph{et~al.}, ``Gut microbiota and tuberculosis,'' \emph{iMeta}, p. e70054.

\bibitem{lee2020single}
J.~Lee, D.~Y. Hyeon, and D.~Hwang, ``Single-cell multiomics: technologies and data analysis methods,'' \emph{Experimental \& Molecular Medicine}, vol.~52, no.~9, pp. 1428--1442, 2020.

\bibitem{baysoy2023technological}
A.~Baysoy, Z.~Bai, R.~Satija, and R.~Fan, ``The technological landscape and applications of single-cell multi-omics,'' \emph{Nature Reviews Molecular Cell Biology}, vol.~24, no.~10, pp. 695--713, 2023.

\bibitem{huynh2024topological}
T.~Huynh and Z.~Cang, ``Topological and geometric analysis of cell states in single-cell transcriptomic data,'' \emph{Briefings in Bioinformatics}, vol.~25, no.~3, p. bbae176, 2024.

\bibitem{rao2021exploring_tissue_architecture}
A.~Rao, D.~Barkley, G.~S. Fran{\c{c}}a, and I.~Yanai, ``Exploring tissue architecture using spatial transcriptomics,'' \emph{Nature}, vol. 596, no. 7871, pp. 211--220, 2021.

\bibitem{longo2021intercellulartissuedynamics}
S.~K. Longo, M.~G. Guo, A.~L. Ji, and P.~A. Khavari, ``Integrating single-cell and spatial transcriptomics to elucidate intercellular tissue dynamics,'' \emph{Nature Reviews Genetics}, vol.~22, no.~10, pp. 627--644, 2021.

\bibitem{sun2022identifyingphenotype-associatedsubpopulations}
D.~Sun, X.~Guan, A.~E. Moran, L.-Y. Wu, D.~Z. Qian, P.~Schedin, M.-S. Dai, A.~V. Danilov, J.~J. Alumkal, A.~C. Adey \emph{et~al.}, ``Identifying phenotype-associated subpopulations by integrating bulk and single-cell sequencing data,'' \emph{Nature biotechnology}, vol.~40, no.~4, pp. 527--538, 2022.

\bibitem{cui2023scgpt}
H.~Cui, C.~Wang, H.~Maan, K.~Pang, F.~Luo, and B.~Wang, ``scgpt: Towards building a foundation model for single-cell multi-omics using generative ai,'' \emph{bioRxiv}, pp. 2023--04, 2023.

\bibitem{theodoris2023geneformer}
C.~V. Theodoris, L.~Xiao, A.~Chopra, M.~D. Chaffin, Z.~R. Al~Sayed, M.~C. Hill, H.~Mantineo, E.~M. Brydon, Z.~Zeng, X.~S. Liu \emph{et~al.}, ``Transfer learning enables predictions in network biology,'' \emph{Nature}, pp. 1--9, 2023.

\bibitem{yang2023genecompass}
X.~Yang, G.~Liu, G.~Feng, D.~Bu, P.~Wang, J.~Jiang, S.~Chen, Q.~Yang, H.~Miao, Y.~Zhang \emph{et~al.}, ``Genecompass: deciphering universal gene regulatory mechanisms with a knowledge-informed cross-species foundation model,'' \emph{Cell Research}, pp. 1--16, 2024.

\bibitem{qin2025scihorizon}
C.~Qin, X.~Chen, C.~Wang, P.~Wu, X.~Chen, Y.~Cheng, J.~Zhao, M.~Xiao, X.~Dong, Q.~Long \emph{et~al.}, ``Scihorizon: Benchmarking ai-for-science readiness from scientific data to large language models,'' in \emph{Proceedings of the 31st ACM SIGKDD Conference on Knowledge Discovery and Data Mining V. 2}, 2025, pp. 5754--5765.

\bibitem{kiselev2019challenges(dimension_curse)}
V.~Y. Kiselev, T.~S. Andrews, and M.~Hemberg, ``Challenges in unsupervised clustering of single-cell rna-seq data,'' \emph{Nature Reviews Genetics}, vol.~20, no.~5, pp. 273--282, 2019.

\bibitem{li2021sparse}
Z.~Li, F.~Nie, J.~Bian, D.~Wu, and X.~Li, ``Sparse pca via l2,p-norm regularization for unsupervised feature selection,'' \emph{IEEE Transactions on Pattern Analysis and Machine Intelligence}, vol.~45, no.~4, pp. 5322--5328, 2021.

\bibitem{mackiewicz1993principal}
A.~Ma{\'c}kiewicz and W.~Ratajczak, ``Principal components analysis (pca),'' \emph{Computers \& Geosciences}, vol.~19, no.~3, pp. 303--342, 1993.

\bibitem{kobak2019art}
D.~Kobak and P.~Berens, ``The art of using t-sne for single-cell transcriptomics,'' \emph{Nature communications}, vol.~10, no.~1, p. 5416, 2019.

\bibitem{becht2019dimensionality}
E.~Becht, L.~McInnes, J.~Healy, C.-A. Dutertre, I.~W. Kwok, L.~G. Ng, F.~Ginhoux, and E.~W. Newell, ``Dimensionality reduction for visualizing single-cell data using umap,'' \emph{Nature biotechnology}, vol.~37, no.~1, pp. 38--44, 2019.

\bibitem{dalman2012fold}
M.~R. Dalman, A.~Deeter, G.~Nimishakavi, and Z.-H. Duan, ``Fold change and p-value cutoffs significantly alter microarray interpretations,'' in \emph{BMC bioinformatics}, vol.~13.\hskip 1em plus 0.5em minus 0.4em\relax Springer, 2012, pp. 1--4.

\bibitem{yip2019evaluation}
S.~H. Yip, P.~C. Sham, and J.~Wang, ``Evaluation of tools for highly variable gene discovery from single-cell rna-seq data,'' \emph{Briefings in bioinformatics}, vol.~20, no.~4, pp. 1583--1589, 2019.

\bibitem{Seuratv3}
T.~Stuart, A.~Butler, P.~Hoffman, C.~Hafemeister, E.~Papalexi, W.~M. Mauck, Y.~Hao, M.~Stoeckius, P.~Smibert, and R.~Satija, ``Comprehensive integration of single-cell data,'' \emph{cell}, vol. 177, no.~7, pp. 1888--1902, 2019.

\bibitem{luecken2022benchmarking}
M.~D. Luecken, M.~B{\"u}ttner, K.~Chaichoompu, A.~Danese, M.~Interlandi, M.~F. M{\"u}ller, D.~C. Strobl, L.~Zappia, M.~Dugas, M.~Colom{\'e}-Tatch{\'e} \emph{et~al.}, ``Benchmarking atlas-level data integration in single-cell genomics,'' \emph{Nature methods}, vol.~19, no.~1, pp. 41--50, 2022.

\bibitem{hetzel2022predicting}
L.~Hetzel, S.~Boehm, N.~Kilbertus, S.~G{\"u}nnemann, F.~Theis \emph{et~al.}, ``Predicting cellular responses to novel drug perturbations at a single-cell resolution,'' \emph{Advances in Neural Information Processing Systems}, vol.~35, pp. 26\,711--26\,722, 2022.

\bibitem{gong2024xtrimogene}
J.~Gong, M.~Hao, X.~Cheng, X.~Zeng, C.~Liu, J.~Ma, X.~Zhang, T.~Wang, and L.~Song, ``xtrimogene: An efficient and scalable representation learner for single-cell rna-seq data,'' \emph{Advances in Neural Information Processing Systems}, vol.~36, 2024.

\bibitem{CellBRF}
\BIBentryALTinterwordspacing
Y.~Xu, H.-D. Li, C.-X. Lin, R.~Zheng, Y.~Li, J.~Xu, and J.~Wang, ``\BIBforeignlanguage{en-US}{Cellbrf: a feature selection method for single-cell clustering using cell balance and random forest},'' \emph{\BIBforeignlanguage{en-US}{Bioinformatics}}, p. i368–i376, Jun 2023. [Online]. Available: \url{http://dx.doi.org/10.1093/bioinformatics/btad216}
\BIBentrySTDinterwordspacing

\bibitem{geneBasis}
A.~Missarova, J.~Jain, A.~Butler, S.~Ghazanfar, T.~Stuart, M.~Brusko, C.~Wasserfall, H.~Nick, T.~Brusko, M.~Atkinson \emph{et~al.}, ``genebasis: an iterative approach for unsupervised selection of targeted gene panels from scrna-seq,'' \emph{Genome biology}, vol.~22, pp. 1--22, 2021.

\bibitem{HRG}
\BIBentryALTinterwordspacing
Y.~Wu, Q.~Hu, S.~Wang, C.~Liu, Y.~Shan, W.~Guo, R.~Jiang, X.~Wang, and J.~Gu, ``\BIBforeignlanguage{en-US}{Highly regional genes: graph-based gene selection for single-cell rna-seq data},'' \emph{\BIBforeignlanguage{en-US}{Journal of Genetics and Genomics}}, p. 891–899, Sep 2022. [Online]. Available: \url{http://dx.doi.org/10.1016/j.jgg.2022.01.004}
\BIBentrySTDinterwordspacing

\bibitem{pont2019single}
F.~Pont, M.~Tosolini, and J.~J. Fourni{\'e}, ``Single-cell signature explorer for comprehensive visualization of single cell signatures across scrna-seq datasets,'' \emph{Nucleic acids research}, vol.~47, no.~21, pp. e133--e133, 2019.

\bibitem{volk2023alphaflow}
A.~A. Volk, R.~W. Epps, D.~T. Yonemoto, B.~S. Masters, F.~N. Castellano, K.~G. Reyes, and M.~Abolhasani, ``Alphaflow: autonomous discovery and optimization of multi-step chemistry using a self-driven fluidic lab guided by reinforcement learning,'' \emph{Nature Communications}, vol.~14, no.~1, p. 1403, 2023.

\bibitem{cai2022survey}
Q.~Cai, C.~Cui, Y.~Xiong, W.~Wang, Z.~Xie, and M.~Zhang, ``A survey on deep reinforcement learning for data processing and analytics,'' \emph{IEEE Transactions on Knowledge and Data Engineering}, vol.~35, no.~5, pp. 4446--4465, 2022.

\bibitem{saadatmand2024many}
H.~Saadatmand and M.~Akbarzadeh-T, ``Many-objective jaccard-based evolutionary feature selection for high-dimensional imbalanced data classification,'' \emph{IEEE Transactions on Pattern Analysis \& Machine Intelligence}, no.~01, pp. 1--16, 2024.

\bibitem{pre}
P.~Wang, W.~Liu, J.~Wang, Y.~Liu, P.~Li, P.~Xu, W.~Cui, R.~Zhang, Q.~Long, Z.~Hu \emph{et~al.}, ``sccompass: An integrated cross-species scrna-seq database for ai-ready,'' \emph{bioRxiv}, pp. 2024--11, 2024.

\bibitem{HVG}
R.~Satija, J.~Farrell, D.~Gennert, A.~Schier, and A.~Regev, ``\BIBforeignlanguage{en-US}{Spatial reconstruction of single-cell gene expression data},'' \emph{\BIBforeignlanguage{en-US}{PMC}}, Apr 2015.

\bibitem{vote}
H.~Zhao, T.~Zhou, G.~Long, J.~Jiang, and C.~Zhang, ``Voting from nearest tasks: Meta-vote pruning of pre-trained models for downstream tasks,'' in \emph{Joint European Conference on Machine Learning and Knowledge Discovery in Databases}.\hskip 1em plus 0.5em minus 0.4em\relax Springer, 2023, pp. 52--68.

\bibitem{schaul2015prioritized}
T.~Schaul, J.~Quan, I.~Antonoglou, and D.~Silver, ``Prioritized experience replay,'' \emph{arXiv preprint arXiv:1511.05952}, 2015.

\bibitem{konda1999actor}
V.~Konda and J.~Tsitsiklis, ``Actor-critic algorithms,'' \emph{Advances in neural information processing systems}, vol.~12, 1999.

\bibitem{sewak2019actor}
M.~Sewak and M.~Sewak, ``Actor-critic models and the a3c: The asynchronous advantage actor-critic model,'' \emph{Deep reinforcement learning: frontiers of artificial intelligence}, pp. 141--152, 2019.

\bibitem{edgar2002gene}
R.~Edgar, M.~Domrachev, and A.~E. Lash, ``Gene expression omnibus: Ncbi gene expression and hybridization array data repository,'' \emph{Nucleic acids research}, vol.~30, no.~1, pp. 207--210, 2002.

\bibitem{brazma2003arrayexpress}
A.~Brazma, H.~Parkinson, U.~Sarkans, M.~Shojatalab, J.~Vilo, N.~Abeygunawardena, E.~Holloway, M.~Kapushesky, P.~Kemmeren, G.~G. Lara \emph{et~al.}, ``Arrayexpress—a public repository for microarray gene expression data at the ebi,'' \emph{Nucleic acids research}, vol.~31, no.~1, pp. 68--71, 2003.

\bibitem{leinonen2010sequence}
R.~Leinonen, H.~Sugawara, M.~Shumway, and I.~N. S.~D. Collaboration, ``The sequence read archive,'' \emph{Nucleic acids research}, vol.~39, no. suppl\_1, pp. D19--D21, 2010.

\bibitem{Cao}
\BIBentryALTinterwordspacing
J.~Cao, J.~S. Packer, V.~Ramani, D.~A. Cusanovich, C.~Huynh, R.~Daza, X.~Qiu, C.~Lee, S.~N. Furlan, F.~J. Steemers, A.~Adey, R.~H. Waterston, C.~Trapnell, and J.~Shendure, ``\BIBforeignlanguage{en-US}{Comprehensive single-cell transcriptional profiling of a multicellular organism},'' \emph{\BIBforeignlanguage{en-US}{Science}}, p. 661–667, Aug 2017. [Online]. Available: \url{http://dx.doi.org/10.1126/science.aam8940}
\BIBentrySTDinterwordspacing

\bibitem{Han}
\BIBentryALTinterwordspacing
X.~Han, R.~Wang, Y.~Zhou, L.~Fei, H.~Sun, S.~Lai, A.~Saadatpour, Z.~Zhou, H.~Chen, F.~Ye, D.~Huang, Y.~Xu, W.~Huang, M.~Jiang, X.~Jiang, J.~Mao, Y.~Chen, C.~Lu, J.~Xie, Q.~Fang, Y.~Wang, R.~Yue, T.~Li, H.~Huang, S.~H. Orkin, G.-C. Yuan, M.~Chen, and G.~Guo, ``\BIBforeignlanguage{en-US}{Mapping the mouse cell atlas by microwell-seq},'' \emph{\BIBforeignlanguage{en-US}{Cell}}, pp. 1091--1107.e17, Feb 2018. [Online]. Available: \url{http://dx.doi.org/10.1016/j.cell.2018.02.001}
\BIBentrySTDinterwordspacing

\bibitem{batch_effect}
M.~Lotfollahi, F.~A. Wolf, and F.~J. Theis, ``scgen predicts single-cell perturbation responses,'' \emph{Nature methods}, vol.~16, no.~8, pp. 715--721, 2019.

\bibitem{Seurat}
R.~Satija, J.~Farrell, D.~Gennert, A.~Schier, and A.~Regev, ``\BIBforeignlanguage{en-US}{Spatial reconstruction of single-cell gene expression data},'' \emph{\BIBforeignlanguage{en-US}{PMC}}, Apr 2015.

\bibitem{NMI}
\BIBentryALTinterwordspacing
A.~Strehl and J.~Ghosh, ``\BIBforeignlanguage{en-US}{10.1162/153244303321897735},'' \emph{\BIBforeignlanguage{en-US}{CrossRef Listing of Deleted DOIs}}, Jan 2000. [Online]. Available: \url{http://dx.doi.org/10.1162/153244303321897735}
\BIBentrySTDinterwordspacing

\bibitem{ARI}
\BIBentryALTinterwordspacing
W.~M. Rand, ``\BIBforeignlanguage{en-US}{Objective criteria for the evaluation of clustering methods},'' \emph{\BIBforeignlanguage{en-US}{Journal of the American Statistical Association}}, p. 846–850, Dec 1971. [Online]. Available: \url{http://dx.doi.org/10.1080/01621459.1971.10482356}
\BIBentrySTDinterwordspacing

\bibitem{SI}
\BIBentryALTinterwordspacing
P.~J. Rousseeuw, ``\BIBforeignlanguage{en-US}{Silhouettes: A graphical aid to the interpretation and validation of cluster analysis},'' \emph{\BIBforeignlanguage{en-US}{Journal of Computational and Applied Mathematics}}, p. 53–65, Nov 1987. [Online]. Available: \url{http://dx.doi.org/10.1016/0377-0427(87)90125-7}
\BIBentrySTDinterwordspacing

\bibitem{CellRanger}
G.~X. Zheng, J.~M. Terry, P.~Belgrader, P.~Ryvkin, Z.~W. Bent, R.~Wilson, S.~B. Ziraldo, T.~D. Wheeler, G.~P. McDermott, J.~Zhu \emph{et~al.}, ``Massively parallel digital transcriptional profiling of single cells,'' \emph{Nature communications}, vol.~8, no.~1, p. 14049, 2017.

\bibitem{PearsonResidualS}
J.~Lause, P.~Berens, and D.~Kobak, ``Analytic pearson residuals for normalization of single-cell rna-seq umi data,'' \emph{Genome biology}, vol.~22, pp. 1--20, 2021.

\bibitem{gpsFISH}
Y.~Zhang, V.~Petukhov, E.~Biederstedt, R.~Que, K.~Zhang, and P.~V. Kharchenko, ``Gene panel selection for targeted spatial transcriptomics,'' \emph{Genome Biology}, vol.~25, no.~1, p.~35, 2024.

\bibitem{scGIST}
M.~A. Yafi, M.~H.~H. Hisham, F.~Grisanti, J.~F. Martin, A.~Rahman, and M.~A.~H. Samee, ``scgist: gene panel design for spatial transcriptomics with prioritized gene sets,'' \emph{Genome Biology}, vol.~25, no.~1, p.~57, 2024.

\bibitem{rf}
L.~Breiman, ``Random forests,'' \emph{Machine learning}, vol.~45, pp. 5--32, 2001.

\bibitem{svm}
D.~A. Pisner and D.~M. Schnyer, ``Support vector machine,'' \emph{Machine learning}, pp. 101--121, 2020.

\bibitem{rfe}
P.~M. Granitto, C.~Furlanello, F.~Biasioli, and F.~Gasperi, ``Recursive feature elimination with random forest for ptr-ms analysis of agroindustrial products,'' \emph{Chemometrics and intelligent laboratory systems}, vol.~83, no.~2, pp. 83--90, 2006.

\bibitem{KBest}
Y.~Yang and J.~O. Pedersen, ``A comparative study on feature selection in text categorization,'' in \emph{Icml}, vol.~97, no. 412-420.\hskip 1em plus 0.5em minus 0.4em\relax Nashville, TN, USA, 1997, p.~35.

\bibitem{M3Drop}
\BIBentryALTinterwordspacing
T.~S. Andrews and M.~Hemberg, ``\BIBforeignlanguage{en-US}{M3drop: dropout-based feature selection for scrnaseq},'' \emph{\BIBforeignlanguage{en-US}{Bioinformatics}}, p. 2865–2867, Aug 2019. [Online]. Available: \url{http://dx.doi.org/10.1093/bioinformatics/bty1044}
\BIBentrySTDinterwordspacing

\bibitem{FEATS}
\BIBentryALTinterwordspacing
E.~Vans, A.~Patil, and A.~Sharma, ``\BIBforeignlanguage{en-US}{Feats: feature selection-based clustering of single-cell rna-seq data},'' \emph{\BIBforeignlanguage{en-US}{Briefings in Bioinformatics}}, Oct 2020. [Online]. Available: \url{http://dx.doi.org/10.1093/bib/bbaa306}
\BIBentrySTDinterwordspacing

\bibitem{FEAST}
\BIBentryALTinterwordspacing
K.~Su, T.~Yu, and H.~Wu, ``\BIBforeignlanguage{en-US}{Accurate feature selection improves single-cell rna-seq cell clustering},'' \emph{\BIBforeignlanguage{en-US}{Briefings in Bioinformatics}}, Sep 2021. [Online]. Available: \url{http://dx.doi.org/10.1093/bib/bbab034}
\BIBentrySTDinterwordspacing

\bibitem{NS-forest}
\BIBentryALTinterwordspacing
B.~Aevermann, Y.~Zhang, M.~Novotny, M.~Keshk, T.~Bakken, J.~Miller, R.~Hodge, B.~Lelieveldt, E.~Lein, and R.~H. Scheuermann, ``\BIBforeignlanguage{en-US}{A machine learning method for the discovery of minimum marker gene combinations for cell type identification from single-cell rna sequencing},'' \emph{\BIBforeignlanguage{en-US}{Genome Research}}, p. 1767–1780, Oct 2021. [Online]. Available: \url{http://dx.doi.org/10.1101/gr.275569.121}
\BIBentrySTDinterwordspacing

\bibitem{sutton2018reinforcement}
R.~S. Sutton, ``Reinforcement learning: An introduction,'' \emph{A Bradford Book}, 2018.

\bibitem{gu2024review}
S.~Gu, L.~Yang, Y.~Du, G.~Chen, F.~Walter, J.~Wang, and A.~Knoll, ``A review of safe reinforcement learning: Methods, theories and applications,'' \emph{IEEE Transactions on Pattern Analysis and Machine Intelligence}, 2024.

\bibitem{li2021structured}
W.~Li, X.~Wang, B.~Jin, D.~Luo, and H.~Zha, ``Structured cooperative reinforcement learning with time-varying composite action space,'' \emph{IEEE Transactions on Pattern Analysis and Machine Intelligence}, vol.~44, no.~11, pp. 8618--8634, 2021.

\bibitem{he2023fear}
X.~He, J.~Wu, Z.~Huang, Z.~Hu, J.~Wang, A.~Sangiovanni-Vincentelli, and C.~Lv, ``Fear-neuro-inspired reinforcement learning for safe autonomous driving,'' \emph{IEEE transactions on pattern analysis and machine intelligence}, 2023.

\bibitem{zhang2022efficient}
T.~Zhang, A.~Hellander, and S.~Toor, ``Efficient hierarchical storage management empowered by reinforcement learning,'' \emph{IEEE Transactions on Knowledge and Data Engineering}, vol.~35, no.~6, pp. 5780--5793, 2022.

\bibitem{guo2022reinforcement}
L.~Guo, J.~Zhang, T.~Chen, X.~Wang, and H.~Yin, ``Reinforcement learning-enhanced shared-account cross-domain sequential recommendation,'' \emph{IEEE Transactions on Knowledge and Data Engineering}, vol.~35, no.~7, pp. 7397--7411, 2022.

\bibitem{du2022learning}
Z.~Du, N.~Yang, Z.~Yu, and S.~Y. Philip, ``Learning from atypical behavior: temporary interest aware recommendation based on reinforcement learning,'' \emph{IEEE Transactions on Knowledge and Data Engineering}, vol.~35, no.~10, pp. 9824--9835, 2022.

\bibitem{zhang2024large}
D.~Zhang, L.~Chen, S.~Zhang, H.~Xu, Z.~Zhao, and K.~Yu, ``Large language models are semi-parametric reinforcement learning agents,'' \emph{Advances in Neural Information Processing Systems}, vol.~36, 2024.

\bibitem{wen2022multi}
M.~Wen, J.~Kuba, R.~Lin, W.~Zhang, Y.~Wen, J.~Wang, and Y.~Yang, ``Multi-agent reinforcement learning is a sequence modeling problem,'' \emph{Advances in Neural Information Processing Systems}, vol.~35, pp. 16\,509--16\,521, 2022.

\bibitem{liu2021automated}
K.~Liu, Y.~Fu, L.~Wu, X.~Li, C.~Aggarwal, and H.~Xiong, ``Automated feature selection: A reinforcement learning perspective,'' \emph{IEEE Transactions on Knowledge and Data Engineering}, vol.~35, no.~3, pp. 2272--2284, 2021.

\bibitem{hu2023mo}
T.~Hu, B.~Luo, C.~Yang, and T.~Huang, ``Mo-mix: Multi-objective multi-agent cooperative decision-making with deep reinforcement learning,'' \emph{IEEE Transactions on Pattern Analysis and Machine Intelligence}, vol.~45, no.~10, pp. 12\,098--12\,112, 2023.

\bibitem{xiao2022self}
M.~Xiao, D.~Wang, M.~Wu, K.~Liu, H.~Xiong, Y.~Zhou, and Y.~Fu, ``Traceable group-wise self-optimizing feature transformation learning: A dual optimization perspective,'' \emph{ACM Transactions on Knowledge Discovery from Data}, vol.~18, no.~4, pp. 1--22, 2024.

\bibitem{zhang2025comprehend}
W.~Zhang, X.~Huang, Y.~Du, Z.~Qiao, Q.~Long, Z.~Meng, Y.~Zhou, and M.~Xiao, ``Comprehend, divide, and conquer: Feature subspace exploration via multi-agent hierarchical reinforcement learning,'' \emph{arXiv preprint arXiv:2504.17356}, 2025.

\end{thebibliography}
\vspace{-1.5cm}
\begin{IEEEbiography}[{\includegraphics[width=1in,height=1.25in,clip,keepaspectratio]{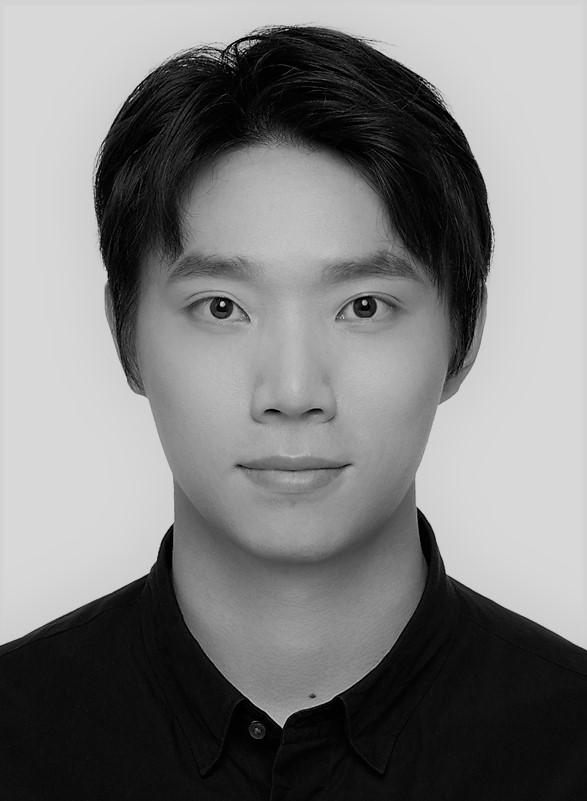}}]{Meng Xiao}
received joint doctoral training from the University of Chinese Academy of Sciences and Singapore's Agency for Science, Technology and Research (ASTAR). 
Currently, he is a research fellow at the DUKE-NUS Medical School, National University of Singapore. 
Meng Xiao has published over 30 papers, including IEEE TKDE, ACM TKDD, AIJ, NeurIPS, ICML, and ICLR. 
His research focuses on Data-Centric AI, AI for Science, and interdisciplinary knowledge modeling. 
He is a Member of IEEE.
\end{IEEEbiography}
\vspace{-1.5cm}
\begin{IEEEbiography}[{\includegraphics[width=1in,height=1.25in,clip,keepaspectratio]{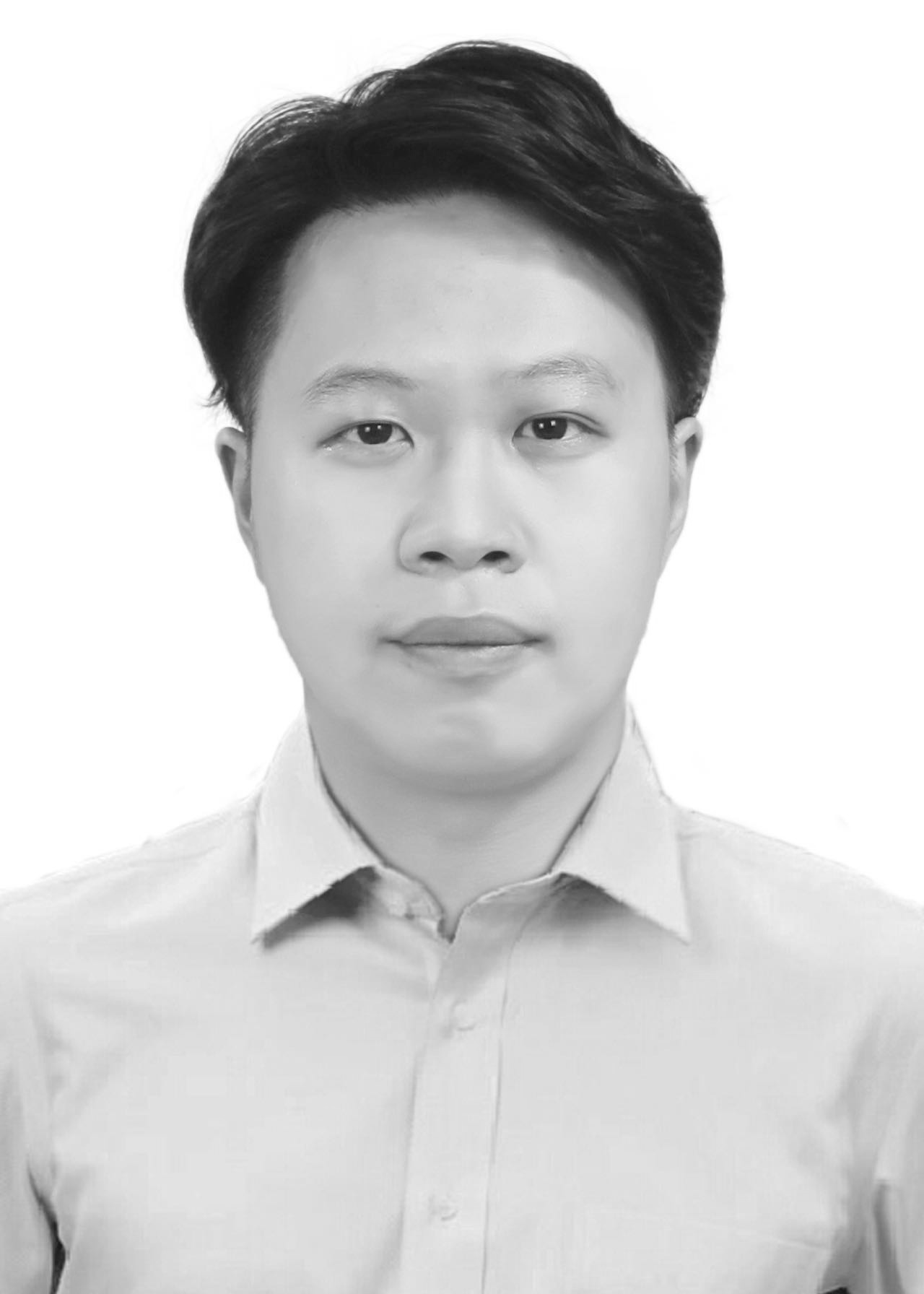}}]{Weiliang Zhang} was born in 2000. He received his B.S. degree in 2023 from Beijing University of Chemical Technology, China. He is currently pursuing a master’s degree with the Computer Network Information Center at the University of Chinese Academy of Sciences. His research interests include Data Mining and Reinforcement Learning.
\end{IEEEbiography}
\vspace{-1.5cm}
\begin{IEEEbiography}[{\includegraphics[width=1in,height=1.25in,clip,keepaspectratio]{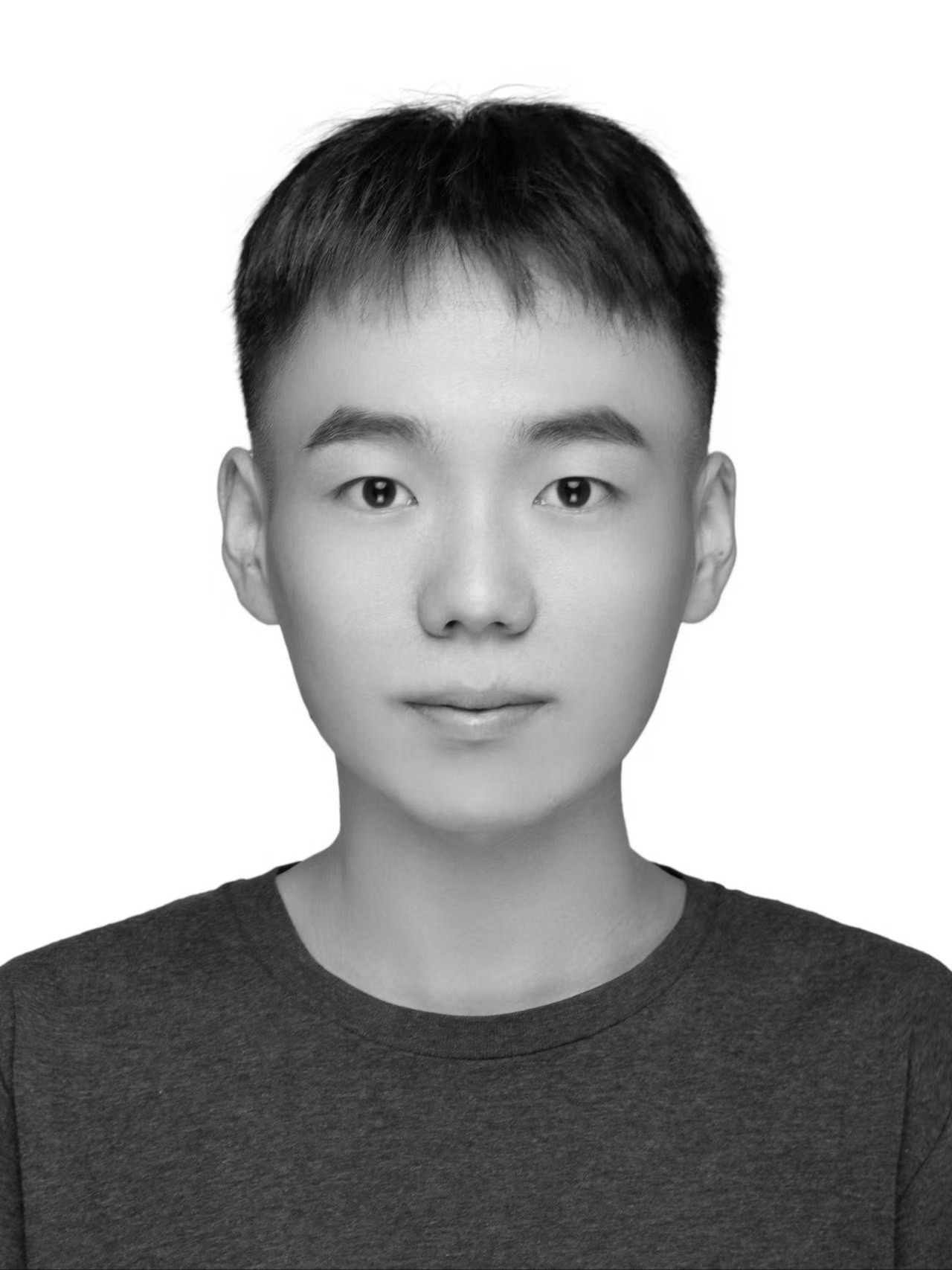}}]{Xiaohan Huang} completed his undergraduate studies in the Department of Computer Technology and Application at Qinghai University. Currently, he is pursuing a master's degree at the Computer Network Information Center, Chinese Academy of Sciences. His research focuses on Data Mining and Artificial Intelligence for Science.
\end{IEEEbiography}
\vspace{-1.5cm}
\begin{IEEEbiography}[{\includegraphics[width=1in,height=1.25in,clip,keepaspectratio]{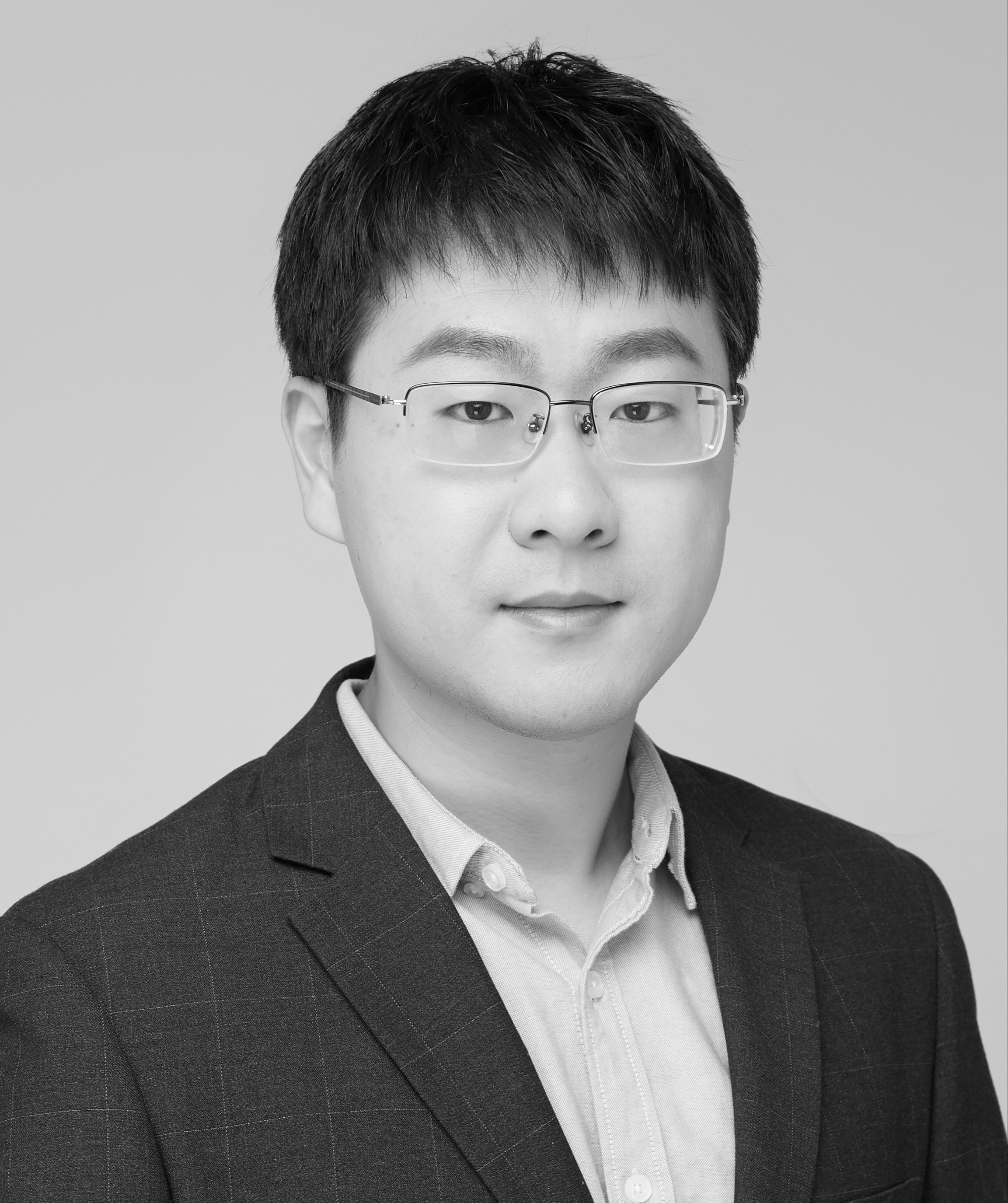}}]{Hengshu Zhu} is currently a full professor at the Computer Network Information Center (CNIC), Chinese Academy of Sciences (CAS). He received a Ph.D. degree in 2014 and a B.E. degree in 2009, both in Computer Science from the University of Science and Technology of China (USTC), China. His general area of research is data mining and machine learning, with a focus on developing advanced data analysis techniques for innovative scientific and business applications. 
He has published prolifically in refereed journals and conference proceedings, such as Nature Cities, Nature Communications, IEEE TKDE, IEEE TMC, ACM TOIS, SIGKDD, SIGIR, and NeurIPS. 
He is a Distinguished Member of CCF and a Senior Member of ACM, CAAI, and IEEE.
\end{IEEEbiography}
\vspace{-1.5cm}
\begin{IEEEbiography}[{\includegraphics[width=1in,height=1.25in,clip,keepaspectratio]{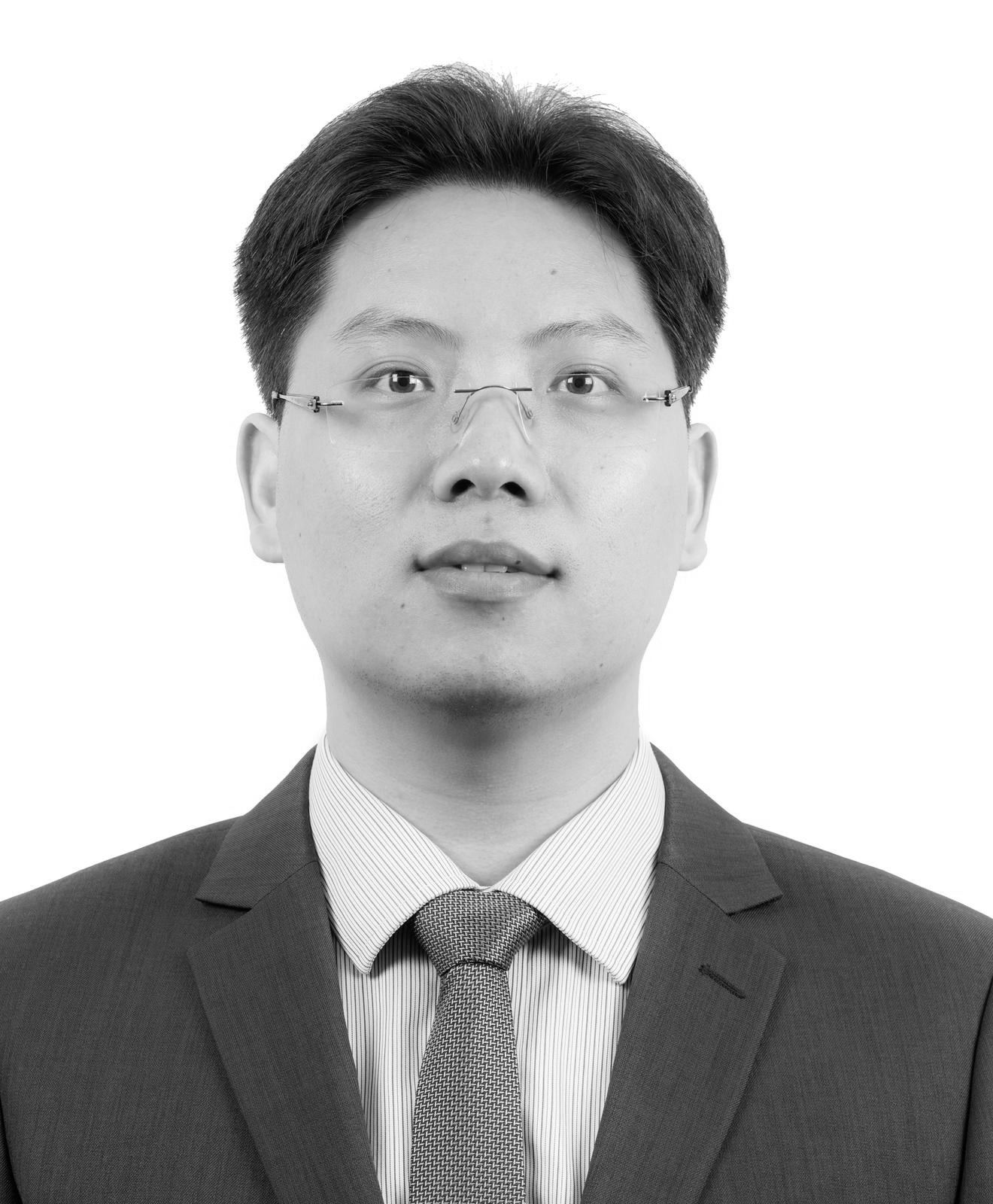}}]{Min Wu}
is currently a Principal Scientist at Institute for Infocomm Research (I2R), Agency for Science, Technology and Research (A*STAR), Singapore. He received his Ph.D. degree in Computer Science from Nanyang Technological University (NTU), Singapore, in 2011 and B.E. degree in Computer Science from University of Science and Technology of China (USTC) in 2006. He received the best paper awards in EMBS Society 2023, IEEE ICIEA 2022, IEEE SmartCity 2022, InCoB 2016 and DASFAA 2015. He also won the CVPR UG2+ challenge in 2021 and the IJCAI competition on repeated buyers prediction in 2015. He has been serving as an Associate Editor for journals like Neurocomputing, Neural Networks and IEEE Transactions on Cognitive and Developmental Systems, as well as area chairs of leading machine learning and data mining conferences, such as ICLR, NeurIPS, AAAI, KDD, etc. His current research interests focus on AI for time series data, graph data, and biological and healthcare data.
\end{IEEEbiography}
\vspace{-1.5cm}
\begin{IEEEbiography}
[{\includegraphics[width=1in,height=1.25in,clip,keepaspectratio]{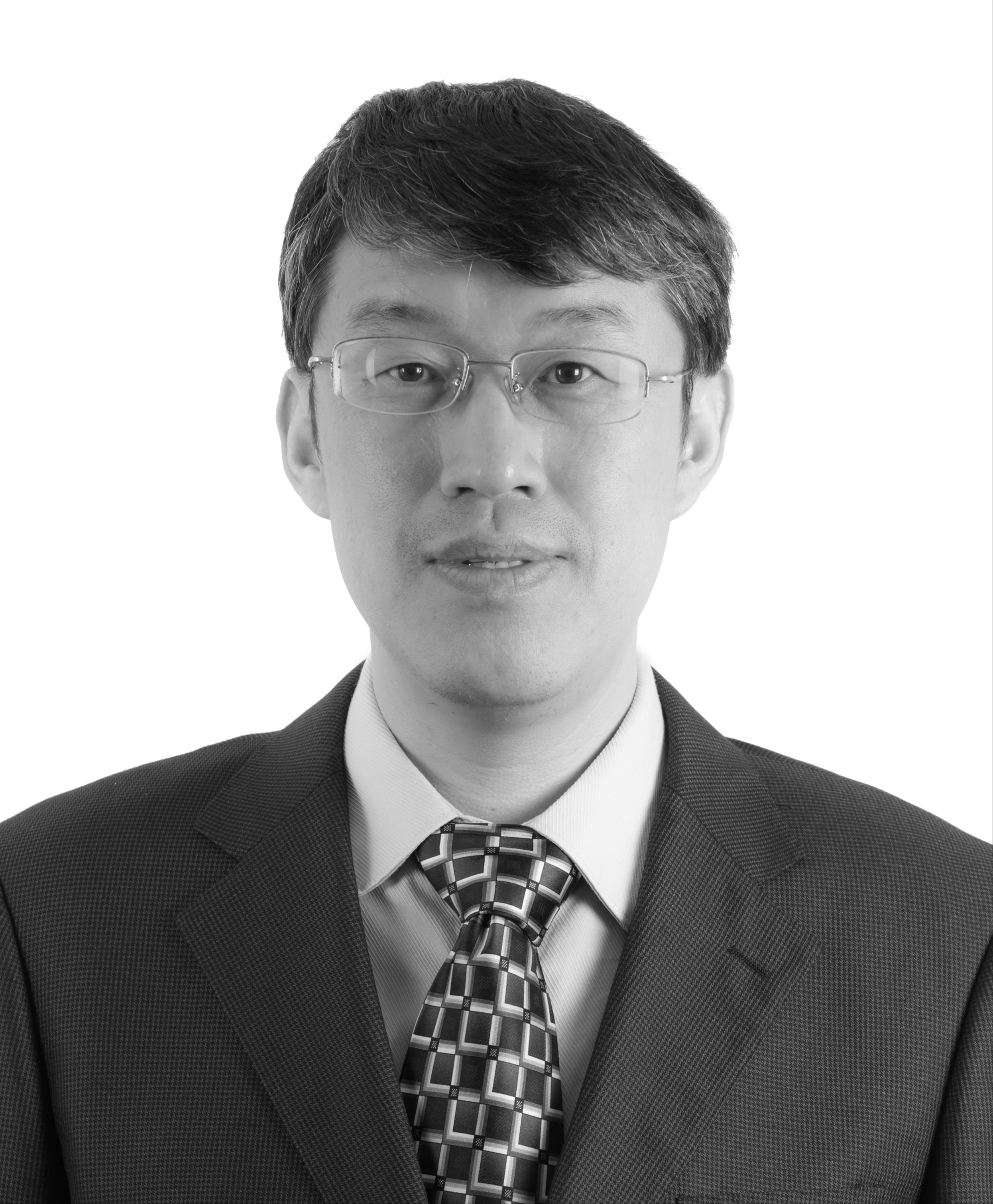}}]{Xiaoli Li} joined SUTD on 15 August 2025 as Head of the Information Systems Technology and Design (ISTD) Pillar. In this role, he provides strategic leadership in academic programme development, research direction, and industry engagement across the Pillar.
Prof Li is internationally recognised in the AI community and has held leadership roles at top-tier conferences including NeurIPS, ICLR, KDD, ICDM, WWW, IJCAI, AAAI, ACL, and EMNLP—serving as Conference Chair, Area Chair, Workshop Chair, and Session Chair.
Xiaoli has published over 300 high-quality papers and won nine best paper/benchmark competition awards. 
He is a Fellow of IEEE.
\end{IEEEbiography}
\vspace{-1.5cm}
\begin{IEEEbiography}[{\includegraphics[width=1in,height=1.25in,clip,keepaspectratio]{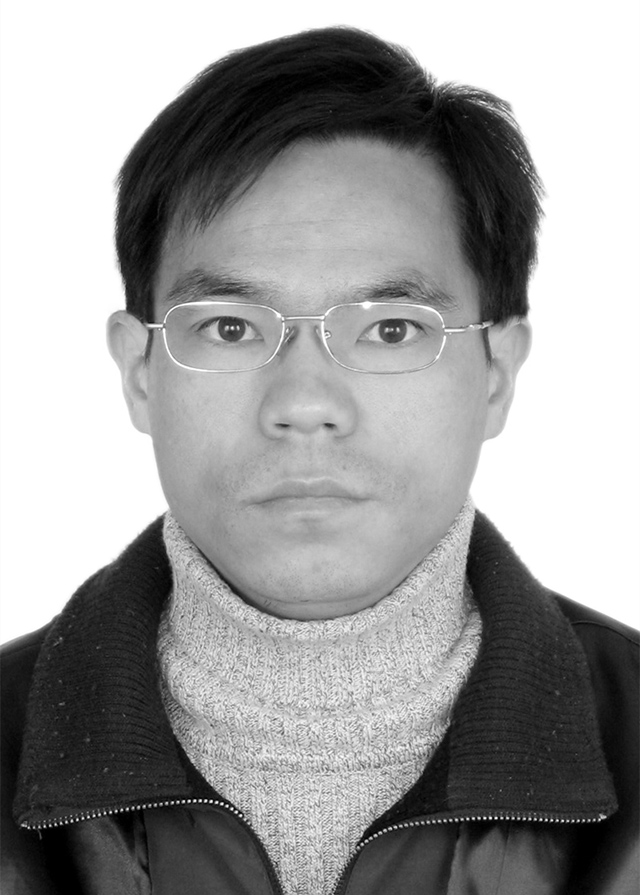}}]{Yuanchun Zhou}
was born in 1975. He received his Ph.D. degree from the Institute of Computing Technology, Chinese Academy of Sciences, in 2006. 
He is a professor, Ph.D. supervisor, the deputy director of  Computer Network Information Center, Chinese Academy of Sciences. 
His research interests include data mining, big data processing, and knowledge graph. 
He has published prolifically in refereed journals and conferences, such as Cell Research, IEEE TKDE, Nucleic Acids Research, NeurIPS, IJCAI, and AAAI. 
\end{IEEEbiography}
\clearpage

\begin{algorithm*}[!h]
\caption{Label-free Biomarker Identification Pipeline}
\label{alg:master}
\textbf{Input}:  
Pre-filtered gene set $\tilde{G}$;  
Single-cell expression matrix $X[\tilde{G}]$;  
Prior knowledge gene subsets $\mathcal{K}=\{G^{f}\}_{f=1}^{m}$;  
Hyper-parameters $\alpha,\lambda,\gamma,N_{\text{explore}},N_{\text{optimize}}$.

\textbf{Initialization}:  
Actors $\{\pi^{i}\}$, Critics $\{V^{\pi^{i}}\}$, Replay buffers $\{\mathcal{B}^{i}\}$, State encoders $\text{Enc}_{\phi}(\cdot)$, Histories of candidate gene subsets $\mathcal{H}\gets\emptyset$.

\textbf{Output}:  
Optimal gene panel $G^{*}$.

\begin{algorithmic}[1]

\STATE {\color{gray}/* ---------- Knowledge Injection ---------- */}
\FOR{each prior subset $G^{f} \in \mathcal{K}$}
    \STATE Compute reward $r^{f}$ via Eq. (4) with $G^{f}$;
    \STATE Encode states $S^{0} \gets \text{Enc}_{\phi}(\tilde{G})$, $S^{f} \gets \text{Enc}_{\phi}(G^{f})$;
    \FOR{gene $i \in \tilde{G}$}
        \STATE $a^{i} \gets \text{Select}$ if $i \in G^{f}$ else $\text{Discard}$;
        \STATE Store transition $(S^{0}, a_{i}^f, r^{f}, S^{f})$ into $\mathcal{B}^{i}$;
    \ENDFOR
\ENDFOR

\STATE {\color{gray}/* ---------- Exploration ---------- */}
\SetKwFunction{FExplore}{Pipeline\_Exploration}
\STATE $\{\mathcal{B}^{i}\},\mathcal{H} \gets$  \FExplore{$\tilde{G}$, $X[\tilde{G}]$, $\alpha$, $\lambda$, $\gamma$, $N_{\text{explore}}$}

\STATE {\color{gray}/* ---------- Exploitation---------- */}
\SetKwFunction{FExploit}{Pipeline\_Exploitation}
\STATE $\mathcal{H} \gets$  \FExploit{$\tilde{G}$, $X[\tilde{G}]$, $\{\mathcal{B}^{i}\}$, $\mathcal{H}$, $\gamma$, $N_{\text{optimize}}$}

\STATE {\color{gray}/* ---------- Optimal Gene Panel Selection ---------- */}
\STATE Initialize $G^{*} \gets \emptyset$, $P^{*} \gets -\infty$;
\FOR{each $G' \in \mathcal{H}$}
    \STATE Evaluate clustering metric $\mathcal{E}\bigl(\mathcal{C}(X[G'])\bigr)$;
    \IF{$\mathcal{E} > P^{*}$}
        \STATE $P^{*} \gets \mathcal{E}$; \quad $G^{*} \gets G'$;
    \ENDIF
\ENDFOR

\end{algorithmic}
\end{algorithm*}
\clearpage

\begin{algorithm*}[!h]
\caption{Pipeline Exploration}
\label{alg:exploration_phase}

\textbf{Input}:  
Pre-filtered gene set $\tilde{G}$;  
Single-cell expression matrix $X[\tilde{G}]$;  
Hyper-parameters $\alpha,\lambda,\gamma,N_{\text{explore}}$.

\textbf{Output}:  
Replay buffers $\{\mathcal{B}^{i}\}$;  
History $\mathcal{H}$.
\clearpage
\begin{algorithmic}[1]
\STATE {\color{gray}/* --- Exploration --- */}
\FOR{step $t=1$ to $N_{\text{explore}}$}
    \STATE {\color{gray}  /* --- Select Gene Subset --- */}
    \STATE Encode state $S_{t} \gets \text{Enc}_{\phi}(G_{t})$;
    \FOR{gene agent $i$}
        \STATE Sample action $a_{t}^{i} \sim \pi^{i}(\cdot|S_{t})$;
    \ENDFOR
    \STATE Form action set $\mathcal{A}_{t} \gets \{a_{t}^{i}\}_{i=1}^{|\tilde{G}|}$;
    \STATE Update subset $G_{t+1} \gets \{i \in \tilde{G} \mid a_{t}^{i} = \text{Select}\}$;
    \STATE Store subset in history $\mathcal{H} \gets \mathcal{H} \cup \{G_{t+1}\}$;
    \STATE {\color{gray} /* --- Assign Pseudo-Label --- */}
    \STATE Assign pseudo-labels $\hat{y}$ via Louvain on $X[G_{t+1}]$;
    \STATE {\color{gray} /* --- Estimate Reward --- */}
    \STATE Compute reward $r_{t}$ in  Eq. (4);
    \STATE Encode next state $S_{t+1} \gets \text{Enc}_{\phi}(G_{t+1})$;
    \STATE {\color{gray} /* --- Store Experience --- */}
    \FOR{gene agent $i$}
        \STATE Store transition $(S_{t}, a_{t}^{i}, r_{t}, S_{t+1})$ into $\mathcal{B}^{i}$;
    \ENDFOR
\ENDFOR
\end{algorithmic}
\end{algorithm*}

\clearpage
\begin{algorithm*}[!h]
\caption{Pipeline Exploitation}
\label{alg:exploitation_phase}

\textbf{Input}:  
Pre-filtered gene set $\tilde{G}$;  
Single-cell expression matrix $X[\tilde{G}]$;  
Replay buffers $\{\mathcal{B}^{i}\}$;  
History $\mathcal{H}$.
Hyper-parameters $\gamma,N_{\text{optimize}}$;

\textbf{Output}:  
History $\mathcal{H}$.


\begin{algorithmic}[1]
\STATE {\color{gray}/* --- Exploitation --- */}
\FOR{step $t=1$ to $N_{\text{optimize}}$}
    \STATE {\color{gray} /* --- Collect Experience --- */}
    \STATE Collect experience $(S_{t}, a_{t}^{i}, r_{t}, S_{t+1})$ \emph{exactly as in Pipeline Exploration};
    \STATE {\color{gray} /* --- Optimize Policy --- */}
    \FOR{gene agent $i$}
        \STATE Sample mini-batch $\mathcal{B}_{m}^{i} \sim \mathcal{B}^{i}$;
        \STATE Update Critic via 
               $L(V^\pi)$ in Eq. (12);
        \STATE Update Actor via 
               $\nabla_{\theta} J(\pi)$ in Eq. (13);
    \ENDFOR
\ENDFOR
\end{algorithmic}
\end{algorithm*}
\clearpage

\begin{table*}[ht]
Accuracy is calculated as follows:
\[
\text{Accuracy} = \frac{\text{TP} + \text{TN}}{\text{TP} + \text{TN} + \text{FP} + \text{FN}}
\]
where:
\begin{itemize}
  \item \textbf{TP}: number of correctly predicted positive cases.
  \item \textbf{TN}: number of correctly predicted negative cases.
  \item \textbf{FP}: number of negative cases incorrectly predicted as positive.
  \item \textbf{FN}: number of positive cases incorrectly predicted as negative.
\end{itemize}
\centering
\caption{Accuracy comparison in cell type annotation task on 24 datasets.} 
\clearpage
\begin{tabular}{lccccccccccc}
\toprule
 Dataaset & Random & CellRanger & PR & Seurat v3 & HRG & GeneBasis & CellBRF & gpsFISH & scGIST & RIGPS \\
\midrule
Chu1 & 33.23 & 97.08 & 87.29 & 97.29 & 26.46 & 84.27 & 54.69 & 44.58 & 45.62 & 96.77 \\
Chung & 88.62 & 90.4 & 90.23 & 91.18 & 88.84 & 90.79 & 86.1 & 63.9 & 85.1 & 77.06 \\
Darmanis & 28.74 & 33.26 & 31.39 & 24.58 & 24.58 & 24.58 & 24.58 & 24.58 & 29.26 & 34.69 \\
Engel & 52.0 & 92.88 & 79.86 & 95.66 & 92.1 & 92.88 & 99.22 & 92.1 & 85.85 & 97.22 \\
Goolam & 32.0 & 30.0 & 74.0 & 46.0 & 36.0 & 48.0 & 62.0 & 40.0 & 20.0 & 64.0 \\
Koh & 24.22 & 96.09 & 82.03 & 14.06 & 10.16 & 94.53 & 23.05 & 10.16 & 59.77 & 43.75 \\
Kumar & 91.91 & 100.0 & 92.83 & 99.48 & 98.96 & 96.48 & 99.48 & 100.0 & 87.87 & 100.0 \\
Leng & 76.07 & 65.02 & 57.99 & 96.88 & 32.52 & 80.62 & 91.03 & 94.67 & 64.62 & 90.38 \\
Li & 91.13 & 93.82 & 89.94 & 91.92 & 75.56 & 86.42 & 98.46 & 66.67 & 94.55 & 98.09 \\
Maria2 & 34.58 & 45.52 & 35.52 & 72.19 & 70.21 & 65.21 & 73.65 & 83.96 & 27.81 & 65.0 \\
Robert & 99.22 & 100.0 & 100.0 & 99.22 & 100.0 & 94.87 & 99.22 & 100.0 & 99.22 & 95.65 \\
Ting & 83.1 & 92.33 & 75.57 & 85.44 & 86.22 & 83.88 & 84.66 & 81.68 & 35.65 & 93.11 \\
Mouse Pancreas1 & 93.23 & 98.18 & 98.7 & 98.44 & 98.44 & 96.33 & 98.18 & 98.18 & 90.36 & 96.88 \\
Cao & 57.11 & 88.38 & 87.16 & 89.65 & 96.36 & 76.18 & 74.0 & 93.88 & 59.32 & 87.94 \\
Chu2 & 34.23 & 81.47 & 82.89 & 98.51 & 99.55 & 96.8 & 99.55 & 54.09 & 94.64 & 99.11 \\
Han & 66.0 & 85.64 & 84.58 & 88.3 & 89.2 & 84.08 & 89.46 & 87.17 & 69.37 & 87.08 \\
MacParland & 77.71 & 83.96 & 74.75 & 93.75 & 93.07 & 87.44 & 93.04 & 92.5 & 62.69 & 92.33 \\
Maria1 & 37.56 & 42.46 & 32.68 & 65.18 & 73.79 & 71.62 & 70.46 & 83.76 & 33.84 & 62.41 \\
Puram & 97.3 & 98.79 & 97.16 & 98.79 & 98.79 & 97.94 & 99.08 & 99.43 & 94.6 & 98.72 \\
Yang & 68.08 & 60.04 & 59.82 & 93.53 & 61.38 & 60.27 & 72.1 & 92.41 & 88.62 & 90.85 \\
Human Pancreas1 & 86.64 & 98.32 & 97.6 & 95.19 & 97.6 & 96.15 & 97.0 & 92.19 & 85.56 & 97.24 \\
Human Pancreas2 & 91.02 & 98.26 & 98.11 & 98.11 & 97.83 & 96.12 & 97.83 & 97.83 & 87.22 & 98.07 \\
Human Pancreas3 & 90.45 & 97.08 & 98.23 & 97.08 & 97.69 & 95.72 & 97.08 & 95.07 & 96.11 & 97.15 \\
Mouse Pancreas2 & 93.96 & 96.64 & 96.98 & 97.1 & 97.77 & 96.53 & 97.2 & 96.42 & 86.48 & 97.99 \\
\bottomrule
\end{tabular}
\end{table*}

\clearpage
\begin{table*}[ht]
Balanced accuracy is calculated as follows:
\[
\text{Balanced Accuracy} = \frac{1}{2}\left(\frac{\text{TP}}{\text{TP} + \text{FN}} + \frac{\text{TN}}{\text{TN} + \text{FP}}\right)
\]
where
where:
\begin{itemize}
  \item \textbf{TP}: number of correctly predicted positive cases.
  \item \textbf{TN}: number of correctly predicted negative cases.
  \item \textbf{FP}: number of negative cases incorrectly predicted as positive.
  \item \textbf{FN}: number of positive cases incorrectly predicted as negative.
\end{itemize}
\centering
\caption{Balanced accuracy comparison in cell type annotation task on 24 datasets.} 

\begin{tabular}{lccccccccccc}
\toprule
Dataset & Random & CellRanger & PR & Seurat v3 & HRG & GeneBasis & CellBRF & gpsFISH & scGIST & RIGPS \\
\midrule
Chu1 & 44.22 & 33.33 & 89.39 & 16.67 & 64.96 & 92.92 & 65.83 & 16.67 & 16.67 & 97.08 \\
Chung & 72.85 & 78.18 & 77.72 & 78.75 & 51.93 & 79.18 & 68.74 & 27.75 & 74.5 & 66.9 \\
Darmanis & 11.11 & 11.11 & 16.12 & 11.11 & 11.51 & 15.31 & 11.11 & 11.11 & 11.11 & 11.11 \\
Engel & 73.35 & 81.89 & 73.6 & 91.67 & 69.02 & 94.55 & 96.96 & 96.09 & 80.24 & 100.0 \\
Goolam & 39.32 & 45.91 & 55.76 & 12.5 & 44.77 & 47.27 & 40.91 & 32.5 & 48.07 & 46.88 \\
Koh & 92.57 & 98.03 & 79.87 & 8.33 & 13.33 & 93.93 & 8.33 & 16.98 & 79.0 & 56.96 \\
Kumar & 91.78 & 100.0 & 91.58 & 99.58 & 61.69 & 97.32 & 100.0 & 100.0 & 89.02 & 100.0 \\
Leng & 33.33 & 33.33 & 51.72 & 86.36 & 65.48 & 86.59 & 89.95 & 95.47 & 56.31 & 88.49 \\
Li & 82.65 & 82.8 & 86.64 & 90.17 & 68.16 & 48.88 & 98.59 & 91.87 & 87.74 & 95.24 \\
Maria2 & 34.81 & 42.44 & 34.37 & 59.96 & 64.77 & 70.37 & 70.01 & 88.01 & 34.34 & 72.6 \\
Robert & 99.17 & 99.17 & 100.0 & 100.0 & 99.17 & 98.53 & 98.33 & 100.0 & 99.17 & 99.17 \\
Ting & 51.65 & 85.55 & 61.98 & 76.74 & 79.99 & 57.44 & 69.79 & 52.24 & 68.33 & 86.26 \\
Mouse Pancreas1 & 76.53 & 96.28 & 94.11 & 98.75 & 98.13 & 92.76 & 95.44 & 93.49 & 70.44 & 91.74 \\
Cao & 50.69 & 89.72 & 87.85 & 89.98 & 96.36 & 75.09 & 75.38 & 93.67 & 59.08 & 85.88 \\
Chu2 & 94.07 & 85.49 & 87.13 & 39.1 & 28.57 & 99.37 & 97.38 & 42.38 & 96.17 & 99.09 \\
Han & 55.39 & 85.58 & 82.23 & 84.62 & 86.48 & 85.33 & 88.42 & 86.59 & 60.94 & 89.34 \\
MacParland & 77.78 & 86.08 & 77.31 & 93.92 & 94.96 & 87.71 & 91.64 & 94.3 & 62.02 & 92.34 \\
Maria1 & 35.99 & 41.95 & 29.45 & 67.51 & 75.71 & 73.65 & 76.57 & 85.87 & 30.23 & 59.21 \\
Puram & 94.6 & 95.79 & 95.54 & 99.21 & 98.72 & 96.26 & 99.31 & 99.38 & 88.02 & 99.05 \\
Yang & 76.75 & 32.21 & 17.14 & 85.35 & 83.69 & 27.13 & 18.02 & 50.67 & 77.72 & 34.06 \\
Human Pancreas1 & 75.25 & 96.4 & 96.32 & 96.88 & 94.45 & 88.79 & 95.07 & 83.2 & 79.25 & 93.85 \\
Human Pancreas2 & 78.15 & 96.79 & 94.39 & 93.97 & 94.25 & 89.49 & 94.47 & 94.48 & 79.94 & 93.69 \\
Human Pancreas3 & 77.37 & 87.55 & 94.37 & 92.64 & 92.5 & 87.64 & 83.43 & 82.63 & 76.63 & 77.76 \\
Mouse Pancreas2 & 83.43 & 95.04 & 94.91 & 95.34 & 96.99 & 93.13 & 95.07 & 93.77 & 85.31 & 95.46 \\
\bottomrule
\end{tabular}
\end{table*}

\clearpage
\begin{table*}[ht]
Macro-F1 is calculated as follows:
\[
\text{Macro-F1} = \frac{1}{K} \sum_{k=1}^{K} F1_{k}
\quad\text{with}\quad
F1_{k} = \frac{2 \cdot \text{Precision}_k \cdot \text{Recall}_k}{\text{Precision}_k + \text{Recall}_k}
\]
where:
\begin{itemize}
  \item $K$ = total number of classes.
  \item $\text{Precision}_k = \dfrac{\text{TP}_k}{\text{TP}_k + \text{FP}_k}$ for class $k$.
  \item $\text{Recall}_k = \dfrac{\text{TP}_k}{\text{TP}_k + \text{FN}_k}$ for class $k$.
  \item $\text{TP}_k$, $\text{FP}_k$, $\text{FN}_k$ are the true-positive, false-positive, and false-negative counts for class $k$.
\end{itemize}
\centering
\caption{Macro-F1 comparison in cell type annotation task on 24 datasets.}
\begin{tabular}{lccccccccccc}
\toprule
Dataset & Random & CellRanger & PR & Seurat v3 & HRG & GeneBasis & CellBRF & gpsFISH & scGIST & RIGPS \\
\midrule
Chu1 & 36.51 & 24.24 & 88.81 & 6.11 & 58.85 & 92.26 & 60.34 & 6.92 & 6.92 & 97.38 \\
Chung & 72.54 & 72.85 & 71.42 & 74.45 & 54.0 & 74.83 & 62.55 & 25.11 & 68.02 & 63.37 \\
Darmanis & 4.4 & 4.38 & 12.3 & 4.38 & 5.2 & 10.66 & 4.38 & 4.38 & 4.38 & 4.39 \\
Engel & 72.43 & 82.6 & 71.7 & 92.2 & 69.25 & 93.56 & 96.71 & 95.36 & 80.95 & 100.0 \\
Goolam & 33.47 & 43.39 & 54.04 & 4.17 & 39.11 & 44.36 & 24.5 & 24.15 & 45.14 & 41.63 \\
Koh & 91.95 & 97.7 & 76.69 & 2.63 & 2.85 & 92.98 & 1.99 & 9.12 & 73.08 & 51.09 \\
Kumar & 91.17 & 100.0 & 91.41 & 99.52 & 51.63 & 96.9 & 100.0 & 100.0 & 88.55 & 100.0 \\
Leng & 16.09 & 16.09 & 47.08 & 86.03 & 52.74 & 86.22 & 89.35 & 95.13 & 55.47 & 87.83 \\
Li & 80.47 & 75.73 & 85.09 & 91.29 & 61.43 & 47.61 & 94.43 & 90.64 & 87.59 & 95.19 \\
Maria2 & 33.58 & 40.09 & 29.32 & 60.22 & 63.97 & 69.72 & 67.23 & 86.94 & 30.22 & 71.22 \\
Robert & 99.21 & 99.21 & 100.0 & 100.0 & 99.21 & 98.44 & 98.42 & 100.0 & 99.21 & 99.21 \\
Ting & 45.28 & 84.48 & 56.99 & 74.6 & 72.86 & 56.17 & 67.62 & 45.68 & 65.26 & 76.96 \\
Mouse Pancreas1 & 74.86 & 94.86 & 93.04 & 97.72 & 96.73 & 89.27 & 91.05 & 88.81 & 65.42 & 86.4 \\
Cao & 48.45 & 89.74 & 87.6 & 90.38 & 96.47 & 73.35 & 72.74 & 93.18 & 57.03 & 84.6 \\
Chu2 & 94.19 & 81.27 & 84.95 & 32.4 & 19.99 & 98.94 & 97.53 & 34.68 & 95.93 & 99.26 \\
Han & 49.29 & 83.36 & 77.29 & 82.3 & 84.5 & 81.49 & 82.94 & 84.62 & 54.46 & 86.11 \\
MacParland & 74.08 & 84.27 & 73.63 & 92.25 & 93.62 & 84.75 & 89.66 & 93.07 & 56.14 & 89.83 \\
Maria1 & 34.22 & 41.73 & 28.38 & 67.15 & 75.17 & 73.36 & 73.9 & 85.76 & 30.75 & 57.42 \\
Puram & 92.99 & 94.8 & 94.62 & 97.12 & 97.44 & 94.46 & 98.98 & 98.65 & 85.35 & 97.76 \\
Yang & 74.27 & 26.72 & 8.75 & 80.9 & 76.48 & 23.03 & 10.59 & 45.45 & 78.56 & 27.85 \\
Human Pancreas1 & 71.44 & 95.17 & 95.34 & 95.95 & 92.54 & 84.52 & 92.85 & 79.42 & 76.36 & 91.19 \\
Human Pancreas2 & 76.55 & 96.28 & 92.91 & 90.99 & 90.15 & 82.33 & 91.73 & 94.47 & 77.54 & 90.8 \\
Human Pancreas3 & 73.27 & 86.32 & 93.9 & 92.43 & 89.99 & 85.03 & 81.49 & 80.46 & 83.74 & 74.58 \\
Mouse Pancreas2 & 81.13 & 93.94 & 93.54 & 94.82 & 95.89 & 91.59 & 94.43 & 90.75 & 84.87 & 94.93 \\
\bottomrule
\end{tabular}
\end{table*}

\clearpage
\begin{table*}[ht]
Micro-F1 is calculated as follows:
\[
\text{Micro-F1} = \frac{2\,\sum_i\mathrm{TP}_i}
                        {2\,\sum_i\mathrm{TP}_i + \sum_i\mathrm{FP}_i + \sum_i\mathrm{FN}_i}
\]
where
\begin{itemize}
  \item $\mathrm{TP}_i$: number of correctly predicted positive case in class 
  \item $\mathrm{FP}_i$: number of negative cases incorrectly predicted as positive in class $i$
  \item $\mathrm{FN}_i$: number of positive cases incorrectly predicted as negative in class $i$.
\end{itemize}
\centering
\caption{Micro-F1 comparison in cell type annotation task on 24 datasets.} 
\begin{tabular}{lcccccccccc}
\toprule
Dataset & Random & CellRanger & PR & Seurat v3 & HRG & GeneBasis & CellBRF & gpsFISH & scGIST & RIGPS \\
\midrule
Chu1 & 21.46 & 96.88 & 86.35 & 95.94 & 26.46 & 92.19 & 94.69 & 88.23 & 26.46 & 94.79 \\
Chung & 84.01 & 87.44 & 85.49 & 73.94 & 78.24 & 86.89 & 88.06 & 87.83 & 89.23 & 89.23 \\
Darmanis & 20.62 & 24.58 & 34.04 & 24.58 & 24.58 & 24.58 & 24.58 & 25.57 & 24.58 & 25.10 \\
Engel & 69.83 & 86.20 & 75.95 & 49.22 & 96.44 & 71.96 & 97.66 & 97.22 & 84.29 & 92.10 \\
Goolam & 47.00 & 54.00 & 80.00 & 24.00 & 46.00 & 56.00 & 44.00 & 62.00 & 60.00 & 65.00 \\
Koh & 5.16 & 38.67 & 79.69 & 12.11 & 56.64 & 90.62 & 50.00 & 55.47 & 11.72 & 93.36 \\
Kumar & 85.87 & 99.48 & 89.83 & 100.00 & 96.48 & 96.48 & 99.48 & 100.00 & 86.83 & 100.00 \\
Leng & 63.15 & 32.52 & 43.84 & 90.89 & 82.83 & 79.98 & 82.70 & 93.37 & 61.76 & 87.39 \\
Li & 52.02 & 86.10 & 89.18 & 84.15 & 97.70 & 95.70 & 98.46 & 98.48 & 92.65 & 92.70 \\
Maria2 & 30.00 & 48.65 & 34.58 & 71.88 & 72.50 & 64.79 & 59.79 & 61.25 & 27.81 & 84.58 \\
Robert & 93.44 & 99.22 & 58.71 & 99.22 & 100.00 & 93.30 & 97.66 & 99.22 & 99.22 & 100.00 \\
Ting & 83.42 & 84.66 & 54.26 & 87.00 & 83.88 & 72.44 & 51.14 & 49.57 & 63.35 & 91.55 \\
Mouse Pancreas1 & 87.19 & 98.70 & 98.70 & 98.96 & 98.70 & 97.92 & 98.44 & 97.66 & 87.50 & 97.66 \\
Cao & 50.16 & 89.88 & 85.66 & 89.94 & 97.06 & 76.24 & 75.72 & 87.65 & 57.82 & 94.51 \\
Chu2 & 87.34 & 97.47 & 82.14 & 62.65 & 64.36 & 96.80 & 99.78 & 99.78 & 78.57 & 81.25 \\
Han & 61.76 & 86.59 & 83.38 & 87.10 & 74.14 & 84.84 & 88.42 & 87.95 & 69.00 & 82.83 \\
MacParland & 72.45 & 83.93 & 75.18 & 93.22 & 93.73 & 87.06 & 92.96 & 92.53 & 64.81 & 93.33 \\
Maria1 & 31.79 & 43.45 & 32.09 & 67.14 & 74.17 & 75.35 & 72.81 & 63.79 & 34.44 & 84.93 \\
Puram & 92.44 & 98.44 & 97.51 & 98.65 & 98.22 & 97.37 & 99.29 & 98.86 & 93.96 & 99.08 \\
Yang & 86.29 & 60.04 & 34.82 & 78.79 & 34.38 & 52.90 & 82.37 & 91.74 & 87.28 & 80.36 \\
Human Pancreas1 & 81.16 & 97.96 & 97.00 & 97.60 & 97.84 & 96.51 & 96.51 & 97.84 & 87.00 & 97.12 \\
Human Pancreas2 & 86.70 & 98.07 & 97.97 & 97.83 & 94.77 & 96.69 & 97.55 & 97.79 & 88.30 & 98.11 \\
Human Pancreas3 & 77.70 & 92.79 & 91.97 & 91.77 & 92.86 & 90.68 & 92.52 & 91.20 & 92.13 & 94.70 \\
Mouse Pancreas2 & 89.29 & 97.20 & 96.31 & 96.98 & 98.21 & 96.98 & 96.98 & 97.99 & 84.58 & 96.31 \\
\bottomrule
\end{tabular}
\end{table*}

\clearpage

\begin{table*}[ht]
\centering
\caption{Detailed information of the datasets used in this study. We divide all datasets into small and large datasets using 1000 cells as the threshold. A size of "S" indicates that the dataset is small, and "L" indicates that the dataset is large.} 
\label{tab:dataset_detailed_information}
\resizebox{0.75\textwidth}{!}{%
\begin{tabular}{ccccccc}
\toprule
Dataset & Size & \#Cells & \#Genes & \#Types & Accession & Description  \\ 
\midrule
\rowcolor[HTML]{EFEFEF} 
{\color[HTML]{333333} Chu1} & S & {\color[HTML]{333333} 758} & {\color[HTML]{333333} 19176} & {\color[HTML]{333333} 6} & {\color[HTML]{333333} GSE75748} & {\color[HTML]{333333} human pluripotent stem cells}  \\
Chung & S & 515 & 20345 & 5 & GSE75688 & human tumor and immune cells \\
\rowcolor[HTML]{EFEFEF} 
Darmanis & S & 466 & 22085 & 9 & GSE67835 & human brain cells \\
Engel & S & 203 & 23337 & 4 & GSE74596 & mouse Natural killer T cells \\
\rowcolor[HTML]{EFEFEF} 
Goolam & S & 124 & 41388 & 8 & E-MTAB-3321 & mouse cells from different stages\\
Koh & S & 498 & 60483 & 9 & GSM2257302 & human embryonic stem cells \\
\rowcolor[HTML]{EFEFEF} 
Kumar & S & 361 & 22394 & 4 & GSE60749 & mouse embryonic stem cells \\
Leng & S & 247 & 19084 & 3 & GSE64016 & human embryonic stem cells \\
\rowcolor[HTML]{EFEFEF} 
Li & S & 561 & 57241 & 7 & GSE81861 & human cell lines \\
Maria2 & S & 759 & 33694 & 7 & GSE124731 & human innate T cells \\
\rowcolor[HTML]{EFEFEF} 
Robert & S & 194 & 23418 & 2 & GSE74923 & \begin{tabular}[c]{@{}c@{}}mouse leukemia cell line\\ and primary CD8+ T-cells\end{tabular} \\
Ting & S & 187 & 21583 & 7 & GSE51372 & mouse circulating tumor cells \\
\rowcolor[HTML]{EFEFEF} 
\begin{tabular}[c]{@{}c@{}}Mouse\\ Pancreas1\end{tabular} & S & 822 & 14878 & 13 & GSE84133 & Mouse Pancreas Islets \\ \midrule
Cao &L & 4186 & 13488 & 10 & \begin{tabular}[c]{@{}c@{}}sci-RNA-seq\\ platform\end{tabular} & worm neuron cells \\
\rowcolor[HTML]{EFEFEF} 
Chu2 &L & 1018 & 19097 & 7 & GSE75748 & human pluripotent stem cells \\
Han & L & 2746 & 20670 & 16 & \begin{tabular}[c]{@{}c@{}}Mouse Cell\\ Atlas project\end{tabular} & mouse bladder cells \\
\rowcolor[HTML]{EFEFEF} 
MacParland & L & 8444 & 5000 & 11 & GSE115469 & human liver cells \\
Maria1 & L & 1277 & 33694 & 7 & GSE124731 & human innate T cells \\
\rowcolor[HTML]{EFEFEF} 
Puram & L & 3363 & 23686 & 8 & GSE103322 & \begin{tabular}[c]{@{}c@{}}non-malignant cells\\ in Head and Neck Cancer\end{tabular} \\
Yang & L & 1119 & 46609 & 6 & GSE90848 & \begin{tabular}[c]{@{}c@{}}mouse bulge hair follicle stem cell, \\ hair germ, basal transient amplifying \\ cells (TACs) and dermal papilla\end{tabular} \\
\rowcolor[HTML]{EFEFEF} 
\begin{tabular}[c]{@{}c@{}}Human\\ Pancreas1\end{tabular} & L & 1937 & 20125 & 14 & GSE84133 & Human Pancreas Islets \\
\begin{tabular}[c]{@{}c@{}}Human\\ Pancreas2\end{tabular} & L & 1724 & 20125 & 14 & GSE84133 & Human Pancreas Islets \\
\rowcolor[HTML]{EFEFEF} 
\begin{tabular}[c]{@{}c@{}}Human\\ Pancreas3\end{tabular} & L & 3605 & 20125 & 14 & GSE84133 & Human Pancreas Islets \\
\begin{tabular}[c]{@{}c@{}}Mouse\\ Pancreas2\end{tabular} & L & 1064 & 14878 & 13 & GSE84133 & Mouse Pancreas Islets\\ \bottomrule
\end{tabular}%
}
\end{table*}
\clearpage



\useunder{\uline}{\ul}{}
\begin{table*}[h]
\centering
\caption{Details of the model performance comparison on each dataset regarding NMI, ARI, and SI. We use light red shade and \textbf{bold font} to highlight the best performance. We use light blue shade and {\ul underline} to highlight the second-best performance.}
\label{tab:main_table}
 \setlength{\tabcolsep}{8pt}
\resizebox{\textwidth}{!}{%

    \begin{tabular}{clllllllllllllll}
    \hline
    \multirow{2}{*}{Dataset}                                  & \multicolumn{3}{c}{Original Dataset}                                                                                                                                               & \multicolumn{3}{c}{CellRanger}                                                                                                                                                                                                             & \multicolumn{3}{c}{PR}                                                                                                                                                                                                                     & \multicolumn{3}{c}{Seurat v3}                                                                                                                                                                                                              & \multicolumn{3}{c}{HRG}                                                                                                                                                                                                                    \\
                                                              & \multicolumn{1}{c}{NMI}                                                   & \multicolumn{1}{c}{\cellcolor[HTML]{EFEFEF}ARI} & \multicolumn{1}{c}{SI}                                                       & \multicolumn{1}{c}{\cellcolor[HTML]{EFEFEF}NMI}                                                      & \multicolumn{1}{c}{ARI}                                                      & \multicolumn{1}{c}{\cellcolor[HTML]{EFEFEF}SI}                                                       & \multicolumn{1}{c}{NMI}                                                      & \multicolumn{1}{c}{\cellcolor[HTML]{EFEFEF}ARI}                                                      & \multicolumn{1}{c}{SI}                                                       & \multicolumn{1}{c}{\cellcolor[HTML]{EFEFEF}NMI}                                                      & \multicolumn{1}{c}{ARI}                                                      & \multicolumn{1}{c}{\cellcolor[HTML]{EFEFEF}SI}                                                       & \multicolumn{1}{c}{NMI}                                                      & \multicolumn{1}{c}{\cellcolor[HTML]{EFEFEF}ARI}                                                      & \multicolumn{1}{c}{SI}                                                       \\ \hline
    Chu1                                                      & 69.38                                                                     & 56.37                   & 2.53                                                                         & 81.82                                                                        & 68.83                                                                        & 14.2                                                                         & 66.54                                                                        & 45.62                                                                        & 6.37                                                                         & 80.6                                                                         & 68.48                                                                        & 7.44                                                                         & 80.71                                                                        & 67.68                                                                        & 13.23                                                                        \\
    Chung                                                     & 41.79                                                                     & 15.88                   & 3.92                                                                         & \cellcolor[HTML]{CBCEFB}{\ul 48.21}    & 18.93                                                                        & 15.21                                                                        & 44.02                                                                        & 17.38                                                                        & 8.02                                                                         & 47.54                                                                        & 17.74                                                                        & 14.42                                                                        & 47.36                                                                        & 17.89                                                                        & 16.86                                                                        \\
    Darmanis                                                  & 12.09                                                                     & 3.77                    & 3.33                                                                         & 16.99                                                                        & 5.97                                                                         & 10.8                                                                         & \cellcolor[HTML]{CBCEFB}{\ul 18.56}    & 4.29                                                                         & 9.4                                                                          & 16.75                                                                        & 4.8                                                                          & 10.36                                                                        & 17.47                                                                        & 4.8                                                                          & 10.32                                                                        \\
    Engel                                                     & 60.45                                                                     & 57.46                   & 1.85                                                                         & 66.05                                                                        & 55.59                                                                        & 2.9                                                                          & 69.29                                                                        & 70.03                                                                        & 1.83                                                                         & 66.95                                                                        & 62.69                                                                        & 4.25                                                                         & 75.72                                                                        & 67.16                                                                        & 8.89                                                                         \\
    Goolam                                                    & 67.98                                                                     & 45.06                   & 5.82                                                                         & 73.21                                                                        & 50.92                                                                        & 15.72                                                                        & \cellcolor[HTML]{CBCEFB}{\ul 73.57}    & 49.47                                                                        & 7.99                                                                         & 73.21                                                                        & 50.93                                                                        & 14.02                                                                        & 67.13                                                                        & 38.11                                                                        & \cellcolor[HTML]{FFCCC9}\textbf{27.97} \\
    Koh                                                       & 70.85                                                                     & 51.8                    & 1.58                                                                         & 87.37                                                                        & 76.42                                                                        & 4.04                                                                         & 47.75                                                                        & 29.14                                                                        & 0.23                                                                         & 76.34                                                                        & 62.86                                                                        & 4.04                                                                         & 93.85                                                                        & 90.52                                                                        & 9.08                                                                         \\
    Kumar                                                     & 68.54                                                                     & 58.28                   & 5.81                                                                         & 93.1                                                                         & 93.55                                                                        & 8.05                                                                         & 76.52                                                                        & 68.47                                                                        & 4.96                                                                         & 91.99                                                                        & 93.46                                                                        & 8.14                                                                         & 96.7                                                                         & 97.05                                                                        & 12.93                                                                        \\
    Leng                                                      & 52.89                                                                     & 55.01                   & 0.38                                                                         & 15.21                                                                        & 11.2                                                                         & -1.74                                                                        & 5.04                                                                         & 2.16                                                                         & -0.42                                                                        & 27.08                                                                        & 26.69                                                                        & -0.81                                                                        & 56.8                                                                         & 59.11                                                                        & 1.01                                                                         \\
    Li                                                        & 79.98                                                                     & 67.31                   & 9.94                                                                         & 87.62                                                                        & 72.68                                                                        & 7.9                                                                          & 88.9                                                                         & \cellcolor[HTML]{CBCEFB}{\ul 82.85}    & 4.43                                                                         & 87.87                                                                        & 72.95                                                                        & 8.11                                                                         & 88.07                                                                        & 78.57                                                                        & 34.26                                                                        \\
    Maria2                                                    & 13.97                                                                     & 9.96                    & 0.11                                                                         & 23.62                                                                        & 17.31                                                                        & 0.36                                                                         & 12.92                                                                        & 7.2                                                                          & 0.09                                                                         & 26.02                                                                        & 18.48                                                                        & 1.24                                                                         & 33.01                                                                        & 21.28                                                                        & 1.45                                                                         \\
    Robert                                                    & 70.26                                                                     & 72.84                   & 16.73                                                                        & 56.3                                                                         & 42.02                                                                        & 6.45                                                                         & 59.05                                                                        & 49.39                                                                        & 4.78                                                                         & 63.46                                                                        & 57.15                                                                        & 24.39                                                                        & 59.84                                                                        & 51.06                                                                        & \cellcolor[HTML]{CBCEFB}{\ul 27.83}    \\
    Ting                                                      & 72.41                                                                     & 50.85                   & 4.72                                                                         & 77.4                                                                         & 53.6                                                                         & 17.01                                                                        & 79.03                                                                        & 58.13                                                                        & 12.44                                                                        & 80.53                                                                        & 59.87                                                                        & 18.79                                                                        & 79.28                                                                        & 58.92                                                                        & 16.9                                                                         \\
    \begin{tabular}[c]{@{}c@{}}Mouse\\ Pancreas1\end{tabular} & 61.41                                                                     & 51.27                   & 3.5                                                                          & 76.47                                                                        & 57.14                                                                        & 11.34                                                                        & 75.31                                                                        & 56.72                                                                        & 13.05                                                                        & \cellcolor[HTML]{CBCEFB}{\ul 80.35}    & \cellcolor[HTML]{CBCEFB}{\ul 67.8}     & 14.39                                                                        & 74.04                                                                        & 62.67                                                                        & 7.23                                                                         \\
    Cao                                                       & 56.23                                                                     & 31.41                   & 0.38                                                                         & 61.18                                                                        & 32.75                                                                        & \cellcolor[HTML]{CBCEFB}{\ul 21.1}     & 57.69                                                                        & 29.3                                                                         & 17.7                                                                         & 61.91                                                                        & 34.62                                                                        & 18.89                                                                        & 56.93                                                                        & 41.67                                                                        & 1.48                                                                         \\
    Chu2                                                      & 90.47                                                                     & 75.78                   & 8.75                                                                         & 94.19                                                                        & 90.67                                                                        & 12.18                                                                        & 91.29                                                                        & 77.35                                                                        & 16.97                                                                        & 95.47                                                                        & 93.9                                                                         & 6.94                                                                         & 96.3                                                                         & 95.95                                                                        & 12.85                                                                        \\
    Han                                                       & 72.46                                                                     & 57.67                   & 1.91                                                                         & 73.45                                                                        & 55.47                                                                        & 9.91                                                                         & 73.27                                                                        & 62.98                                                                        & \cellcolor[HTML]{CBCEFB}{\ul 13.64}    & 73.73                                                                        & 54.57                                                                        & 9.64                                                                         & \cellcolor[HTML]{CBCEFB}{\ul 76.54}    & \cellcolor[HTML]{FFCCC9}\textbf{68.41} & 3.15                                                                         \\
    MacParland                                                & 67.59                                                                     & 39.77                   & 2.31                                                                         & 77.45                                                                        & 56.01                                                                        & 6.37                                                                         & 67.08                                                                        & 50.22                                                                        & \cellcolor[HTML]{CBCEFB}{\ul 11.45}    & 79.84                                                                        & 61.04                                                                        & 5.11                                                                         & 73.22                                                                        & 48.05                                                                        & 2.32                                                                         \\
    Maria1                                                    & 19.91                                                                     & 13.98                   & -0.11                                                                        & 19.61                                                                        & 15.35                                                                        & 0.12                                                                         & 6.91                                                                         & 3.95                                                                         & -0.6                                                                         & 36.59                                                                        & 27.8                                                                         & 1.09                                                                         & 37.93                                                                        & 22.95                                                                        & 0.97                                                                         \\
    Puram                                                     & 65.71                                                                     & 51.15                   & 1.68                                                                         & 63.6                                                                         & 30.17                                                                        & 9.59                                                                         & 70.1                                                                         & 46                                                                           & 4.99                                                                         & 71.58                                                                        & 42.61                                                                        & 5.92                                                                         & 71.28                                                                        & 43.69                                                                        & 3.65                                                                         \\
    Yang                                                      & 42.29                                                                     & 38.13                   & 4.88                                                                         & 64.01                                                                        & 52.95                                                                        & 10.32                                                                        & 54.9                                                                         & 35.65                                                                        & 6.62                                                                         & 59.88                                                                        & 36.47                                                                        & 8.34                                                                         & 62.06                                                                        & 42.25                                                                        & 12.15                                                                        \\
    \begin{tabular}[c]{@{}c@{}}Human\\ Pancreas1\end{tabular} & 67.44                                                                     & 60.25                   & 2.4                                                                          & 81.22                                                                        & 58.3                                                                         & 11.92                                                                        & 78.97                                                                        & 49.81                                                                        & 13.63                                                                        & \cellcolor[HTML]{CBCEFB}{\ul 83.67}    & 59.69                                                                        & 13.62                                                                        & 67.87                                                                        & 60.55                                                                        & 2.52                                                                         \\
    \begin{tabular}[c]{@{}c@{}}Human\\ Pancreas2\end{tabular} & 85.1                                                                      & 89.19                   & 5                                                                            & 87.22                                                                        & 82.97                                                                        & 14.67                                                                        & 81.91                                                                        & 65.63                                                                        & 17.96                                                                        & \cellcolor[HTML]{CBCEFB}{\ul 88.12}    & 84.64                                                                        & 14.1                                                                         & 85.3                                                                         & \cellcolor[HTML]{CBCEFB}{\ul 89.35}    & 5.26                                                                         \\
    \begin{tabular}[c]{@{}c@{}}Human\\ Pancreas3\end{tabular} & 79.94                                                                     & 84.17                   & 5.55                                                                         & \cellcolor[HTML]{CBCEFB}{\ul 89.52}    & \cellcolor[HTML]{CBCEFB}{\ul 92.9}     & 23.7                                                                         & 89.11                                                                        & 92.8                                                                         & \cellcolor[HTML]{CBCEFB}{\ul 26.86}    & 84.21                                                                        & 78.18                                                                        & 18.07                                                                        & 84.17                                                                        & 87.76                                                                        & 11.47                                                                        \\
    \begin{tabular}[c]{@{}c@{}}Mouse\\ Pancreas2\end{tabular} & 45.46                                                                     & 37.02                   & 0.85                                                                         & 68.37                                                                        & 40.55                                                                        & 8.97                                                                         & 66.66                                                                        & 38.72                                                                        & 12.68                                                                        & 72.77                                                                        & 45.06                                                                        & 12.05                                                                        & \cellcolor[HTML]{CBCEFB}{\ul 74.56}    & \cellcolor[HTML]{CBCEFB}{\ul 48.43}    & 8.56                                                                         \\ \hline
    \multirow{2}{*}{Dataset}                                  & \multicolumn{3}{c}{GeneBasis}                                                                                                                                                      & \multicolumn{3}{c}{CellBRF}                                                                                                                                                                                                                & \multicolumn{3}{c}{gpsFISH}                                                                                                                                                                                                                & \multicolumn{3}{c}{scGIST}                                                                                                                                                                                                                 & \multicolumn{3}{c}{\model}                                                                                                                                                                                  \\
                                                              & \multicolumn{1}{c}{NMI}                                                   & \multicolumn{1}{c}{ARI\cellcolor[HTML]{EFEFEF}} & \multicolumn{1}{c}{SI}                                                       & \multicolumn{1}{c}{NMI\cellcolor[HTML]{EFEFEF}}                                                      & \multicolumn{1}{c}{ARI}                                                      & \multicolumn{1}{c}{SI\cellcolor[HTML]{EFEFEF}}                                                       & \multicolumn{1}{c}{NMI}                                                      & \multicolumn{1}{c}{ARI\cellcolor[HTML]{EFEFEF}}                                                      & \multicolumn{1}{c}{SI}                                                       & \multicolumn{1}{c}{NMI\cellcolor[HTML]{EFEFEF}}                                                      & \multicolumn{1}{c}{ARI}                                                      & \multicolumn{1}{c}{SI\cellcolor[HTML]{EFEFEF}}                                                       & \multicolumn{1}{c}{NMI}                                                      & \multicolumn{1}{c}{ARI\cellcolor[HTML]{EFEFEF}}                                                      & \multicolumn{1}{c}{SI}                                                       \\ \hline
    Chu1                                                      & 74.53                                                                     & 62.29                   & 15.32                                                                        & 84.87                                                                        & \cellcolor[HTML]{CBCEFB}{\ul 78.25}    & 16.39                                                                        & \cellcolor[HTML]{CBCEFB}{\ul 86.09}    & 74.56                                                                        & 11.69                                                                        & 78.11                                                                        & 65.12                                                                        & \cellcolor[HTML]{CBCEFB}{\ul 16.84}    & \cellcolor[HTML]{FFCCC9}\textbf{88.83} & \cellcolor[HTML]{FFCCC9}\textbf{82.71} & \cellcolor[HTML]{FFCCC9}\textbf{17.9}  \\
    Chung                                                     & 47.43                                                                     & 19.1                    & \cellcolor[HTML]{FFCCC9}\textbf{18.97} & 46.73                                                                        & 19.24                                                                        & 16.7                                                                         & 45.58                                                                        & \cellcolor[HTML]{FFCCC9}\textbf{24.47} & 9.96                                                                         & 46.55                                                                        & \cellcolor[HTML]{CBCEFB}{\ul 20.81}    & 12.81                                                                        & \cellcolor[HTML]{FFCCC9}\textbf{48.29} & 18.31                                                                        & \cellcolor[HTML]{CBCEFB}{\ul 18.0}     \\
    Darmanis                                                  & 13.04                                                                     & 4.79                    & \cellcolor[HTML]{CBCEFB}{\ul 13.64}    & 17.79                                                                        & 5.24                                                                         & \cellcolor[HTML]{FFCCC9}\textbf{13.86} & 18.11                                                                        & \cellcolor[HTML]{FFCCC9}\textbf{14.45} & 10.02                                                                        & 16.53                                                                        & \cellcolor[HTML]{CBCEFB}{\ul 7.0}      & 6.86                                                                         & \cellcolor[HTML]{FFCCC9}\textbf{19.65} & 6.75                                                                         & 11.54                                                                        \\
    Engel                                                     & 68.2                                                                      & 64.01                   & 7.3                                                                          & \cellcolor[HTML]{FFCCC9}\textbf{80.85} & \cellcolor[HTML]{CBCEFB}{\ul 73.64}    & \cellcolor[HTML]{FFCCC9}\textbf{20.18} & 68.28                                                                        & 61.06                                                                        & 11.48                                                                        & 80.19                                                                        & \cellcolor[HTML]{FFCCC9}\textbf{85.05} & 8.16                                                                         & \cellcolor[HTML]{CBCEFB}{\ul 80.49}    & 72.55                                                                        & \cellcolor[HTML]{CBCEFB}{\ul 14.8}     \\
    Goolam                                                    & 60.59                                                                     & 36.92                   & 10.7                                                                         & 71.27                                                                        & 46.31                                                                        & \cellcolor[HTML]{CBCEFB}{\ul 22.82}    & 70.38                                                                        & \cellcolor[HTML]{FFCCC9}\textbf{56.49} & 14.06                                                                        & 67.44                                                                        & 41.88                                                                        & 22.17                                                                        & \cellcolor[HTML]{FFCCC9}\textbf{74.45} & \cellcolor[HTML]{CBCEFB}{\ul 51.6}     & 18.36                                                                        \\
    Koh                                                       & 89.38                                                                     & 85.76                   & 10.05                                                                        & \cellcolor[HTML]{CBCEFB}{\ul 98.44}    & \cellcolor[HTML]{CBCEFB}{\ul 98.28}    & \cellcolor[HTML]{FFCCC9}\textbf{23.59} & 55.5                                                                         & 42.25                                                                        & 7.43                                                                         & 86.92                                                                        & 77.08                                                                        & 7.27                                                                         & \cellcolor[HTML]{FFCCC9}\textbf{99.09} & \cellcolor[HTML]{FFCCC9}\textbf{99.21} & \cellcolor[HTML]{CBCEFB}{\ul 20.58}    \\
    Kumar                                                     & 88.29                                                                     & 87.66                   & 13.62                                                                        & 90.3                                                                         & 86.85                                                                        & 22.65                                                                        & 94.22                                                                        & 79.99                                                                        & 21.51                                                                        & \cellcolor[HTML]{CBCEFB}{\ul 98.07}    & \cellcolor[HTML]{CBCEFB}{\ul 98.51}    & \cellcolor[HTML]{FFCCC9}\textbf{24.68} & \cellcolor[HTML]{FFCCC9}\textbf{98.07} & \cellcolor[HTML]{FFCCC9}\textbf{98.51} & \cellcolor[HTML]{CBCEFB}{\ul 24.3}     \\
    Leng                                                      & 6.97                                                                      & 2.79                    & 0.93                                                                         & 70.37                                                                        & 71.18                                                                        & \cellcolor[HTML]{FFCCC9}\textbf{8.77}  & 15.57                                                                        & 13.88                                                                        & 5.91                                                                         & \cellcolor[HTML]{CBCEFB}{\ul 82.62}    & \cellcolor[HTML]{CBCEFB}{\ul 83.02}    & 5.97                                                                         & \cellcolor[HTML]{FFCCC9}\textbf{82.82} & \cellcolor[HTML]{FFCCC9}\textbf{85.72} & \cellcolor[HTML]{CBCEFB}{\ul 7.97}     \\
    Li                                                        & 89.06                                                                     & 78.98                   & 25.74                                                                        & 89.03                                                                        & 78.92                                                                        & \cellcolor[HTML]{CBCEFB}{\ul 40.63}    & \cellcolor[HTML]{CBCEFB}{\ul 92.52}    & 82.51                                                                        & 17.12                                                                        & 88.19                                                                        & 77.34                                                                        & 17.75                                                                        & \cellcolor[HTML]{FFCCC9}\textbf{93.41} & \cellcolor[HTML]{FFCCC9}\textbf{83.33} & \cellcolor[HTML]{FFCCC9}\textbf{41.63} \\
    Maria2                                                    & 35.37                                                                     & 29.79                   & 1.64                                                                         & \cellcolor[HTML]{FFCCC9}\textbf{53.73} & \cellcolor[HTML]{FFCCC9}\textbf{43.87} & \cellcolor[HTML]{CBCEFB}{\ul 7.0}      & 20.53                                                                        & 14.84                                                                        & \cellcolor[HTML]{FFCCC9}\textbf{7.17}  & \cellcolor[HTML]{CBCEFB}{\ul 53.59}    & \cellcolor[HTML]{CBCEFB}{\ul 43.6}     & 3.74                                                                         & 43.2                                                                         & 33.14                                                                        & 2.9                                                                          \\
    Robert                                                    & 52.57                                                                     & 36.83                   & 9.58                                                                         & 55.38                                                                        & 38.49                                                                        & 17.21                                                                        & \cellcolor[HTML]{CBCEFB}{\ul 70.53}    & \cellcolor[HTML]{FFCCC9}\textbf{73.82} & 26.23                                                                        & 52.62                                                                        & 35.05                                                                        & 13.13                                                                        & \cellcolor[HTML]{FFCCC9}\textbf{71.13} & \cellcolor[HTML]{CBCEFB}{\ul 73.1}     & \cellcolor[HTML]{FFCCC9}\textbf{52.86} \\
    Ting                                                      & 81.48                                                                     & 60.58                   & 18.37                                                                        & 77.63                                                                        & 58.34                                                                        & \cellcolor[HTML]{FFCCC9}\textbf{28.36} & 81.86                                                                        & \cellcolor[HTML]{FFCCC9}\textbf{68.91} & 13.97                                                                        & \cellcolor[HTML]{CBCEFB}{\ul 82.02}    & 61.28                                                                        & 16.33                                                                        & \cellcolor[HTML]{FFCCC9}\textbf{83.17} & \cellcolor[HTML]{CBCEFB}{\ul 62.3}     & \cellcolor[HTML]{CBCEFB}{\ul 25.78}    \\
    \begin{tabular}[c]{@{}c@{}}Mouse\\ Pancreas1\end{tabular} & 75.93                                                                     & 54.28                   & \cellcolor[HTML]{FFCCC9}\textbf{16.57} & 77.19                                                                        & 60.87                                                                        & \cellcolor[HTML]{CBCEFB}{\ul 15.54}    & 73.63                                                                        & 59.51                                                                        & 10.87                                                                        & 78.49                                                                        & 63.8                                                                         & 14.45                                                                        & \cellcolor[HTML]{FFCCC9}\textbf{84.57} & \cellcolor[HTML]{FFCCC9}\textbf{78.14} & 14.17                                                                        \\
    Cao                                                       & 49.75                                                                     & 32.23                   & 7.2                                                                          & 47.12                                                                        & 26.53                                                                        & 10.54                                                                        & 46.72                                                                        & 30.64                                                                        & \cellcolor[HTML]{FFCCC9}\textbf{30.28} & \cellcolor[HTML]{CBCEFB}{\ul 62.45}    & \cellcolor[HTML]{FFCCC9}\textbf{64.72} & 8.85                                                                         & \cellcolor[HTML]{FFCCC9}\textbf{63.66} & \cellcolor[HTML]{CBCEFB}{\ul 50.53}    & 8.13                                                                         \\
    Chu2                                                      & 99.05                                                                     & 99.23                   & 20.08                                                                        & \cellcolor[HTML]{CBCEFB}{\ul 99.4}     & 99.64                                                                        & \cellcolor[HTML]{FFCCC9}\textbf{33.92} & 94.21                                                                        & \cellcolor[HTML]{FFCCC9}\textbf{102.1} & 13.81                                                                        & 92.56                                                                        & 78.14                                                                        & \cellcolor[HTML]{CBCEFB}{\ul 32.84}    & \cellcolor[HTML]{FFCCC9}\textbf{100.0} & \cellcolor[HTML]{CBCEFB}{\ul 100.0}    & 29.81                                                                        \\
    Han                                                       & 68.94                                                                     & 51.47                   & 11.26                                                                        & 76.05                                                                        & 66.38                                                                        & 10.08                                                                        & 62.05                                                                        & 45.39                                                                        & \cellcolor[HTML]{FFCCC9}\textbf{15.73} & 76.21                                                                        & 66.09                                                                        & 10.03                                                                        & \cellcolor[HTML]{FFCCC9}\textbf{78.87} & \cellcolor[HTML]{CBCEFB}{\ul 66.57}    & 10.33                                                                        \\
    MacParland                                                & 70.89                                                                     & 51.93                   & 8.91                                                                         & \cellcolor[HTML]{CBCEFB}{\ul 82.74}    & \cellcolor[HTML]{CBCEFB}{\ul 70.48}    & 5.71                                                                         & 65.33                                                                        & 42.72                                                                        & \cellcolor[HTML]{FFCCC9}\textbf{12.89} & 82.18                                                                        & 64.8                                                                         & 6.76                                                                         & \cellcolor[HTML]{FFCCC9}\textbf{83.94} & \cellcolor[HTML]{FFCCC9}\textbf{75.51} & 5.75                                                                         \\
    Maria1                                                    & \cellcolor[HTML]{CBCEFB}{\ul 45.52} & 37                      & 2.66                                                                         & \cellcolor[HTML]{FFCCC9}\textbf{51.22} & \cellcolor[HTML]{FFCCC9}\textbf{40.16} & \cellcolor[HTML]{CBCEFB}{\ul 5.34}     & 27.13                                                                        & 23.98                                                                        & \cellcolor[HTML]{FFCCC9}\textbf{8.27}  & 42.07                                                                        & \cellcolor[HTML]{CBCEFB}{\ul 38.52}    & 3.15                                                                         & 43.77                                                                        & 30.86                                                                        & 3.25                                                                         \\
    Puram                                                     & 65.75                                                                     & 37.69                   & 11.06                                                                        & 79.51                                                                        & 65.57                                                                        & \cellcolor[HTML]{FFCCC9}\textbf{17.04} & \cellcolor[HTML]{FFCCC9}\textbf{83.48} & \cellcolor[HTML]{FFCCC9}\textbf{74.47} & 9.36                                                                         & 75.07                                                                        & 52.83                                                                        & 7.87                                                                         & \cellcolor[HTML]{CBCEFB}{\ul 80.16}    & \cellcolor[HTML]{CBCEFB}{\ul 67.58}    & \cellcolor[HTML]{CBCEFB}{\ul 15.26}    \\
    Yang                                                      & 62.72                                                                     & 42.43                   & 15.99                                                                        & \cellcolor[HTML]{CBCEFB}{\ul 66.58}    & 53.6                                                                         & 13.96                                                                        & 63.01                                                                        & \cellcolor[HTML]{FFCCC9}\textbf{64.45} & \cellcolor[HTML]{FFCCC9}\textbf{17.49} & \cellcolor[HTML]{FFCCC9}\textbf{69.15} & \cellcolor[HTML]{CBCEFB}{\ul 56.19}    & 9.51                                                                         & 66.36                                                                        & 54.5                                                                         & \cellcolor[HTML]{CBCEFB}{\ul 16.39}    \\
    \begin{tabular}[c]{@{}c@{}}Human\\ Pancreas1\end{tabular} & 80.74                                                                     & 55.38                   & 19.65                                                                        & 83.29                                                                        & \cellcolor[HTML]{CBCEFB}{\ul 66.37}    & \cellcolor[HTML]{CBCEFB}{\ul 19.81}    & 68.33                                                                        & 42.35                                                                        & \cellcolor[HTML]{FFCCC9}\textbf{20.17} & 82.22                                                                        & 57.12                                                                        & 17.88                                                                        & \cellcolor[HTML]{FFCCC9}\textbf{86.6}  & \cellcolor[HTML]{FFCCC9}\textbf{73.78} & 18.5                                                                         \\
    \begin{tabular}[c]{@{}c@{}}Human\\ Pancreas2\end{tabular} & 82.7                                                                      & 68.12                   & \cellcolor[HTML]{FFCCC9}\textbf{22.84} & 79.14                                                                        & 59.97                                                                        & 15.71                                                                        & 75.65                                                                        & 66.62                                                                        & 12.37                                                                        & 79.96                                                                        & 61.25                                                                        & 18.8                                                                         & \cellcolor[HTML]{FFCCC9}\textbf{89.58} & \cellcolor[HTML]{FFCCC9}\textbf{91.55} & \cellcolor[HTML]{CBCEFB}{\ul 21.37}    \\
    \begin{tabular}[c]{@{}c@{}}Human\\ Pancreas3\end{tabular} & 84.88                                                                     & 88.05                   & 20.65                                                                        & 80.47                                                                        & 63.87                                                                        & 10.56                                                                        & 85.59                                                                        & 91.71                                                                        & 26.32                                                                        & 86.53                                                                        & 80.77                                                                        & 20.31                                                                        & \cellcolor[HTML]{FFCCC9}\textbf{89.98} & \cellcolor[HTML]{FFCCC9}\textbf{93.14} & \cellcolor[HTML]{FFCCC9}\textbf{28.97} \\
    \begin{tabular}[c]{@{}c@{}}Mouse\\ Pancreas2\end{tabular} & 67.8                                                                      & 37.35                   & \cellcolor[HTML]{FFCCC9}\textbf{14.45} & 69.51                                                                        & 38.33                                                                        & 12.16                                                                        & 57.57                                                                        & 38.77                                                                        & 12.36                                                                        & 69.6                                                                         & 37.7                                                                         & 9.5                                                                          & \cellcolor[HTML]{FFCCC9}\textbf{77.92} & \cellcolor[HTML]{FFCCC9}\textbf{60.07} & \cellcolor[HTML]{CBCEFB}{\ul 13.04}    \\ \hline
    \end{tabular}
 
}
\end{table*}

\clearpage

\begin{figure*}[htbp]
\centering
\renewcommand{\thesubfigure}{\arabic{subfigure}}
\subfloat[Cao]{
        \includegraphics[width=0.16\textwidth]{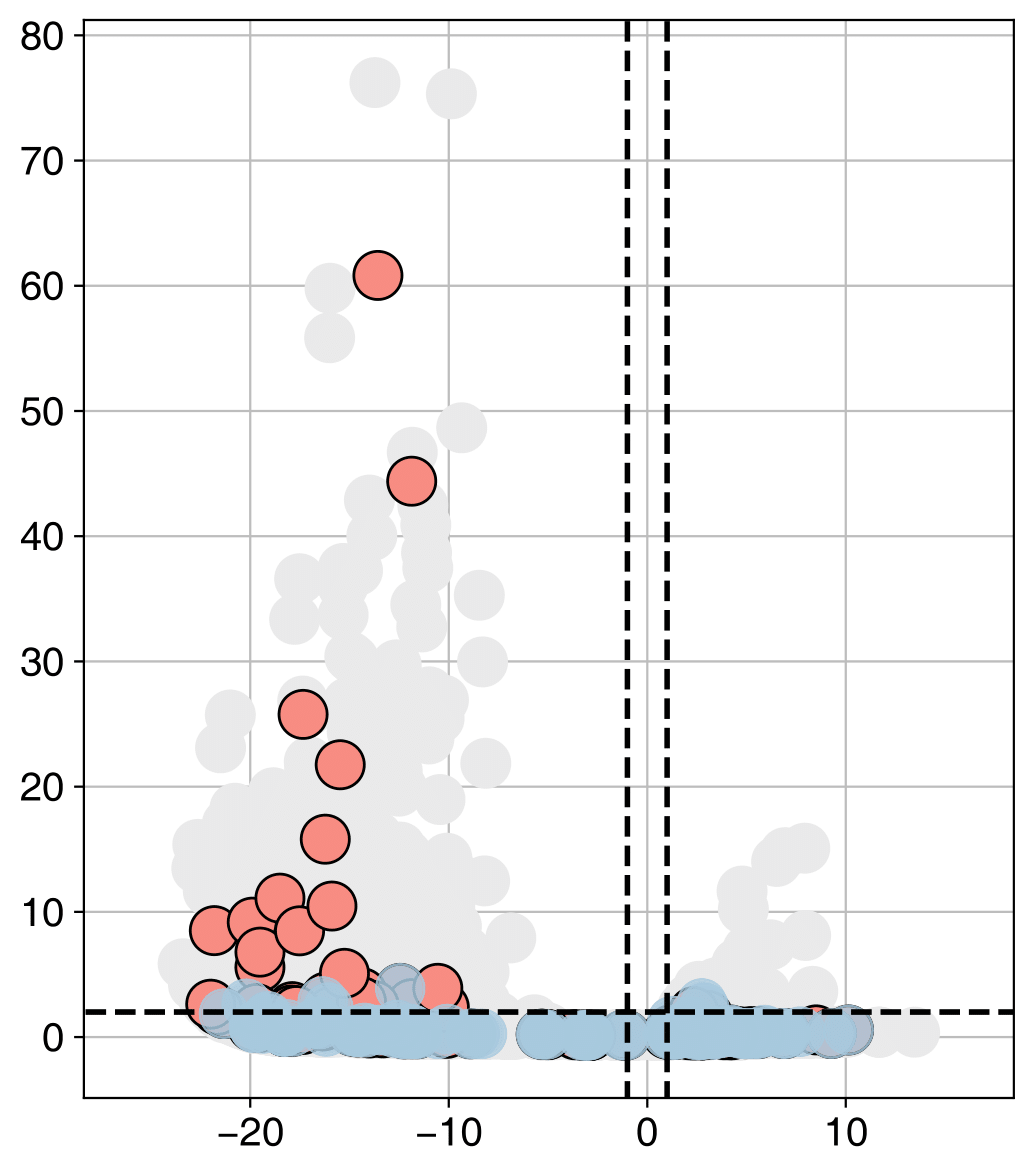}}
\subfloat[Chu2]{
		\includegraphics[width=0.16\textwidth]{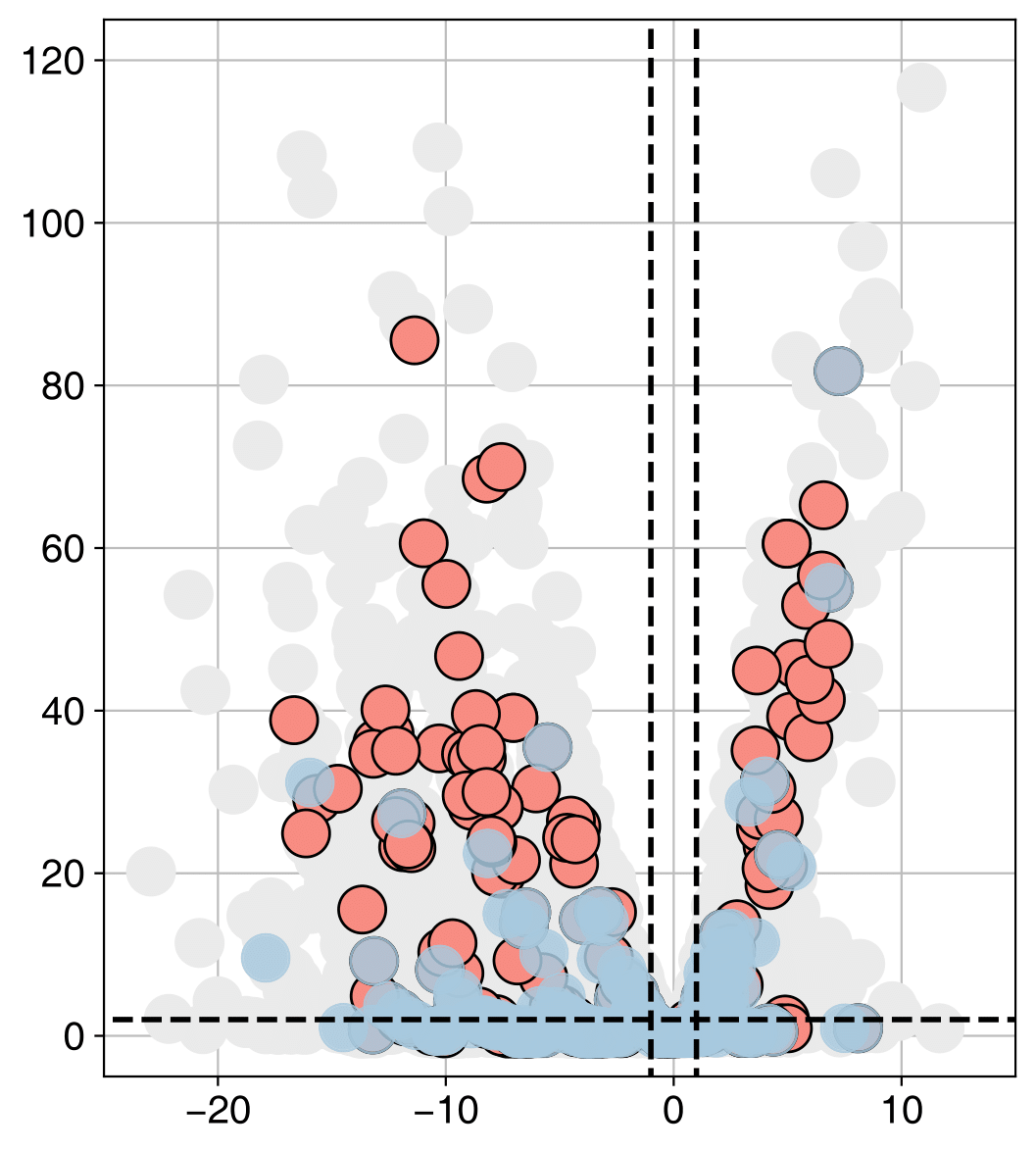}}
\subfloat[Han]{
		\includegraphics[width=0.16\textwidth]{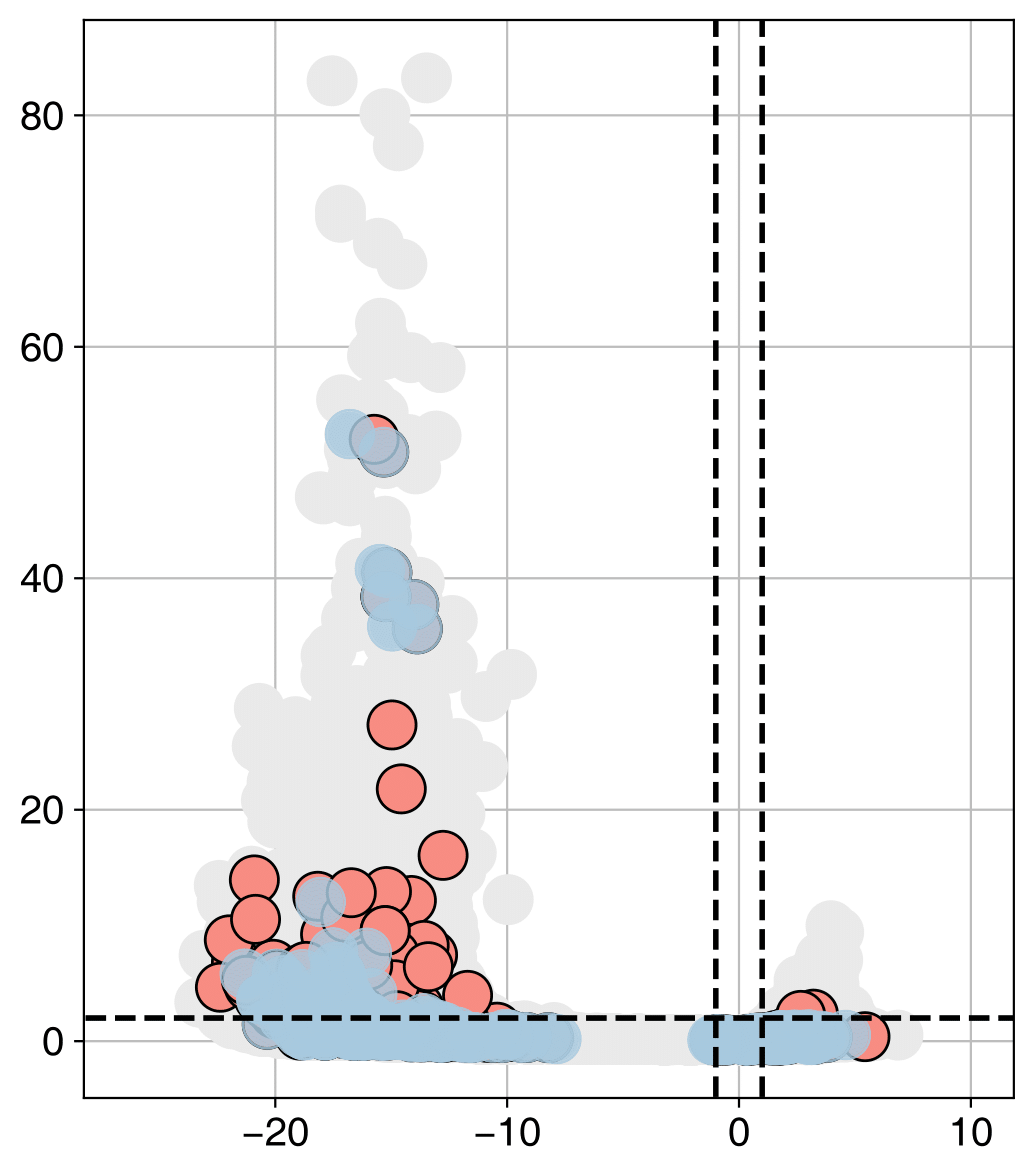}}
\subfloat[Human Pancreas1]{
		\includegraphics[width=0.16\textwidth]{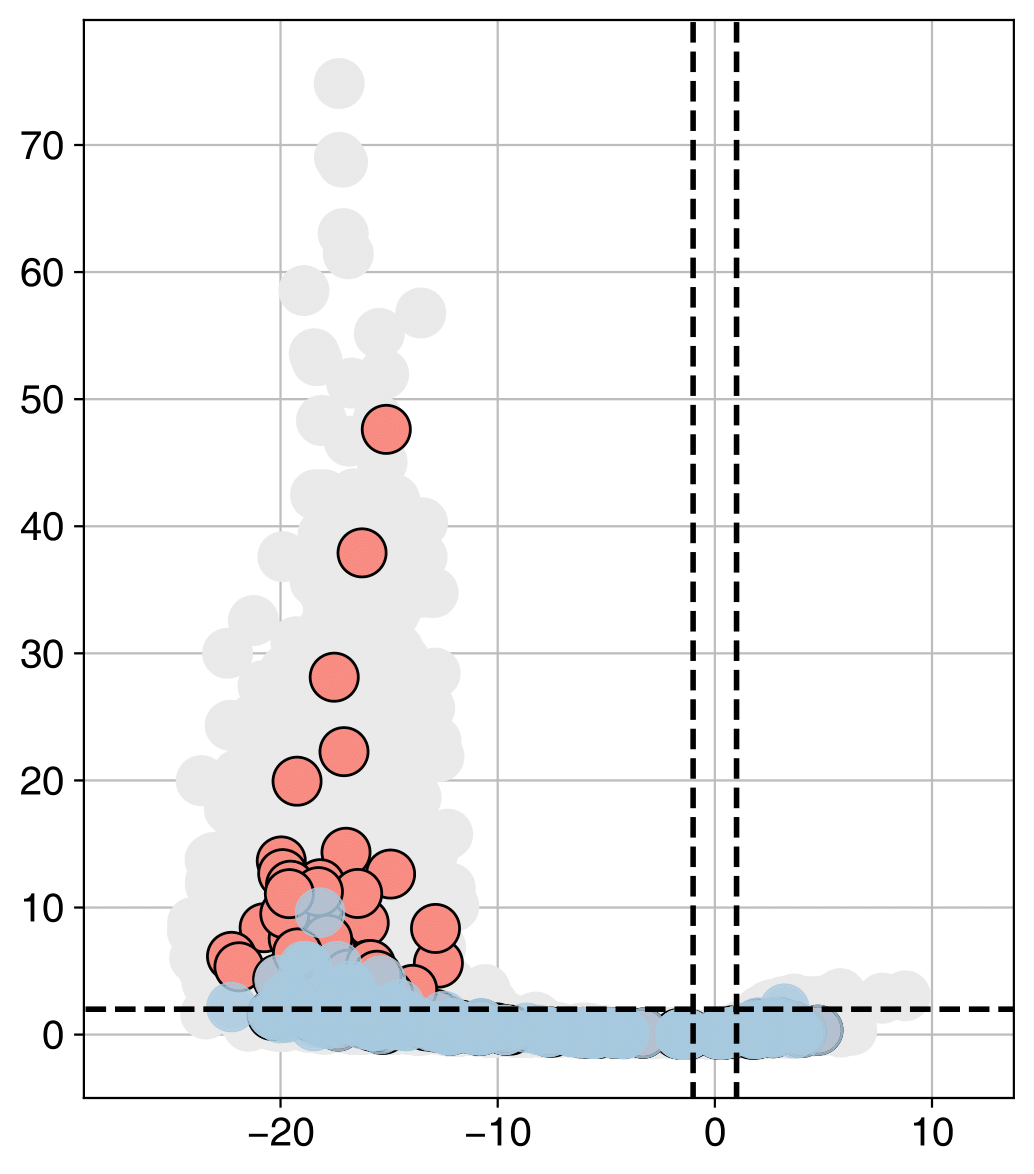}}
\subfloat[Human Pancreas2]{
		\includegraphics[width=0.16\textwidth]{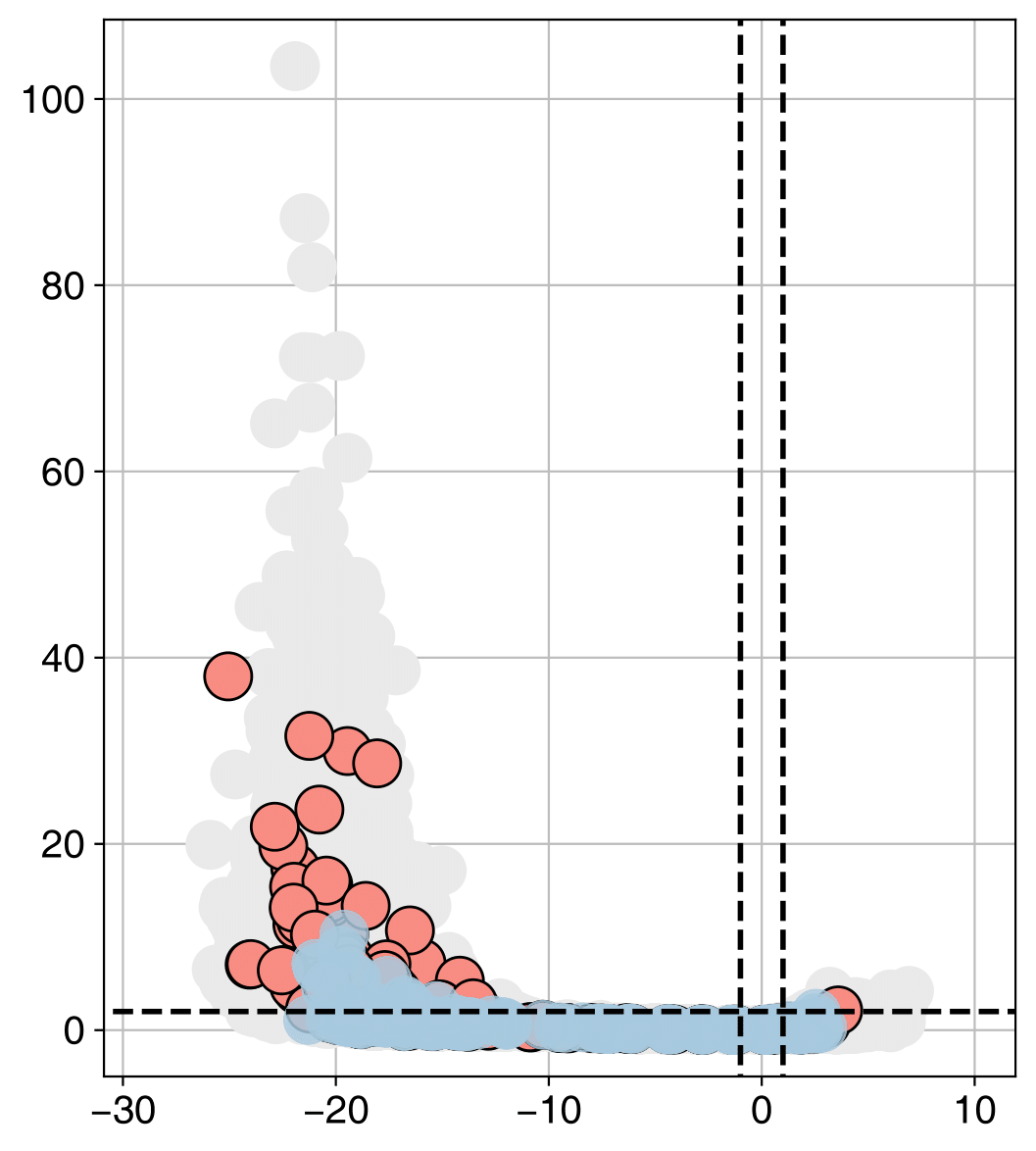}}
\subfloat[Human Pancreas3]{
		\includegraphics[width=0.16\textwidth]{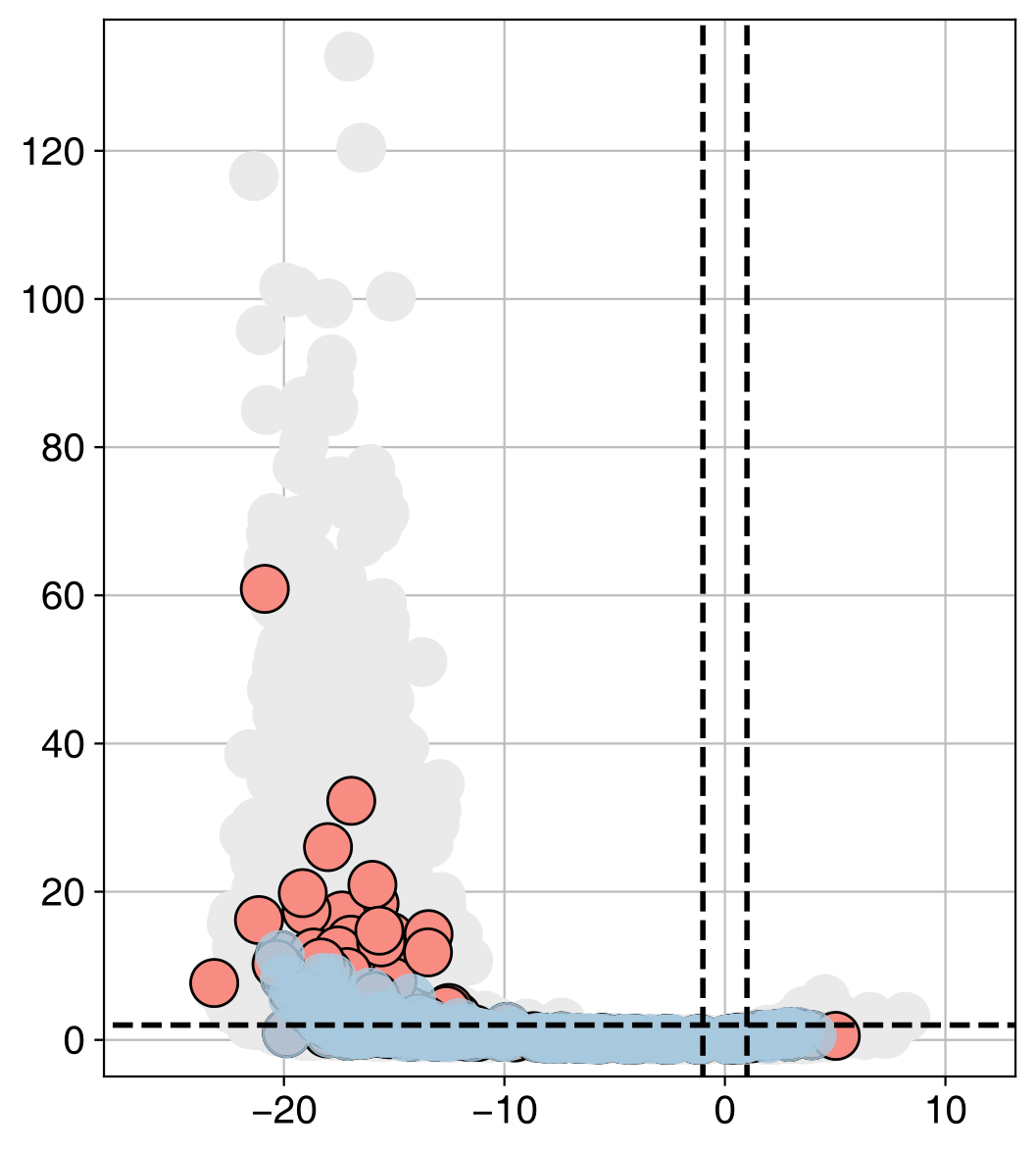}}
\\
\subfloat[Chung]{
		\includegraphics[width=0.16\textwidth]{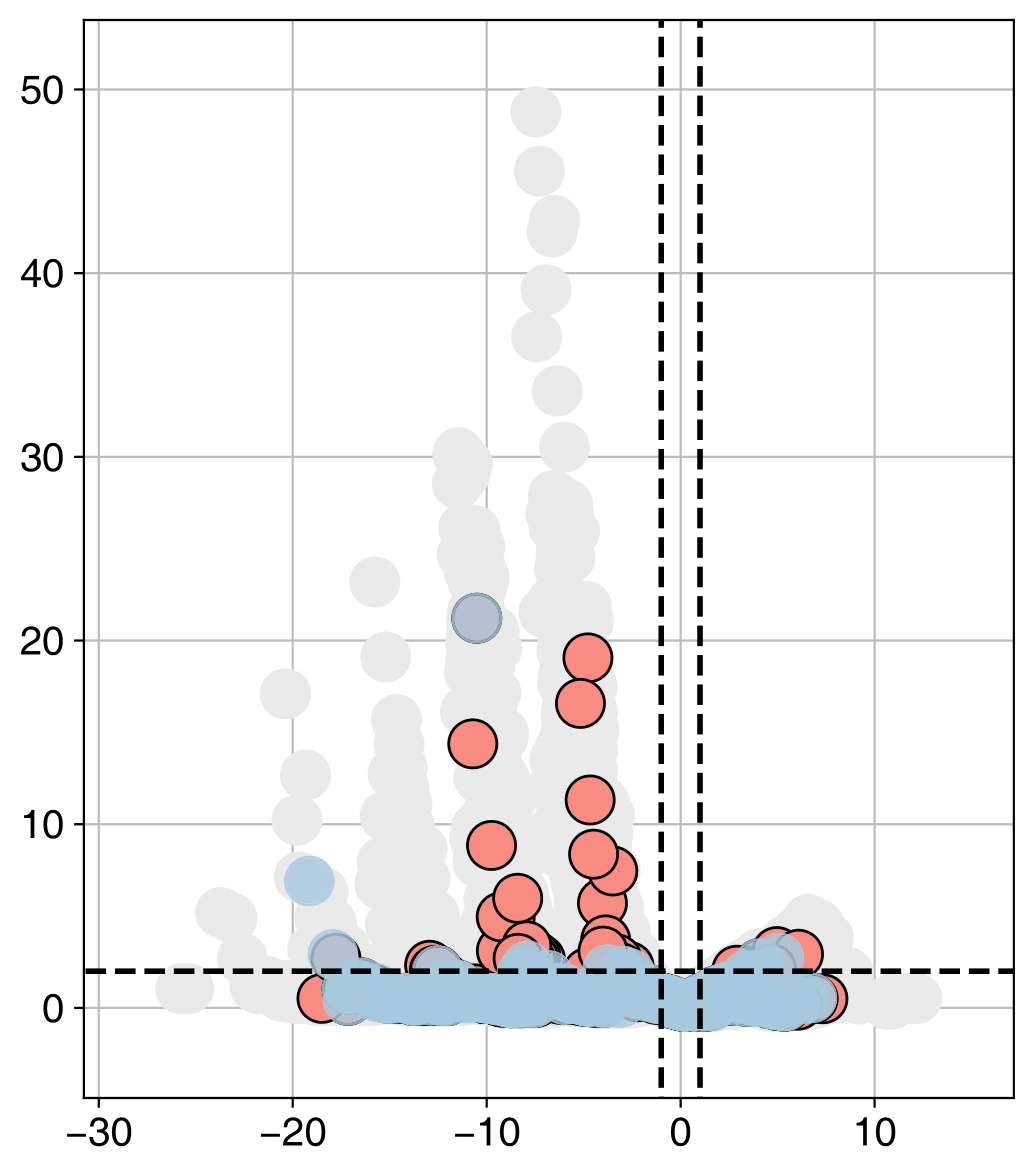}}
\ContinuedFloat
\subfloat[Darmanis]{
		\includegraphics[width=0.16\textwidth]{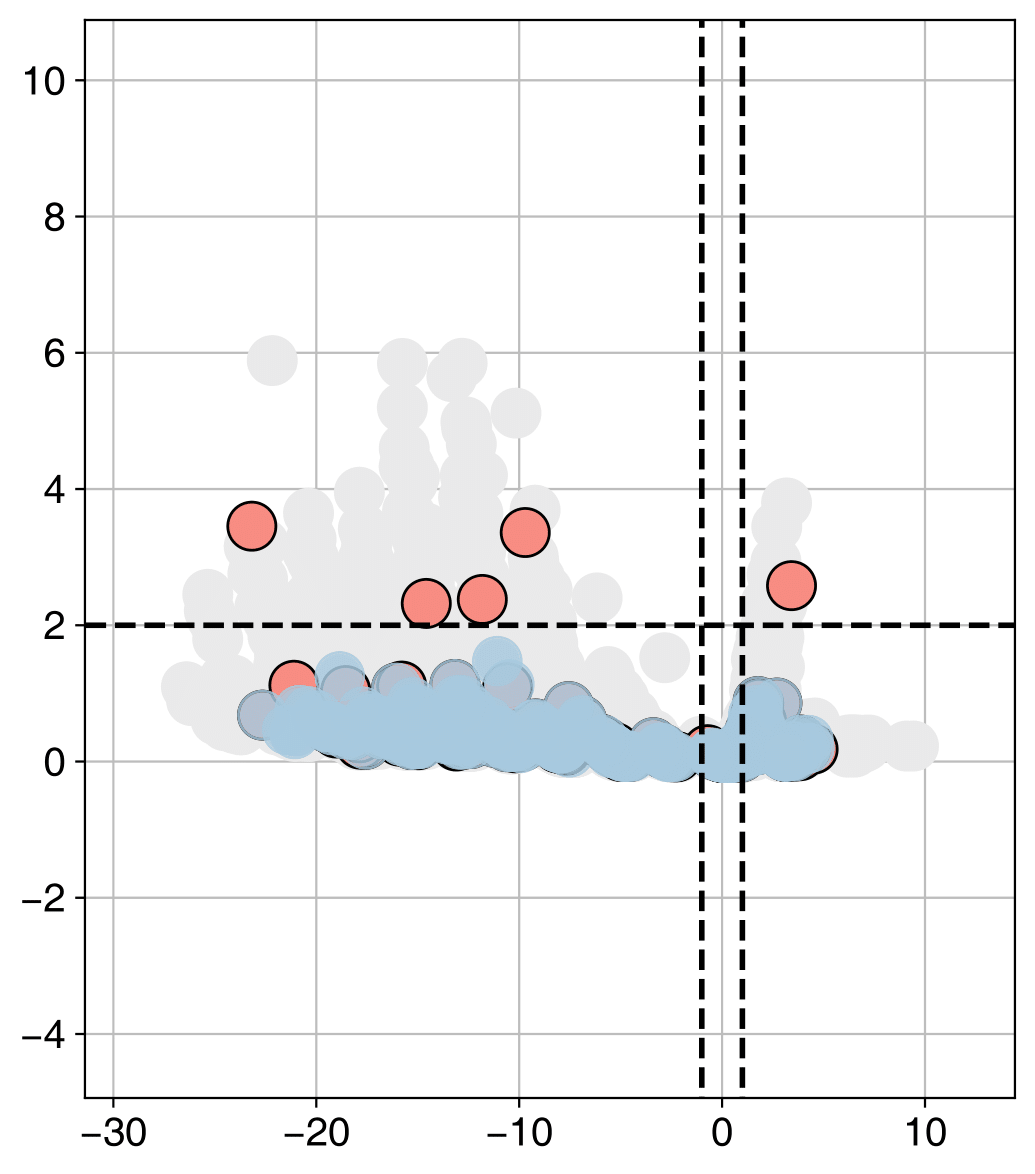}}
\subfloat[Engel]{
		\includegraphics[width=0.16\textwidth]{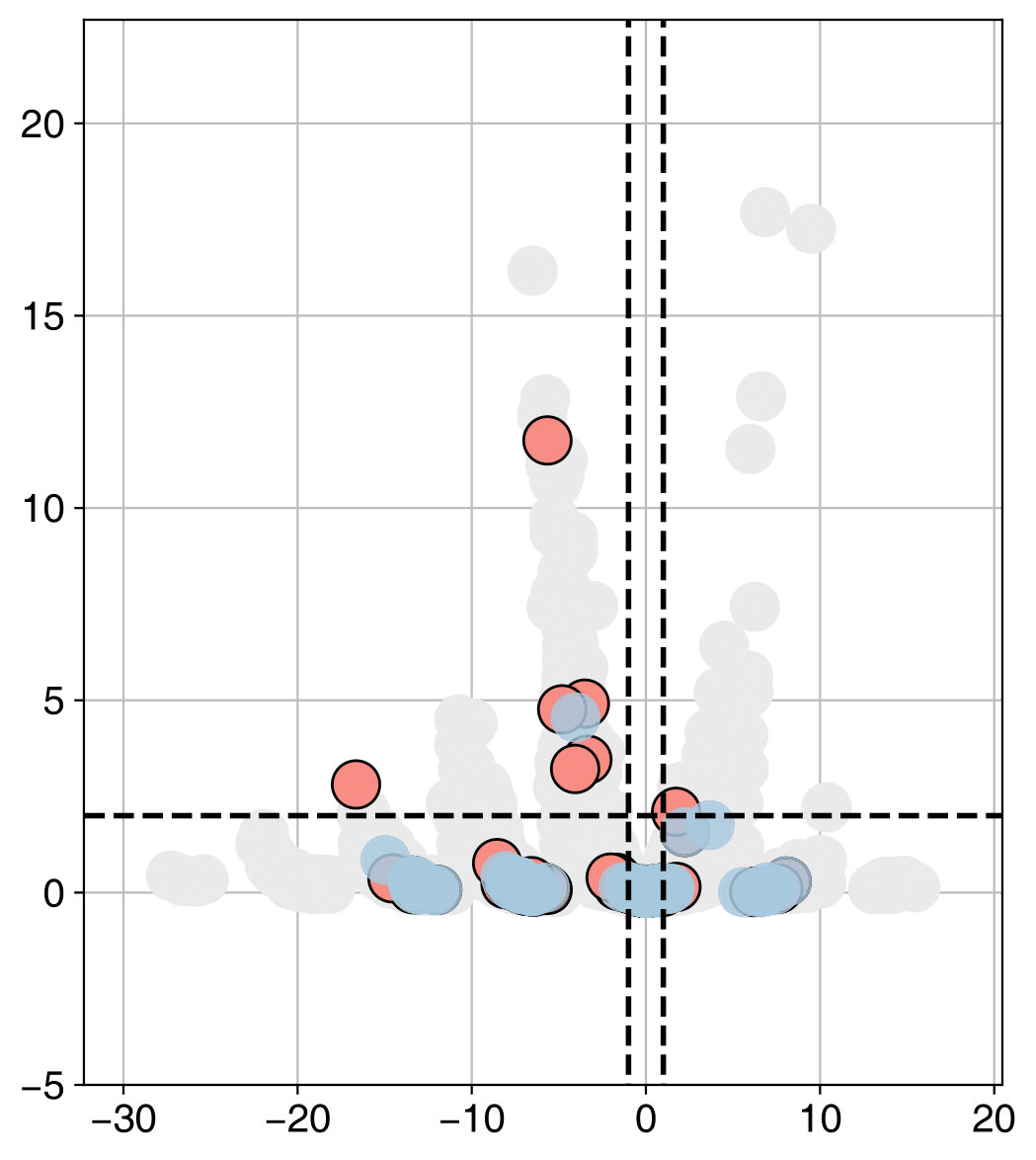}}
\subfloat[Goolam]{
		\includegraphics[width=0.16\textwidth]{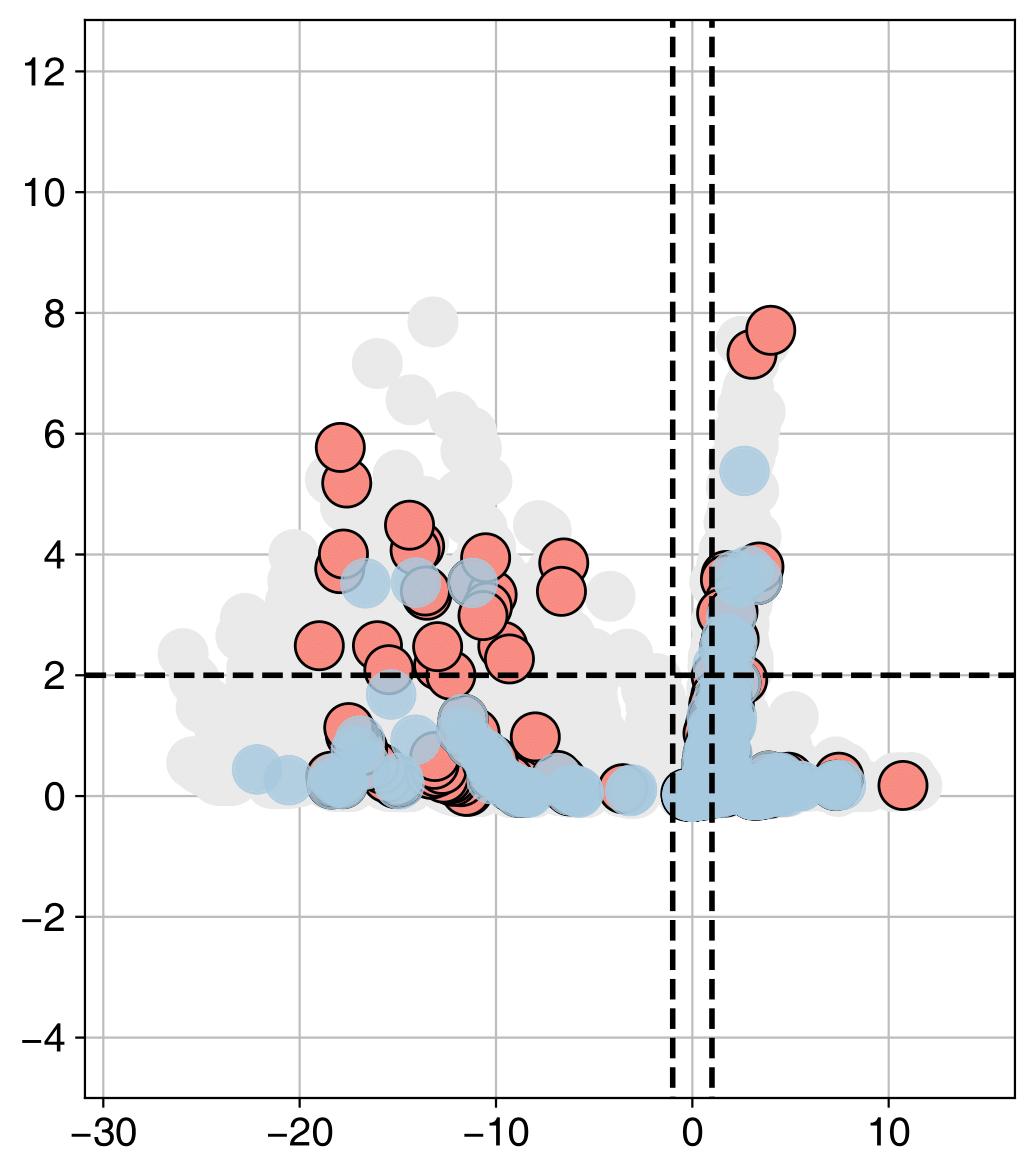}}
\subfloat[Koh]{
		\includegraphics[width=0.16\textwidth]{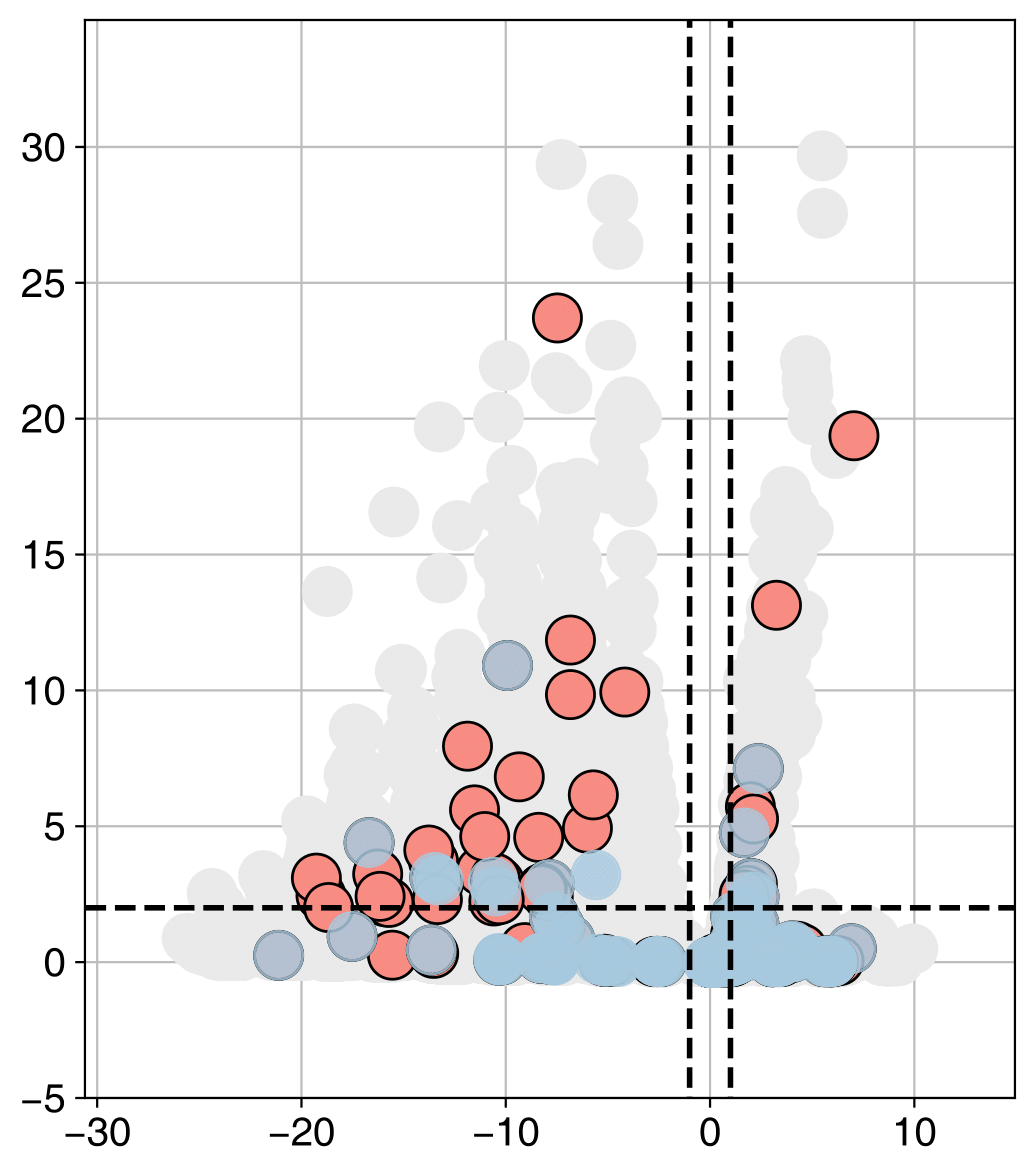}}	
\subfloat[Kumar]{
		\includegraphics[width=0.16\textwidth]{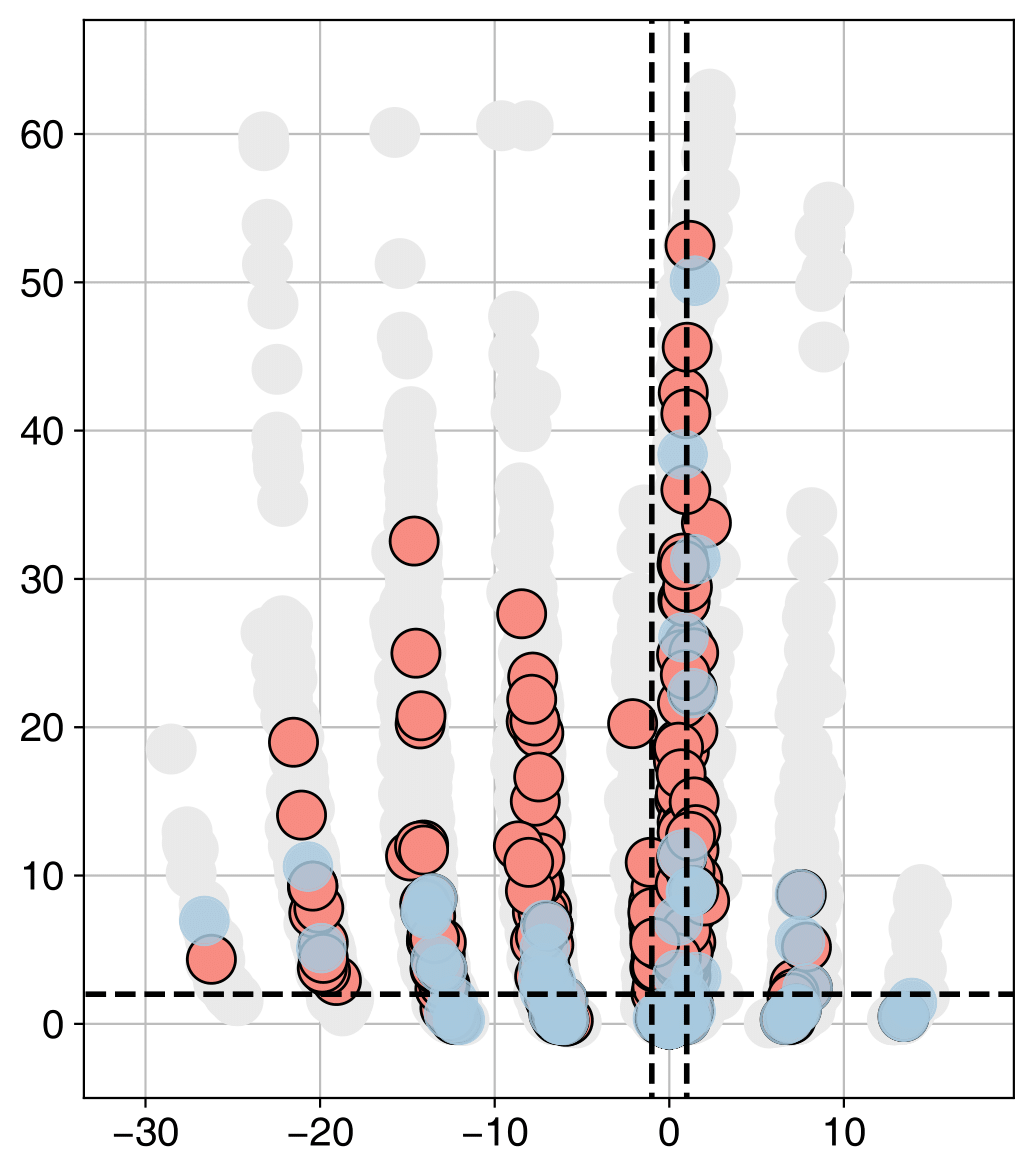}}
\\
\subfloat[Leng]{
		\includegraphics[width=0.16\textwidth]{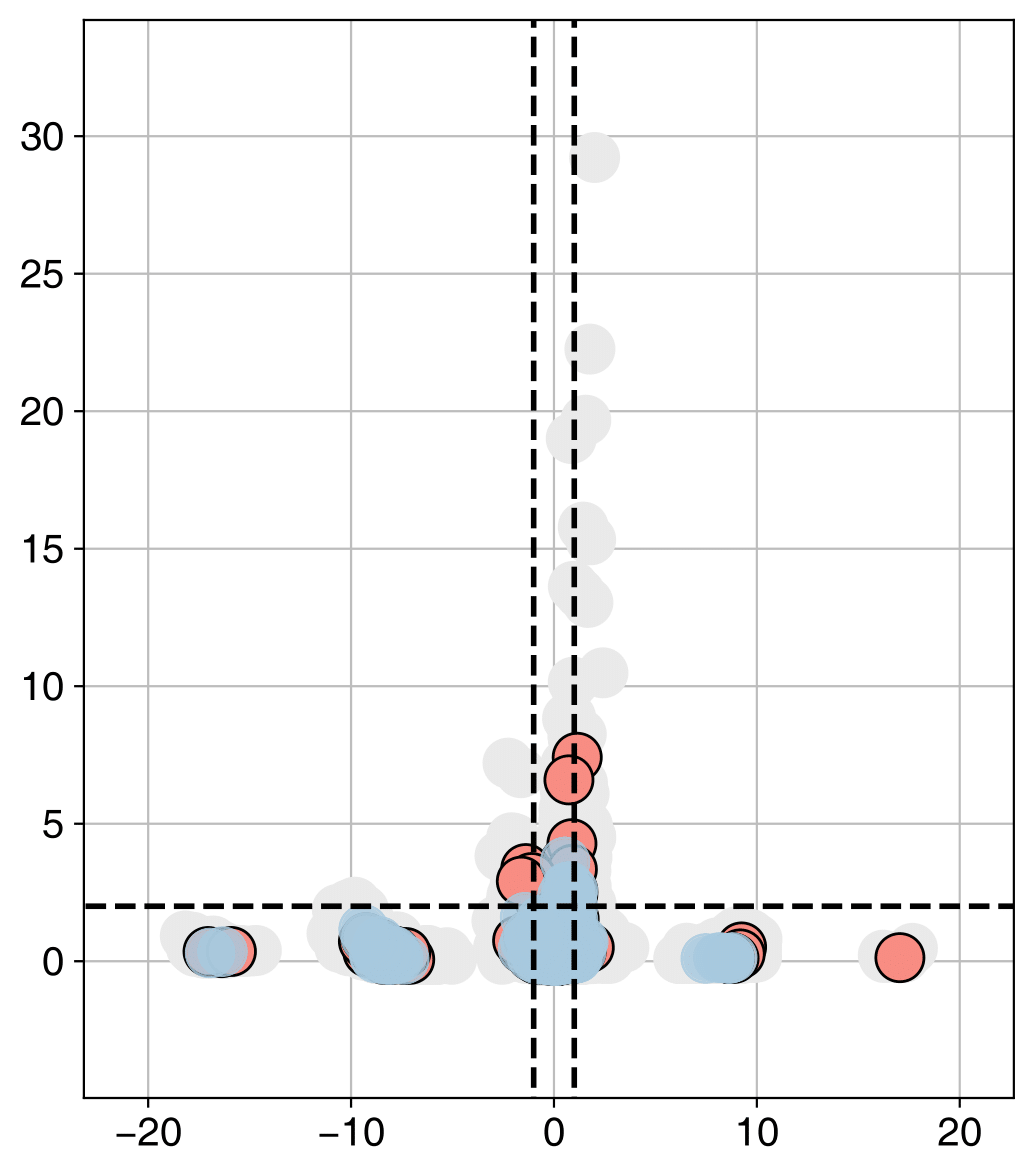}}
\subfloat[Li]{
		\includegraphics[width=0.16\textwidth]{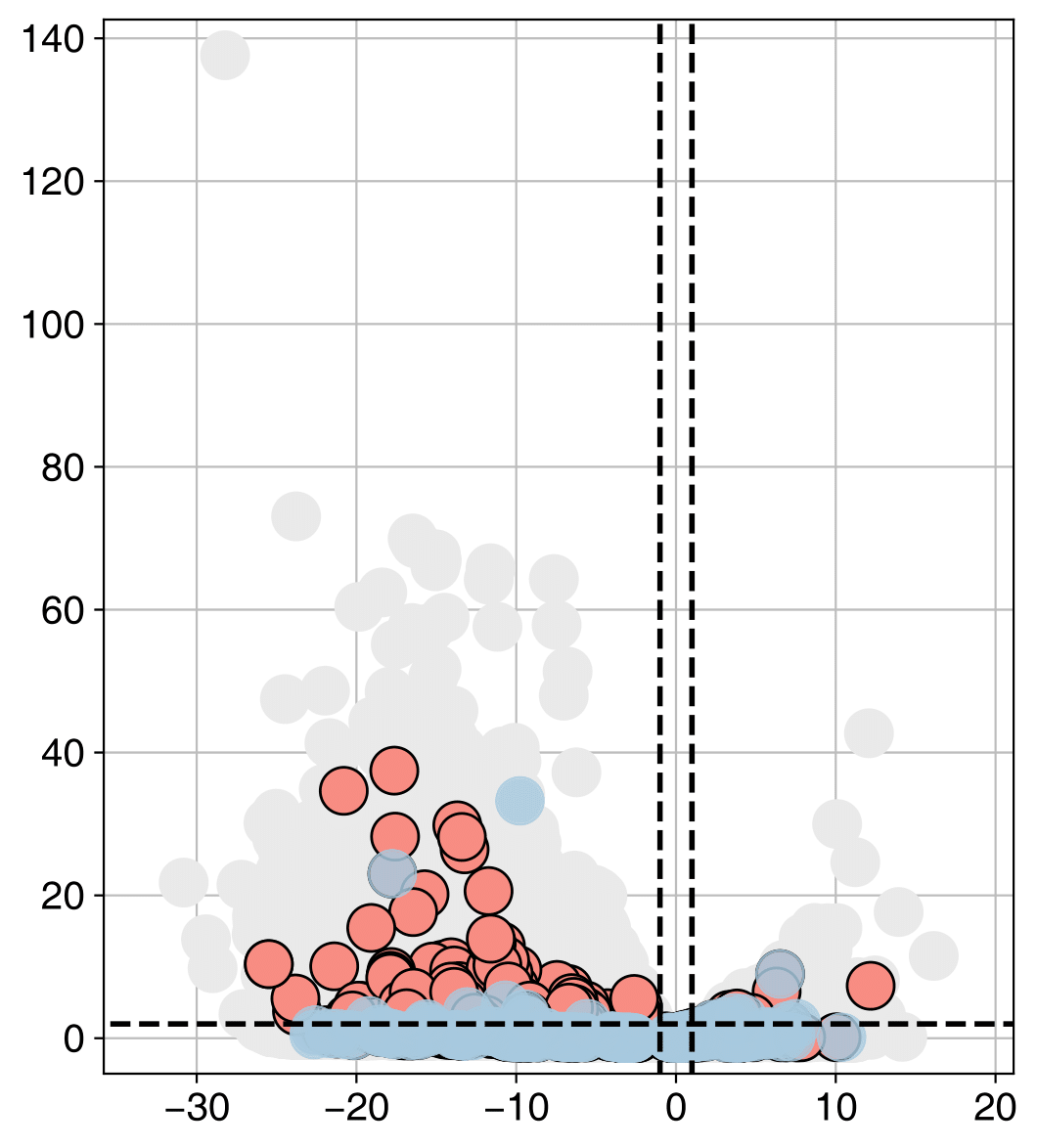}}
\subfloat[Maria1]{
		\includegraphics[width=0.16\textwidth]{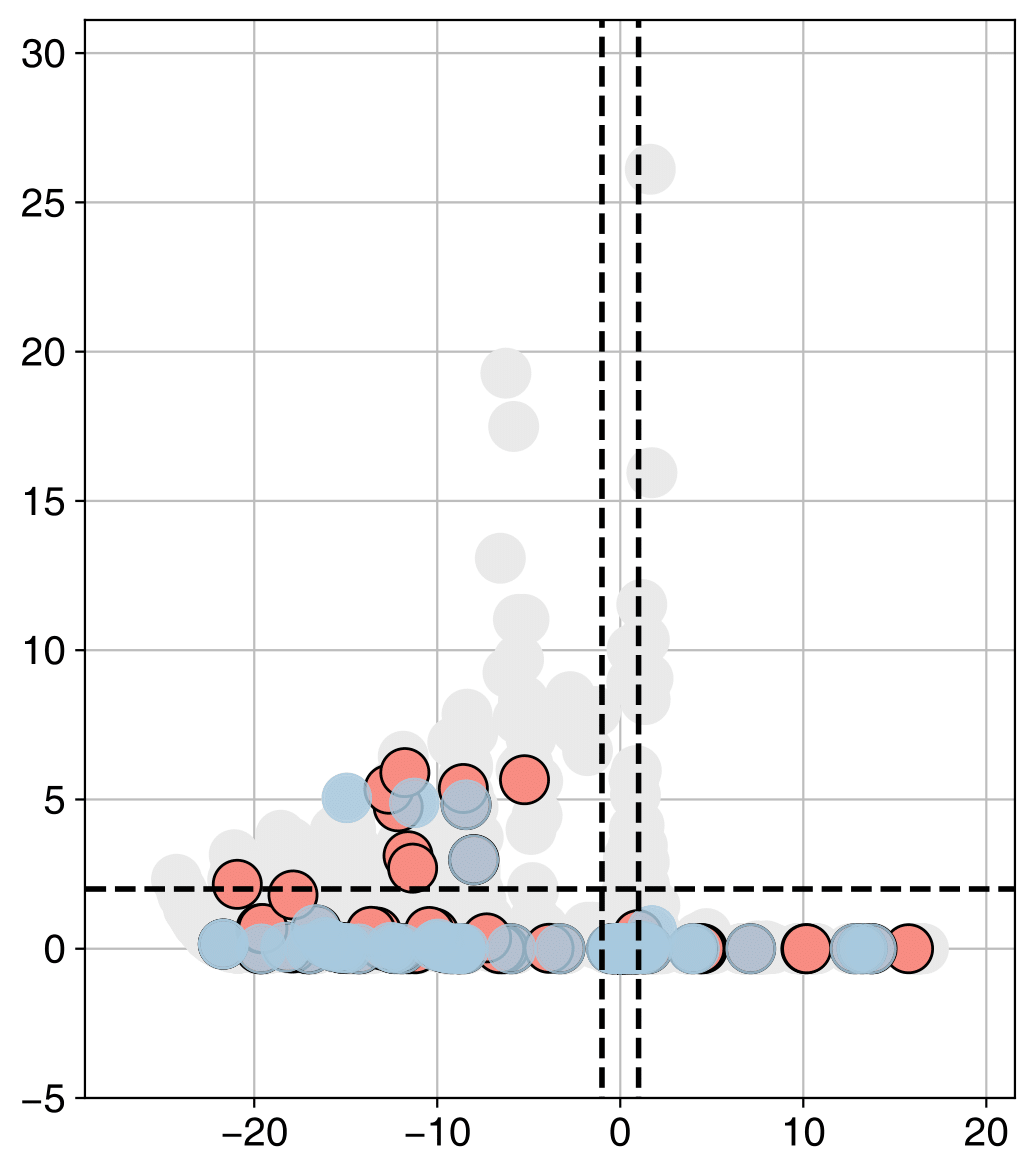}}
\subfloat[Maria2]{
		\includegraphics[width=0.16\textwidth]{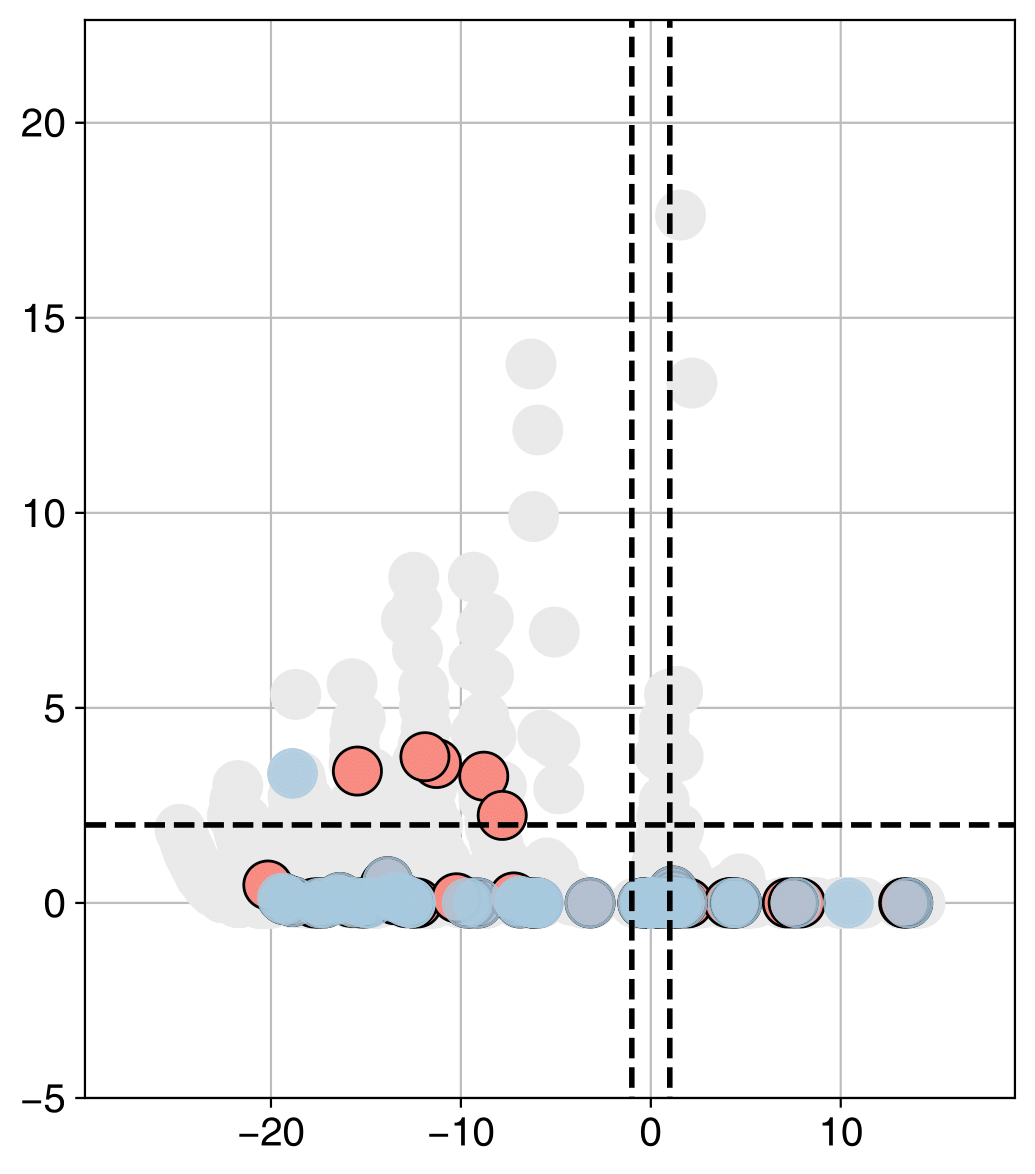}}
\subfloat[Mouse Pancreas1]{
		\includegraphics[width=0.16\textwidth]{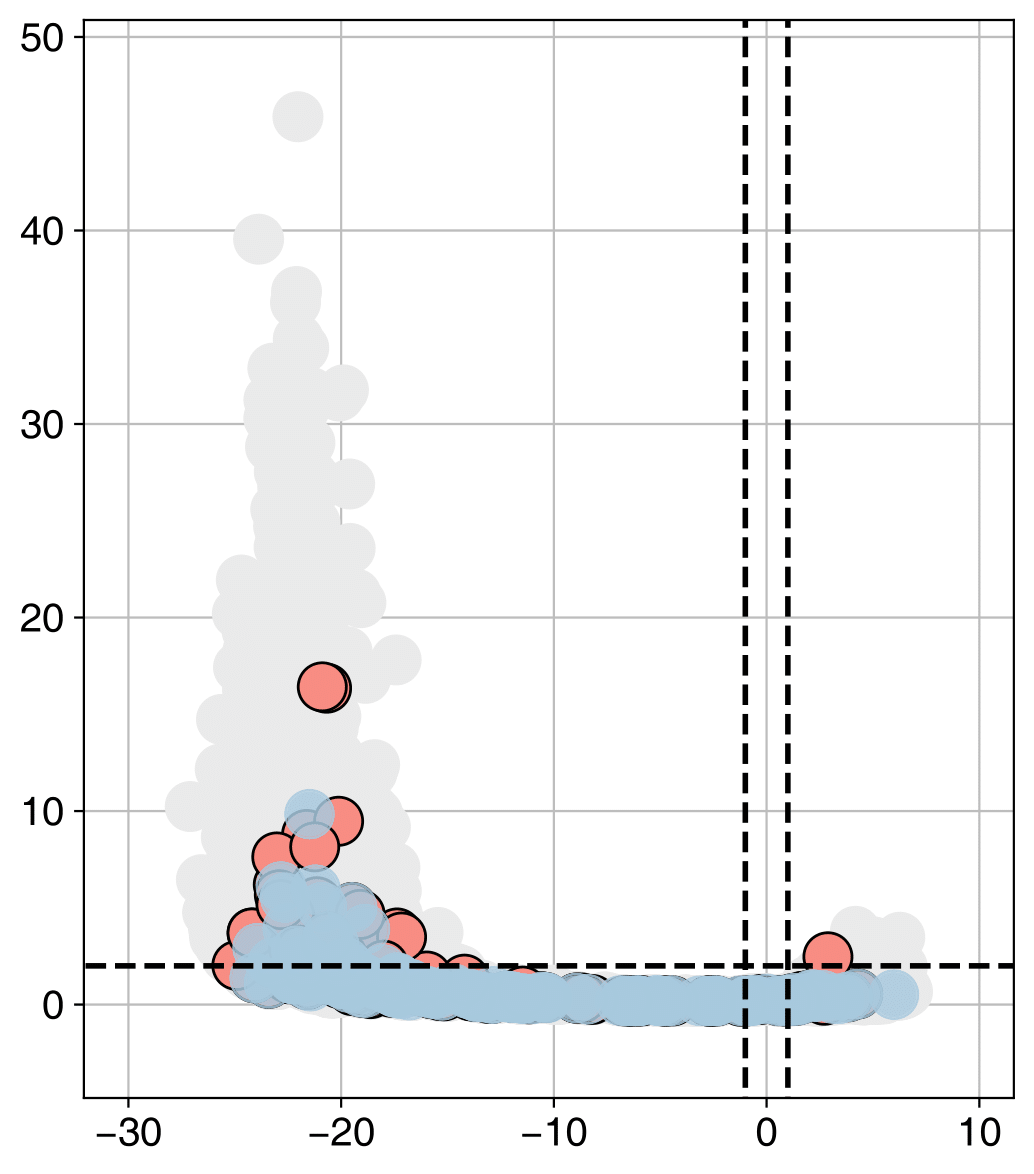}}
\subfloat[MacParland]{
		\includegraphics[width=0.16\textwidth]{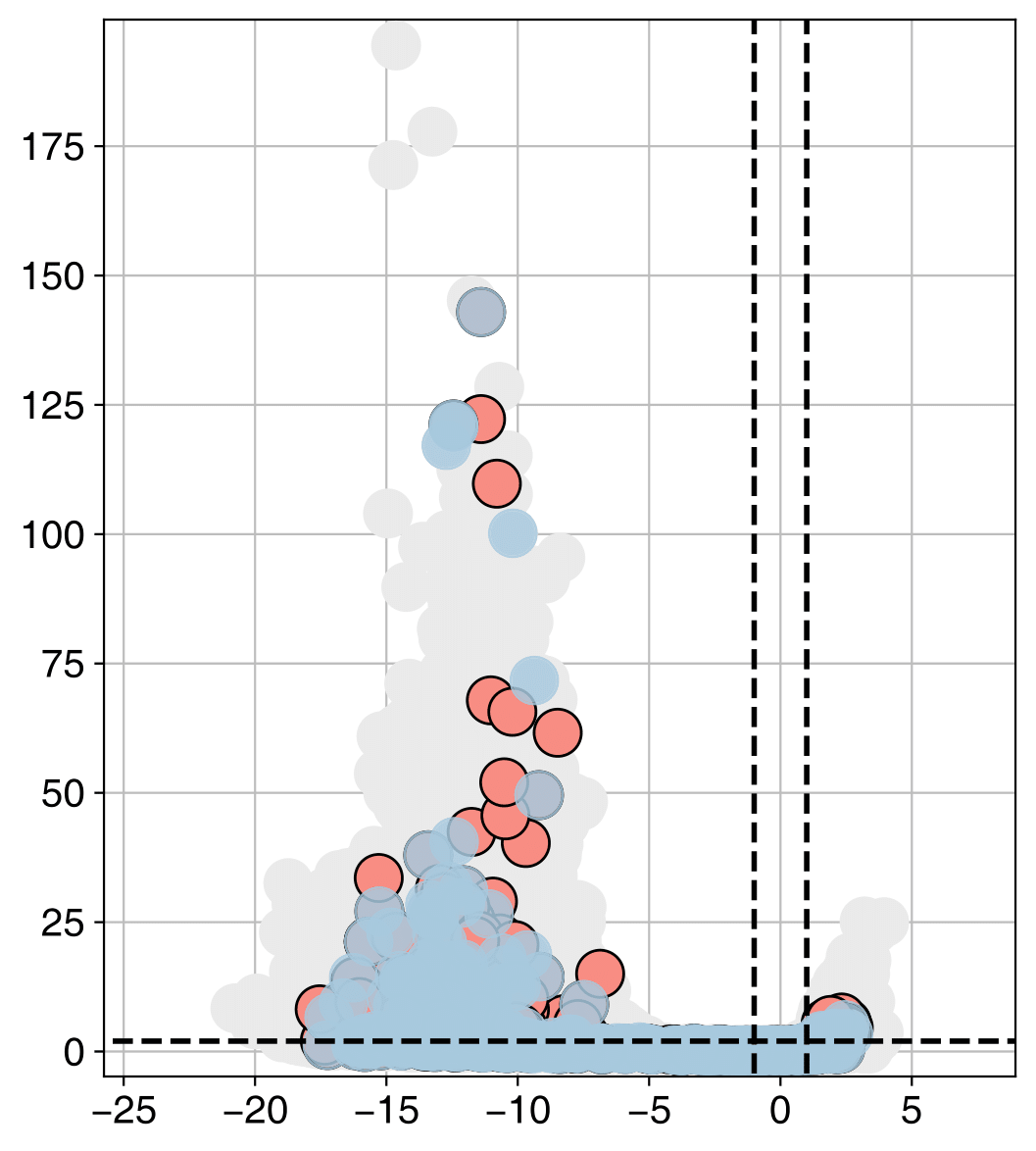}}
\\
\subfloat[Mouse Pancreas2]{
		\includegraphics[width=0.16\textwidth]{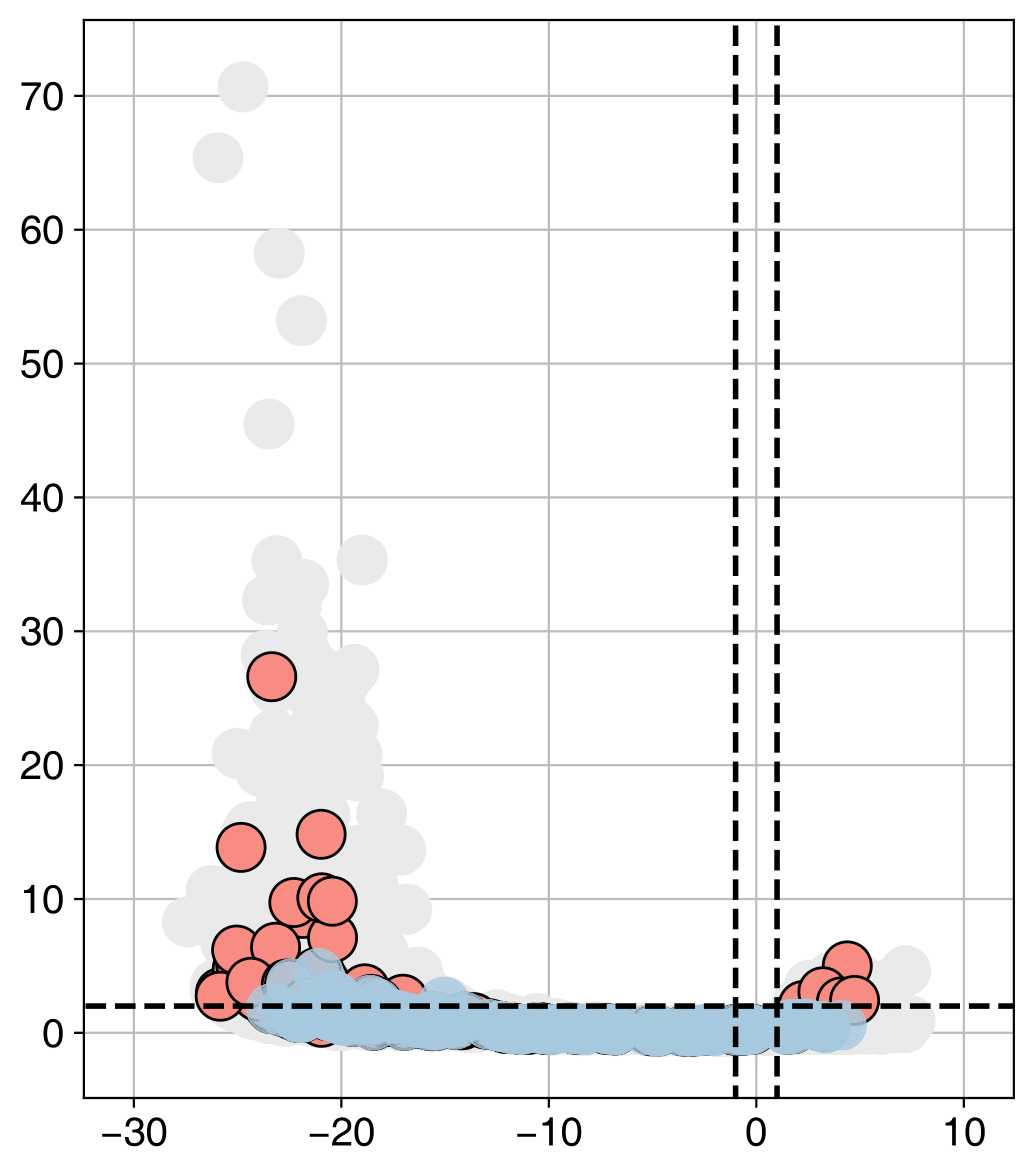}}
\subfloat[Robert]{
		\includegraphics[width=0.16\textwidth]{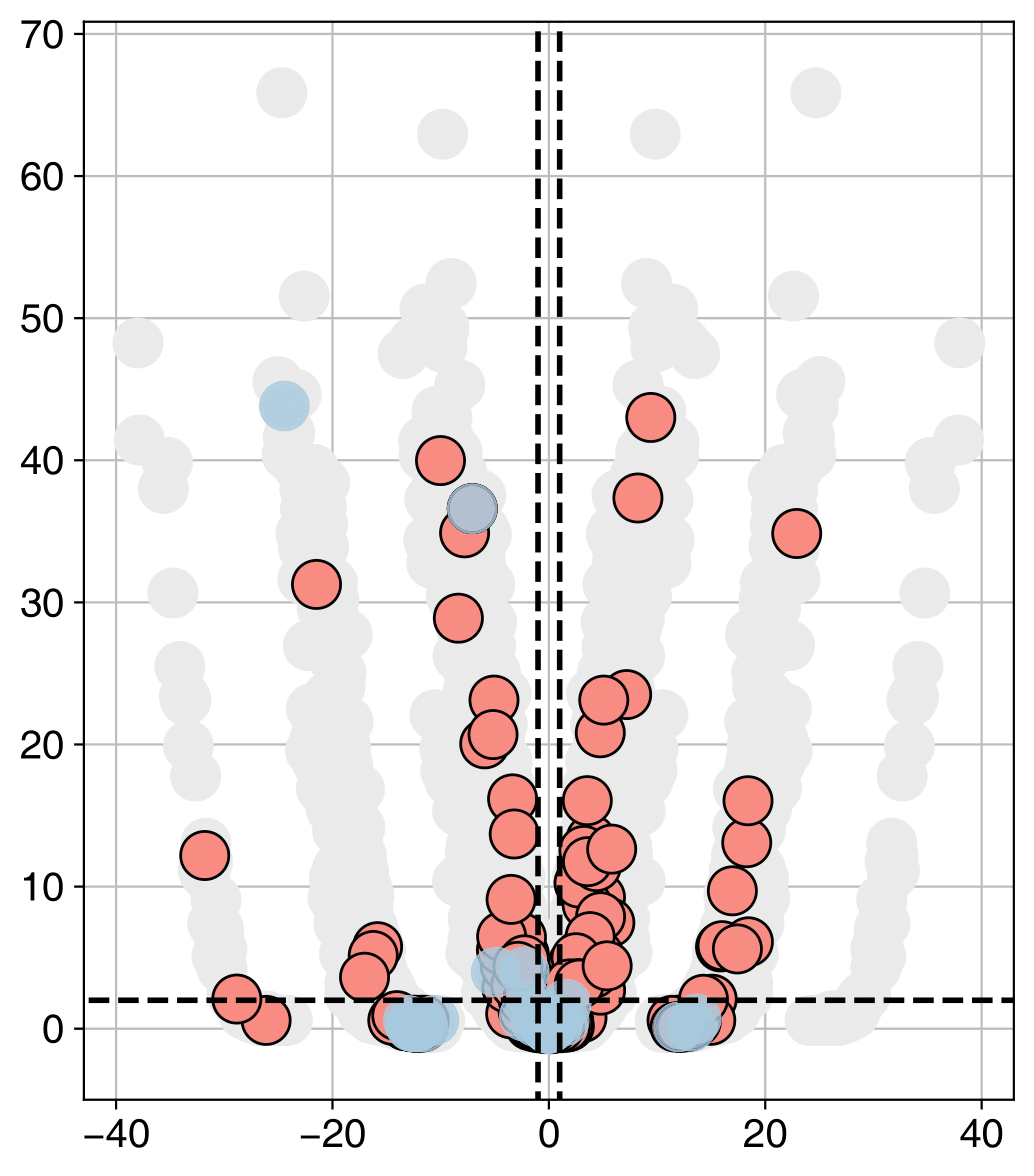}}
\subfloat[Ting]{
		\includegraphics[width=0.16\textwidth]{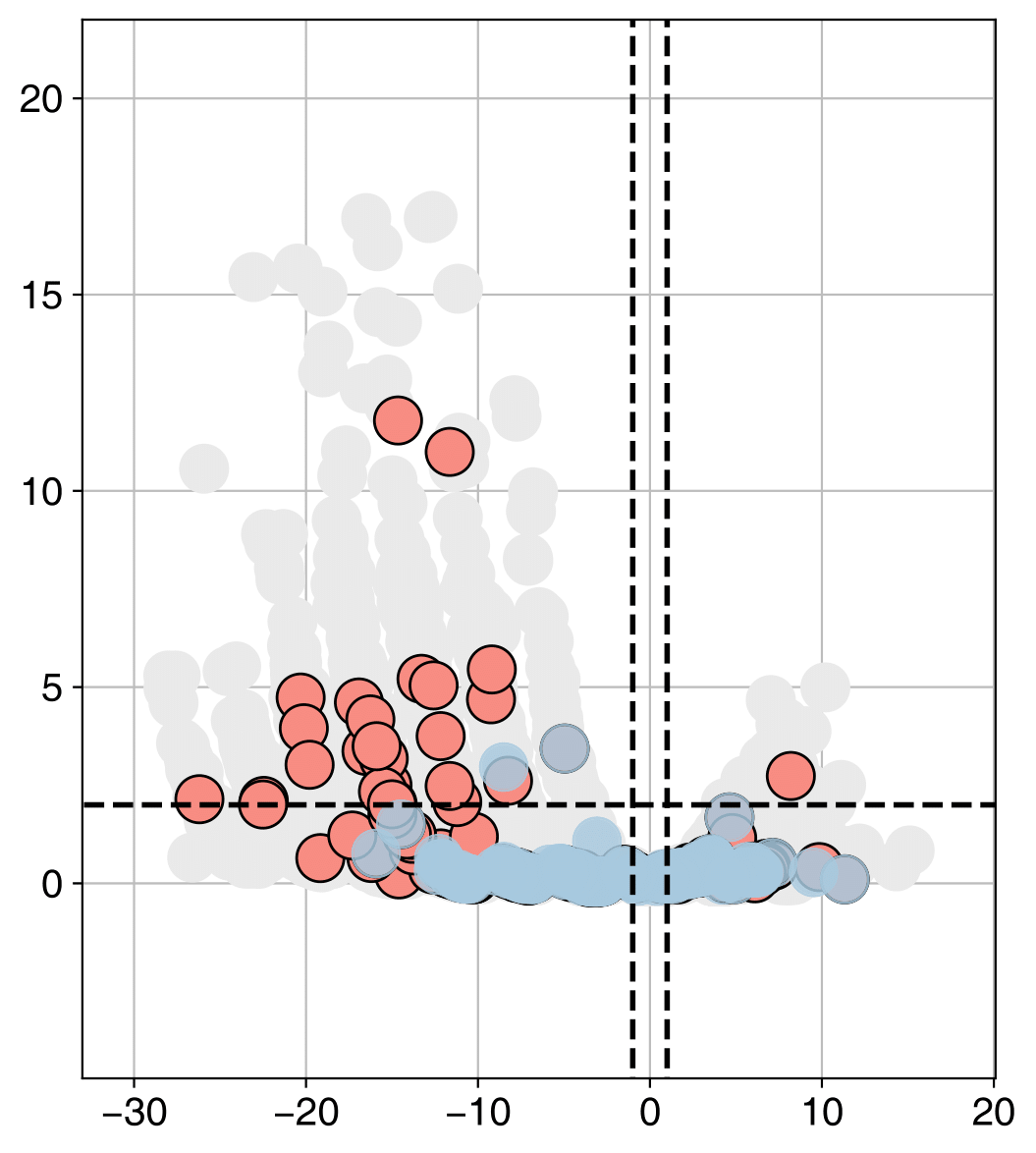}}
\subfloat[Yang]{
		\includegraphics[width=0.16\textwidth]{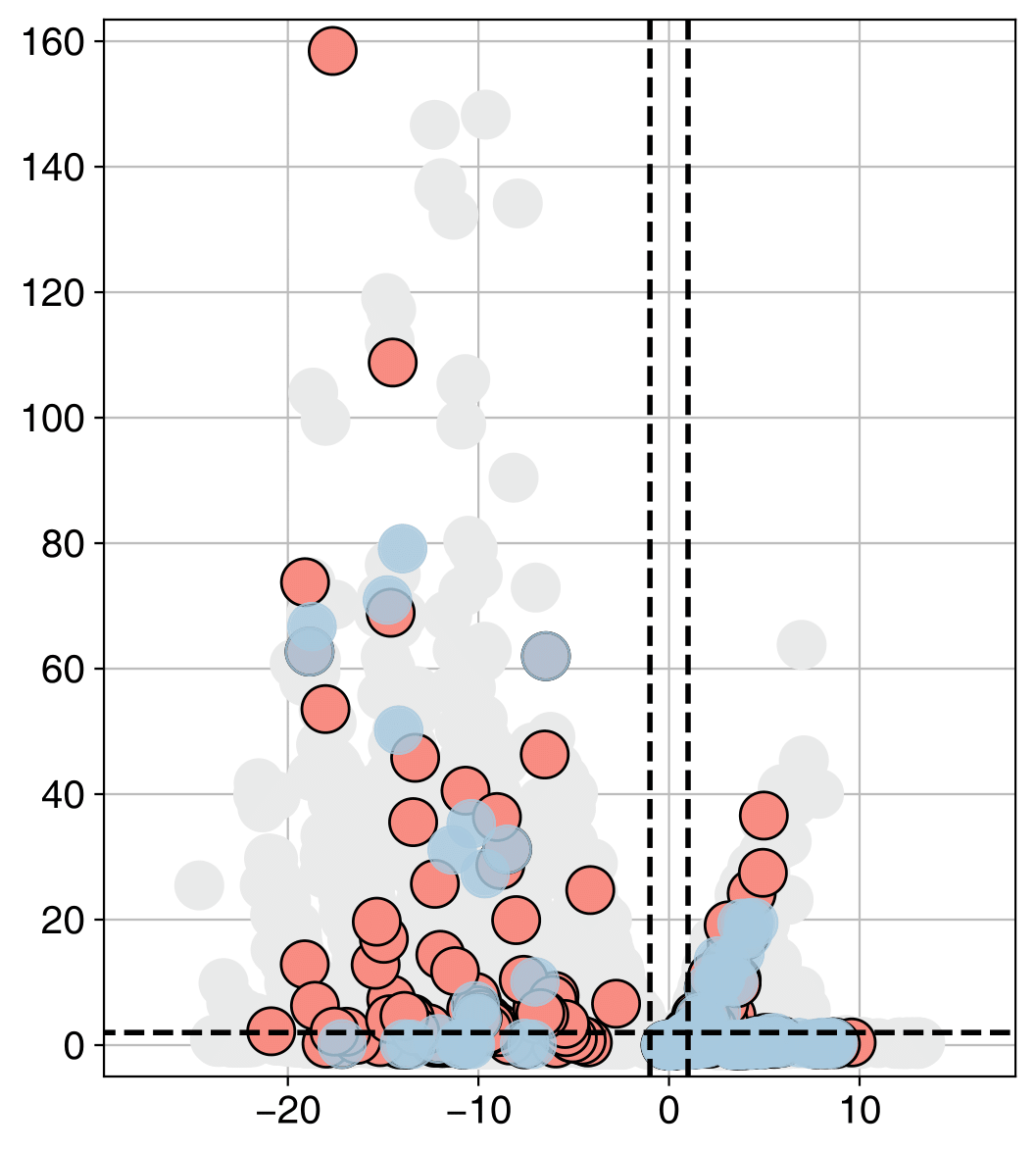}}

\caption{Expression differential analysis of the rest datasets.}
\label{deg_sup}
\end{figure*}

\clearpage

\begin{figure*}[htbp]
\centering
\renewcommand{\thesubfigure}{\arabic{subfigure}}
\subfloat[Cao]{
		\includegraphics[width=0.25\textwidth]{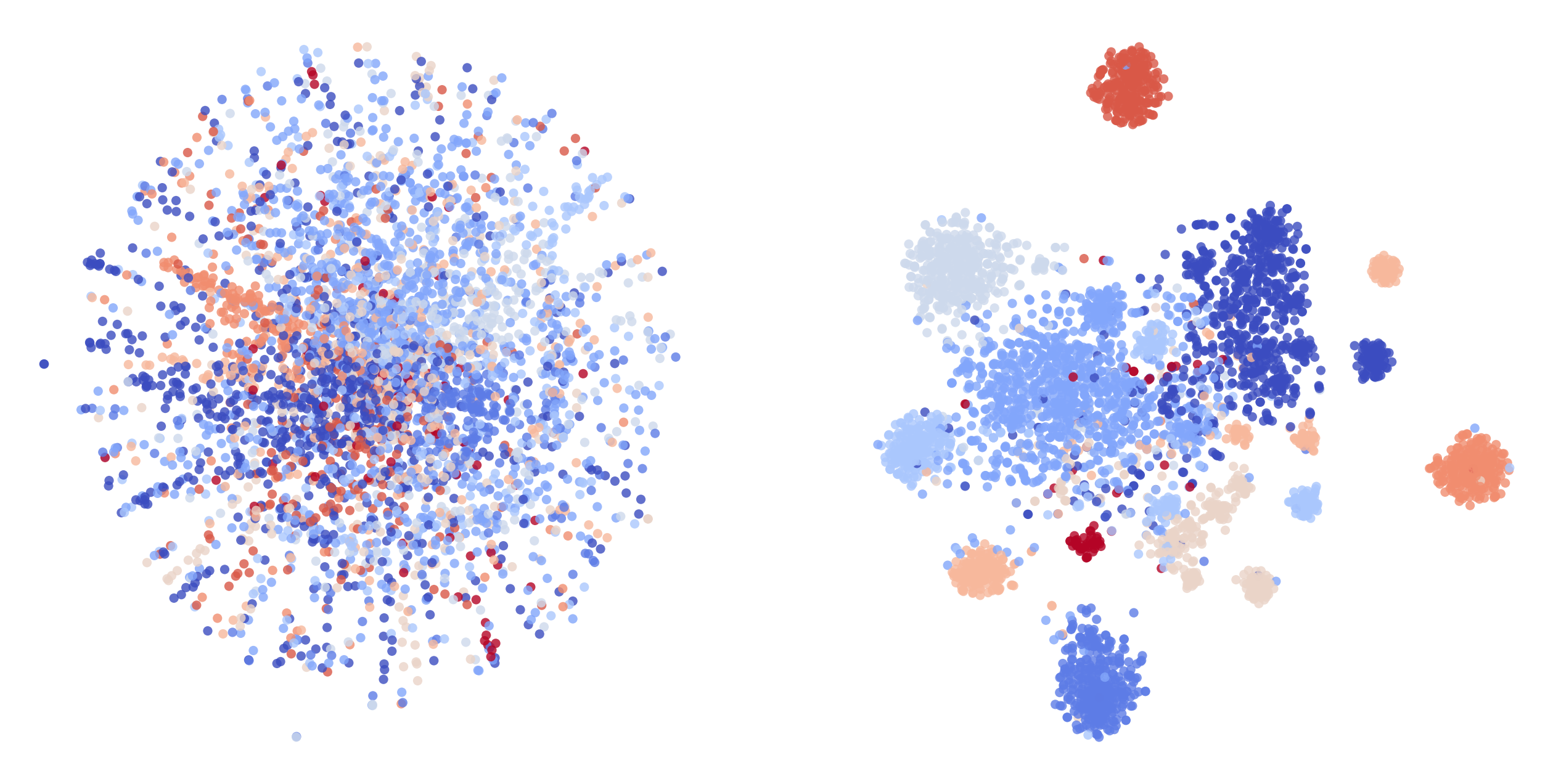}}
\subfloat[Chu1]{
		\includegraphics[width=0.25\textwidth]{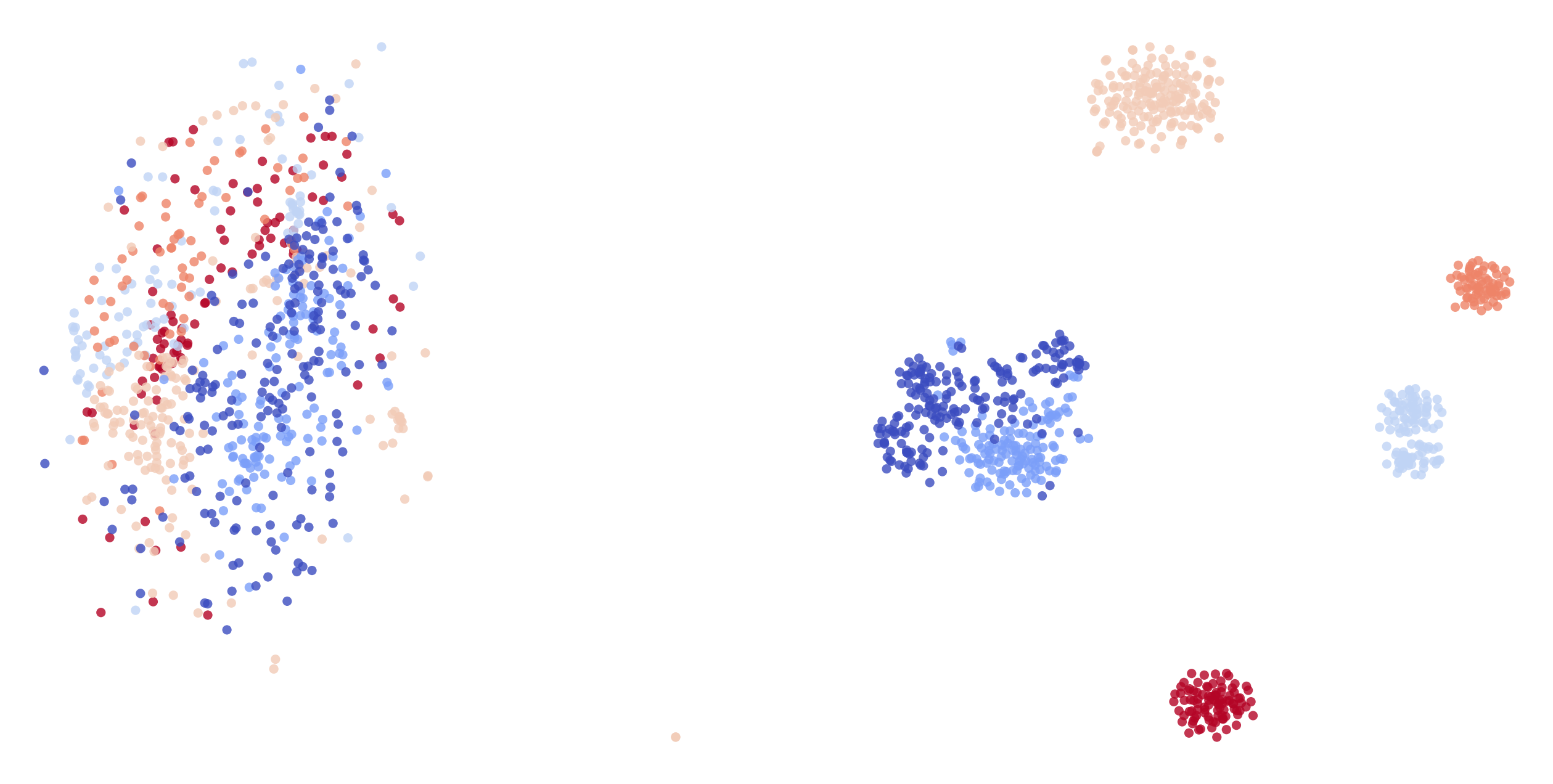}}
\subfloat[Chu2]{
		\includegraphics[width=0.25\textwidth]{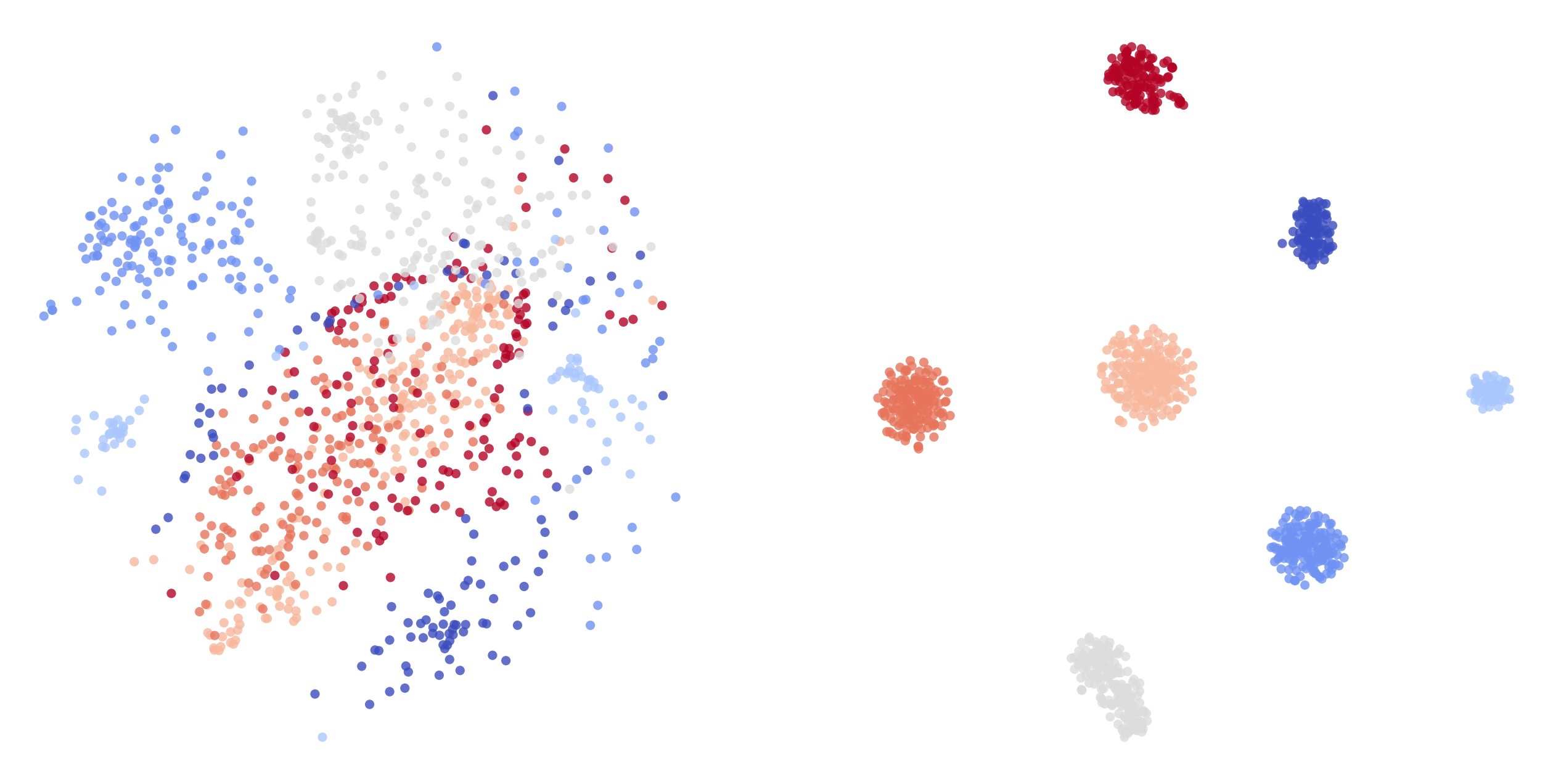}}
\subfloat[Han]{
		\includegraphics[width=0.25\textwidth]{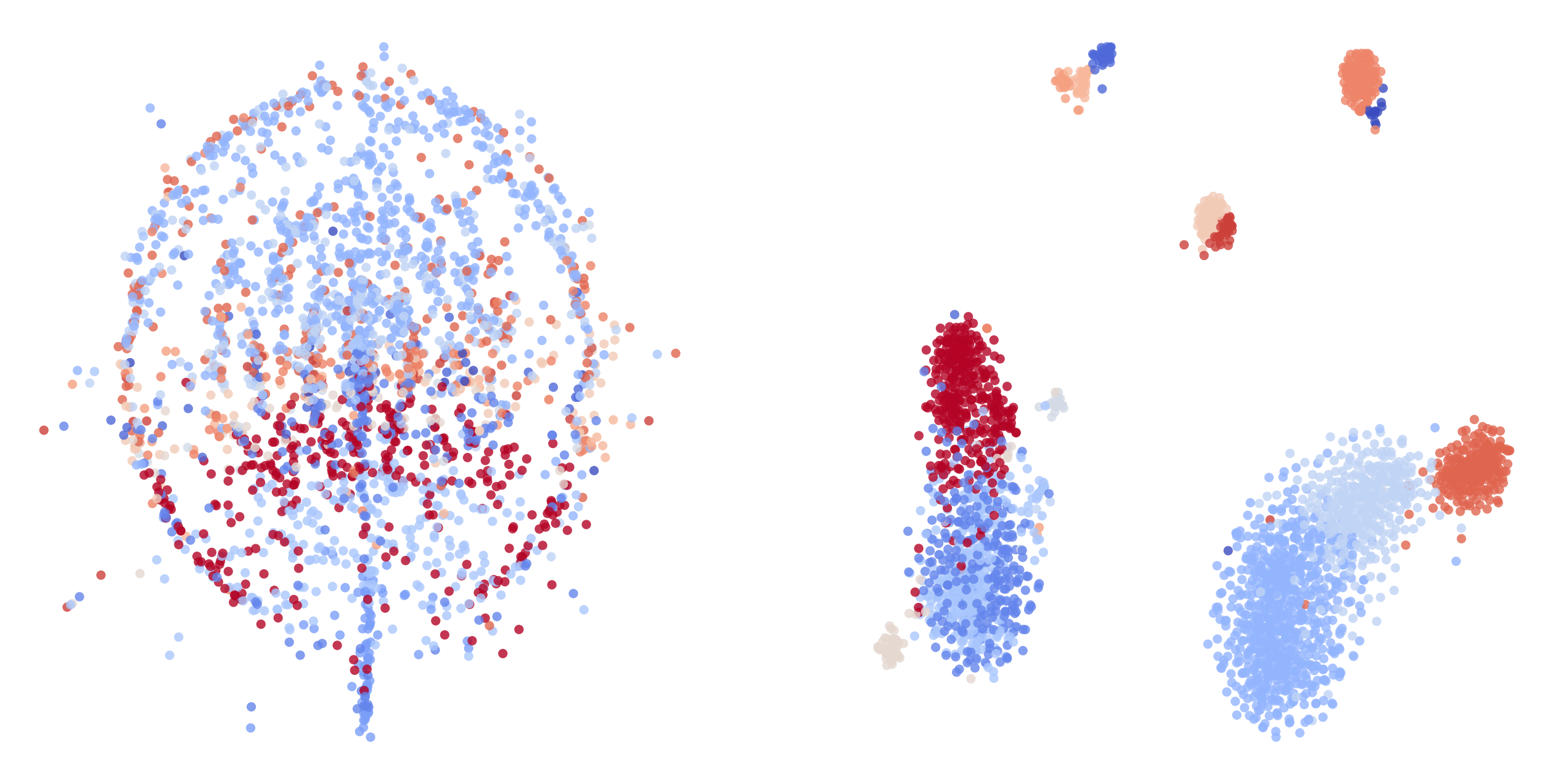}}
\\
\subfloat[Human Pancreas1]{
		\includegraphics[width=0.25\textwidth]{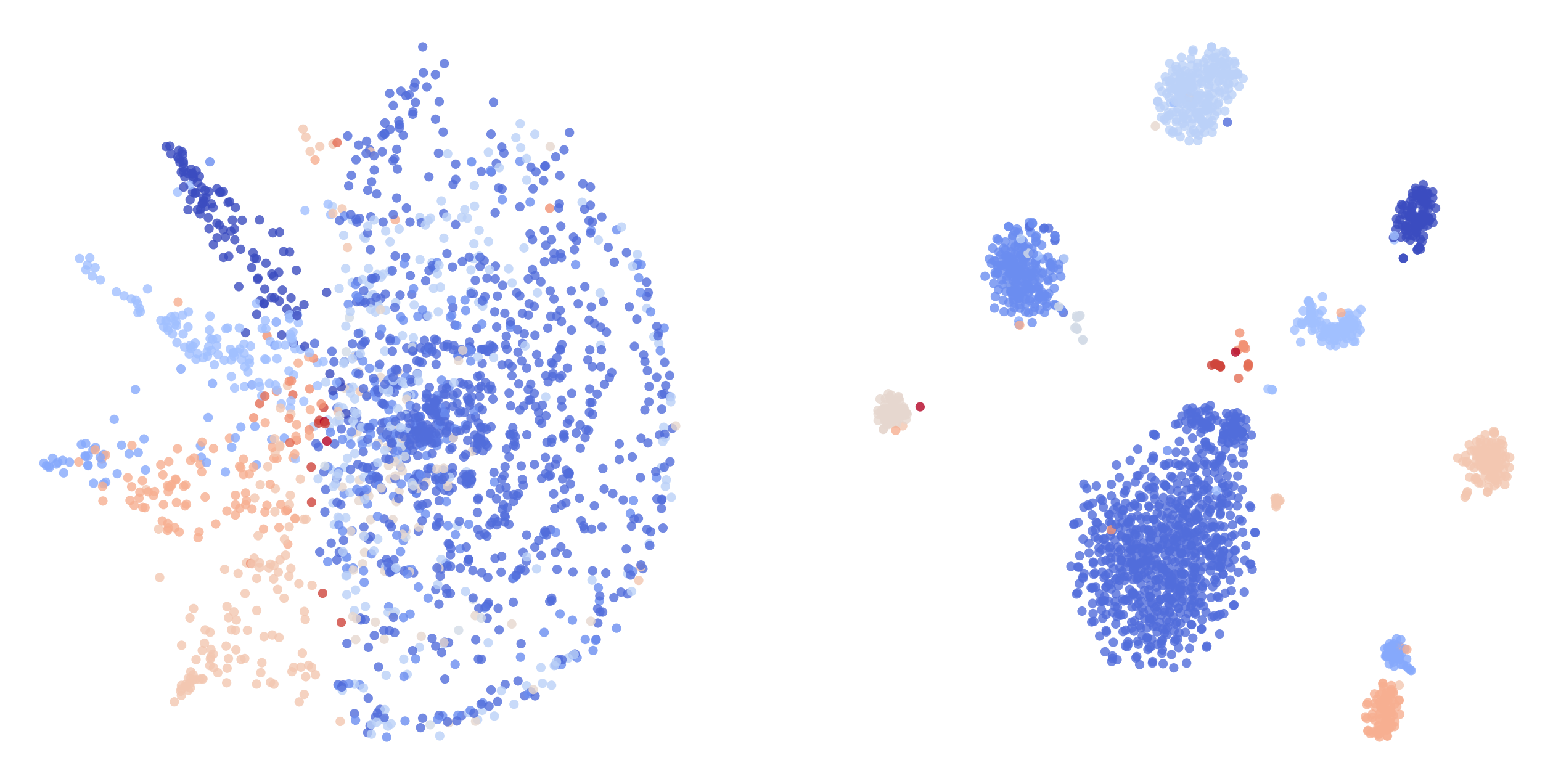}}
\subfloat[Human Pancreas2]{
		\includegraphics[width=0.25\textwidth]{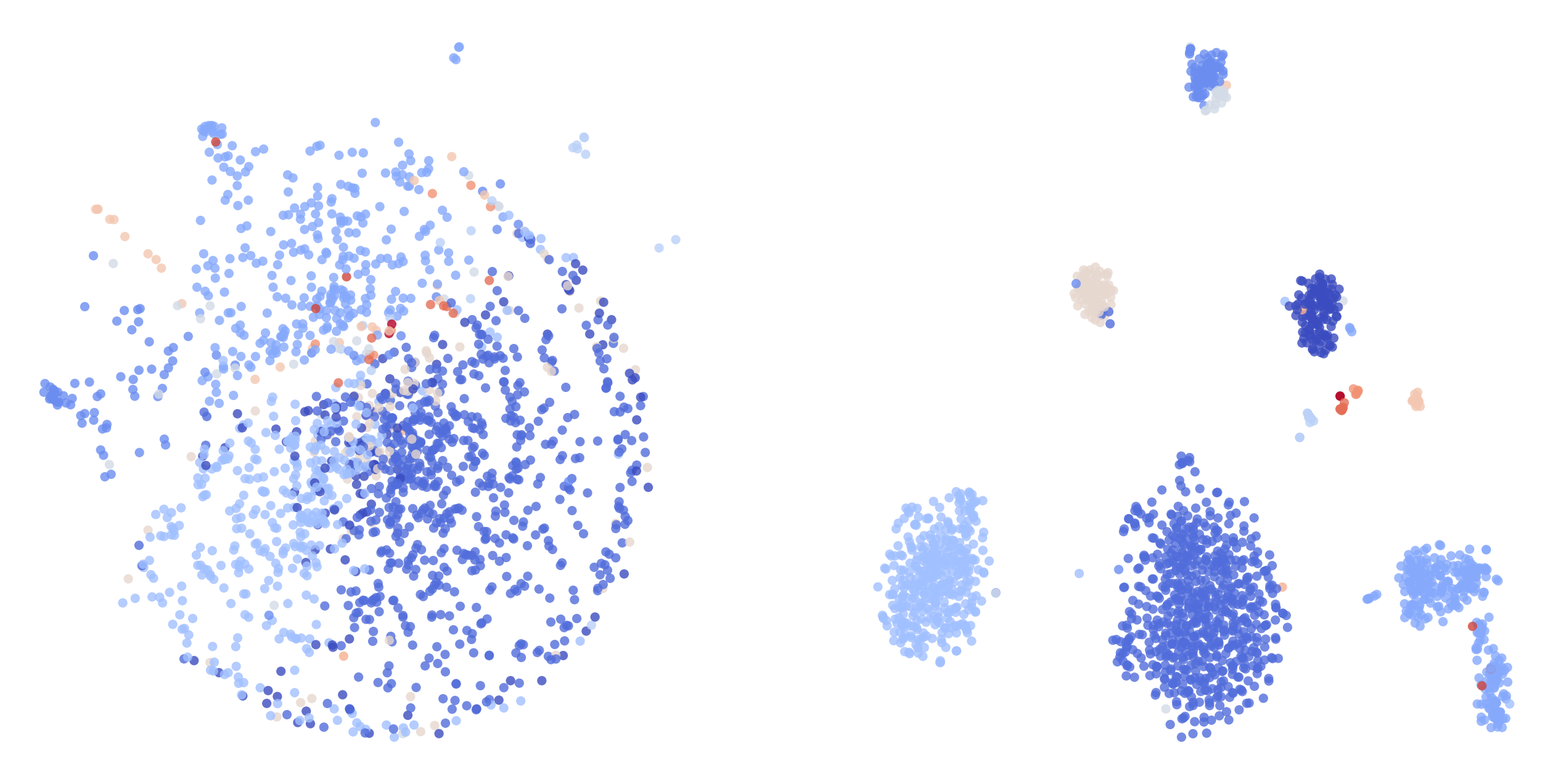}}
\subfloat[Human Pancreas3]{
		\includegraphics[width=0.25\textwidth]{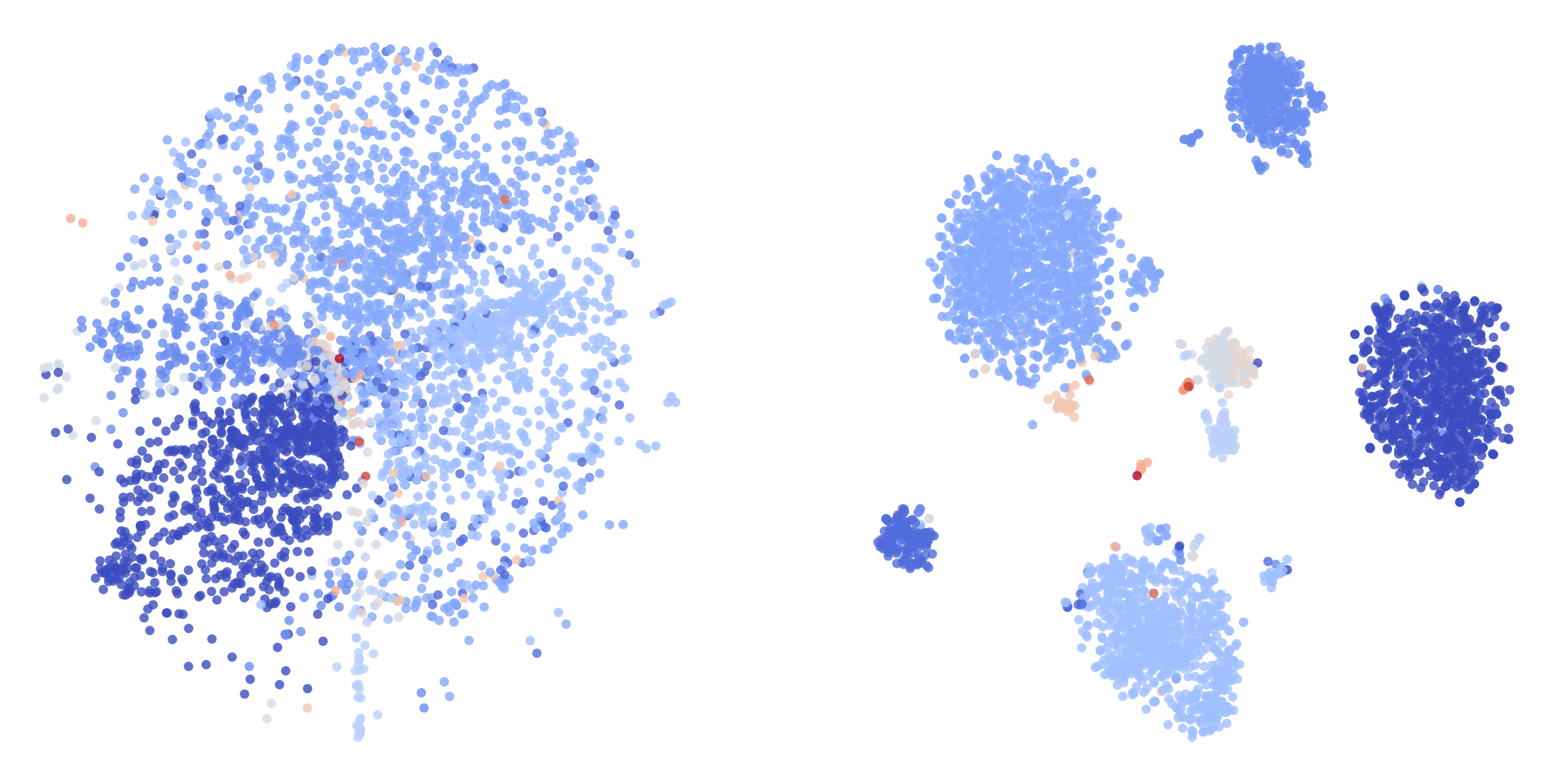}}
\subfloat[Chung]{
		\includegraphics[width=0.25\textwidth]{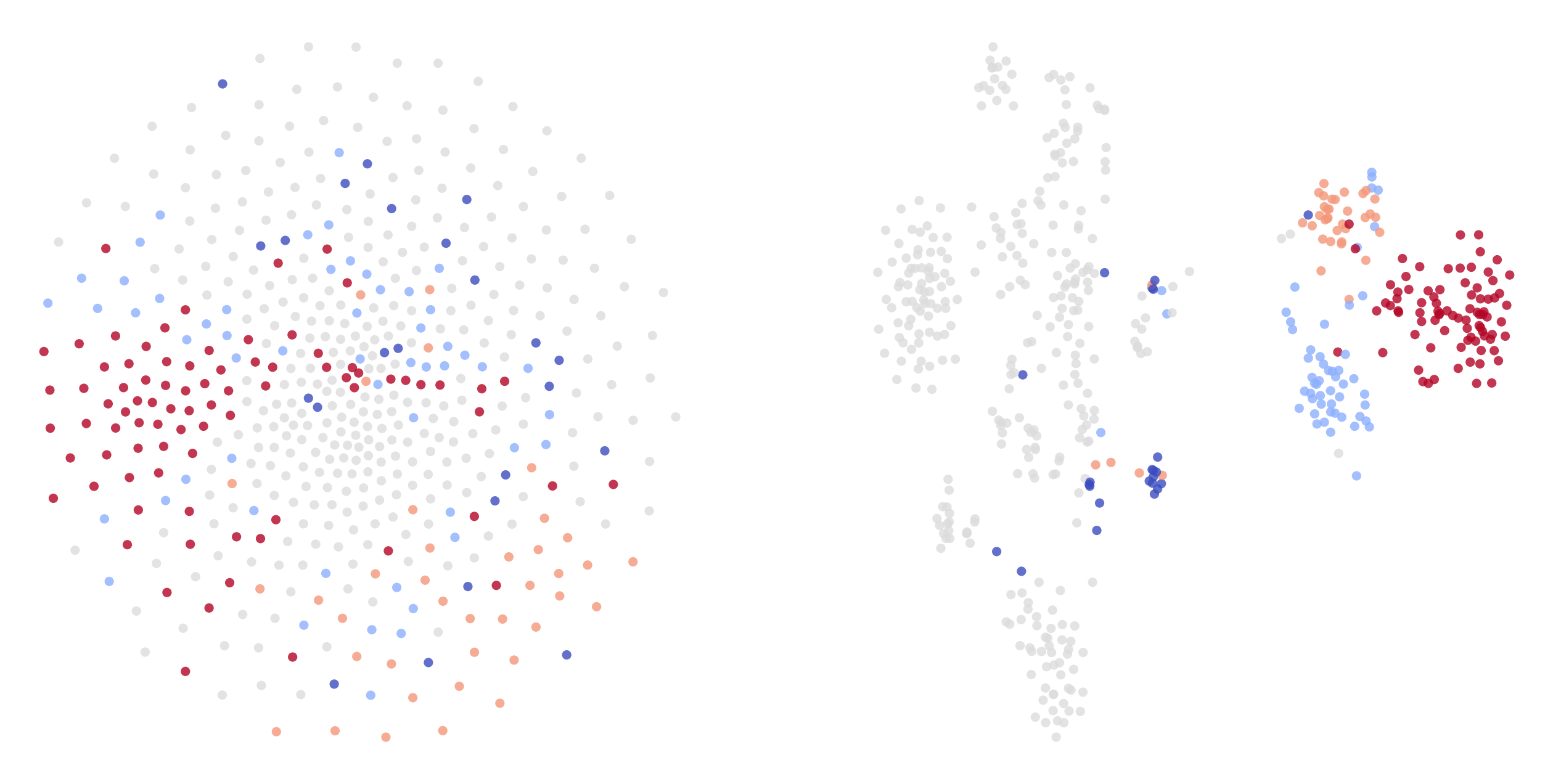}}
\\
\subfloat[Darmanis]{
		\includegraphics[width=0.25\textwidth]{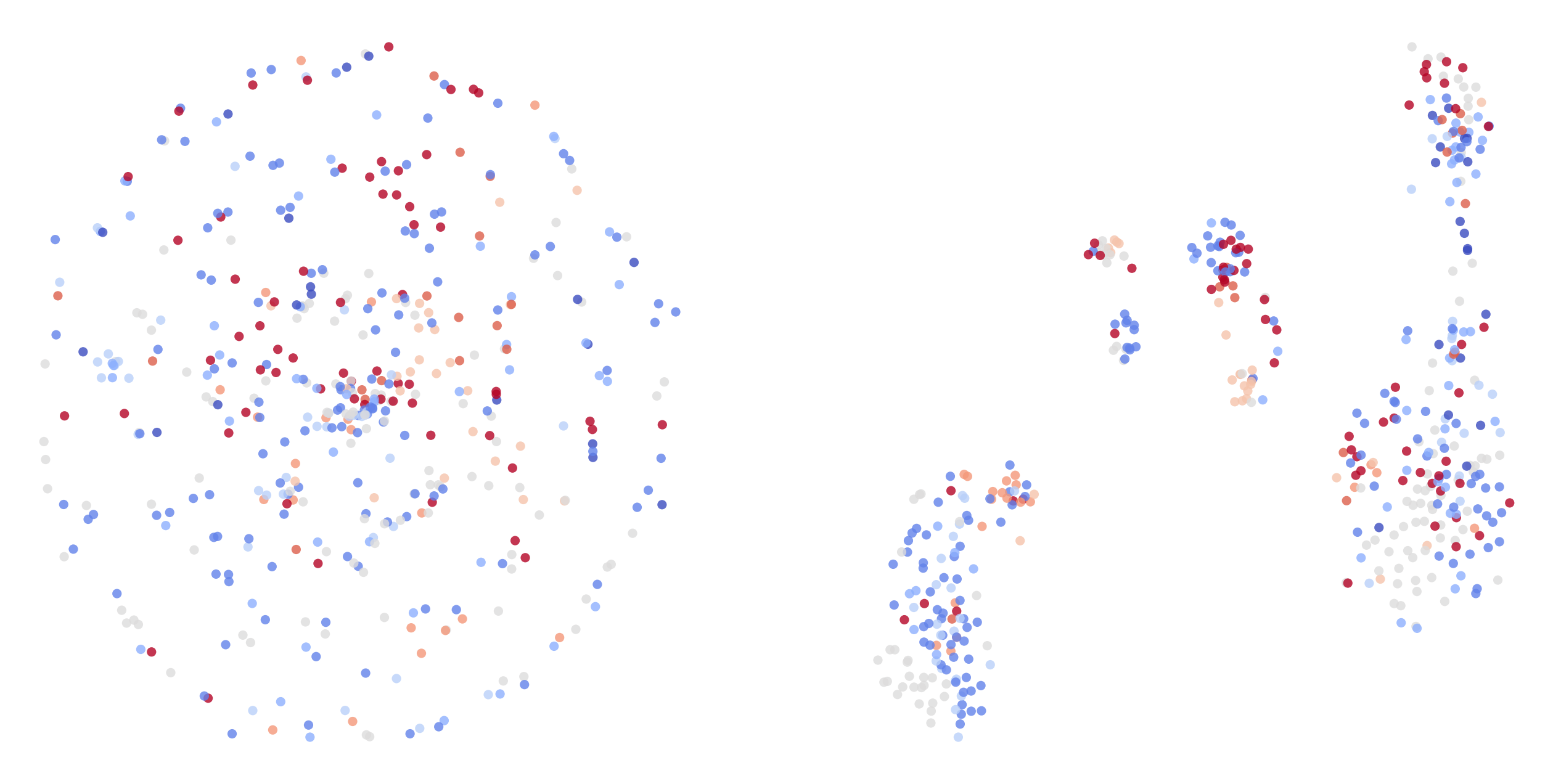}}
\subfloat[Engel]{
		\includegraphics[width=0.25\textwidth]{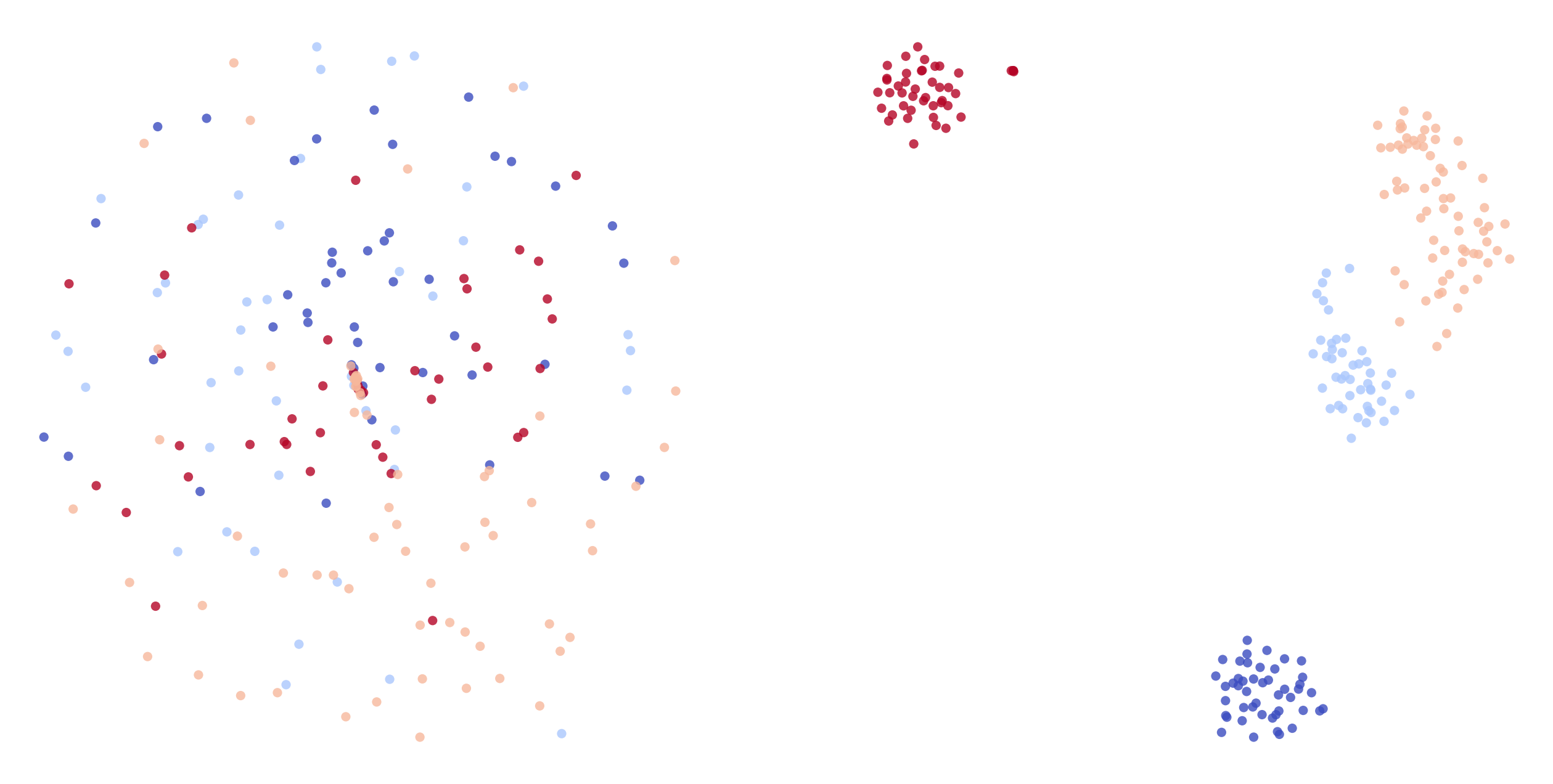}}
\subfloat[Goolam]{
		\includegraphics[width=0.25\textwidth]{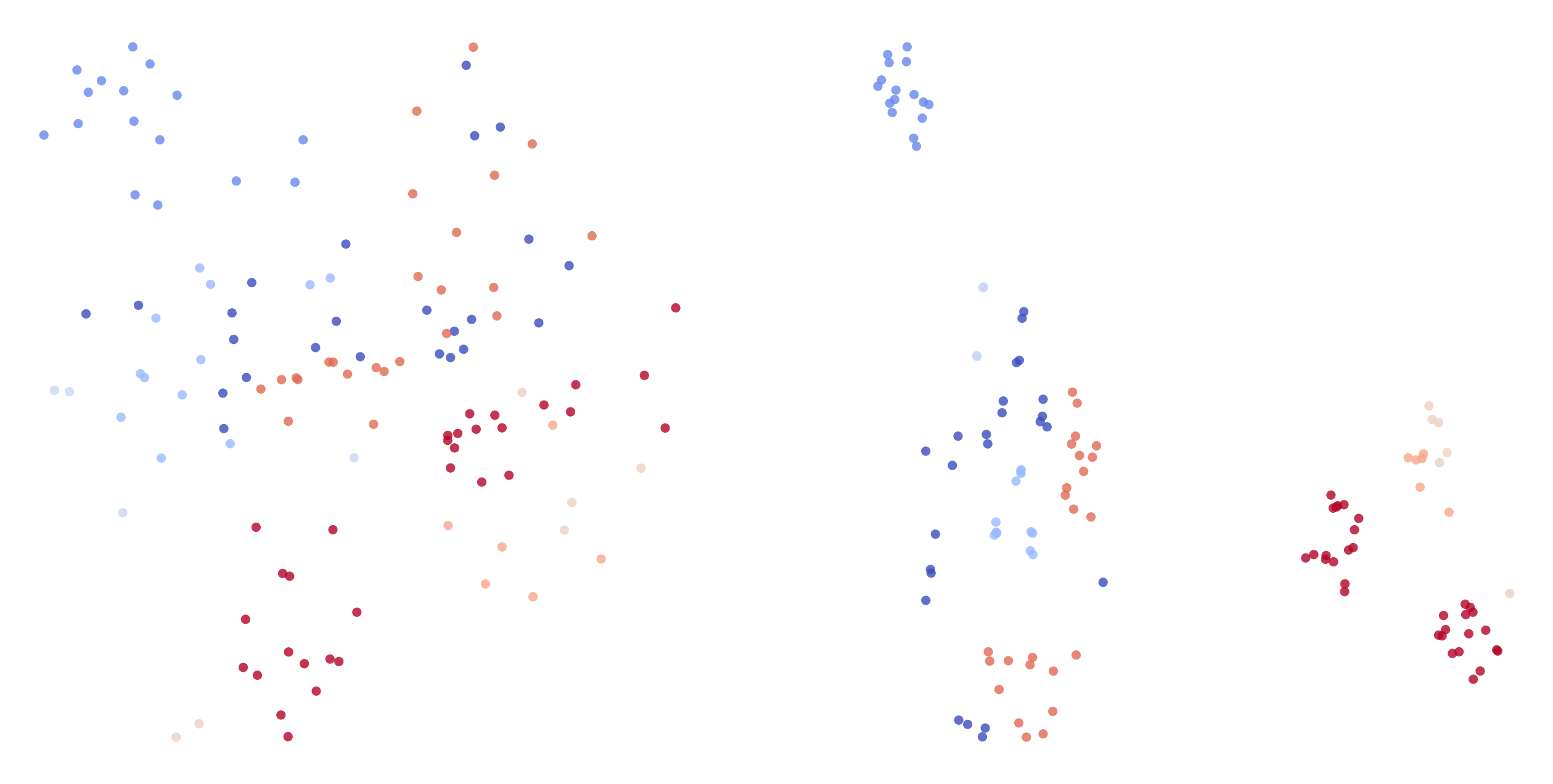}}
\subfloat[Koh]{
		\includegraphics[width=0.25\textwidth]{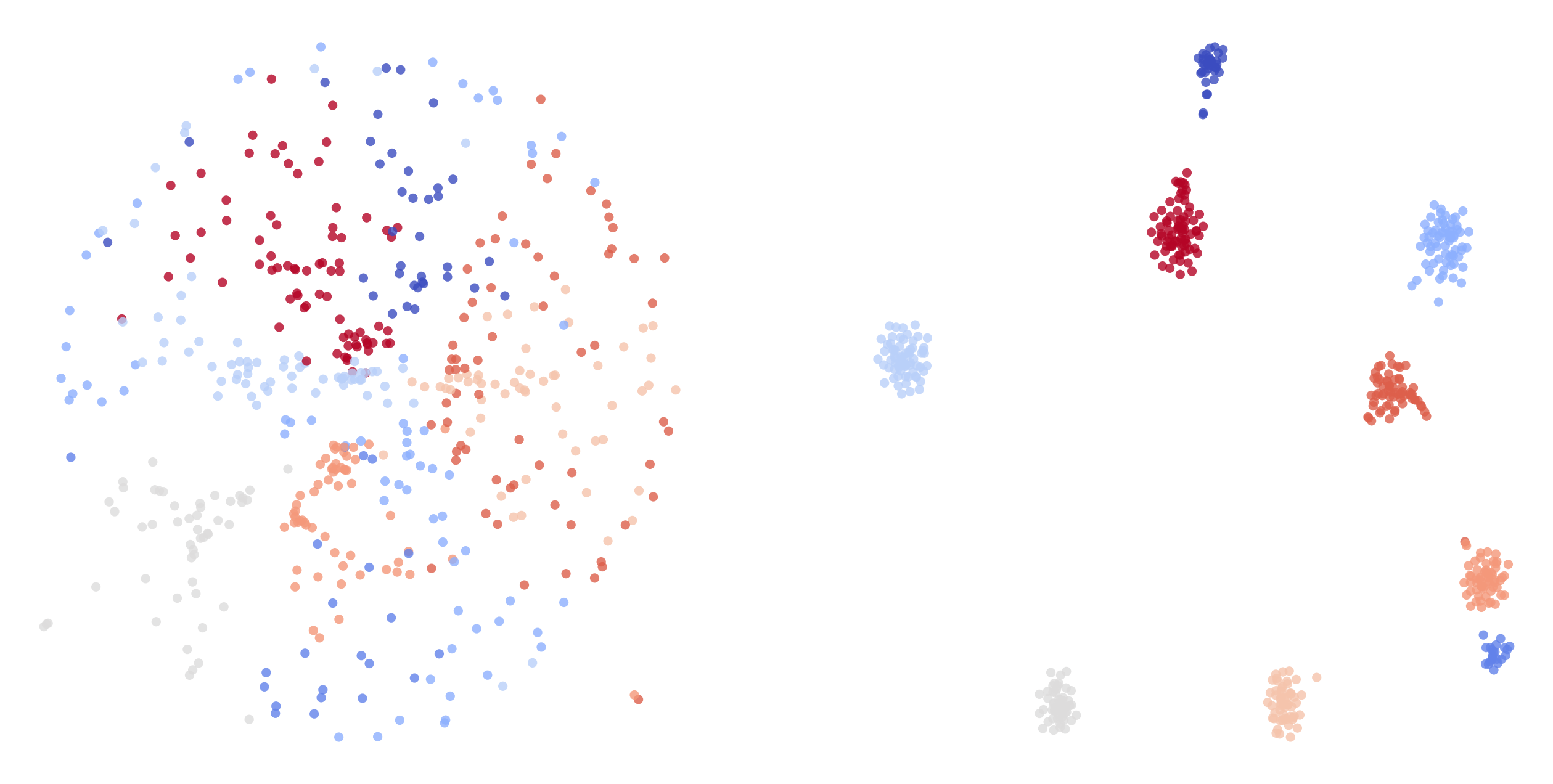}}
\\
\subfloat[Kumar]{
		\includegraphics[width=0.25\textwidth]{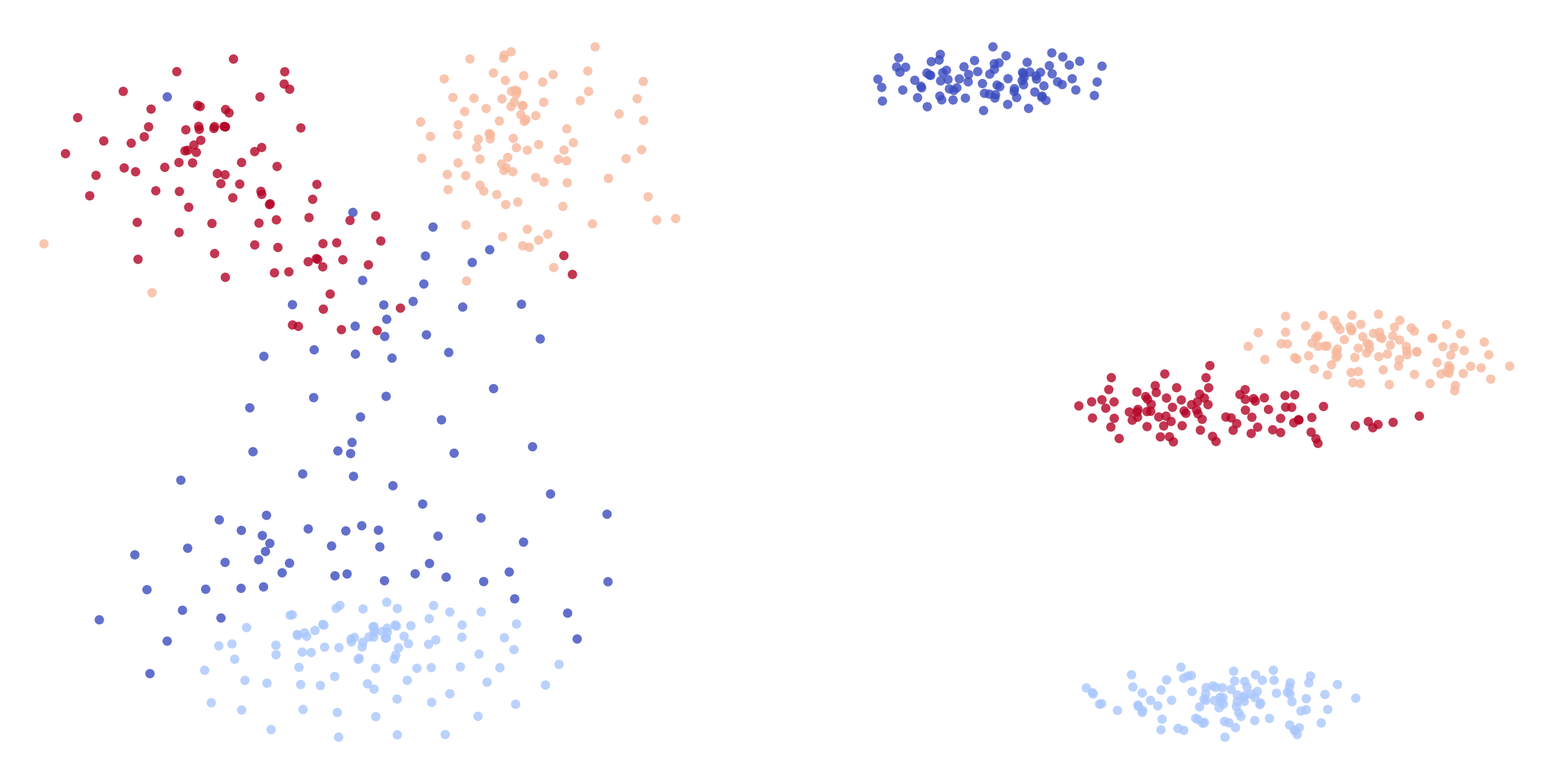}}
\subfloat[Leng]{
		\includegraphics[width=0.25\textwidth]{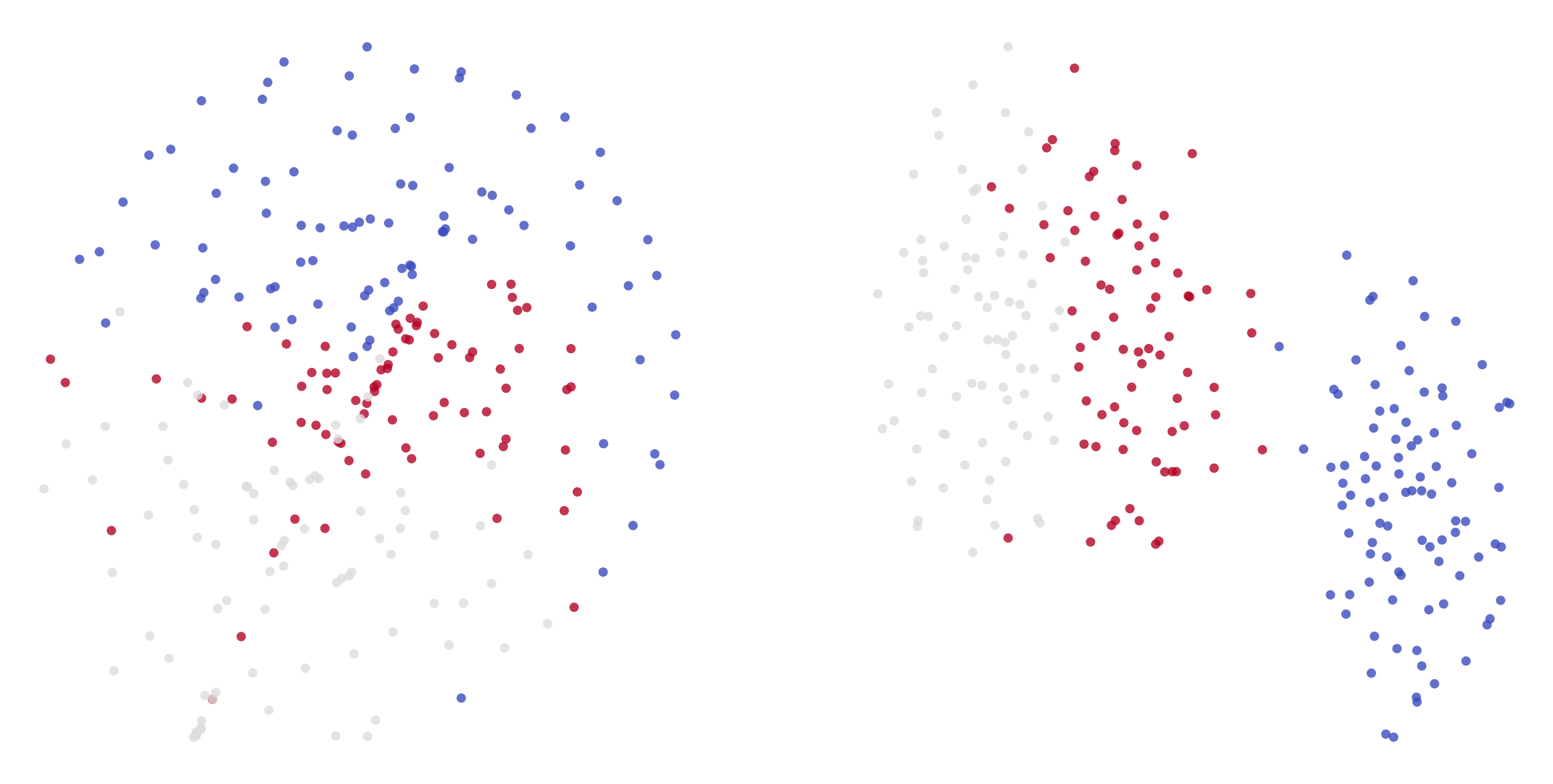}}
\subfloat[Li]{
		\includegraphics[width=0.25\textwidth]{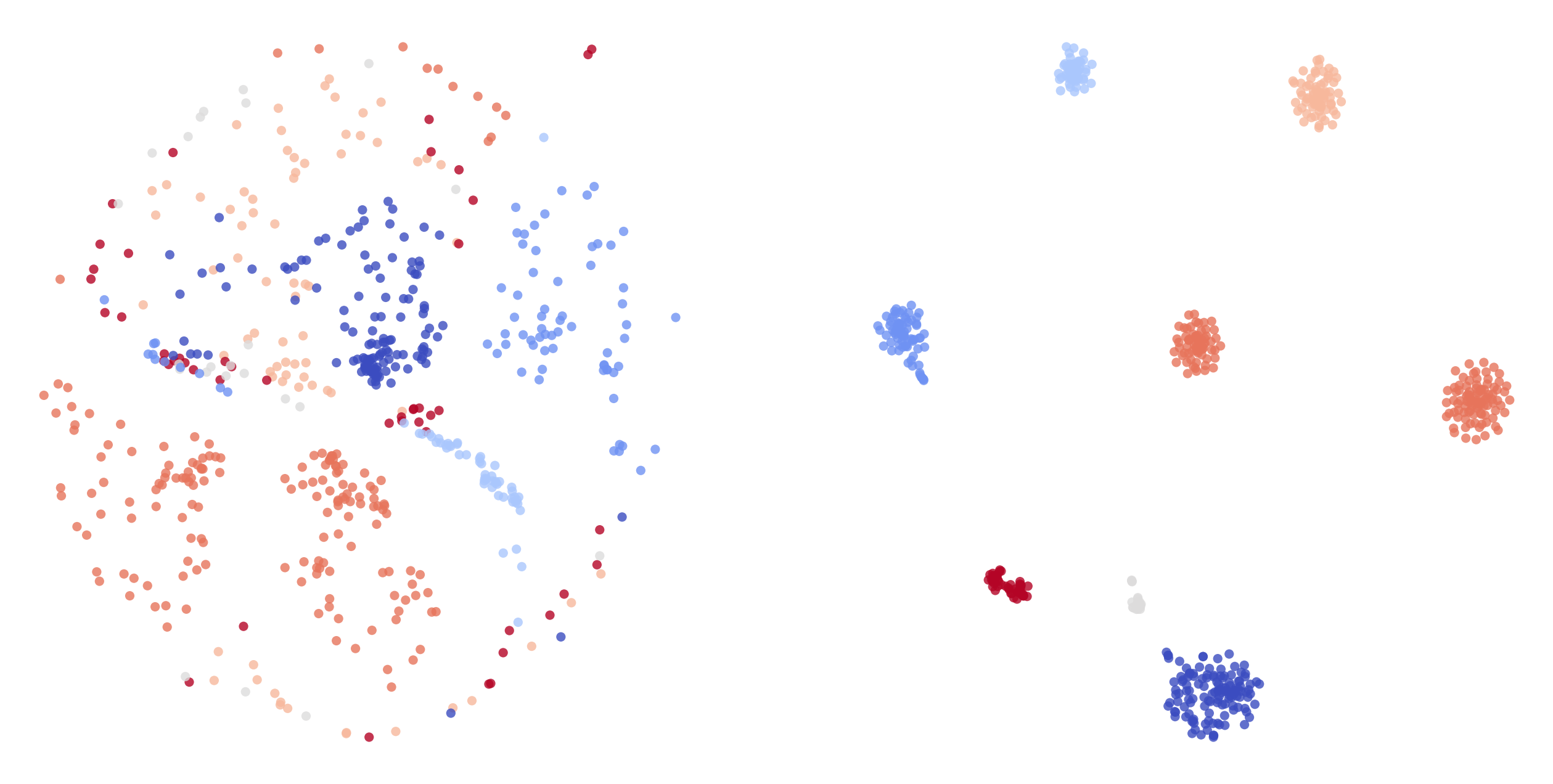}}
\subfloat[Maria1]{
		\includegraphics[width=0.25\textwidth]{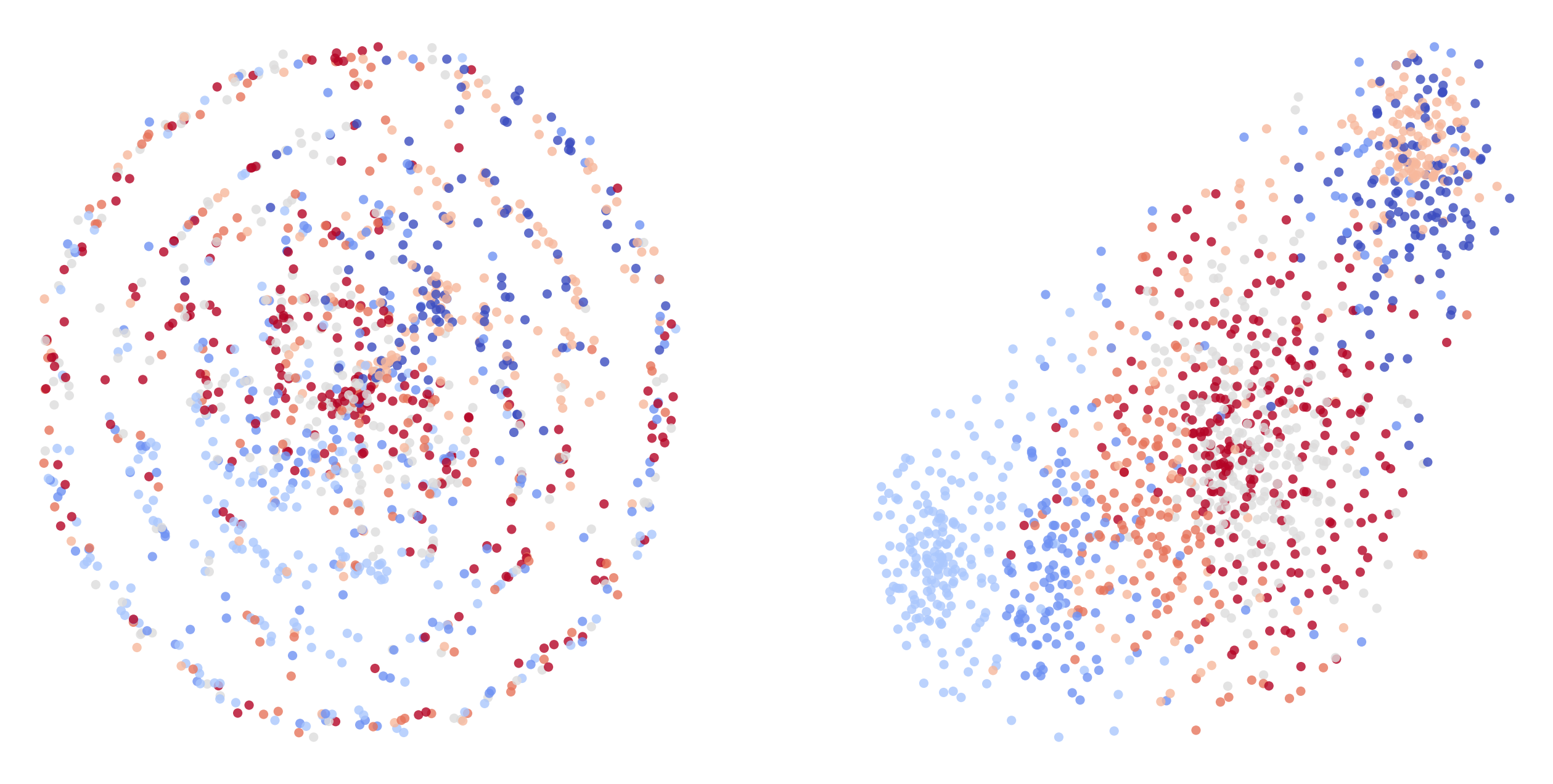}}
\\
\subfloat[Maria2]{
		\includegraphics[width=0.25\textwidth]{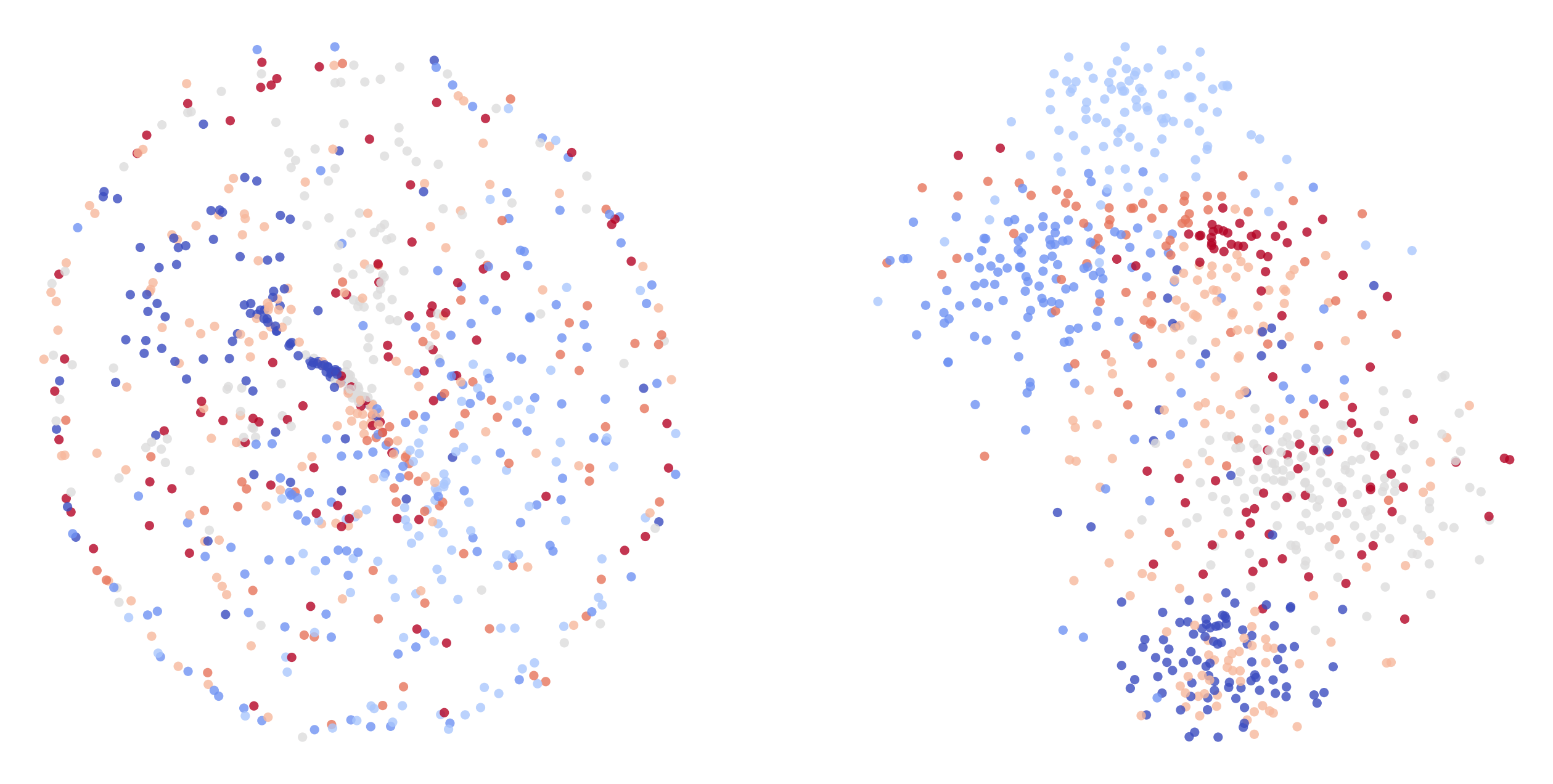}}
\subfloat[Mouse Pancreas1]{
		\includegraphics[width=0.25\textwidth]{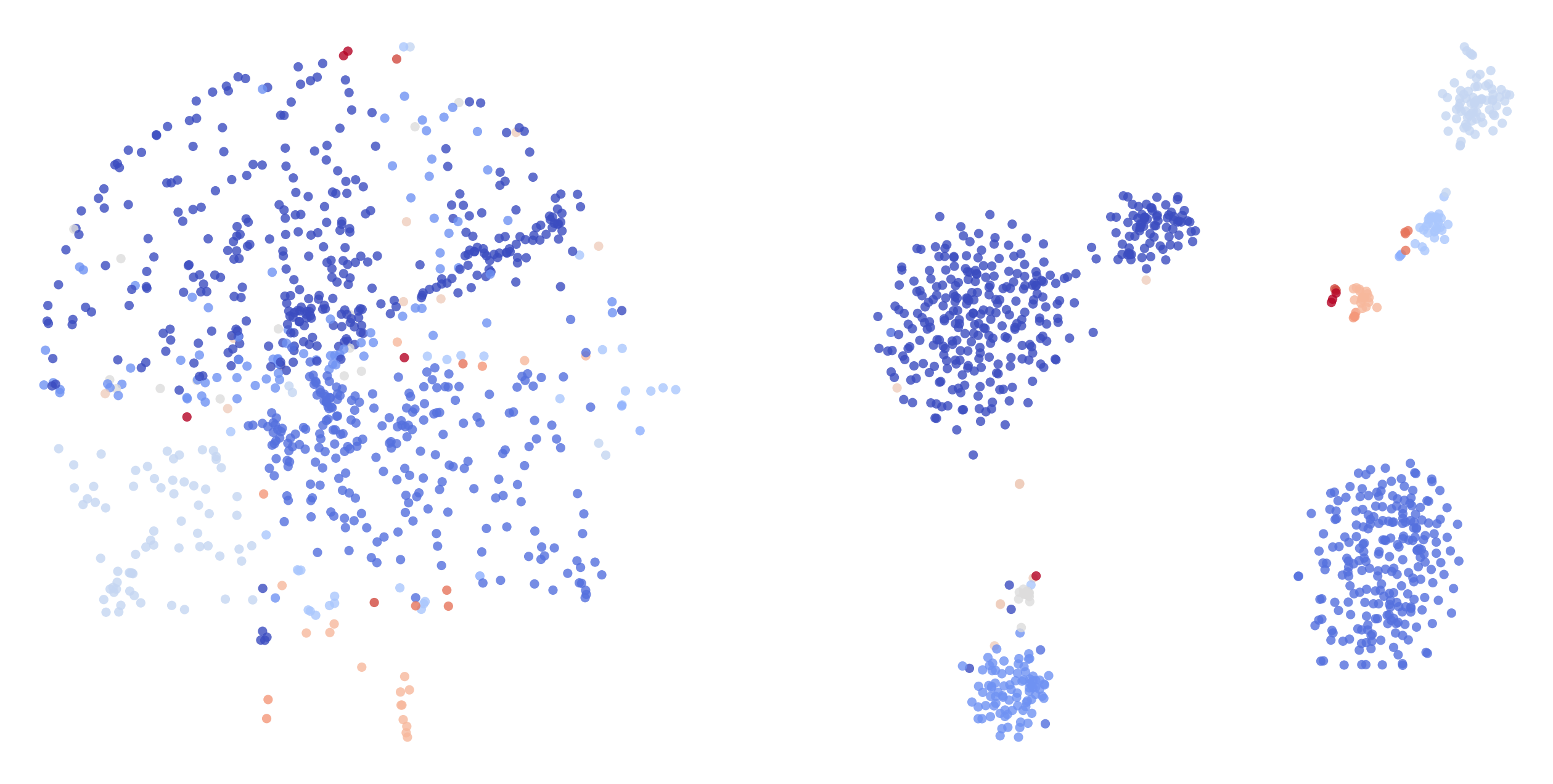}}
\subfloat[MacParland]{
		\includegraphics[width=0.25\textwidth]{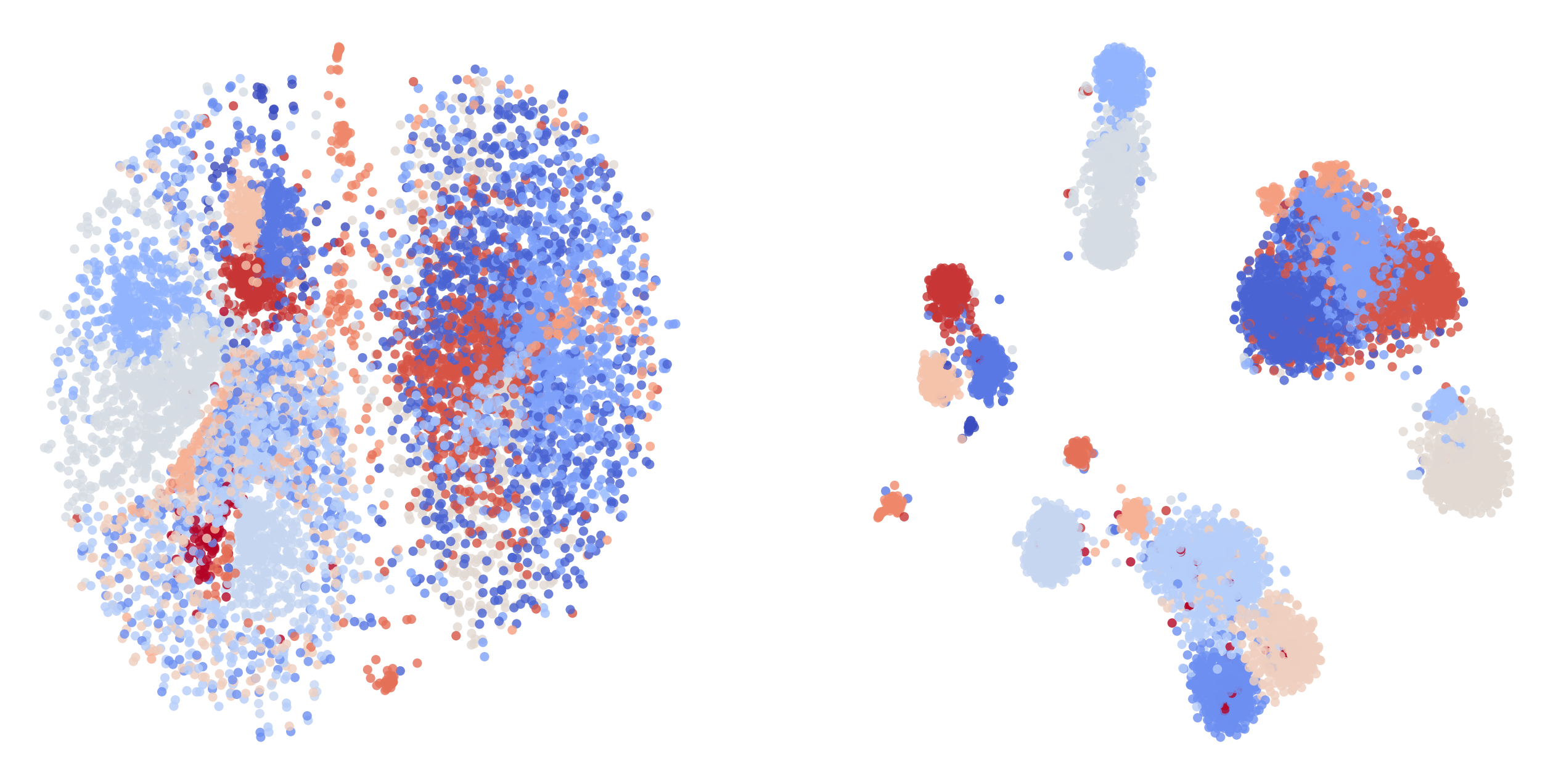}}
\subfloat[Mouse Pancreas2]{
		\includegraphics[width=0.25\textwidth]{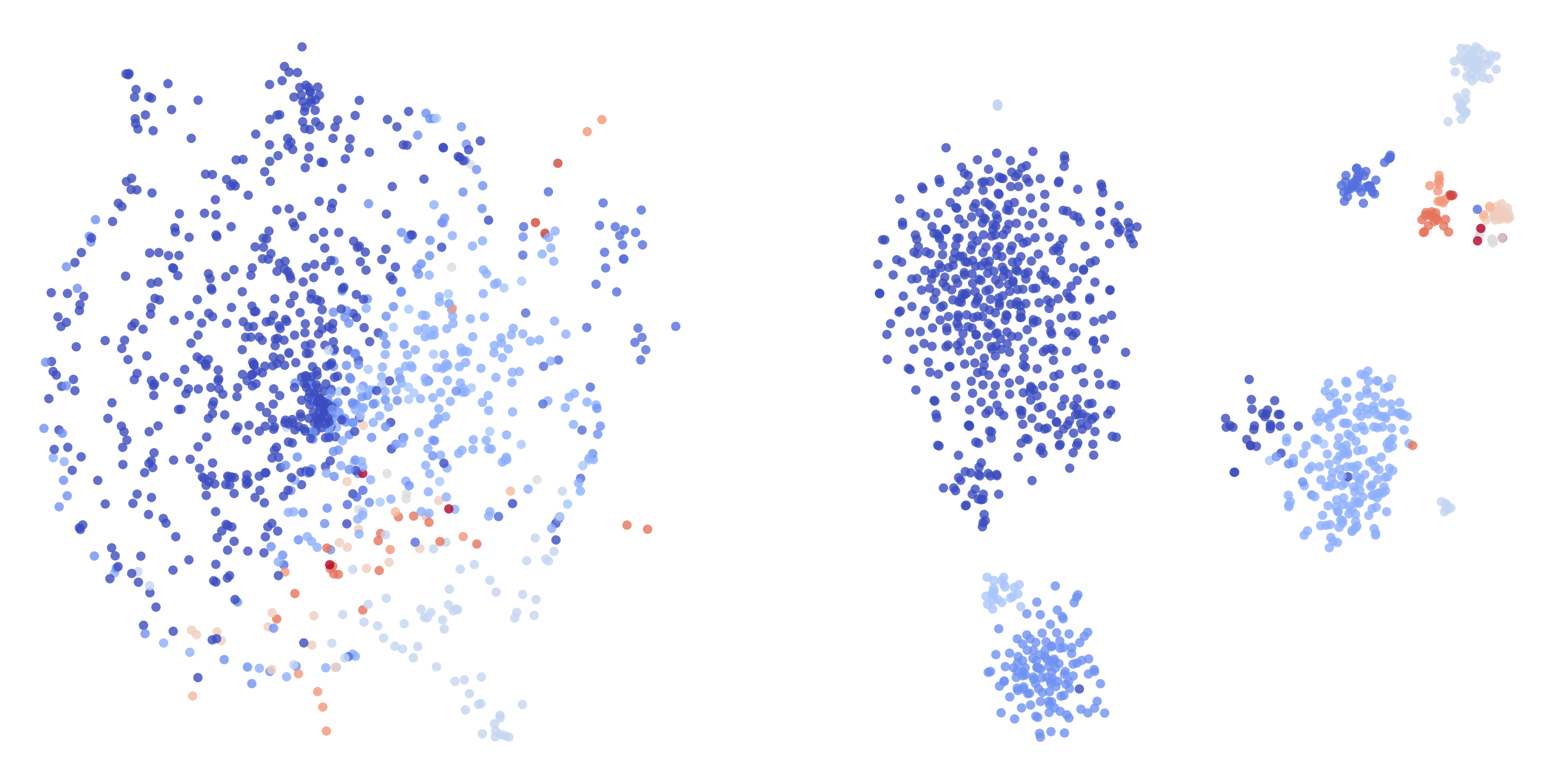}}
\\
\subfloat[Robert]{
		\includegraphics[width=0.25\textwidth]{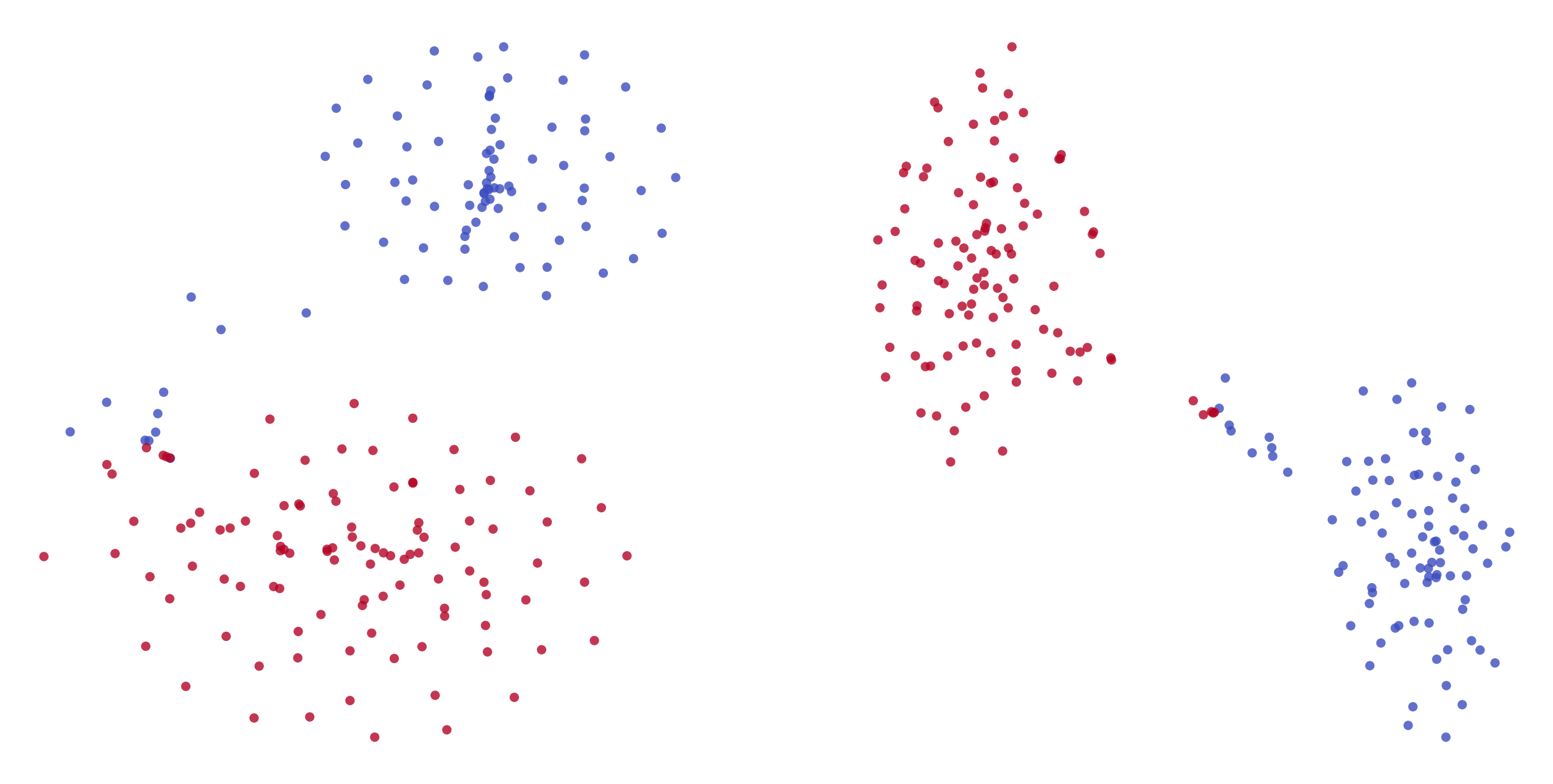}}
\subfloat[Ting]{
		\includegraphics[width=0.25\textwidth]{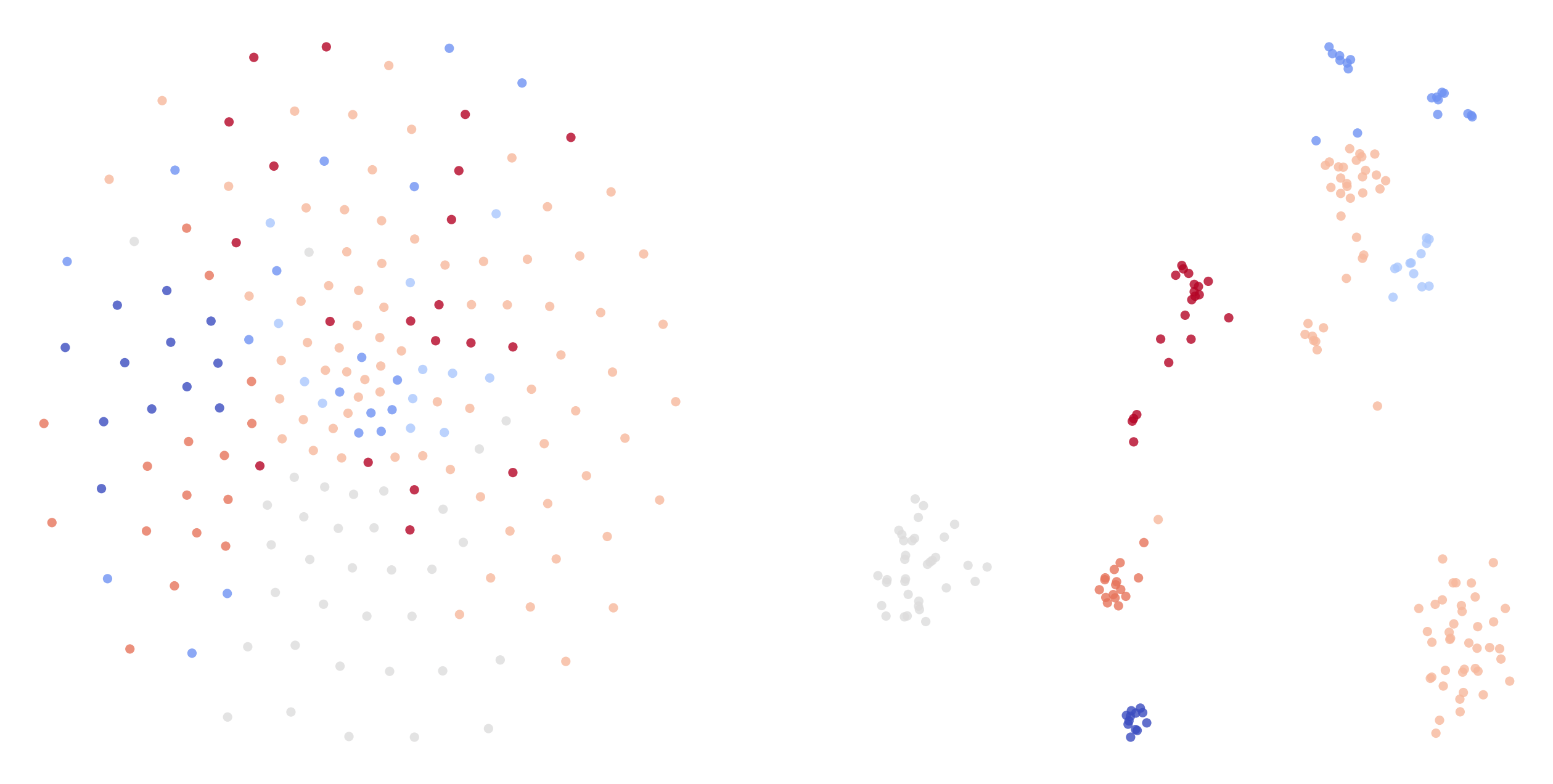}}
\subfloat[Yang]{
		\includegraphics[width=0.25\textwidth]{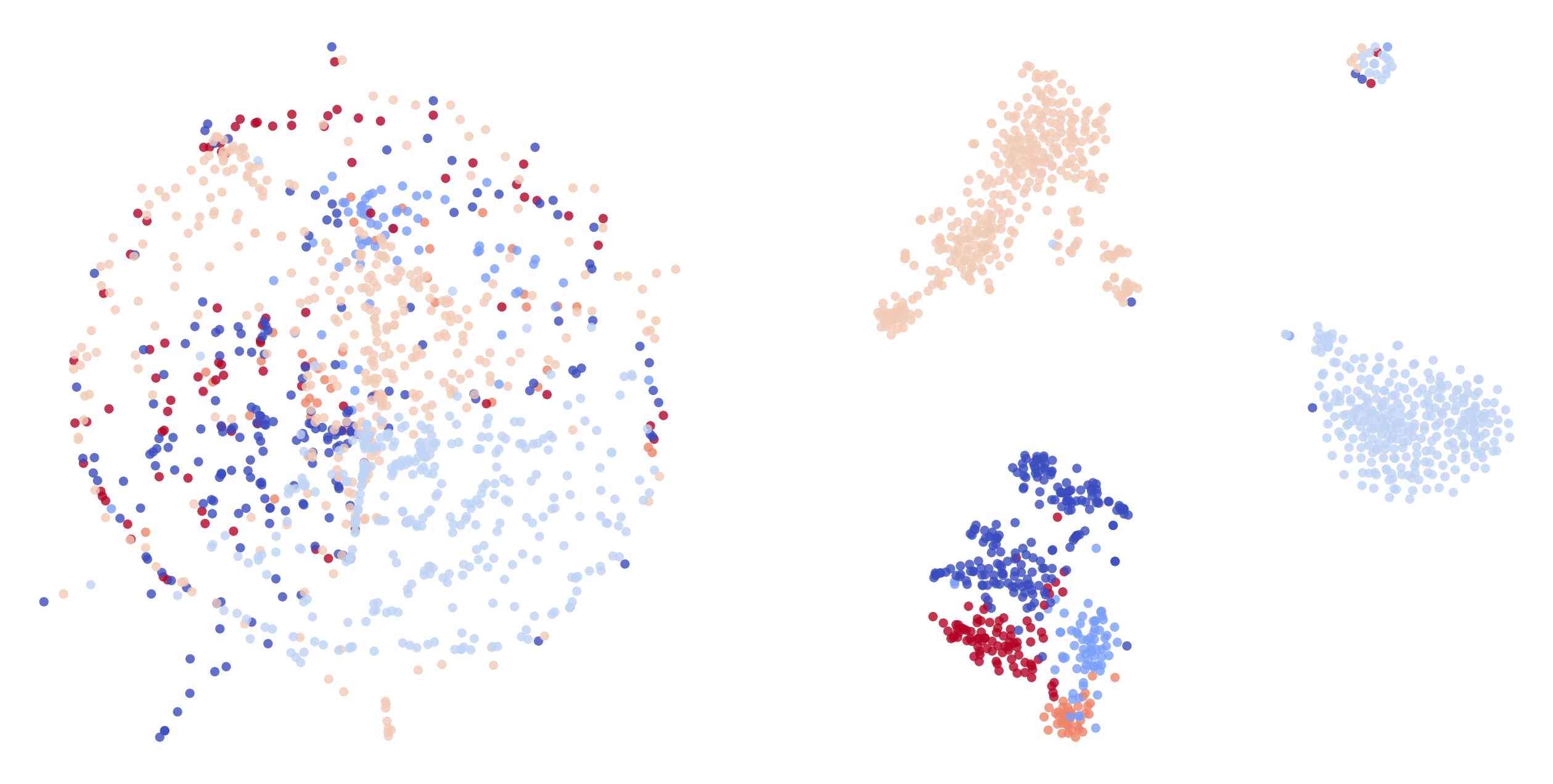}}
\caption{t-SNE visualization of the rest datasets, where the figure in the left panel is visualized from the original dataset.}
\label{tsne_vis}
\end{figure*}

\clearpage
\begin{figure*}[htbp]
\centering
\renewcommand{\thesubfigure}{\arabic{subfigure}}
\subfloat[Cao]{
		\includegraphics[width=0.32\textwidth]{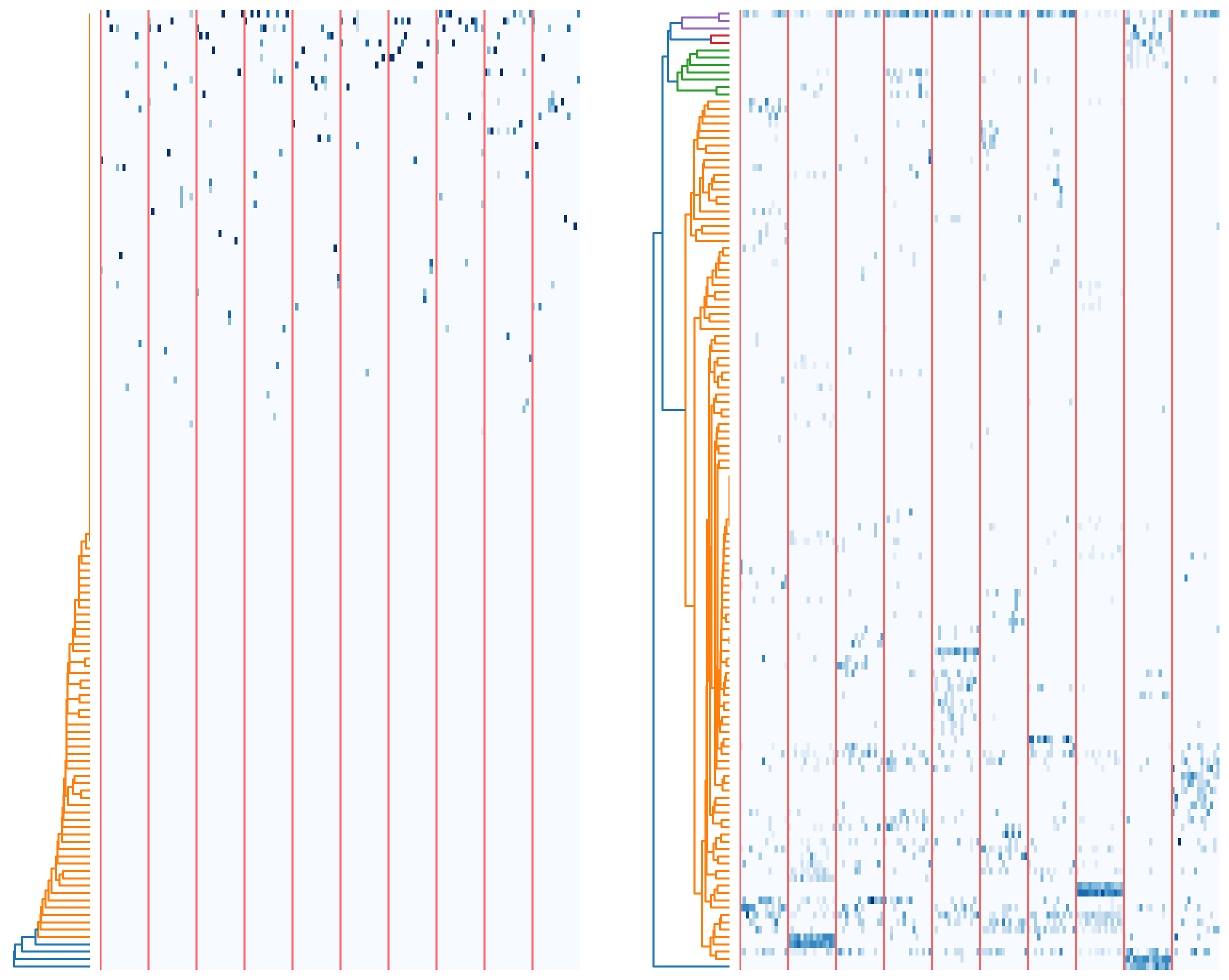}}
\subfloat[Chu1]{
		\includegraphics[width=0.32\textwidth]{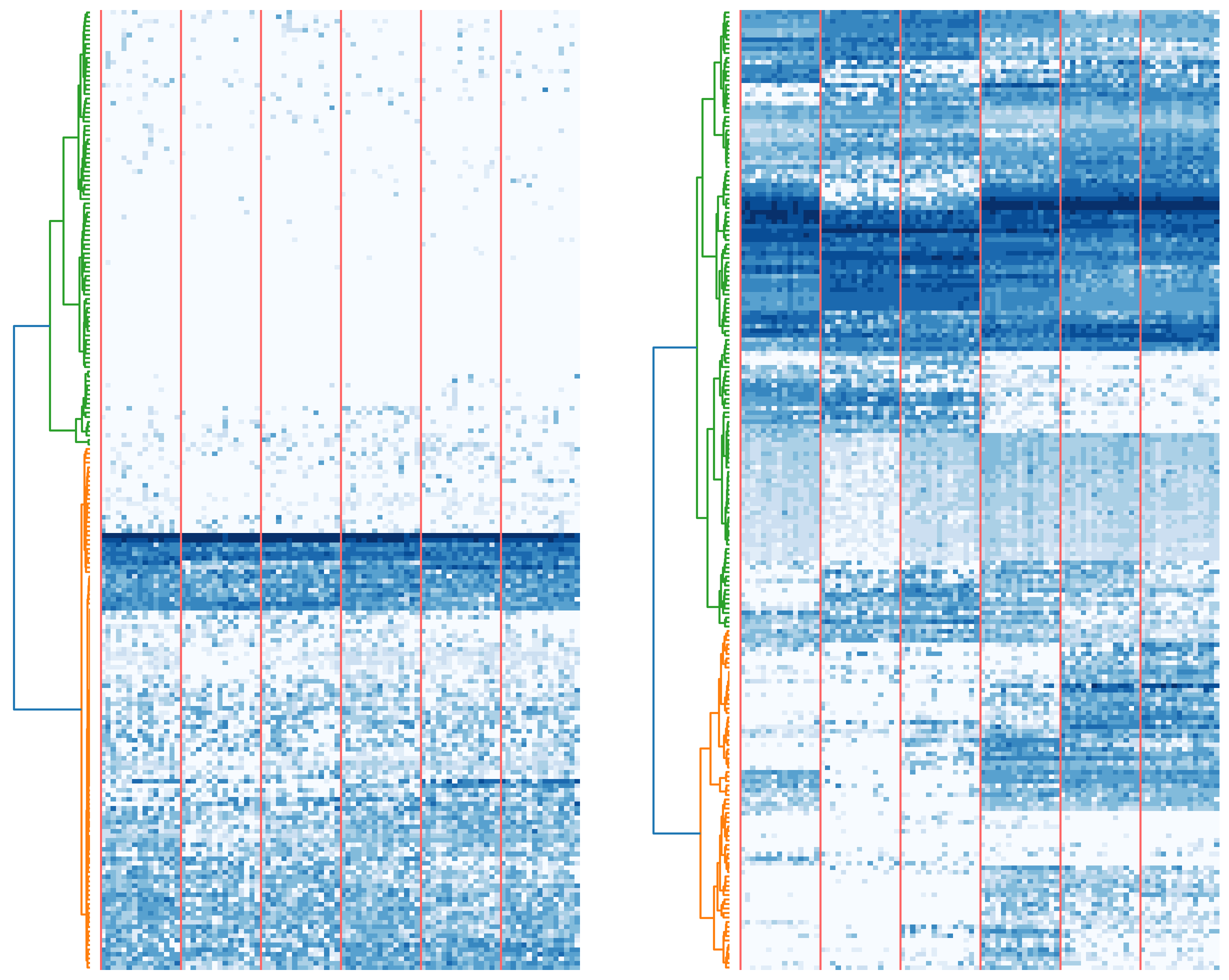}}
\subfloat[Chu2]{
		\includegraphics[width=0.32\textwidth]{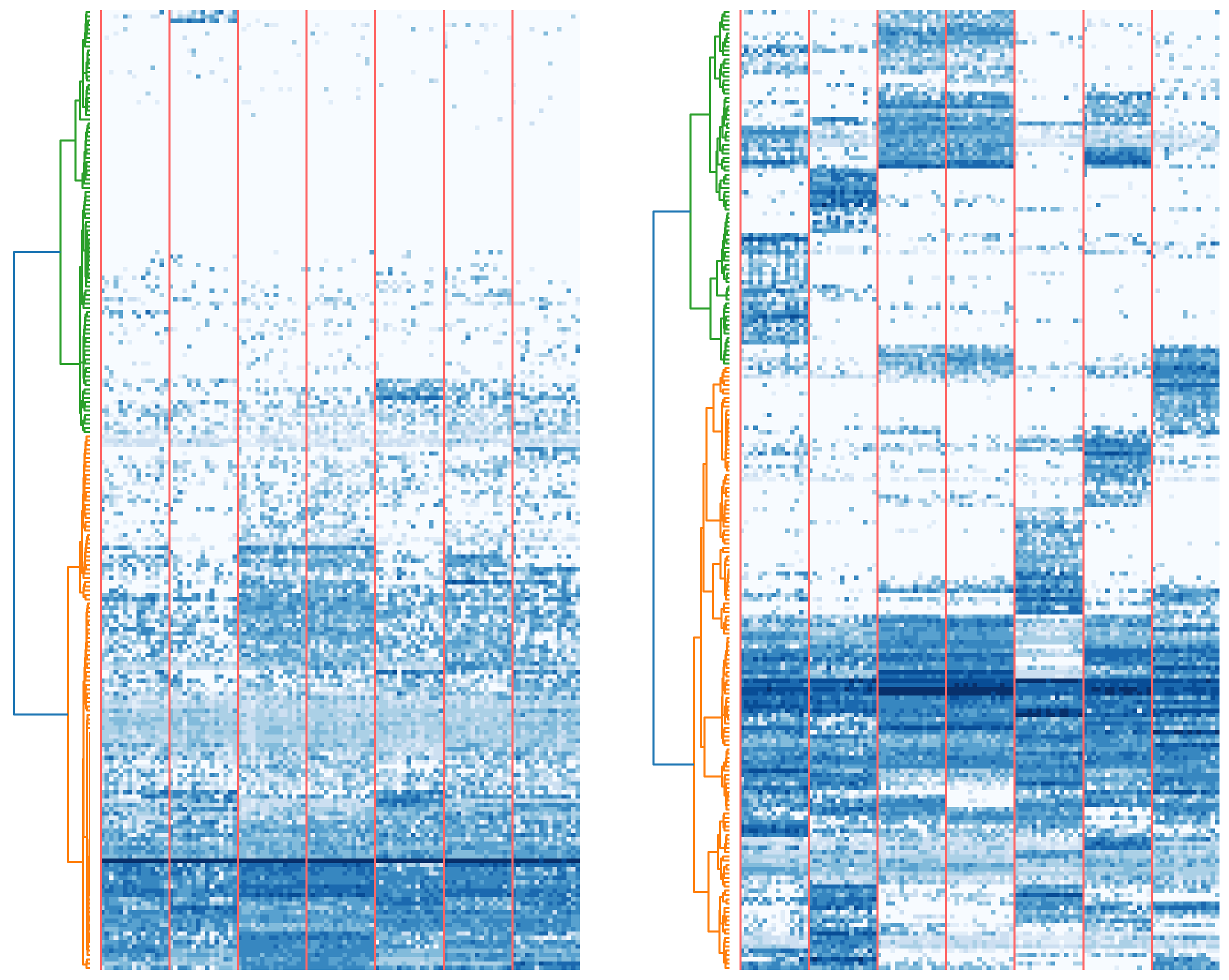}}
\\
\subfloat[Han]{		\includegraphics[width=0.32\textwidth]{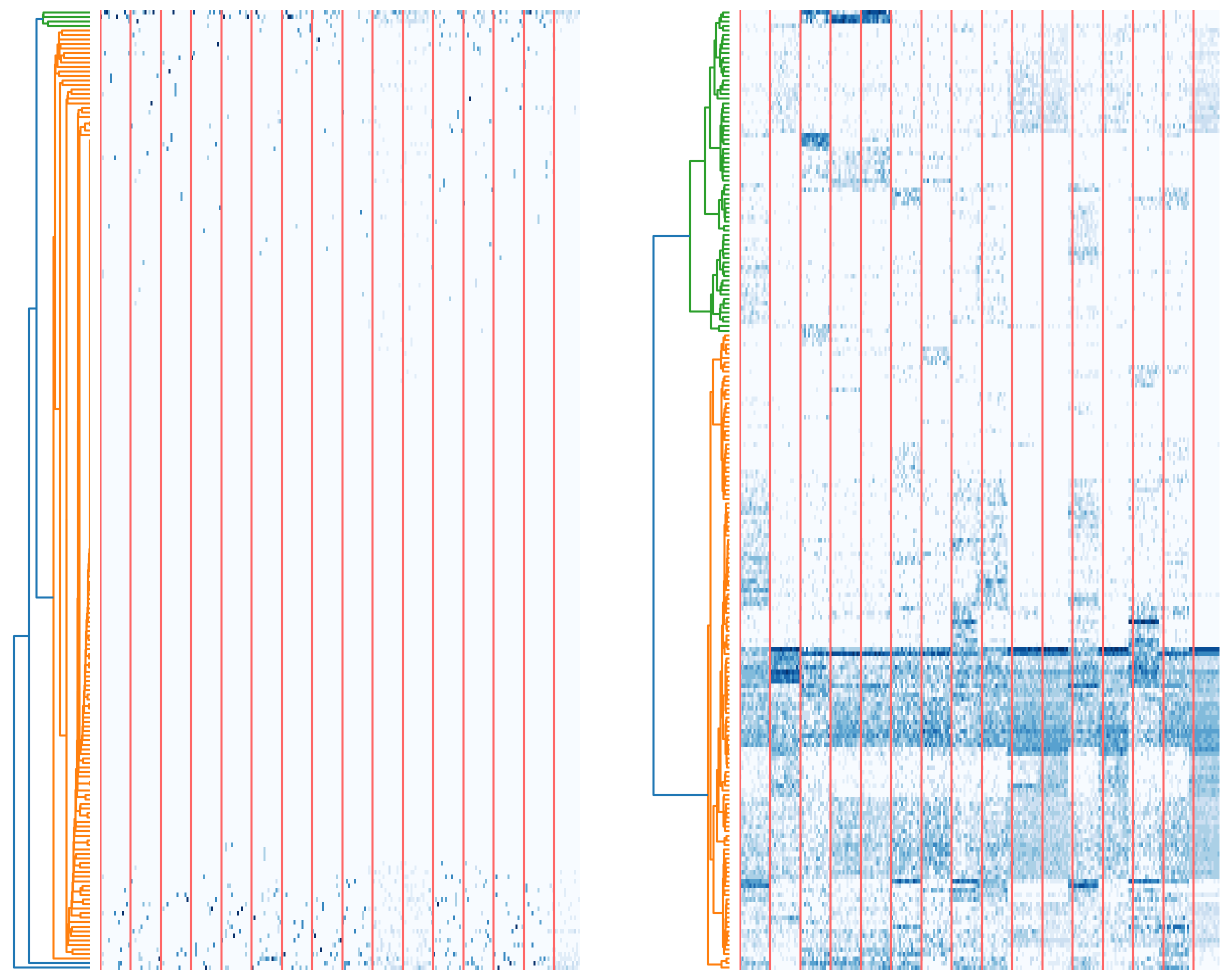}}
\subfloat[Human Pancreas1]{
\includegraphics[width=0.32\textwidth]{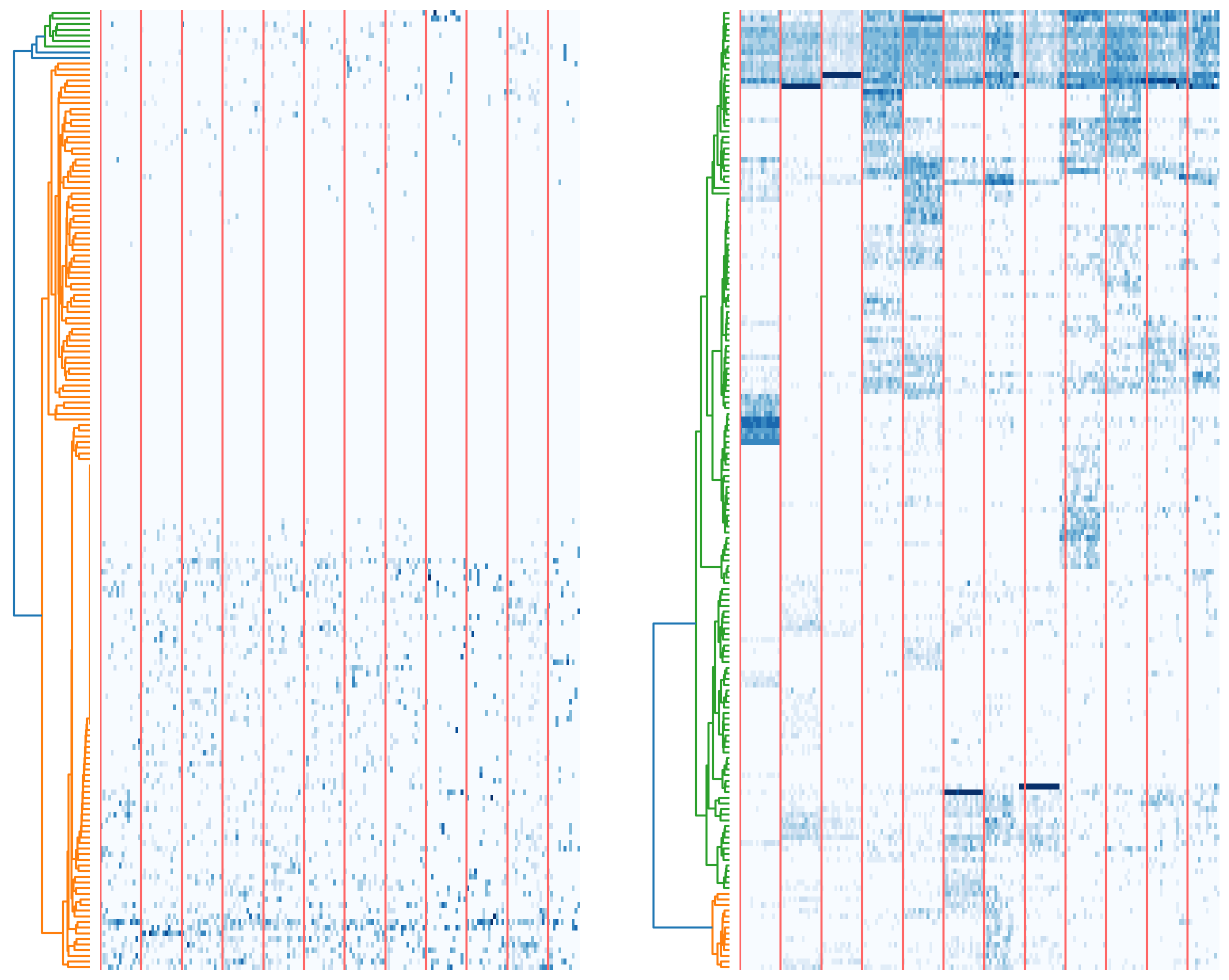}}
\subfloat[Human Pancreas2]{
		\includegraphics[width=0.32\textwidth]{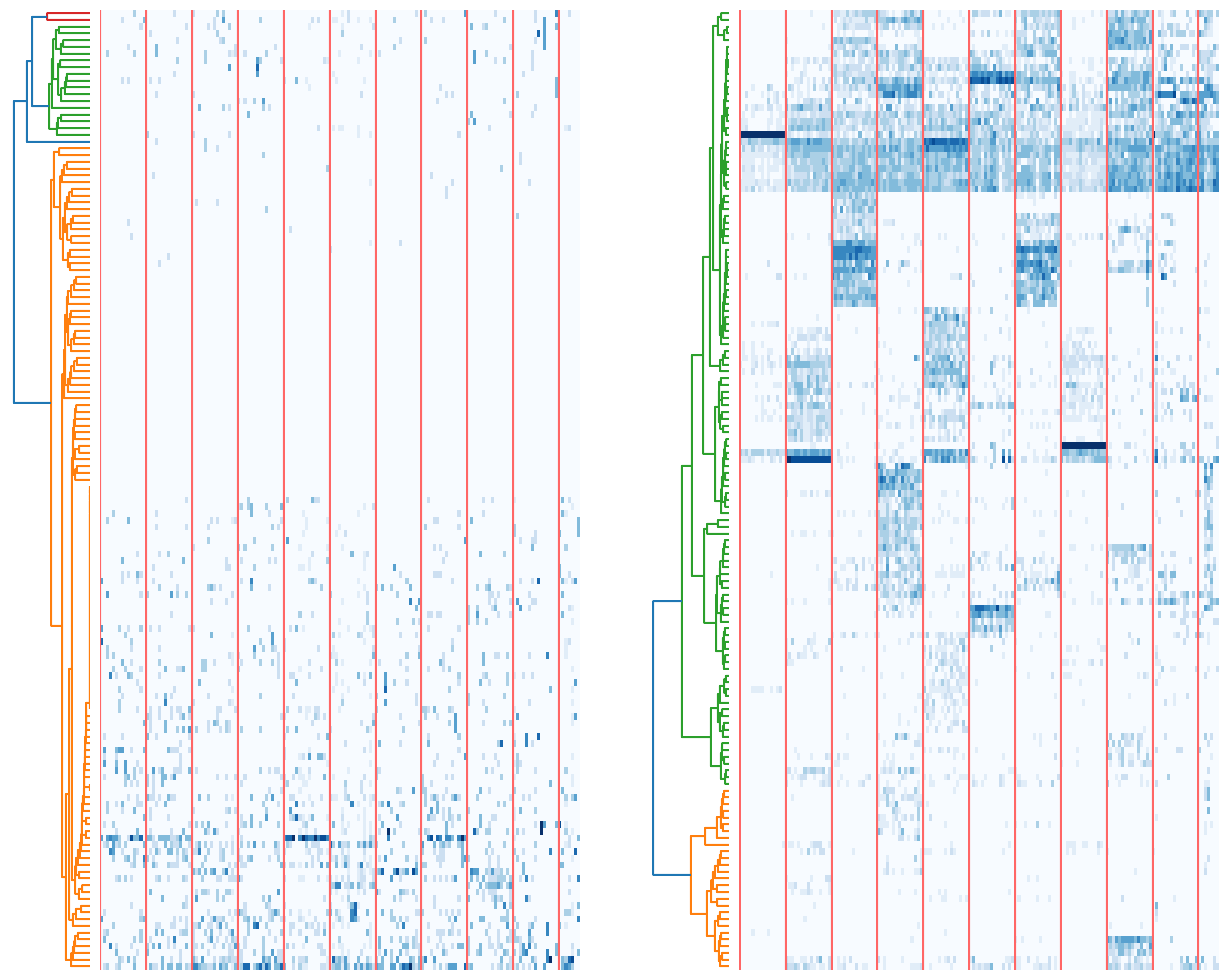}}\\
\subfloat[Human Pancreas3]{
		\includegraphics[width=0.32\textwidth]{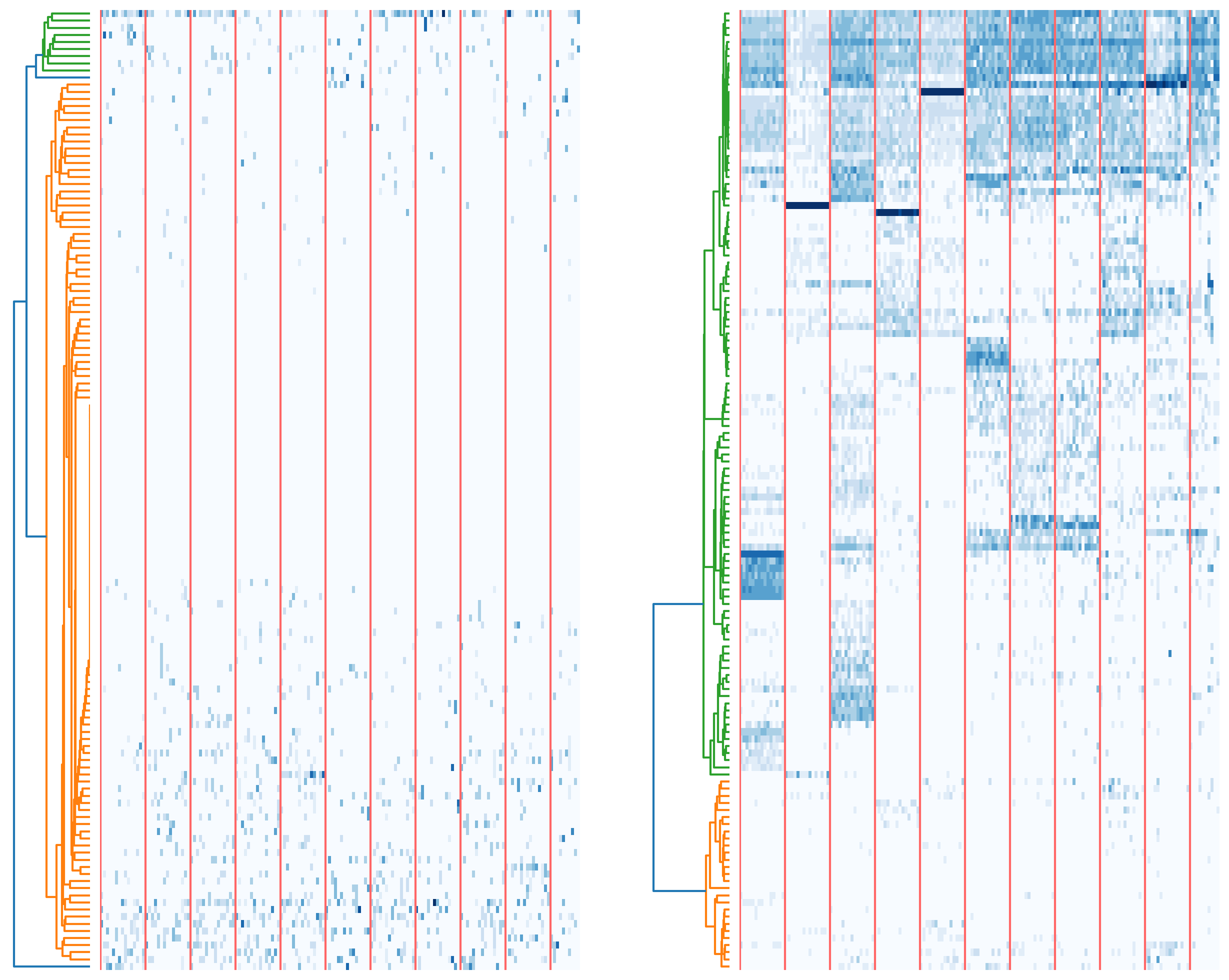}}
\subfloat[Chung]{
		\includegraphics[width=0.32\textwidth]{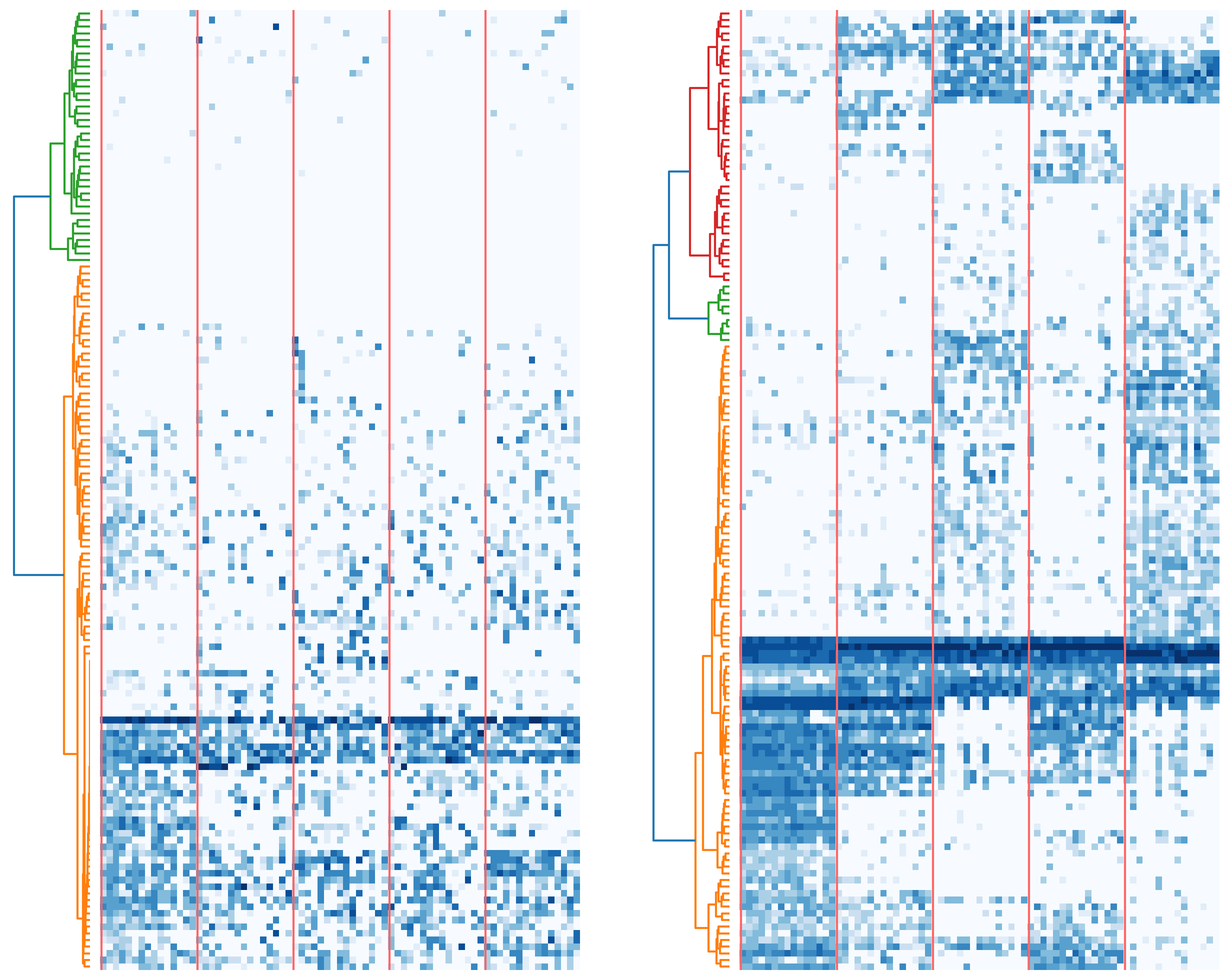}}
\ContinuedFloat
\subfloat[Darmanis]{
\includegraphics[width=0.32\textwidth]{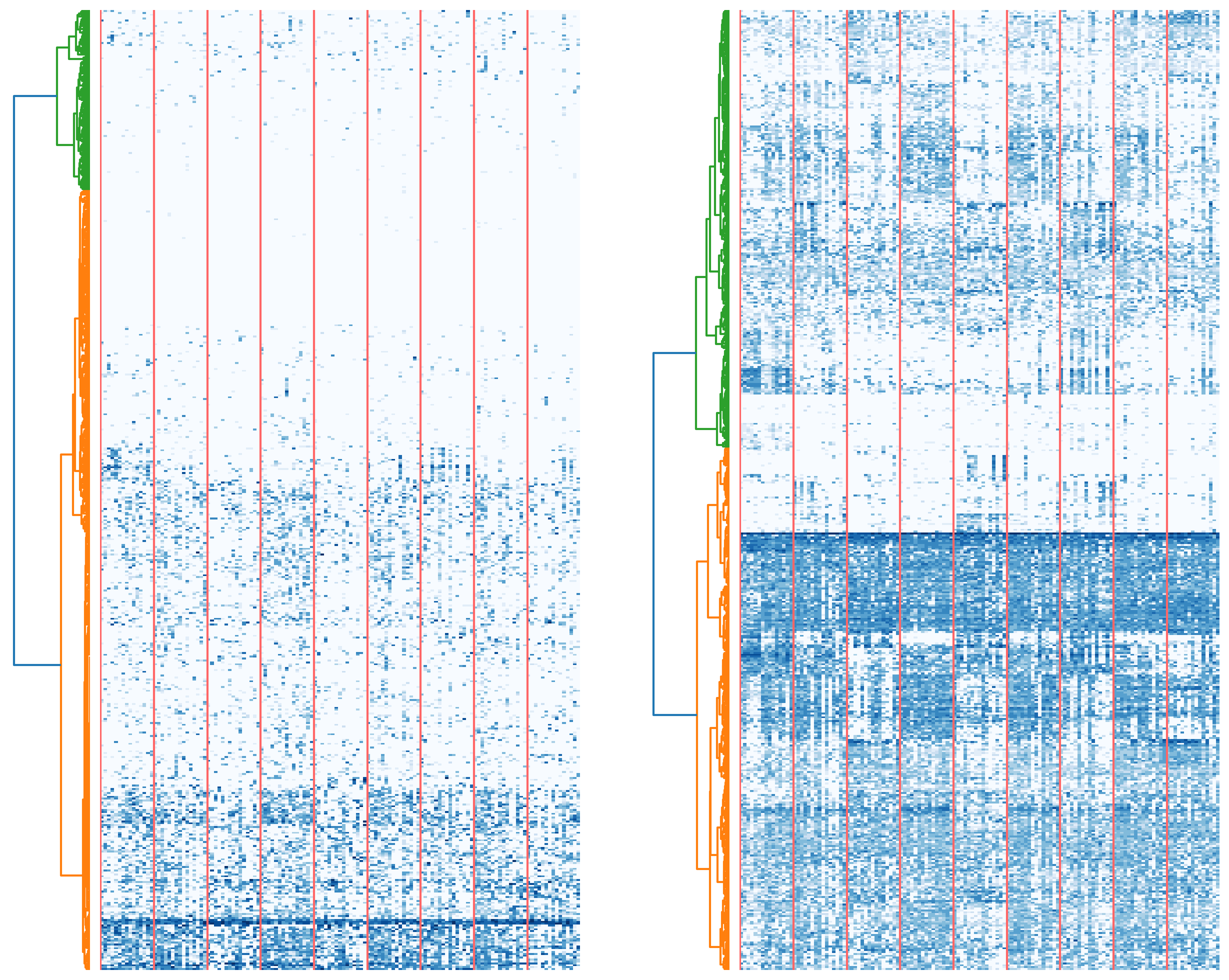}}\\
\subfloat[Engel]{
\includegraphics[width=0.32\textwidth]{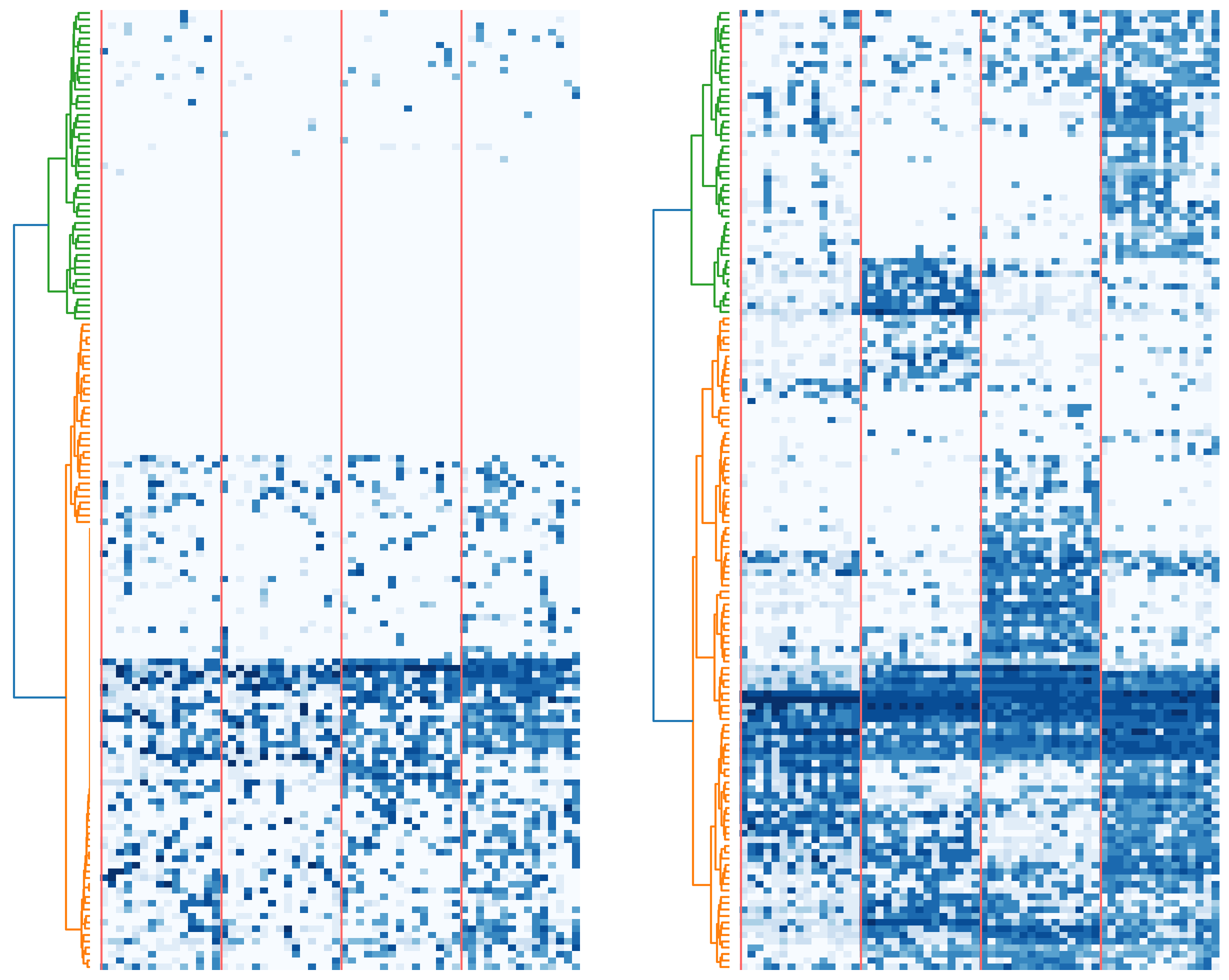}}
\subfloat[Goolam]{
\includegraphics[width=0.32\textwidth]{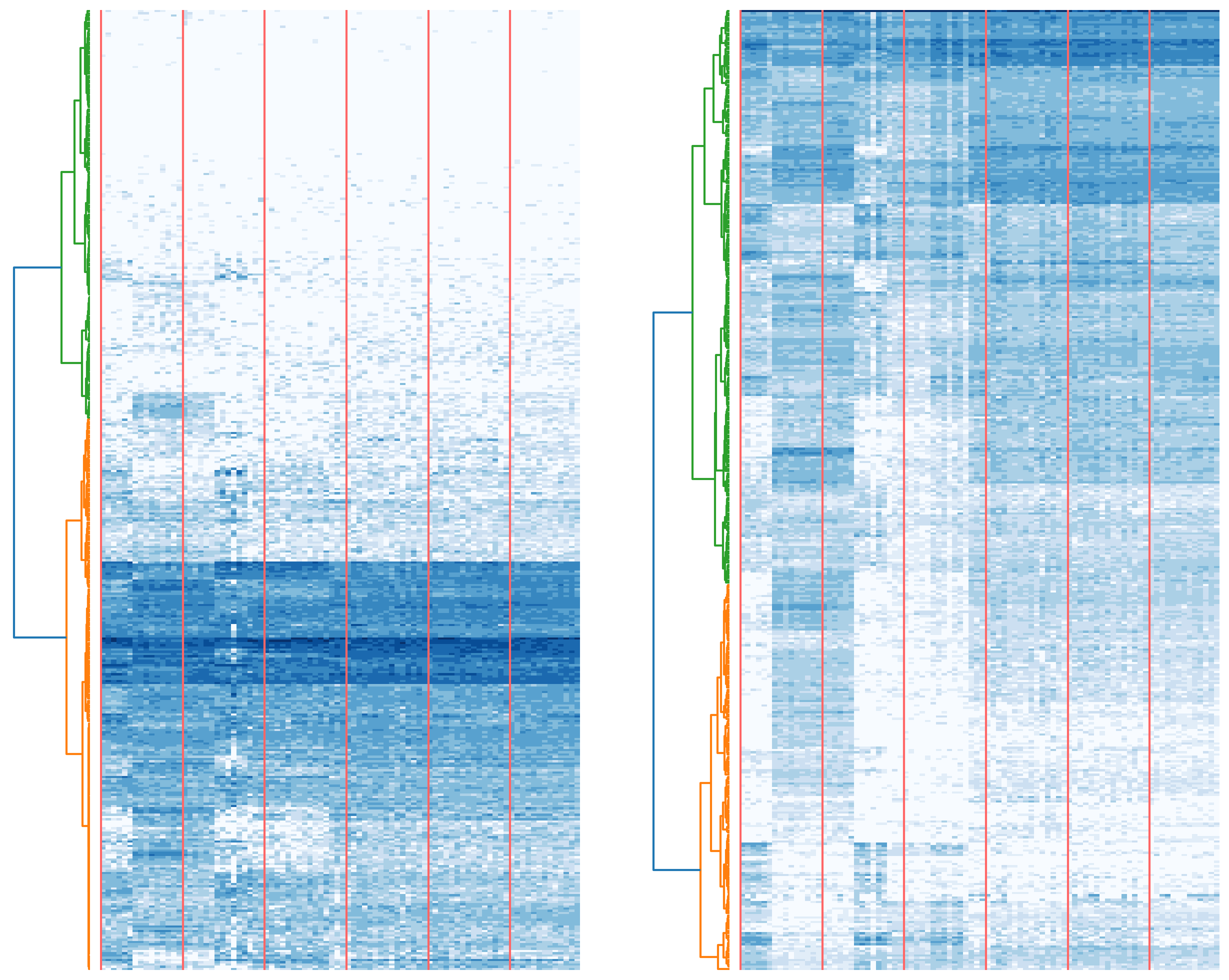}}
\subfloat[Koh]{
\includegraphics[width=0.32\textwidth]{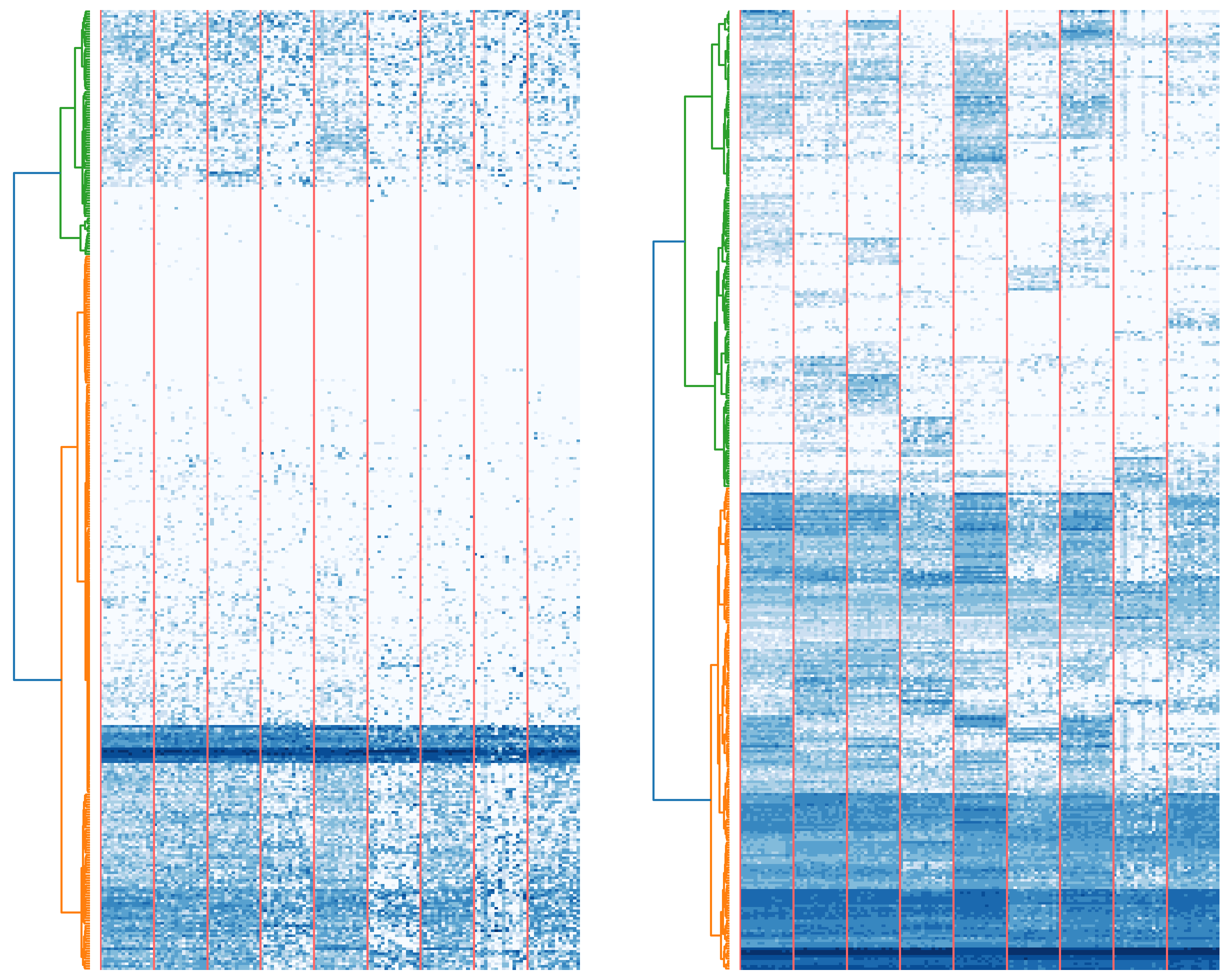}}
\caption{(1/2) Expression heatmap of the rest datasets, where the figure in the left panel is visualized from the original dataset.}
\label{heat_map}
\end{figure*}
\begin{figure*}[htbp]
\renewcommand{\thesubfigure}{\arabic{subfigure}}
\centering
\subfloat[Kumar]{
		\includegraphics[width=0.32\textwidth]{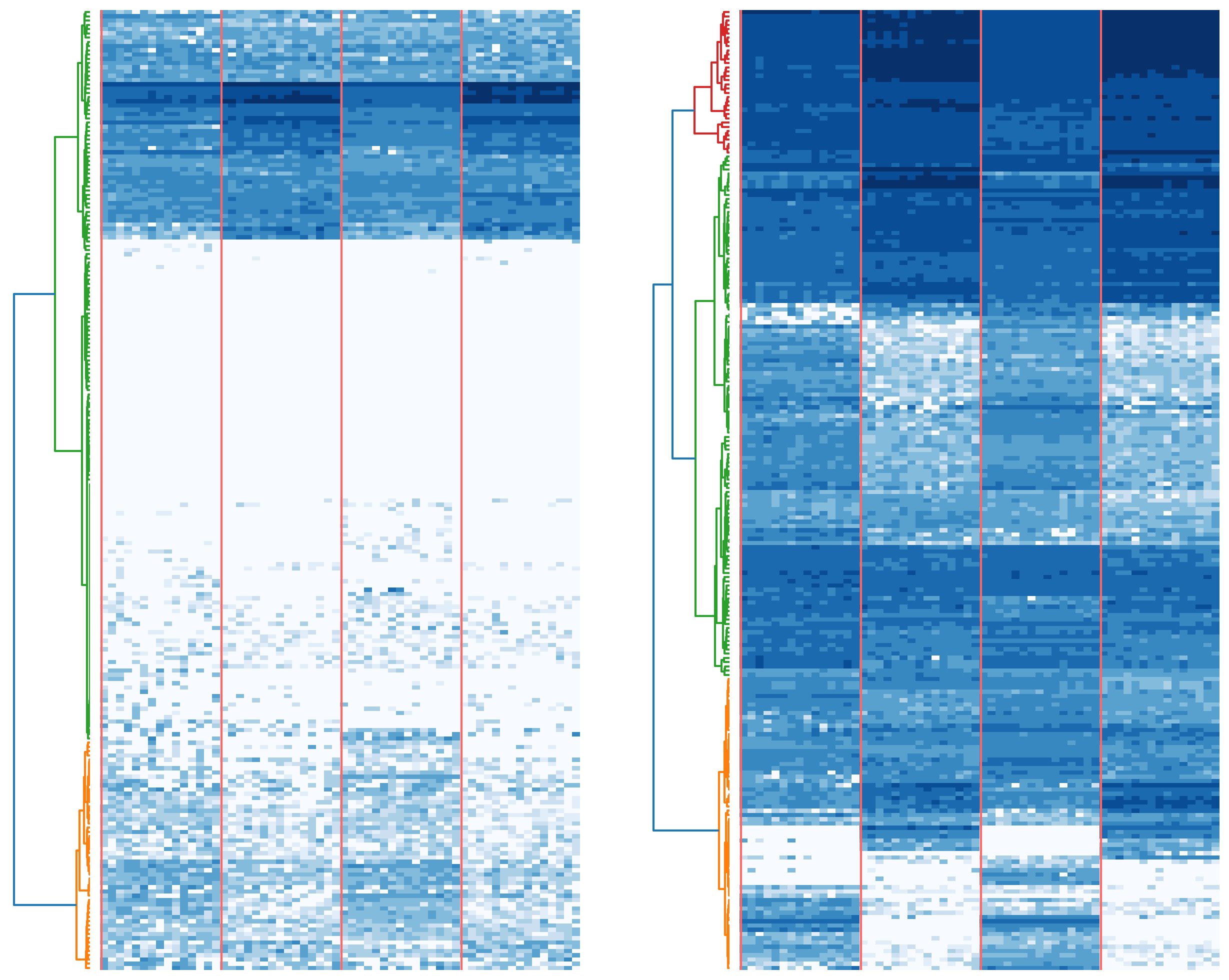}}
\subfloat[Leng]{
		\includegraphics[width=0.32\textwidth]{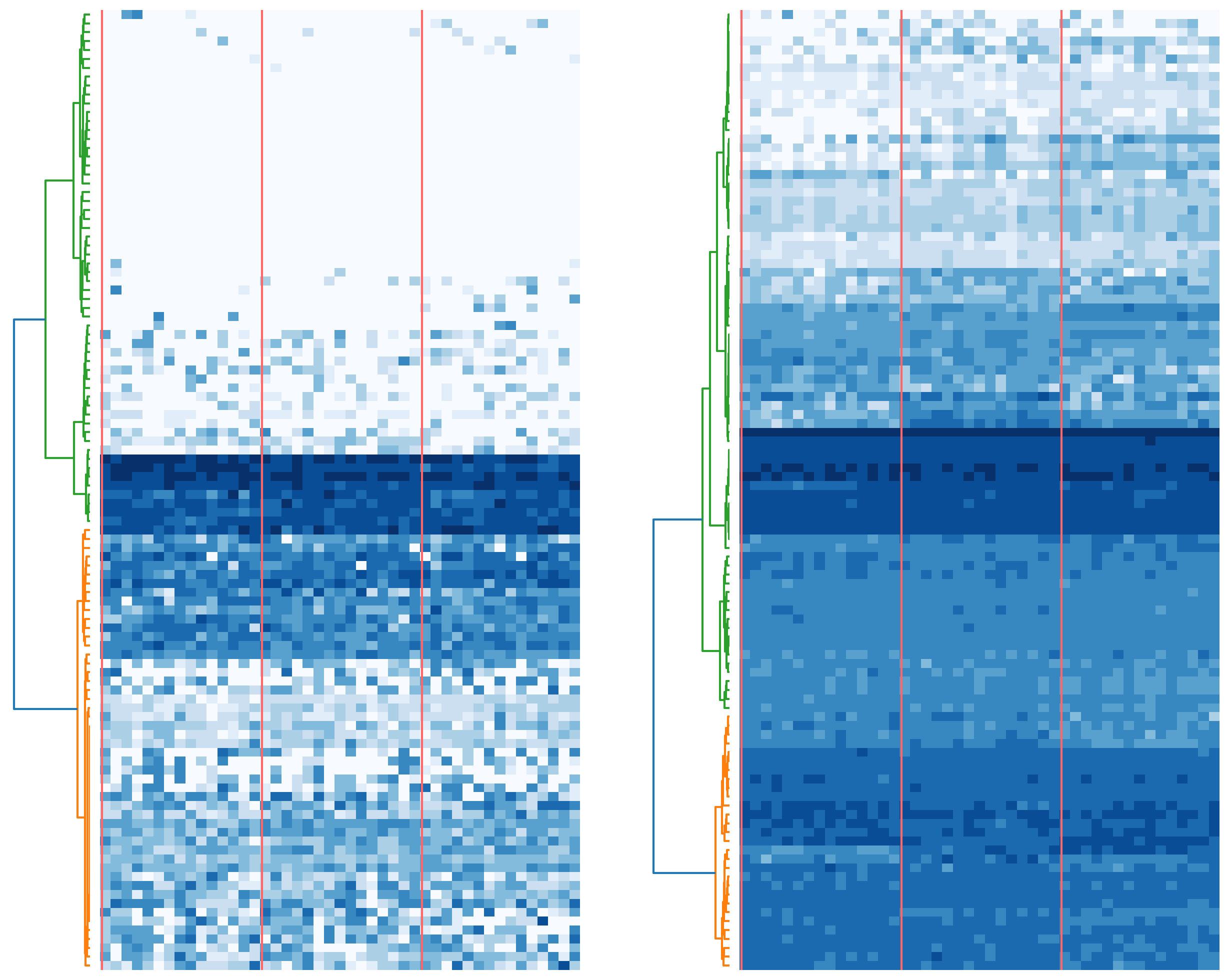}}
\subfloat[Li]{
		\includegraphics[width=0.32\textwidth]{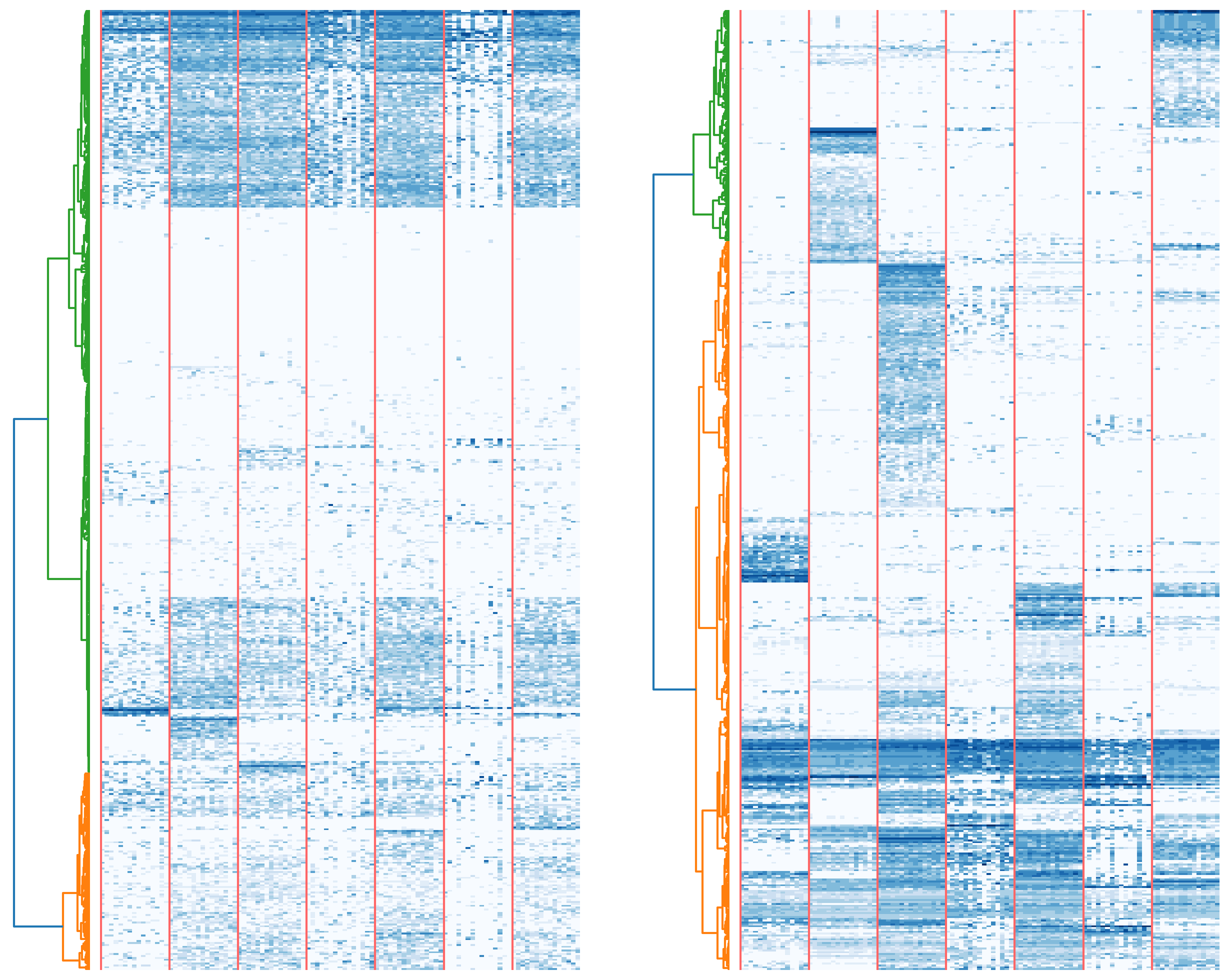}}\\
\subfloat[Maria1]{
	\includegraphics[width=0.32\textwidth]{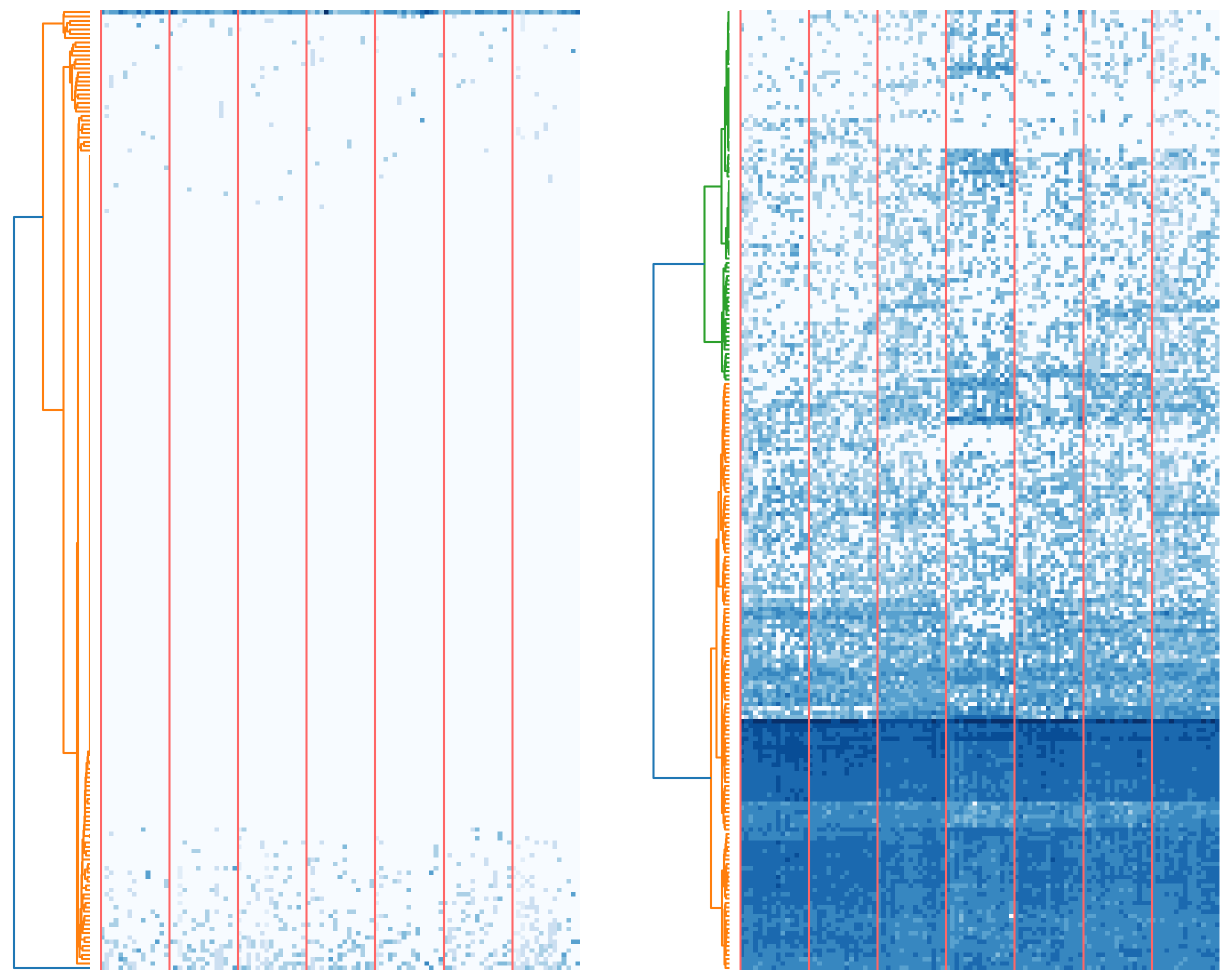}}
\subfloat[Maria2]{
\includegraphics[width=0.32\textwidth]{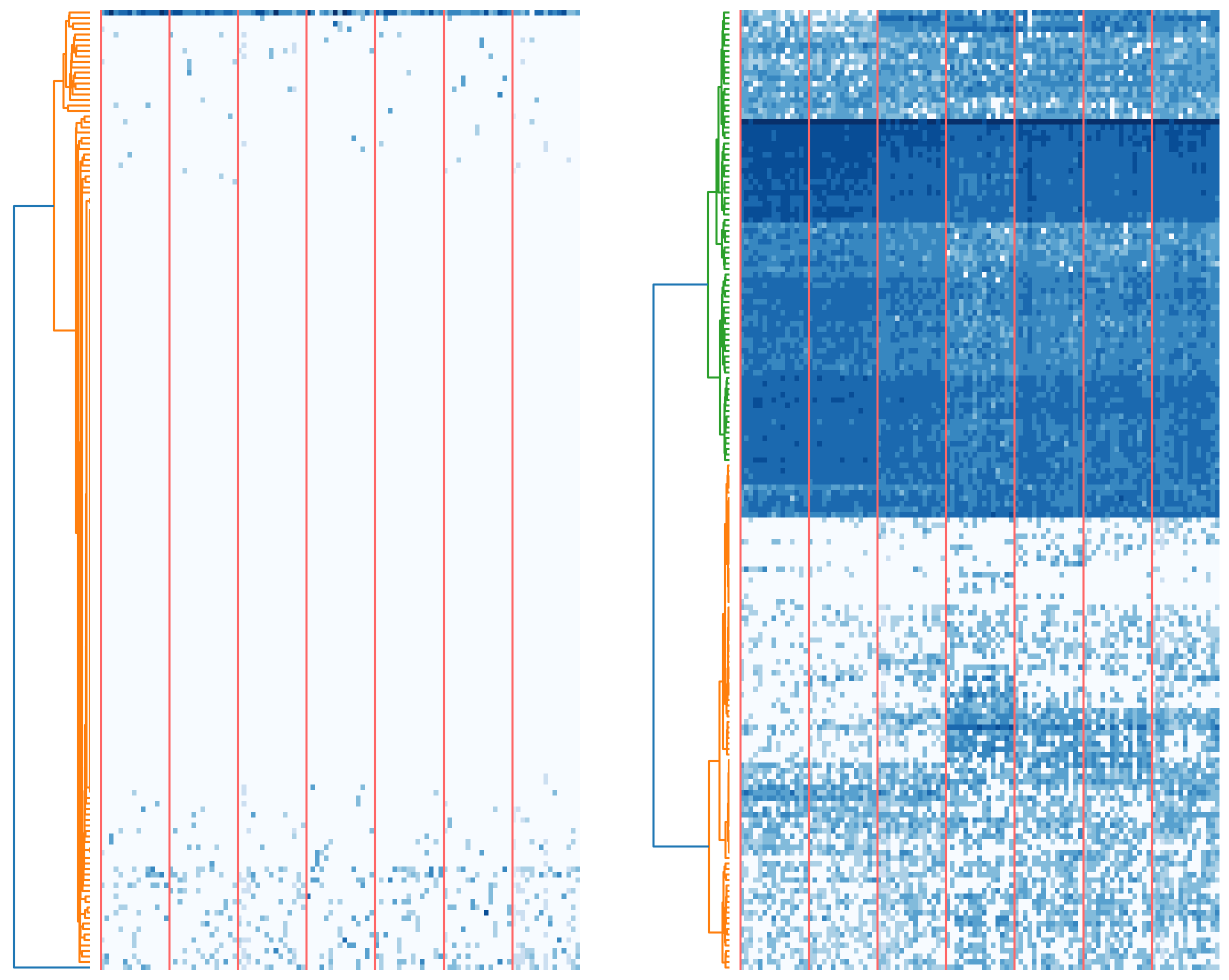}}
\subfloat[Mouse Pancreas1]{
\includegraphics[width=0.32\textwidth]{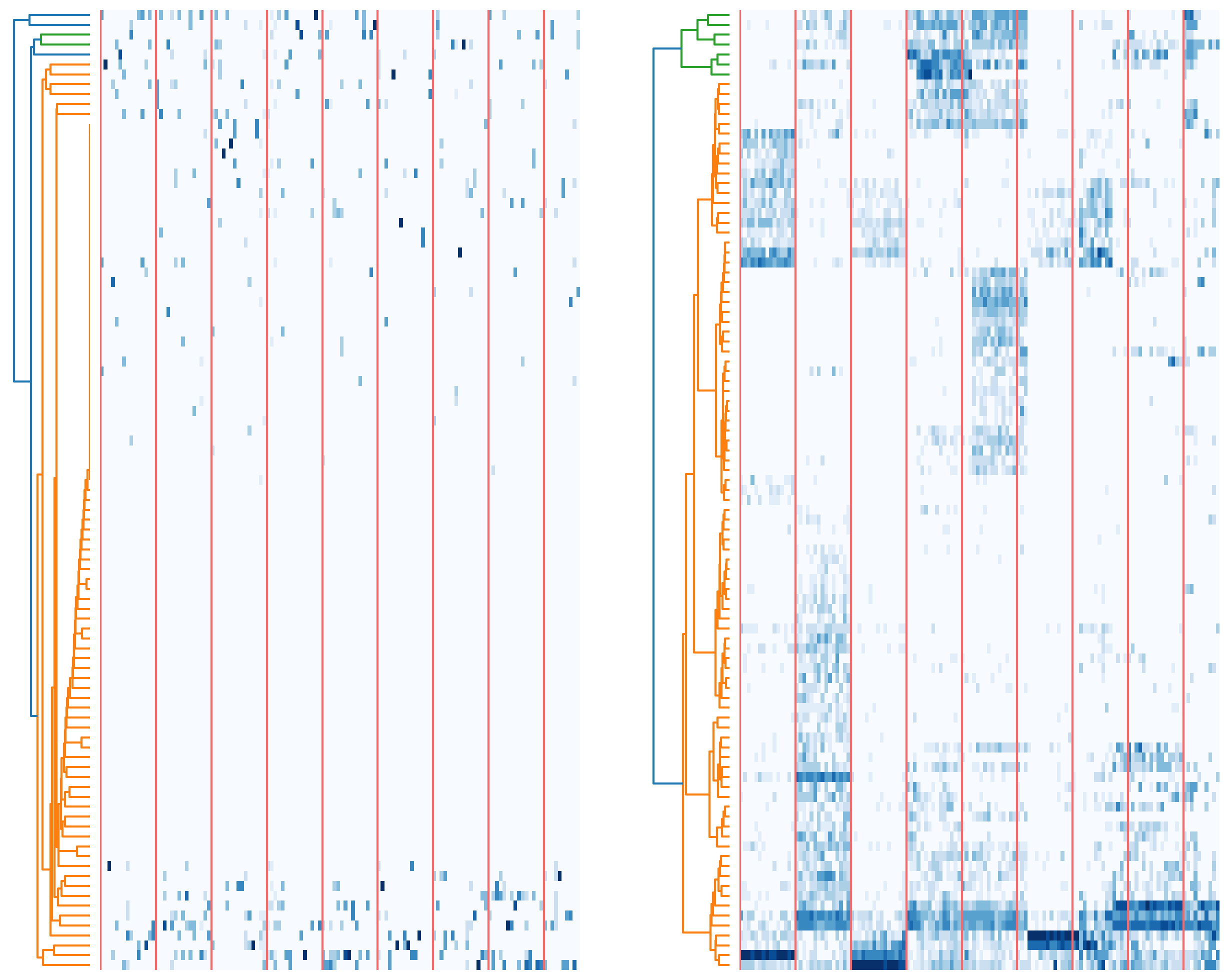}}\\
\subfloat[MacParland]{
\includegraphics[width=0.32\textwidth]{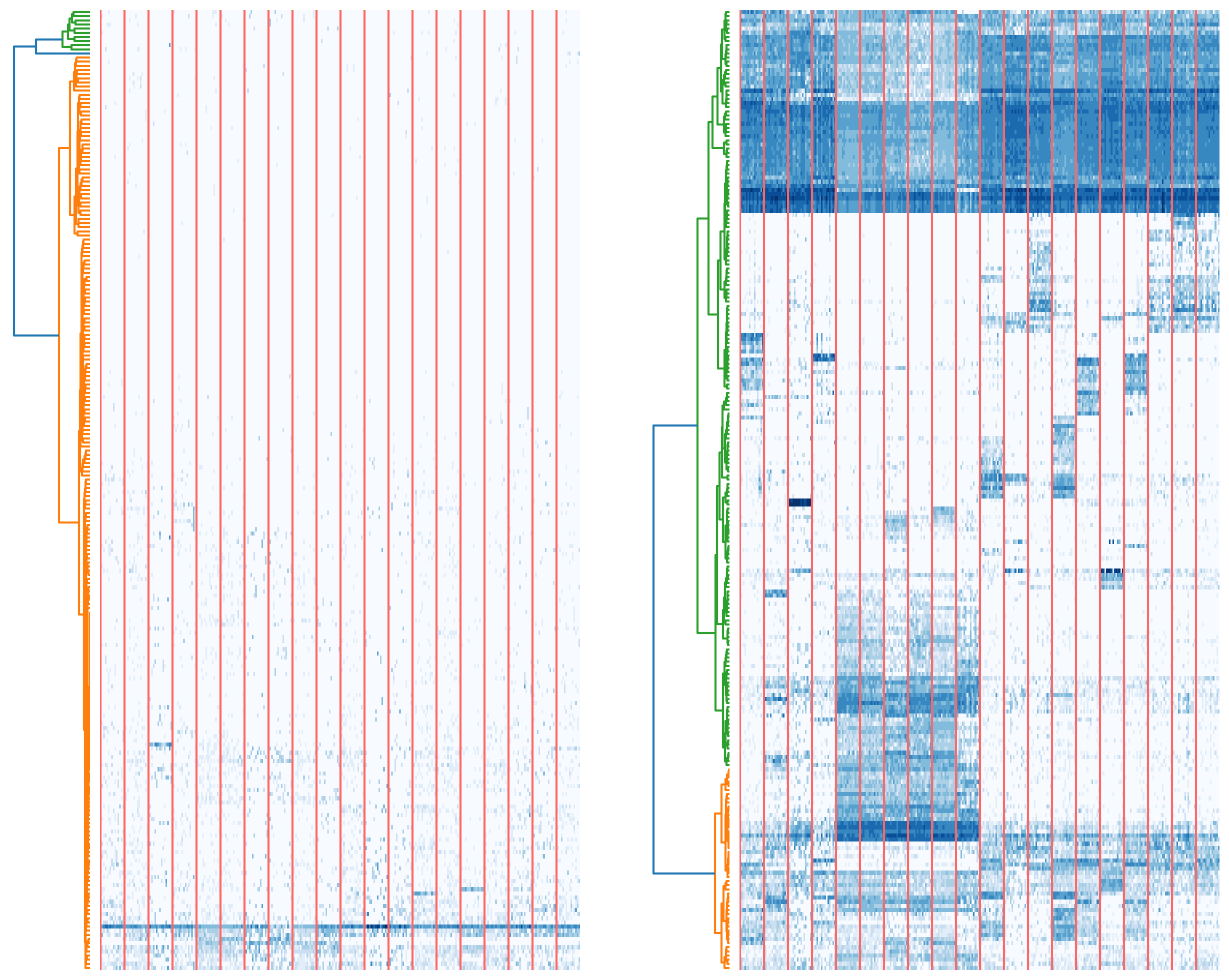}}
\subfloat[Mouse Pancreas2]{
\includegraphics[width=0.32\textwidth]{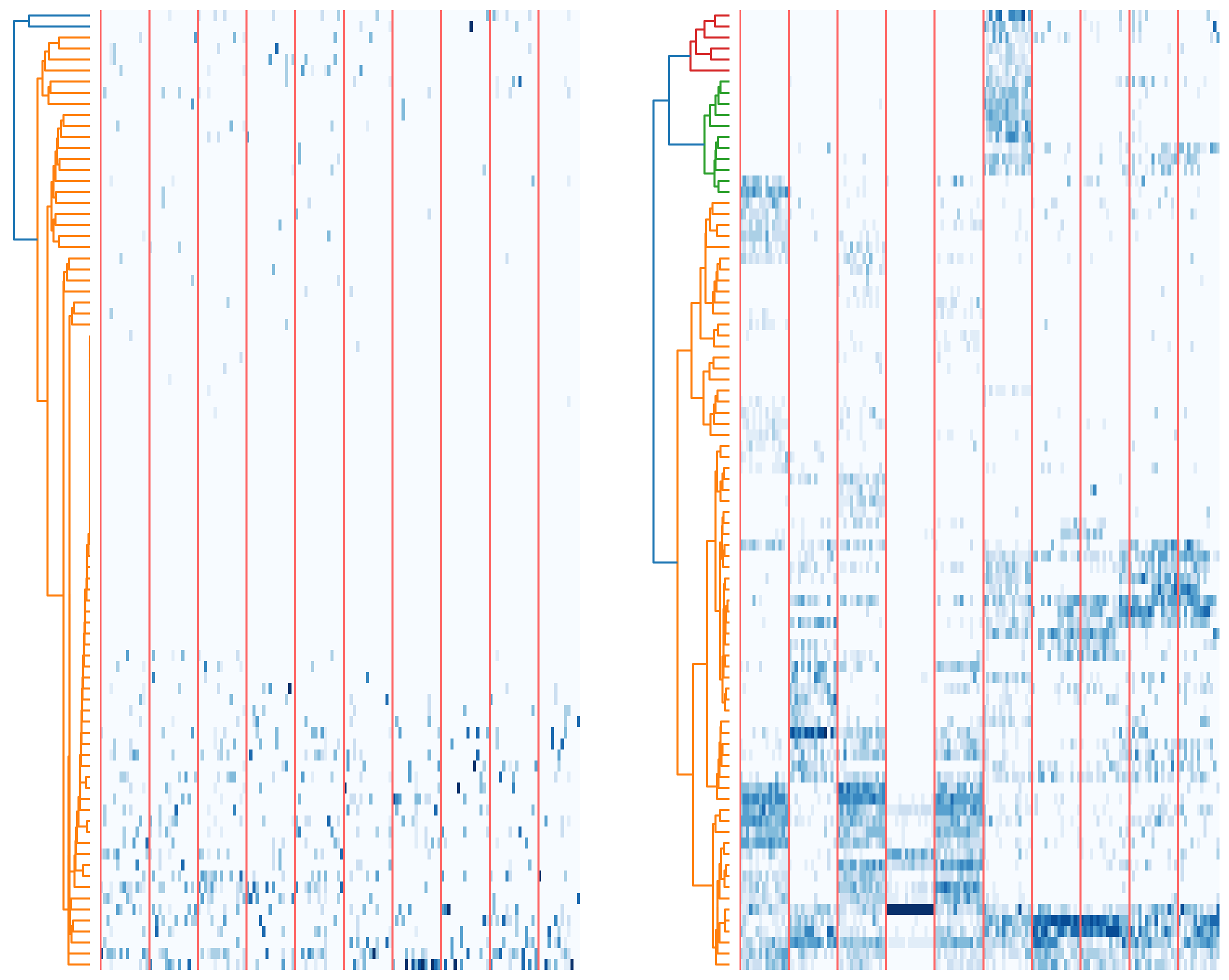}}
\subfloat[Robert]{
\includegraphics[width=0.32\textwidth]{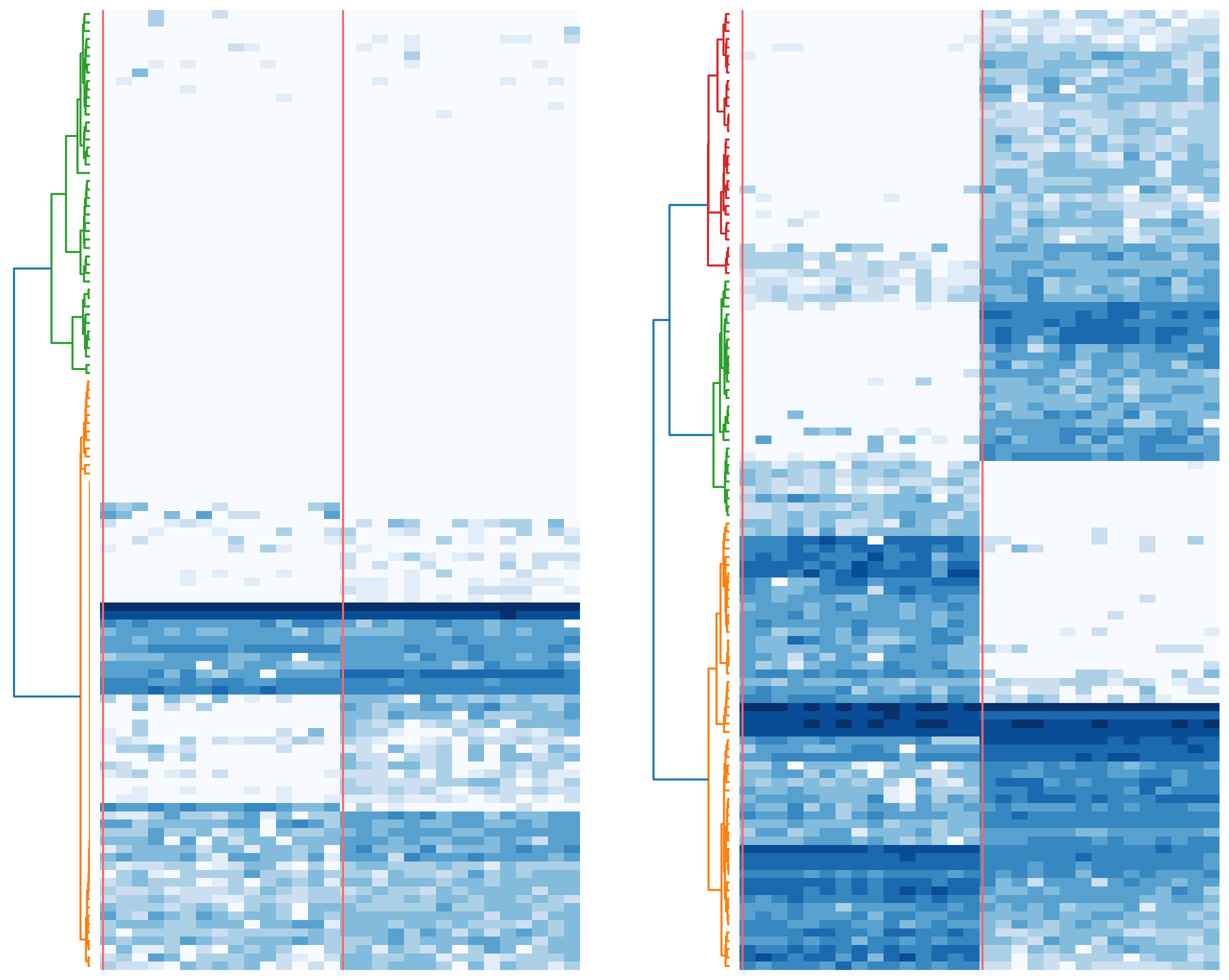}}\\
\subfloat[Ting]{
\includegraphics[width=0.32\textwidth]{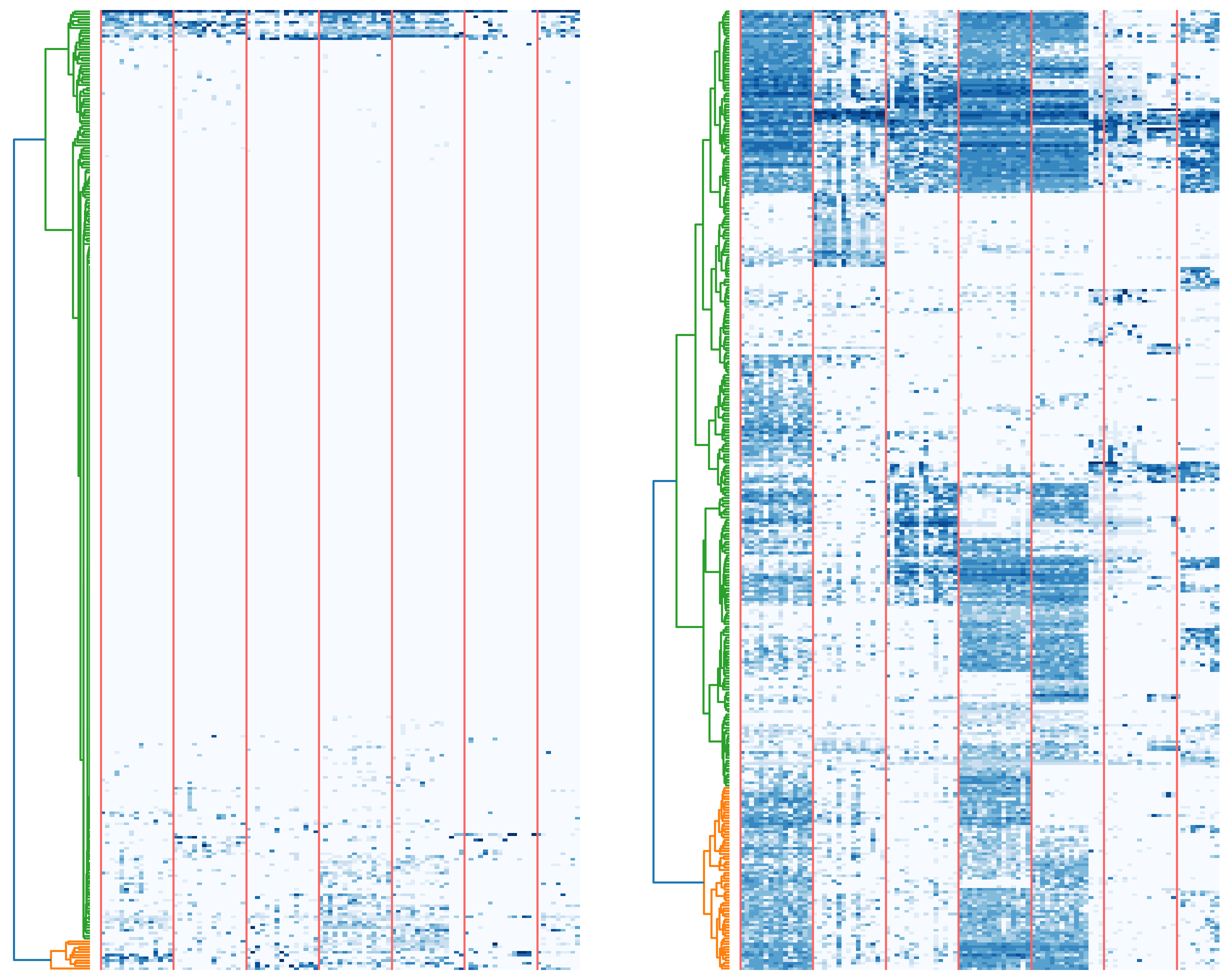}}
\subfloat[Yang]{
\includegraphics[width=0.32\textwidth]{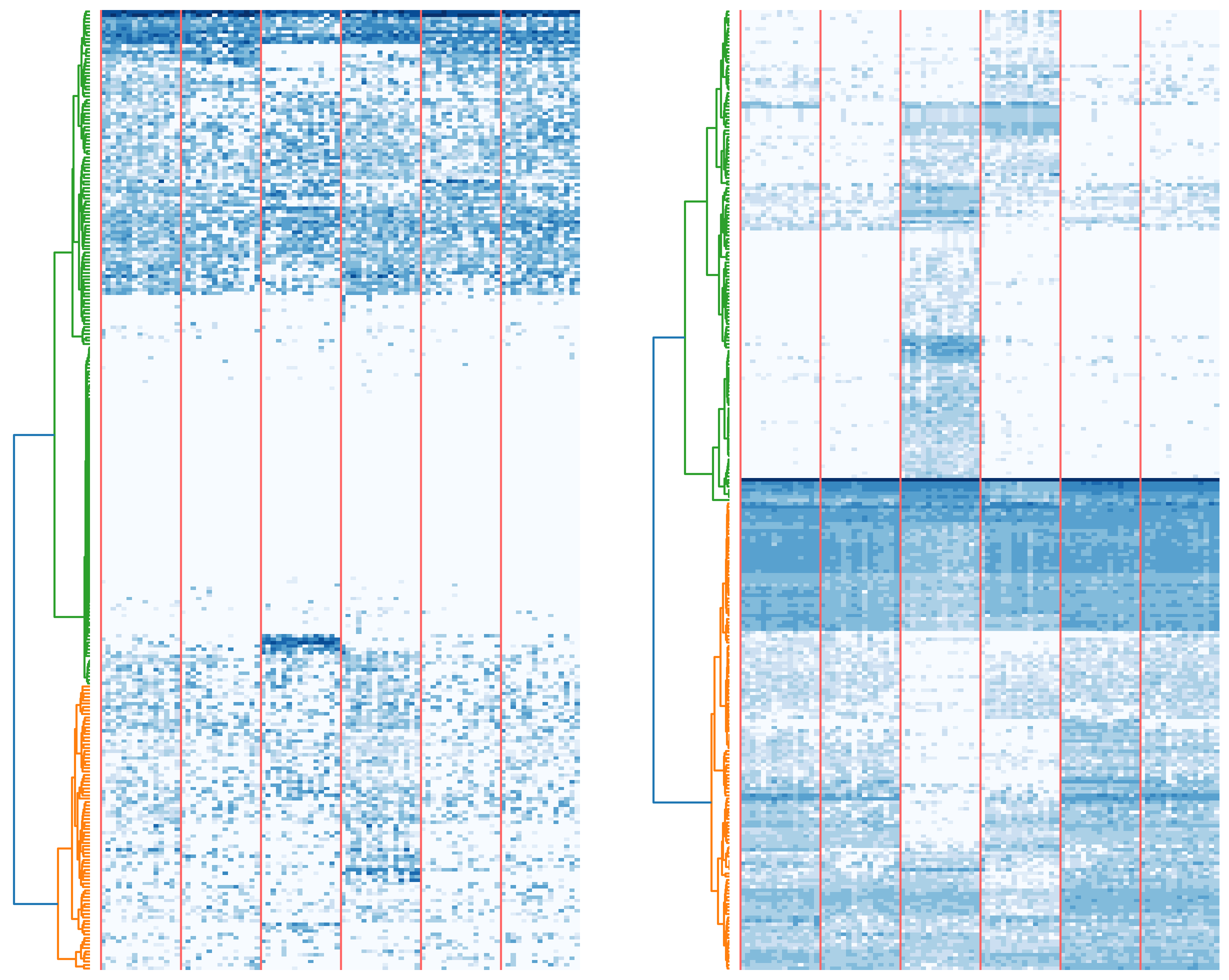}}

\caption{(2/2) Expression heatmap of the rest datasets, where the figure in the left panel is visualized from the original dataset.}
\label{heat_map2}
\end{figure*}
\end{document}